\documentclass[aps,prl,twocolumn,superscriptaddress]{revtex4-1}

\usepackage{advdate}
\usepackage{autobreak}
\usepackage{amsmath,amssymb}
\usepackage{braket}
\usepackage{mathtools}
\usepackage{amsfonts}
\usepackage{bm}
\usepackage{ascmac}
\usepackage{graphicx}
\usepackage{color}
\usepackage[colorlinks=true,citecolor=black,linkcolor=black,urlcolor=black]{hyperref} %linktoc=page
\usepackage{adjustbox}
\usepackage{floatrow}
\usepackage{here}
\usepackage{multirow}
\usepackage{afterpage}
\definecolor{dartmouthgreen}{rgb}{0.05, 0.5, 0.06}
\definecolor{deepjunglegreen}{rgb}{0.0, 0.29, 0.29}

\newcommand{\Z}{$\mathbb{Z}$}
\newcommand{\Zt}{$\mathbb{Z}_{2}$}

\begin{document}

\title{Bulk-boundary correspondence in point-gap topological phases}

\author{Daichi Nakamura}
\email{daichi.nakamura@yukawa.kyoto-u.ac.jp}
\affiliation{Center for Gravitational Physics and Quantum Information, Yukawa Institute for Theoretical Physics, Kyoto University, Kyoto 606-8502, Japan}

\author{Takumi Bessho}
\email{takumi1.bessho@toshiba.co.jp}
\affiliation{Corporate Research and Development Center, Toshiba Corporation, Kawasaki, Japan}

\author{Masatoshi Sato}
\email{msato@yukawa.kyoto-u.ac.jp}
\affiliation{Center for Gravitational Physics and Quantum Information, Yukawa Institute for Theoretical Physics, Kyoto University, Kyoto 606-8502, Japan}

%\date{\AdvanceDate[-1]\today}

\begin{abstract}
A striking feature of non-Hermitian systems is the presence of two different types of topology. One generalizes Hermitian topological phases, and the other is intrinsic to non-Hermitian systems, which are called line-gap topology and point-gap topology, respectively. Whereas the bulk-boundary correspondence is a fundamental principle in the former topology, its role in the latter has not been clear yet. This paper establishes the bulk-boundary correspondence in the point-gap topology in non-Hermitian systems.
After revealing the requirement for point-gap topology in the open boundary conditions, we clarify that the bulk point-gap topology in open boundary conditions can be different from that in periodic boundary conditions. 
On the basis of real space topological invariants and the $K$-theory, we give a complete classification of the open boundary point-gap topology with symmetry and  
show that the non-trivial open boundary topology results in robust and exotic surface states.  
\end{abstract}

\maketitle

%\section{Introduction}
Recently, non-Hermitian topological phases have attracted much attention~\cite{Rudner-09,Sato-11,Esaki-11,Hu-11,Schomerus-13,Malzard-15,Lee-16,Leykam-17,Xu-17,Xiong-18,Shen-18,Kozii-17,Takata-18,MartinezAlvarez-18,KAKU-18,Gong-18,Kawabata-19,YW-18-SSH,YSW-18-Chern,Kunst-18,KSU-18,McDonald-18,Carlstrom-18,Carlstrom-19,Lee-19,Jin-19,Budich-19,Okugawa-19,Liu-19,Yoshida-19,Zhou-19,Lee-Li-Gong-19,Kunst-19,Longhi-19,Edvardsson-19,KSUS-19,ZL-19,Herviou-19,Zeng-20,Hirsbrunner-19,Zirnstein-19,Pocock-19,Kimura-19,Borgnia-19,KBS-19,YM-19,Song-real-space-19,McClarty-19,Okuma-19,Song-19-Lindblad,Bergholtz-19,Lee-Vishwanath-19,Guo-19,Yoshida-19-FQH,Rui-19,Brzezicki-19,Schomerus-20,Imura-19,Herviou-19-ES,Chang-19,Zhang-20,Zhang-19,Yang-19,OKSS-20,Longhi-20,Wang-19,Matsumoto-19,Yoshida-19-mirror,Yokomizo-20,Li-20,KOS-20,Terrier-20,Bessho-Sato-20,Claes-20,Zirnstein-20,Denner-20,Okugawa-takahashi-yokomizo,KSS-20,Fu-20,Okuma-Sato-21,KSR-21,Yi-Yang-20,Zhang-21,Sun-21,Shiozaki-Ono-21,Okugawa-21,Vecsei-21,Hu-21,Ghosh-22,Poli-15,Zeuner-15,Zhen-15,Weimann-17,Xiao-17,St-Jean-17,Parto-17,Bahari-17,Zhao-18,Zhou-18,Harari-18,Bandres-18,Cerjan-19,Zhao-19,Brandenbourger-19-skin-exp,Ghatak-19-skin-exp,Helbig-19-skin-exp,Hofmann-19-skin-exp,Xiao-19-skin-exp,Weidemann-20-skin-exp,Wang-21-1Dwind-exp,Zhang-21-1Dwind-exp,Zhang-21-skin-exp,Zou-21,Palacios-21-skin-exp,Wang-21-Braid-exp}. Non-Hermitian systems differ essentially from Hermitian ones: 
%First, the distinction between transpose and complex conjugation in Hamiltonians ramifies possible symmetries~\cite{Bernard-LeClair-02,KSUS-19}.
%Second, 
The complex-valued energy spectra  of non-Hermitian systems allow two types of the gap structure, {\it i.e.}, line-gap and point-gap~\cite{KSUS-19}. 
Whereas the line-gap is a relatively straightforward generalization of a gap in Hermitian systems, the point-gap is intrinsic in non-Hermitian systems. The multiple gap structures enable corresponding topological phases of non-Hermitian systems, line-gap, and point-gap topological phases~\cite{Gong-18,KSUS-19}. Both topological phases are indispensable to understanding non-Hermitian topological phenomena.

%(ii) The bulk spectra of non-Hermitian systems can be very sensitive to the boundary condition.

%One-dimensional point-gap topology makes the difference between the bulk spectra under the open boundary conditions (OBC) and that under the periodic boundary conditions (PBC)~[] due to the non-Hermitian skin effect (NHSE)~[].

A central character of topological phases is the bulk-boundary correspondence (BBC): the bulk topology causes anomalous gapless boundary modes in the open boundary conditions (OBCs). For example, the quantum Hall systems support chiral edge modes from the nontrivial bulk Chern number~\cite{Hatsugai-93}. 
The exact quantization of the Hall conductance is a consequence of 
the dissipationless current of the chiral edge modes. 

Despite the importance, recent studies have shown that non-Hermiticity obscures the BBC~\cite{Xiong-18,YW-18-SSH,Kunst-18,Lee-19,Helbig-19-skin-exp,Weidemann-20-skin-exp,Borgnia-19,MartinezAlvarez-18,Shen-18,Lee-16,YSW-18-Chern,Leykam-17,YM-19,Zirnstein-19,Zirnstein-20,Jin-19,Herviou-19,Pocock-19,Brzezicki-19,KOS-20,Yang-19}.
A class of non-Hermitian systems 
shows completely different bulk spectra in the OBCs than in the periodic boundary conditions (PBCs).
Because of this phenomenon-{\it the non-Hermitian skin effect} (NHSE)~\cite{YW-18-SSH,Kunst-18}, the bulk energy gap in OBCs would be closed in PBCs. 
Therefore, a bulk topological number in PBCs can be ill-defined even when a gapless boundary mode exists in a gap in OBCs. 
%
%
%On the other hand, the BBC may break down for non-Hermitian systems~[].
%
%
%
%One is the apparent breakdown of the BBC. The changes in the bulk spectra under OBC owing to the NHSE mean that 
%
%In other words, topology under PBC cannot capture topology under OBC~[]. 
Yao and Wang~\cite{YW-18-SSH} solved this problem in the case of line-gap topological phases. 
Using  the generalized Brillouin zone (GBZ)~\cite{YW-18-SSH,YSW-18-Chern,YM-19,Yang-19,KOS-20},  
they define a topological number in OBCs, and show that the new bulk topological number correctly recovers the BBC. 
Furthermore, later, the BBC in OBCs was also formulated in terms of real-space topological invariants, enabling the application in the study of the BBC in higher dimensions and higher-order line-gap topological phases~\cite{Song-real-space-19,Tang-20,Liu-20,Liu-21,Ghorashi-HOD-21,Ghorashi-HOW-21}.

Whereas the above prescription settled the BBC in line-gap topological phases, the BBC in point-gap topological phases has not been clear yet.
In one dimension, point-gap topological numbers in PBCs result in NHSEs in OBCs, as a result of the BBC~\cite{Zhang-19,OKSS-20}.
However, there remains uncertainty of the BBC in higher dimensions:
It has been suggested that topological surface states originate from the three-dimensional winding number in PBCs~\cite{Denner-20,Vecsei-21,Hu-21}, which is the point-gap topological number for general three-dimensional systems. Nonetheless, these surface states can disappear without changing the bulk topological number. Thus, the relation between the bulk 
% topological number
topology
and the surface states is ambiguous. 

%Furthermore, the BBC with symmetry has not been clarified yet.
%While a particular class of point-gap topological numbers in PBCs are responsible to skin effects in higher dimensions, others are not. 

%other is the absence of topologically protected surface states for point-gap topology. Namely, the surface states don't appear under OBC even if the point-gap topological phase transition point seems not to shift from in the case under PBC~[]. The reason is unknown, and one of the difficulties is that the definition of the point-gap topological number using GBZ is an open problem. Recently, M. M. Denner et al.~[] showed the model that the 

%induces topologically protected surface states, suggesting the existence of the BBC for point-gap topology. However, since they have been discussing using point-gap topological number defined under PBC, many points remain unclear, including the relationship with the NHSE. For these reasons, the BBC for the point-gapped topological phases is still unclear.

In this paper, we establish the BBC in point-gapped topological phases. Following the strategy learned from line-gap topological phases, our arguments rely on topological numbers in OBCs. Remarkably, there appears an essentially new feature intrinsic to point-gap topological phases. We find that a particular class of non-Hermitian skin effects, which we dub in-gap skin effects,  ruins point-gap topological numbers in OBCs. As a result, the topological classification in OBCs can be different
from that in PBCs. Based on this result, we resolve the uncertainty of the BBC in point-gap topological phases, and show that non-trivial topological numbers in OBCs result in robust and exotic surface states. Using the $K$-theory, we also give a complete classification table for point-gap topological phases under OBCs in the presence of symmetry.  

\begin{figure}[h]
\centering
\includegraphics[keepaspectratio,width=\linewidth]{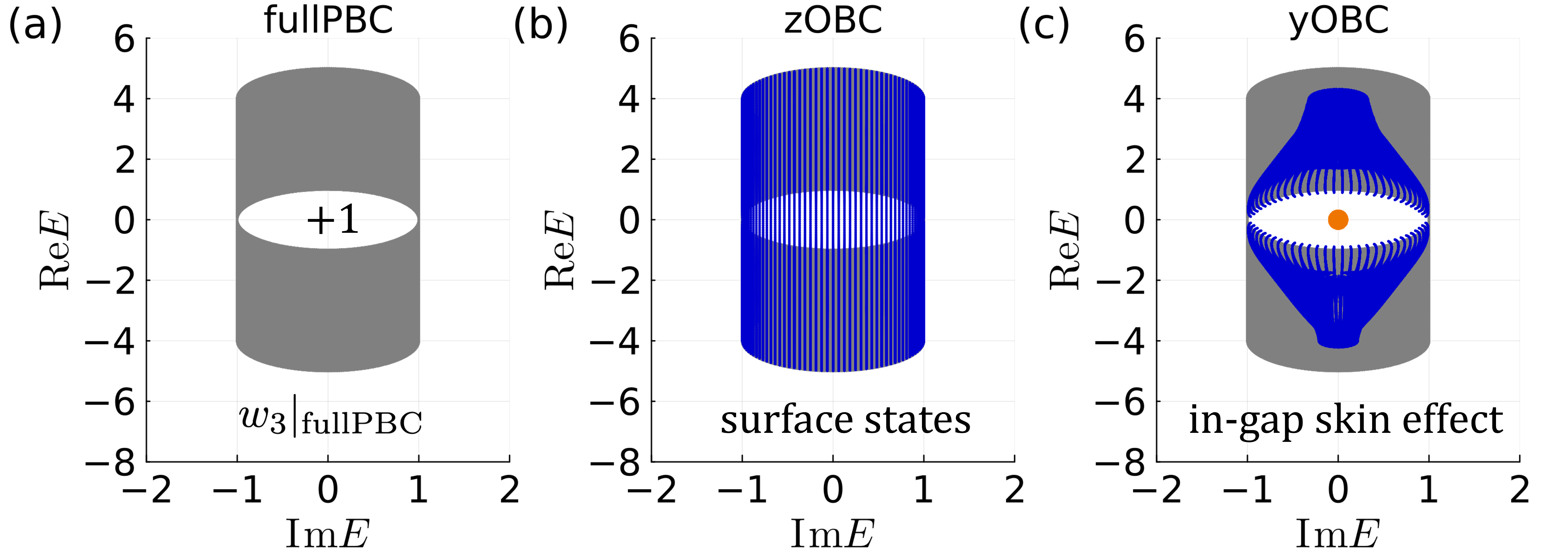}
\caption{The energy spectra of ETI in Eq.~(\ref{eq:ETI}). (a)~The full PBC spectrum (gray). The point gap is open around $E=0$, with the nontrivial 3D winding number $+1$. (b)~The zOBC spectrum (blue) and the full PBC spectrum (gray) in comparison.  No NHSE occurs but surface states appear in the region where the 3D winding number takes the nontrivial value. The system size is $L_x=L_y=100$ and $L_z=30$.
(c)~The yOBC spectrum (blue) and the full PBC spectrum (gray) in comparison. NHSEs occur, and in-gap skin modes (orange) appear. The system size in the $y$ direction is $L_y=30$ and the momentum resolutions in both $k_{x,z}$-directions are taken as $\Delta k_{x,z} = 2\pi/100$.}
\label{fig:ETI}
\end{figure}

{\it Uncertainty of BBC in point-gap topological phases.--}
First, let us see the fore-mentioned uncertainty of the BBC in point-gap topological phases. We start with a model of exceptional topological insulators (ETIs) \cite{Denner-20},
\begin{align}
  H_{\rm ETI}(\bm{k}) &= \sin{k_x}\sigma_{x} + \sin{k_z}\sigma_{y} + \left(2 - \sum_{i=x,y,z}\cos{k_i}\right)\sigma_{z}
\nonumber 
 \\ 
 &- i\sin{k_y}\sigma_{0},
\label{eq:ETI}
\end{align}
where $\sigma_{i=x,y,z}$ are the Pauli
matrices and $\sigma_{0}$ is the 2×2 identity matrix.
As shown in Fig.~\ref{fig:ETI}(a), the ETI has a point gap at $E=0$ in the complex energy plane, {\it i.e.}, no complex spectrum crossing the reference point $E=0$, under PBCs in all directions (full PBC).    
Therefore, we can define the three-dimensional (3D) winding number
over the 3D Brillouin zone (BZ)~\cite{Gong-18,KSUS-19}, 
\begin{align}
  w_{3}|_\text{fullPBC} &= -\dfrac{1}{24\pi^{2}}\int_{\text{ BZ}}\text{Tr}[(H-E)^{-1}d(H-E)]^3,
\label{eq:w3}
\end{align}
which takes $+1$ for $H=H_{\rm ETI}$ and $E=0$. 
%\footnote{When we consider the point gap at $E$, we make the substitution $H \to H - E$.}.

%The great features of this model are as follows.
%(i) Unit 3D winding number under full PBC: 

%We can easily check Eq.~(1) has a point-gap at $E=0$ and 
%the winding number defined on the 3D Brillouin zone is $+1$ (Figure 1(a)).
Interestingly, the ETI hosts surface states once we impose the OBC in the $z$-direction (zOBC). See Fig.~\ref{fig:ETI}(b). %For fixed $k_y$,
Because the non-Hermitian term $i\sin {k_y}\sigma_0$ in Eq.~(\ref{eq:ETI}) only gives a complex shift of the energy under the zOBC, 
%Because of $[H(\bm{k}),i\sin{k_y}\sigma_{0}]=0$,
the surface states of the ETI are equivalent to those of a Weyl semimetal,  $H_{\text{WSM}}(\bm{k}) = H_{\rm ETI}(\bm{k})+i\sin{k_y}\sigma_{0}=\sin{k_x}\sigma_{x} + \sin{k_z}\sigma_{y} + \left(2 - \sum_{i=x,y,z}\cos{k_i}\right)\sigma_{z}$.
For a fixed $k_y$ with $-\pi/2<k_y<\pi/2$,
the Weyl semimetal supports a non-zero Chern number in the PBCs, and thus it has corresponding chiral edge modes under the zOBC. 
By taking into account the complex energy shift from the non-Hermitian term $i\sin {k_y}\sigma_0$, the chiral modes give surface states filling the point-gapped region in Fig.~\ref{fig:ETI}(b).

%The system size is $L_x=L_z=100$ and $L_y=30$.
%The system size in the $y$ direction is $L_y=30$ and the momentum resolutions in both $k_{x,z}$-directions are taken as $\Delta k_{x,z} = 2\pi/100$.
% \begin{figure}[h]
% \RawFloats
% \begin{minipage}[t]{0.32\hsize}
% \centering
% \includegraphics[keepaspectratio,width=\linewidth]{triv1.pdf}
% \end{minipage}\hspace{0mm}
% \begin{minipage}[t]{0.32\hsize}
% \centering
% \includegraphics[keepaspectratio,width=\linewidth]{triv2.pdf}
% \end{minipage}\hspace{0mm}
% \begin{minipage}[t]{0.32\hsize}
% \centering
% \includegraphics[keepaspectratio,width=\linewidth]{triv3.pdf}
% \end{minipage}\hspace{0mm}
% \caption{The energy spectra of ETI in Eq.~(\ref{eq:ETI}). (a)~The full PBC spectrum (gray). The point gap is open in the region containing $E=0$, with the nontrivial 3D winding number $+1$. (b)~The zOBC spectrum (blue).  No NHSE occurs and surface states appear in the region where the 3D winding number takes the nontrivial value. The system sizes are $L_x=L_y=100,L_z=30$. 
% (c)~The yOBC spectrum (blue). NHSEs occur, and in-gap skin modes (orange) appear. The system size in the $y$ direction is $L_y=30$ and the momentum resolutions in both $k_{x,z}$-directions are taken as $\Delta k_{x,z} = 2\pi/100$.}
% \label{fig:ETI}
% \end{figure}

These surface states have a topological number with the same value as the bulk 3D winding number: The effective Hamiltonian of the surface states around $E=0$ takes the form of $h_{\rm surface}(k_x,k_y)=v_xk_x - ik_y$ as the chiral edge mode of the Weyl semimetal has $v_xk_x$ with a real positive constant $v_x>0$ and the non-Hermitian term of $H_{\rm ETI}$ gives $-ik_y$. 
The 1D winding number $w_1$ on a circle $S_1$ around $(k_x,k_y)=(0,0)$ on the surface BZ~\cite{Denner-20}:
\begin{align}
  w_{1} &= -\oint_{S^{1}}\dfrac{\text{d}{\bm k}}{2\pi i}
  \cdot
  h^{-1}_{\rm surface}(k_x,k_y)
  {\bm \nabla}_{k}h_{\rm surface}(k_x,k_y)=1
\label{eq:w1}
\end{align}
measures the topology of the surface states, of which value coincides with the 3D winding number. 

The coincidence of the topological numbers suggests the BBC in point-gap topological phases \cite{Denner-20}. However, there is an uncertainty in this interpretation.
%However, this interpretation has an uncertainty.  
In Fig.~\ref{fig:ETI}(c), we show the spectrum of the same model under the OBC in the $y$-direction (yOBC). 
Whereas the bulk topological number in Eq.(\ref{eq:w3}) remains the same, no surface state covering the point-gapped region appears. 
Instead, we have skin modes in the gap.
Since the skin modes are localized bulk modes \footnote{skin modes are obtained using GBZ and thus bulk modes \cite{YW-18-SSH}, while surface states are not obtained using GBZ and thus not bulk modes}, and do not have 1D winding number~\cite{OKSS-20},
the simple BBC does not hold under the yOBC.

{\it Point-gap topological number under OBCs and BBC. --}
%{\it Open bulk point gap topological number and BBC --}
The solution of the problem is to use topological numbers in OBCs.
%To solve the inconsistency in the above, we introduce topological numbers in OBCs. 
Let us consider a 3D Hamiltonian $\mathcal{H}(k_x,k_y)$ with momentum-space representation in the $x$- and $y$-directions and real-space representation in the $z$-direction under the zOBC. 
Then, we construct the  bulk Hamiltonian $\mathcal{H}_{\text{bulk}}(k_x,k_y)$ by the projection of $\mathcal{H}(k_x,k_y)$ onto the bulk~\cite{Song-real-space-19}.  
When the bulk Hamiltonian has a point gap at $E$ ($\text{det}[\mathcal{H}_{\text{bulk}}-E]\neq0$), we can define the real-space 3D winding number $w_{3}$ under the zOBC,
\begin{align}
  w_{3}|_{\rm zOBC} = -\dfrac{i}{12\pi}\int_{\text{BZ}}d^{2}\bm{k}\hspace{0.5mm}\mathcal{T}_z[\varepsilon^{ijk}Q_iQ_jQ_k],
\label{eq:w3zOBC}
\end{align}
with $Q_{i=x,y} = i(\mathcal{H}_{\text{bulk}}-E)^{-1}\partial_{k_{i}}(\mathcal{H}_{\text{bulk}}-E)$ and $Q_z = (\mathcal{H}_{\text{bulk}}-E)^{-1}[Z,\mathcal{H}_{\text{bulk}}-E]$,
where 
%the summation of repeated indices $i, j, k$ is implied, $\varepsilon^{ijk}=\pm1$ is the Levi-Civita symbol, 
$Z$ is the position operator in the $z$-coordinate, and $\mathcal{T}_z$
stands for the trace per unit length in the $z$-direction. 
This is a non-Hermitian generalization of the real-space topological number in Hermitian systems~\cite{Mondragon-Shem-13,Prodan-14,Prodan-16,Katsura-Koma-18}. 
For the full PBCs, this quantity reproduces Eq.~(\ref{eq:w3}) with the identification
\begin{align}
  \int\dfrac{dk_z}{2\pi}\text{Tr}[A(k_z)] \leftrightarrow \mathcal{T}_z[\mathcal{A}], 
  \quad
  i\partial_{k_z} \leftrightarrow [Z,\cdot],
\end{align}
where $A(k_z)$ is a function of $k_z$ and $\mathcal{A}$ is the real-space representation of $A(k_z)$~\cite{Bellissard-94,Kitaev-06}.
%{\footnote{(cite to footnote wo doujini yaruhouhou ga wakaranai.) 
%The identity between $w_{3}|_\text{fullPBC}$ and $w_{3}|_\text{zOBC}$ is up to sign, as it relies on the definition of the Fourier transform.
%. 
%$w_{3}|_\text{zOBC}$ becomes $1$ under zOBC and ill-defined under yOBC respectively for the present model in Eq.~(\ref{eq:ETI}).
%Therefore, the coincidence between the bulk topological number and the surface one observed in the ETI still holds under zOBC.
Thus, for the ETI in Eq.~(\ref{eq:ETI}), the coincidence between $w_3|_{\rm fullPBC}$ with $E=0$ in Eq.~(\ref{eq:w3}) and $w_1$ in Eq.~(\ref{eq:w1}) 
results in the correspondence between 
$w_3|_{\rm zOBC}$ with $E=0$
in Eq.~(\ref{eq:w3zOBC}) and $w_1$ in Eq.~(\ref{eq:w1}).
Since any nontrivial 3D winding number under zOBC can be produced by stacking the ETIs in Eq.~(\ref{eq:ETI}) up to continuous deformations,
we generally have the BBC
%Applying the real-space 3D winding number to the case with zOBC, we find that when $w_{3}|_\text{zOBC}$ is nontrivial, surface localized states appear at $z=0,L_z$, and their total number is given by $w_{3}|_\text{zOBC}$.
%
%\begin{itembox}[l]{BBC for $w_{3}|_\text{zOBC}$}
\begin{align}
    w_{3}|_\text{zOBC} = \sum_{\bm{k}_p} w_{1}(\bm{k}_p),
    \label{eq:BBC}
\end{align}
with
\begin{align}
    w_{1}(\bm{k}_p) = -\oint_{S^{1}_p}\dfrac{d\bm{k}}{2\pi i}\cdot\text{Tr}[(h_{\text{surface}}-E)^{-1}\nabla (h_{\text{surface}}-E)],
  \end{align}
%\end{itembox}
where $h_{\text{surface}}(k_x,k_y)$ is the surface effective Hamiltonian, $\bm{k}_p$ is the Fermi point satisfying $\text{det}[h_{\text{surface}}({\bm k}_p)-E]=0$, and $S^{1}_p$ stands for the counter clockwise circle around $\bm{k}=\bm{k}_p$ on the surface Brillouin zone.
%and $w_{1}(\bm{k}_p)$ means the surface chirality around the Fermi point $\bm{k}=\bm{k}_p$.
As we shall show later, the bulk topological number under the zOBC can be different from that under the yOBC. 
Therefore, the above BBC does not require surface states under the yOBC.

%\section{nontrivial example}

\begin{figure}[htbp]
\RawFloats
\begin{minipage}[t]{0.49\hsize}
\centering
\includegraphics[keepaspectratio,width=\linewidth]{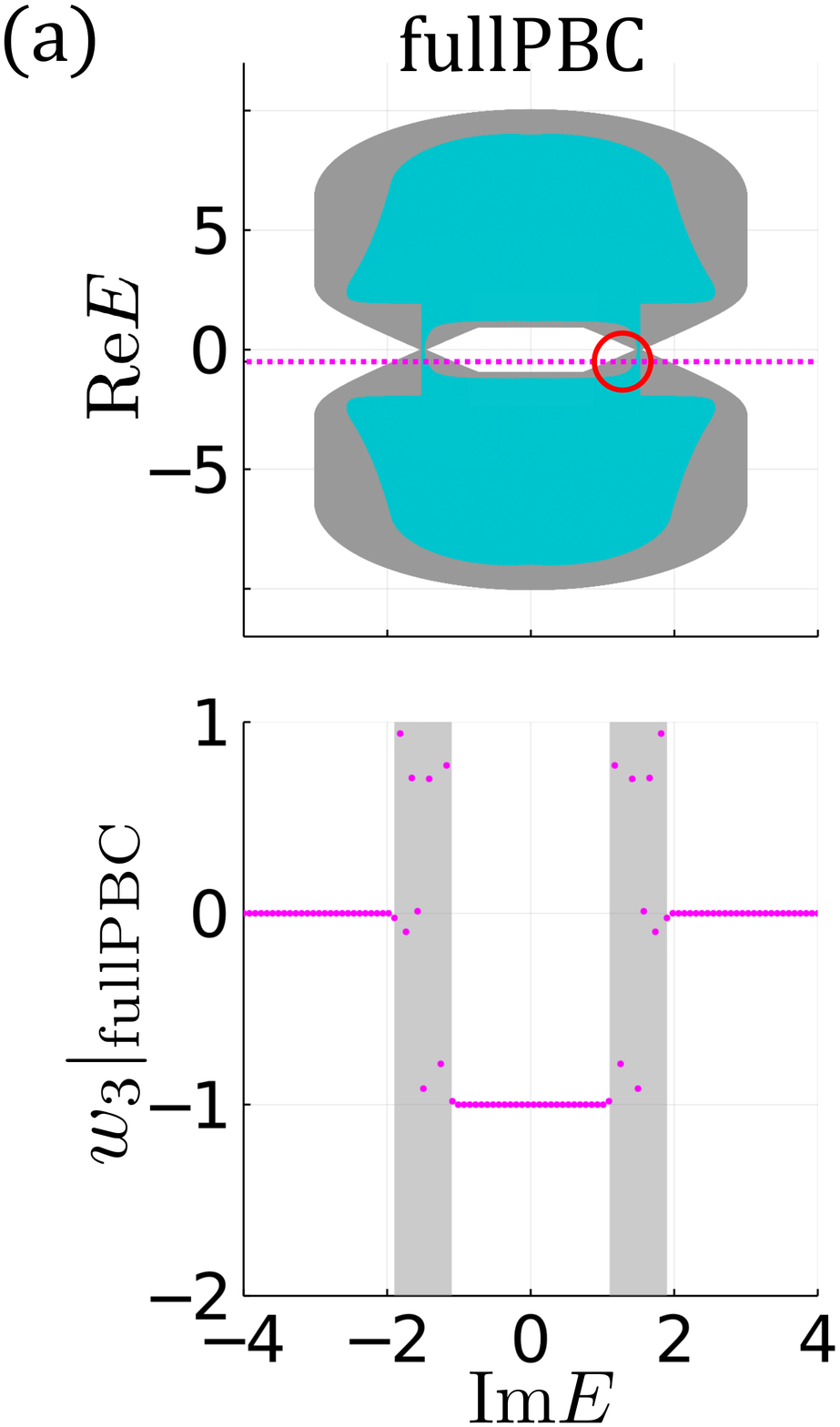}
\end{minipage}\hspace{0mm}
\begin{minipage}[t]{0.49\hsize}
\centering
\includegraphics[keepaspectratio,width=\linewidth]{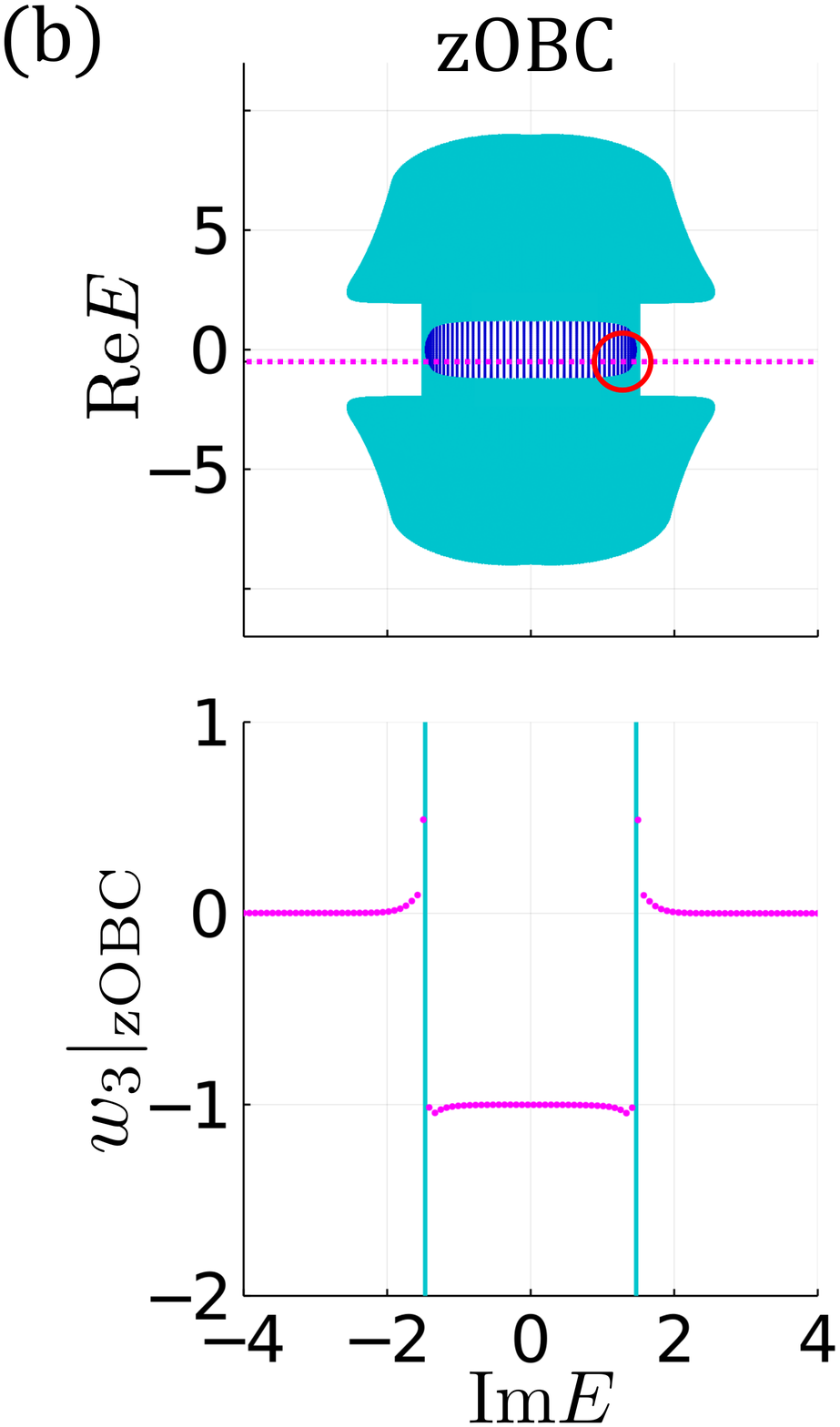}
\end{minipage}\hspace{0mm}
\caption{
The energy spectra~(top) and the 3D winding numbers~(bottom) of the model in Eq.~(\ref{eq:ETI2}) under different boundary conditions.
The magenta lines in the top figures represent $E=-0.5+i{\rm Im}E$ with $-4<{\rm Im}E<4$. 
(a)(top) The bulk spectra under the full PBC (gray) and the zOBC (turquoise). The zOBC spectrum is calculated by using the non-Bloch theory in Ref.\cite{YM-19}. The point-gapless region under the zOBC is wider than that under the full PBCs. (bottom) The 3D winding number under the full PBC along the magenta line in the top figure.
%The 3D winding number at $E$ is defined by Eq.~(\ref{eq:w3}) with the replacement $H \to H - E$. 
The gray shadings represent point-gapless regions. %$E=-0.5+i\mathrm{Im}E$ with $-1.9<\mathrm{Im}E<-1.1$ and $1.1<\mathrm{Im}E<1.9$.
(b)(top) The energy spectrum under the zOBC. The blue modes are surface states which are calculated with the system size $L_x=L_y=100,L_z=20$. Surface states appear even in the regions where $w_{3}|_\text{fullPBC}$ is ill-defined (inside the red circle). 
(bottom) The real-space 3D winding number under the zOBC along the magenta line in the top figure.
%The real-space 3D winding number is defined by Eq.~(\ref{eq:w3zOBC}) with $\mathcal{H}_{\text{bulk}} \to \mathcal{H}_{\text{bulk}} - E$. 
The turquoise segments represent point-gapless regions.
}
\label{fig:ETI2}
\end{figure}

The necessity of the OBC bulk topological number becomes obvious once we consider another model. 
Figure~\ref{fig:ETI2}(a) is the bulk spectra of the following model under different boundary conditions:
\begin{align}
  H(\bm{k}) &= \sin{k_x}\sigma_{x} + 2\sin{k_z}\sigma_{y} + 2\left(2 - \sum_{i=x,y,z}\cos{k_i}\right)\sigma_{z}
\nonumber\\
  &+ \dfrac{3}{2}i(\sin{k_y}+\sin{k_z})\sigma_{0}.
\label{eq:ETI2}
\end{align}
%The minimal model Eq.~(1) is a trivial model in the sense that no NHSE occurs. In this section, we show numerically that even in the presence of the NHSE, when $w_{3}|_\text{zOBC}$ is nontrivial, topologically protected surface states appear. We consider the following model.
Because of NHSEs in the bulk spectrum, the OBC spectrum has a wider point-gapped region than the PBC spectrum. 
Therefore, there is a region where only the topological number under the zOBC is well-defined, as shown in Fig.~\ref{fig:ETI2}.
We find that the BBC holds for the topological number under the zOBC, not for that under the PBC.

{\it In-gap skin effects and absence of surface states.--}
Now we show that the BBC in Eq.~(\ref{eq:BBC}) also explains the absence of surface states in Fig.~\ref{fig:ETI}(c) under the yOBC. 
A key observation is the presence of in-gap skin modes.
In a manner similar to Eq.~(\ref{eq:w3zOBC}), we can introduce the 3D winding number 
under the yOBC, but the in-gap bulk skin modes make the topological number ill-defined.
Consequently, the BBC under the yOBC does not require any surface states. 

\begin{figure}[htbp]
\centering
\includegraphics[keepaspectratio,width=\linewidth]{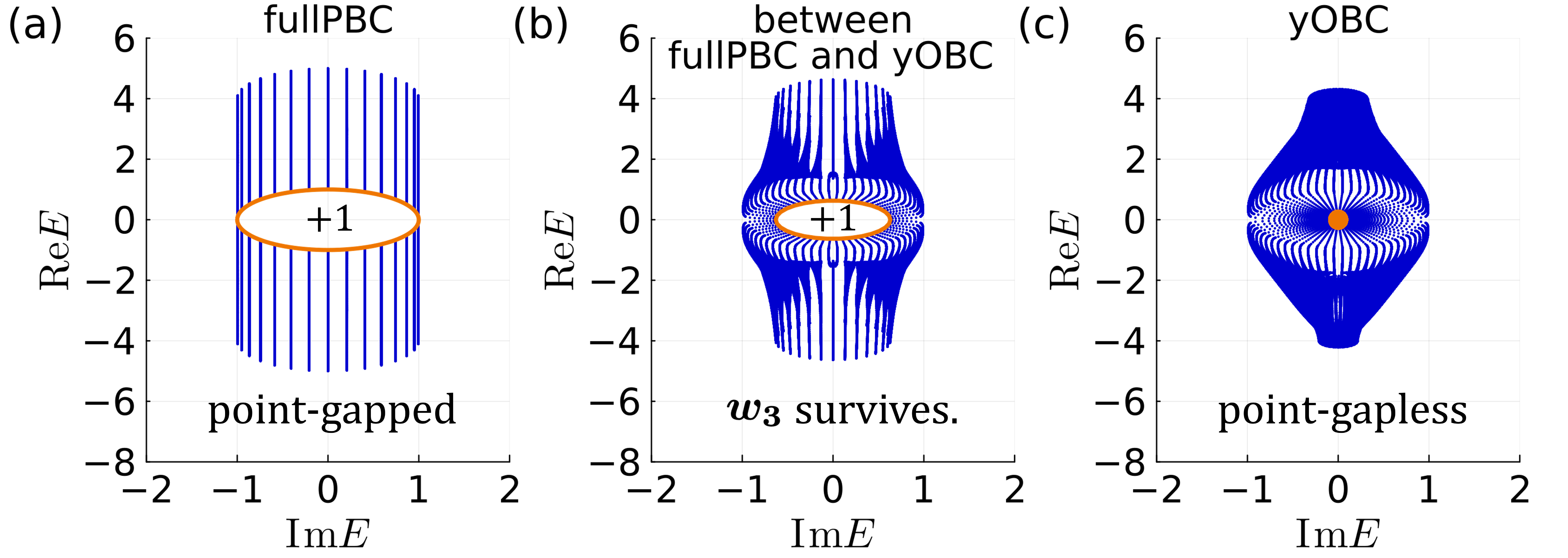}
\caption{The changes of the spectra of ETI in Eq.~(\ref{eq:ETI}) from under the full PBC to the yOBC. $L_y$ is the same as that in Fig.~\ref{fig:ETI}(c) but the momentum resolution around $(k_x,k_z)=(0,0)$ is much finer. The orange modes represent the modes with $k_x=k_z=0$. 
(a) The full PBC spectrum. A point gap is open in the region containing $E=0$ with the nontrivial 3D winding number $+1$. The modes at $k_x=k_z=0$ form a loop and have a nontrivial 1D winding number in each eigensector of $\sigma_z=\pm 1$. 
(b) The spectrum under a boundary condition between the full PBC and the yOBC. Here the hopping amplitude between the $y=1$ sites and the $y=L_y$ sites is $10^{-6}$. The point-gapped region with the non-zero 3D winding number shrinks.
%ut the 3D winding number is preserved in the region where the point gap remains open. 
(c) The yOBC spectrum. The region including $E=0$ is completely closed by the in-gap skin effect of the modes at $k_x=k_z=0$. Comparing with Fig.~\ref{fig:ETI}(c), we can see that the point-gapped region is completely collapsed by the modes near $k_x=k_z=0$.}
\label{fig:yOBC2}
\end{figure}

Let us see how this happens in detail.
First, we note that the in-gap skin modes originate from modes with $(k_x, k_z)=(0,0)$, where
%Along the $k_y$ axis with $k_x=k_z=0$, 
the Hamiltonian in Eq.~(\ref{eq:ETI}) becomes
\begin{align}
  H_{\rm ETI}(k_x=0,k_y,k_z=0) = -\cos{k_y}\sigma_{z} -i\sin{k_y}\sigma_{0}.
\end{align}
This 1D Hamiltonian gives the complex spectra $\mp e^{\pm ik_y}$ in the eigensector of $\sigma_z=\pm 1$, which have the energy winding numbers $\pm 1$ along the $k_y$ direction, as illustrated in Fig.~\ref{fig:yOBC2}(a).
Thus, from the general theory of NHSE~\cite{OKSS-20}, we have the skin modes inside the point gap when one imposes the yOBC~\footnote{Note that the isotropic structure of $\mp e^{\pm ik_y}$ in the complex energy plane causes highly degenerated in-gap skin modes at $E=0$.}.
%The in-gap skin modes originate from 
%As shown in Fig.~\ref{fig:ETI2}, the ETI in Eq.~(\ref{eq:ETI}) does not host surface states under yOBC. 

%The in-gap skin modes make the OBC topological number ill-defined.

At first glance, the in-gap skin modes appear to be isolated from the other bulk modes, but this is not the case.
%At first sight, the in-gap skin modes looks isolated from other bulk modes, but it is not the case.
%For finite systems, in fact, this modes are only apparently isolated. 
For a finite $L_y$, %which is the system size in the $y$-direction, 
the Hamiltonian under the yOBC is a continuous function with respect to $k_x$ and $k_z$, so is its eigenvalues. 
Therefore, there must be ordinary bulk modes nearby the skin modes. 
%in a sufficient neighborhood of $k_x=k_z=0$, there exist many modes that are continuously connected to the in-gap skin modes. In fact, 
Figure~\ref{fig:yOBC2}(c) shows the energy spectrum of Eq.~(\ref{eq:ETI}) under the yOBC, 
with a high momentum resolution. 
%in sufficient neighborhood of $k_x=k_z=0$. 
The bulk modes around the in-gap skin modes are now evident.
Importantly, the point-gapped region disappears due to these bulk modes.
Therefore, 
we do not have a well-defined 3D winding number and surface states under the yOBC.

The disappearance of surface modes can be regard as a result of a topological phase transition under continuous change of the boundary conditions. 
Decreasing the hopping terms between the $y=1$ sites and $y=L_y$ sites, 
we can smoothly change the boundary condition from the full PBC to the yOBC.
According to the deformation, the modes at $(k_x,k_z)=(0,0)$ shrink to the in-gap skin modes as shown in Fig.~\ref{fig:yOBC2}(b).
%As shown in Fig.~\ref{fig:yOBC2} (b), the modes at $(k_x,k_z)=(0,0)$ shrink to the in-gap skin modes accordingly.
Finally, the originally point-gapped region is fully covered by bulk modes under the yOBC.
We also show the change of the 3D winding number throughout the topological phase transition for the model in Eq.~(\ref{eq:ETI2}) in the Supplemental Material \footnote{See Supplemental Material, which includes Refs.~\cite{Prodan-10,Katsura-Koma-16,Akagi-17,Yang-Schnyder-Hu-Chiu}, for a demonstration of how the real-space 3D winding number changes with different boundary conditions, a proof of the BBC for point-gap topological phases in all 38-fold symmetry classes, and a connection between intrinsic point-gap topological phases in the OBC and a single exceptional point.}.

% \begin{figure}[h]
% \RawFloats
% \begin{minipage}[t]{0.32\hsize}
% \centering
% \includegraphics[keepaspectratio,width=\linewidth]{yOBC1.pdf}
% \end{minipage}\hspace{0mm}
% \begin{minipage}[t]{0.32\hsize}
% \centering
% \includegraphics[keepaspectratio,width=\linewidth]{yOBC2.pdf}
% \end{minipage}\hspace{0mm}
% \begin{minipage}[t]{0.32\hsize}
% \centering
% \includegraphics[keepaspectratio,width=\linewidth]{yOBC3.pdf}
% \end{minipage}\hspace{0mm}
% \caption{The changes of the spectra of ETI in Eq.~(\ref{eq:ETI}) from under the full PBC to the yOBC. $L_y$ is the same as that in Fig.~\ref{fig:ETI}(c) but the momentum resolution around $(k_x,k_z)=(0,0)$ is much finer. The orange modes represent the modes with $k_x=k_z=0$. 
% (a) The full PBC spectrum. A point gap is open in the region containing $E=0$ with the nontrivial 3D winding number $+1$. The modes at $k_x=k_z=0$ form a loop and have a nontrivial 1D winding number in each eigensector of $\sigma_z=\pm 1$. 
% (b) The spectrum under a boundary condition between the full PBC and the yOBC. Here the hopping amplitude between the $y=1$ sites and the $y=L_y$ sites is $10^{-6}$. The point-gapped region with the non-zero 3D winding number shrinks.
% %ut the 3D winding number is preserved in the region where the point gap remains open. 
% (c) The yOBC spectrum. The region including $E=0$ is completely closed by the in-gap skin effect of the modes at $k_x=k_z=0$. Comparing with Fig.~\ref{fig:ETI}(c), we can see that the point-gapped region is completely collapsed by the modes near $k_x=k_z=0$.}
% \label{fig:yOBC2}
% \end{figure}

{\it BBC of point-gap topology with symmetry.--}
The above arguments can also apply to point-gapped systems under symmetries.
Namely,
{\it when a symmetry-protected  point-gap topological number under the OBC is non-zero, then the corresponding boundary states appear.}

It should be noted here that  possible point-gap topological numbers under OBCs can be different from those under PBCs.
The disagreement stems from the property that some point-gap topological numbers in PBCs are always accompanied by in-gap skin modes, which spoil the corresponding topological numbers in OBCs.

Whereas we have considered the full general symmetries in the Supplemental Material, 
we focus here on a particular class of symmetries, which we call AZ$^\dagger$ symmetry. 
The AZ$^\dagger$ symmetries are a non-Hermitian generalization of the Altland-Zirnbauer (AZ) symmetry \cite{KSUS-19}:
It consists of non-Hermitian versions of time-reversal symmetry (TRS$^\dagger$),
${\cal C} H^T({\bm k}) {\cal C}^{-1}=H(-{\bm k})$, 
particle-hole symmetry (PHS$^\dagger$), ${\cal T} H^*({\bm k}) {\cal T}^{-1}=-H(-{\bm k})$, 
and chiral symmetry (CS) $\Gamma H^\dagger({\bm k}) \Gamma^{-1}=-H({\bm k})$, where 
${\cal C}$, ${\cal T}$, and $\Gamma$ are unitary operators.
The AZ$^\dagger$ symmetry naturally arises in the non-Hermitian Hamiltonian of the retarded Green function, and thus it governs non-Hermitian topological phases in materials \cite{Yoshida-dagger-symmetry}. 
The presence and/or absence of these symmetries define ten symmetry classes, and
the topological classification in these classes under the full PBC has been known \cite{KSUS-19}.

In Table \ref{table:PBCtoOBC}, we show how the point-gap topological classification under the PBCs changes under OBCs:
In one dimension, all the point-gap topological numbers in the AZ$^\dagger$ classes become trivial under the OBC, 
%The $\mathbb{Z}_2$ topological number in 1D class AII$^\dagger$ is also a topological number in class DII$^\dagger$, and the non-trivial $\mathbb{Z}_2$ topological number is known to induce symmetry-protected skin modes inside the point-gap under OBCs \cite{OKSS-20}, 
because their non-trivial values in the PBC always result in in-gap skin modes under the OBC. 
Actually, all the $\mathbb{Z}$ indices in 1D AZ$^\dagger$ classes reduce to the 1D winding number~\cite{KSUS-19}, which gives in-gap skin modes under the OBC~\cite{Zhang-19,OKSS-20}.
Furthermore, the $\mathbb{Z}_2$ topological number in 1D classes AII$^\dagger$ and DIII$^\dagger$ under the PBC causes symmetry-protected skin modes inside the point gap under the OBC \cite{OKSS-20}. 
%Furthermore, other $\mathbb{Z}$ and $2\mathbb{Z}$ topological number reduce to the 1D winding number in class A~\cite{KSUS-19}. Non-zero 1D winding number is known to results in the in-gap skin modes under OBCs~\cite{Zhang-19,OKSS-20}.
Therefore, no 1D point-gap topological number survives under the OBC.

The reduction of topological numbers in Table \ref{table:PBCtoOBC} in two and three dimensions occurs as a result of the dimensional reduction \cite{QHZ-08}. 
%%%%%%%%%%%%%%%%%%%%%%%
First, we focus on class AII$^\dagger$. From a Hamiltonian $H({\bm k})$ with a point gap at $E$ in class AII$^\dagger$, we can obtain a topologically equivalent gapped Hermitian Hamiltonian $\tilde{H}({\bm k})$: 
\begin{align}
\tilde{H}({\bm k})=
\begin{pmatrix}
0&H({\bm k})-E\\
H^{\dagger}({\bm k})-E^* & 0
\end{pmatrix},
\end{align}
which belongs to class DIII as it has an additional CS $\Sigma \tilde{H}({\bm k}) \Sigma^{-1}=-\tilde{H}({\bm k})$ ($\Sigma=\sigma_z\otimes 1$) together with TRS $\tilde{\cal C}\tilde{H}^*({\bm k})\tilde{\cal C}^{-1}=\tilde{H}(-{\bm k})$ $(\tilde{C}=\sigma_x\otimes {\cal C})$~\cite{KSUS-19}.  
Using the dimensional reduction in class DIII~\cite{QHZ-08}, one can show that 
the parity of the 3D $\mathbb{Z}$ index for $H(k_x,k_y,k_z)$ equals to the product of the 2D $\mathbb{Z}_2$ indices of $H(k_x,k_y,k_z^\text{0})$ with $k_z^0=0$ and $\pi$, and  
similarly, the 2D $\mathbb{Z}_2$ index of $H(k_x,k_y,k_z^\text{0})$ equals to the product of the 1D $\mathbb{Z}_2$ indices of $H(k_x,k_y^0,k_z^\text{0})$ with $k_y^0=0$ and $\pi$.
%Using the dimensional reduction in class DIII~\cite{QHZ-08}, one can show that for class AII$^\dagger$ under full PBCs, a non-trivial 2D $\mathbb{Z}_2$ index or an odd parity of the 3D $\mathbb{Z}$ index at $E=E_{\rm P}$ yield a non-zero 1D $\mathbb{Z}_2$ index at $E=E_{\rm P}$ along a high symmetric line in the BZ.
Thus, for class AII$^\dagger$ under full PBCs, a non-trivial 2D $\mathbb{Z}_2$ index or an odd parity of the 3D $\mathbb{Z}$ index at $E$ yield a non-zero 1D $\mathbb{Z}_2$ index at $E$ along a high symmetric line in the BZ.
Therefore, they always accompany symmetry-protected in-gap skin modes~\cite{OKSS-20}, trivializing the corresponding topological numbers in OBCs.  
As a result, only the even part of the $\mathbb{Z}$ index in three dimensions survives in the OBC. 
We can also show the reduction $\mathbb{Z}\to2\mathbb{Z}$ in 2D class DIII$^\dagger$, using a similar dimensional reduction. 
%The reduction in 2D class DIII$^\dagger$ comes from the reduction $\mathbb{Z}_2\to 0$ in 2D class AII$^\dagger$ since the parity of the $\mathbb{Z}$ index in 2D class DIII$^\dagger$ coincides with the $\mathbb{Z}_2$ index in 2D class AII$^\dagger$.
On the basis of the $K$-theory, we prove the BBC for point-gap topological phases under the OBC in all 38-fold symmetry classes in non-Hermitian systems, including AZ$^\dagger$ ones, in the Supplemental Material~\cite{Note3}.

\begin{center}
\begin{table}[htbp]
    \caption{Classification of point-gap topological phases. For topological numbers with arrows, the left specifies topological numbers under PBCs and the right specifies those under OBCs. For topological numbers without arrows, the classification under OBCs coincides with that under PBCs. We consider the $\mathrm{AZ}^{\dag}$ symmetry classes with the spatial dimension $d =1,2$ and $3$.}
     \begin{adjustbox}{width=\linewidth}
     \begin{tabular}{ccccccc} \hline \hline
    Symmetry class & ~TRS$^\dagger$~ & ~PHS$^\dagger$~ & ~~CS~~ & ~$d=1$~ & ~$d=2$~ & ~$d=3$~ \\ \hline
    \multirow{1}{*}{A}
    & $0$ & $0$ & $0$ & $\mathbb{Z}\to 0$ & $0$ & $\mathbb{Z}$ \\ %\hline
    \multirow{1}{*}{AIII}
    & $0$ & $0$ & $1$ & $0$ & $\mathbb{Z}$ & $0$ \\ \hline
    \multirow{1}{*}{$\text{AI}^{\dag}$}
    & $+1$ & $0$ & $0$ &$0$ & $0$ & $2\mathbb{Z}$ \\ %\hline
    \multirow{1}{*}{$\text{BDI}^{\dag}$}
    & $+1$ & $+1$ & $1$ &$0$ & $0$ & $0$ \\ %\hline
    \multirow{1}{*}{$\text{D}^{\dag}$}
    & $0$ & $+1$ & $0$ & $\mathbb{Z}\to 0$ & $0$ & $0$ \\ %\hline
    \multirow{1}{*}{$\text{DIII}^{\dag}$}
    & $-1$ & $+1$ & $1$ & $\mathbb{Z}_2\to 0$ & $\mathbb{Z}\to2\mathbb{Z}$ & $0$ \\ %\hline
    \multirow{1}{*}{$\text{AII}^{\dag}$}
    & $-1$ & $0$ & $0$ & $\mathbb{Z}_2\to 0$ & $\mathbb{Z}_2\to 0$ & $\mathbb{Z}\to 2\mathbb{Z}$ \\ %\hline
    \multirow{1}{*}{$\text{CII}^{\dag}$}
    & $-1$ & $-1$ & $1$ &$0$ & $\mathbb{Z}_2$ & $\mathbb{Z}_2$ \\ %\hline
    \multirow{1}{*}{$\text{C}^{\dag}$}
    & $0$ & $-1$ & $0$ & $2\mathbb{Z}\to 0$ & $0$ & $\mathbb{Z}_2$ \\ %\hline
    \multirow{1}{*}{$\text{CI}^{\dag}$}
    & $+1$ & $-1$ & $1$ &$0$ & $2\mathbb{Z}$ & $0$ \\ \hline \hline
  \end{tabular}
  \end{adjustbox}
 \label{table:PBCtoOBC} 
\end{table}
\end{center}

Remarkably, we can predict novel topological phase transitions intrinsic to non-Hermitian systems,
using the reduction of the point-gap topological numbers in the presence of symmetry:
The symmetry-protected in-gap skin modes in 3D class AII$^\dagger$ systems may disappear suddenly once one breaks TRS$^\dagger$ by an infinitesimal perturbation.
The disappearance of the in-gap skin modes allows the well-defined 3D winding number under the OBC, and thus we have an abrupt transmutation from the topologically trivial state with the in-gap skin modes to a topologically non-trivial one with surface states. 
Infinitesimal instability \cite{Okuma-19} intrinsic to non-Hermitian systems induces this topological phase transition, and thus it never happens in Hermitian systems.
We also find that boundary states in intrinsic point-gap topological phases may support a single exceptional point~\cite{Note3}, which is also unique to non-Hermitian systems.
%\footnote{See Sec. S6 of Supplemental Material}

{\it Conclusions.--}
%In this paper we have discussed the BBC for the point-gap topology in the three-dimensional class A. We have demonstrated analytically and numerically that the BBC for the point-gap topology is valid by introducing the real-space 3D winding number that can be easily used even under the open boundary. We also have investigated the in-gap skin effect, which causes the BBC to break in detail. Our study has revealed that the BBC is universally valid even in non-Hermitian systems. In the following paper, we give the BBC for the point-gap topology for all other classes and a classification table of point-gapless phases.
In this paper, we establish the BBC for the point-gap topology in non-Hermitian systems. We demonstrate that  real-space topological numbers under OBCs result in robust and exotic surface states.
We find that the bulk point-gap topology in OBCs can be different from that in PBCs, and give a complete classification of the OBC point-gap topology in the presence of symmetry.
Our finding reveals a novel universal property of non-Hermitian topological phases of matters.

\smallskip
We are grateful to Nobuyuki Okuma, Ken Shiozaki, Shuhei Ohyama, Yusuke Nakai, and Hiroto Oka for valuable discussions. 
This work was supported by JST CREST Grant No.JPMJCR19T2, the establishment of university fellowships towards the creation of science technology innovation, Grant No.JPMJFS2123, and KAKENHI Grant No.JP20H00131.
This work was done while Takumi Bessho was at Yukawa Institute for Theoretical Physics, Kyoto University. 

\smallskip
{\it Appendix on the BBC for point-gap topological phases in general symmetry classes.--}
Our argument is applicable to any symmetry classes in non-Hermitian systems.
We summarize here our results for point-gap topological phases in general symmetry classes in non-Hermitian systems.
For the details of the derivation, see Supplemental Material~\cite{Note3}. 

In addition to AZ$^\dagger$ symmetries discussed in the main text, non-Hermitian systems allow the original AZ symmetries defined by the following equations:
\begin{align}
  \label{eq:sym-AZ}
  {\cal T} H^{*}({\bm k}) {\cal T}^{-1} &= H(-{\bm k}),\quad {\cal T}{\cal T}^{*}=\pm 1,\nonumber
  \\
  {\cal C} H^T({\bm k}) {\cal C}^{-1} &= -H(-{\bm k}), \quad {\cal C}{\cal C}^{*}=\pm 1,
\end{align}
where ${\cal T}$ and ${\cal C}$ are unitary operators corresponding to time-reversal symmetry (TRS) and particle-hole symmetry (PHS), respectively. Furthermore, as an additional general symmetry, one can also introduce sublattice symmetry (SLS),
%\begin{align}
%{\cal S}H{\cal S}^{-1}=-H, \quad {\cal S}^2=1,    
%\end{align}
\begin{align}
  {\cal S}H(\bm{k}){\cal S}^{-1} = -H(\bm{k}), \quad {\cal S}^{2}=1,
  \label{sym:SLS}
\end{align}
or pseudo-Hermiticity (pH)
\begin{align}
 \eta H^\dagger({\bm k})\eta^{-1}=H({\bm k}), \quad \eta^2=1   
\end{align}
with unitary operators ${\cal S}$ and $\eta$.  The presence and/or absence of these symmetries define symmetry classes intrinsic to non-Hermitian systems \cite{KSUS-19}.

Tables \ref{table:realAZ+SLS} and \ref{table:realAZdag+SLS} summarize our results on point-gap topological phases in general symmetry classes.
Here, the presence or absence of AZ or AZ$^\dagger$ symmetries define AZ or AZ$^\dagger$ symmetry classes in Table \ref{table:AZ_AZdag_class}. 
Moreover, each AZ or AZ$^\dagger$ class can host SLS (pH) additionally, 
%which defines AZ class+${\cal S}$ (AZ class+$\eta$) and AZ$^\dagger$ class+${\cal S}$ (AZ$^\dagger$ class+$\eta$). 
where the subindex $+(-)$ of ${\cal S}/\eta$ in Tables \ref{table:realAZ+SLS} and \ref{table:realAZdag+SLS} specifies the commutation (anti-commutation) relation between SLS/pH and AZ or AZ$^\dagger$ symmetries. 
For an AZ (AZ$^\dagger$) class having both TRS (TRS$^\dagger$) and PHS (PHS$^\dagger$), ${\cal S}$ or $\eta$ has a double subindex, where the first index specifies the commutation or anticommutation relation between SLS and TRS (TRS$^\dagger$), and the second one specifies those between SLS and PHS (PHS$^\dagger$), respectively.
Tables \ref{table:realAZ+SLS} and \ref{table:realAZdag+SLS} show how the point-gap topological classification under the PBCs changes under OBCs. For the topological numbers with arrows, the classification under the PBCs shown on the left changes to that under OBCs on the right, while the topological numbers without arrows remain the same under both boundary conditions.
We also find that the BBC holds for the point-gap topological classification under OBCs:
Topologically protected boundary states appear when the bulk point-gap topological numbers under OBCs are non-trivial. 

\begin{center}
\begin{table}[htbp]
\caption{AZ and AZ$^\dagger$ symmetry classes for non-Hermitian Hamiltonians. Here, $``0"$ denotes the absence of symmetries, while $``\pm1"$ indicates the presence of each symmetry, dependent on whether its operator squares to $\pm1$.}
     \begin{tabular}{lcccccc} \hline \hline
    Sym. class & &TRS & PHS & CS & TRS$^\dagger$ & PHS$^\dagger$ \\ \hline 
	Complex AZ &\multirow{1}{*}{A}
		& $0$ & $0$ & $0$ & $0$ & $0$ \\
		&\multirow{1}{*}{AIII}
		& $0$ & $0$ & $1$ & $0$ & $0$  \\ \hline
    Real AZ &\multirow{1}{*}{$\text{AI}$}
    & $+1$ & $0$ & $0$ &$0$ & $0$  \\ 
    &\multirow{1}{*}{$\text{BDI}$}
    & $+1$ & $+1$ & $1$ &0 & 0 \\ 
    &\multirow{1}{*}{$\text{D}$}
    & $0$ & $+1$ & $0$ &0 & 0  \\ 
    &\multirow{1}{*}{$\text{DIII}$}
    & $-1$ & $+1$ & $1$ & $0$ & 0  \\ 
    &\multirow{1}{*}{$\text{AII}$}
    & $-1$ & $0$ & $0$ &0 & $0$  \\ 
    &\multirow{1}{*}{$\text{CII}$}
    & $-1$ & $-1$ & $1$ &$0$ & 0  \\ 
    &\multirow{1}{*}{$\text{C}$}
    & $0$ & $-1$ & $0$ & $0$ & $0$  \\ 
    &\multirow{1}{*}{$\text{CI}$}
    & $+1$ & $-1$ & $1$ &$0$ & $0$  \\ \hline
    Real AZ$^\dagger$ &\multirow{1}{*}{$\text{AI}^\dagger$}
    & $0$ & $0$ & $0$ &$+1$ & $0$  \\ 
    &\multirow{1}{*}{$\text{BDI}^\dagger$}
    & $0$ & $0$ & $1$ &$+1$ & $+1$ \\ 
    &\multirow{1}{*}{$\text{D}^\dagger$}
    & $0$ & $0$ & $0$ &0 & $+1$  \\ 
    &\multirow{1}{*}{$\text{DIII}^\dagger$}
    & $0$ & $0$ & $1$ & $-1$ & $+1$  \\ 
    &\multirow{1}{*}{$\text{AII}^\dagger$}
    & $0$ & $0$ & $0$ &$-1$ & $0$  \\ 
    &\multirow{1}{*}{$\text{CII}^\dagger$}
    & $0$ & $0$ & $1$ &$-1$ & $-1$  \\ 
    &\multirow{1}{*}{$\text{C}^\dagger$}
    & $0$ & $0$ & $0$ & $0$ & $-1$  \\ 
    &\multirow{1}{*}{$\text{CI}^\dagger$}
    & $0$ & $0$ & $1$ &$+1$ & $-1$  \\ \hline \hline  
    \end{tabular}
 \label{table:AZ_AZdag_class} 
\end{table}
\end{center}

\clearpage

\begin{table}[H]
  \centering
  \begingroup
  \renewcommand{\arraystretch}{1.1}
    \begin{adjustbox}{width=\linewidth}
    \begin{tabular}{ccccc} \hline \hline
        AZ class & ~Add. sym.~ & ~$d=1$~ & ~$d=2$~ & ~$d=3$~ \\ \hline %\hline
        \multirow{1}{*}{A}
        & -&$\mathbb{Z}\to 0$ & $0$ & \Z \\ 
        \multirow{1}{*}{AIII}
        &  -&$0$ &\Z & $0$ \\ \hline %\hline
        \multirow{1}{*}{$\text{A}$}&{$\mathcal{S}$}
        &  ${\begin{aligned}[t]&\mathbb{Z}\oplus\mathbb{Z}\\\to &\mathbb{Z}[1,-1]\end{aligned}}$ & $0$ & $\mathbb{Z}\oplus\mathbb{Z}$ \\ 
        \multirow{1}{*}{$\text{AIII}$}&{$\mathcal{S}_{-},\eta_{-}$}
        &  $0$ & $\mathbb{Z}\oplus\mathbb{Z}$ & $0$ \\ \hline %\hline
        \multirow{1}{*}{$\text{A}$}&{$\eta$}
        &  $0$ & \Z & $0$ \\ 
        \multirow{1}{*}{$\text{AIII}$}&{$\mathcal{S}_{+},\eta_{+}$}
        &  \Z & $0$ & \Z \\ \hline
        \multirow{1}{*}{$\text{AI}$}
        & -&$\mathbb{Z}\to 0$ & $0$ & $0$ \\ 
        \multirow{1}{*}{$\text{BDI}$}
        & -&\Zt & \Z & $0$ \\ 
        \multirow{1}{*}{$\text{D}$}
        & -&\Zt & \Zt & \Z \\ 
        \multirow{1}{*}{$\text{DIII}$}
        & -&$0$ &\Zt & \Zt \\ 
        \multirow{1}{*}{$\text{AII}$}
        & -&$2\mathbb{Z}\to 0$ & $0$ & \Zt \\ 
        \multirow{1}{*}{$\text{CII}$}
        & -&$0$ & 2\Z & $0$ \\ 
        \multirow{1}{*}{$\text{C}$}
        & -&$0$ & $0$ & 2\Z \\ 
        \multirow{1}{*}{$\text{CI}$}
        & -&$0$ & $0$ & $0$ \\ \hline %\hline
        \multirow{1}{*}{$\text{AI}$}&{$\mathcal{S}_{+}$}
        & ${\begin{aligned}[t]&\mathbb{Z}\oplus\mathbb{Z}\\\to &\mathbb{Z}[1,-1]\end{aligned}}$ & $0$ & $0$ \\ 
        \multirow{1}{*}{$\text{BDI}$}&{$\mathcal{S}_{+-},\eta_{-+}$}
        & ${\begin{gathered}[t]\mathbb{Z}_{2}\oplus\mathbb{Z}_{2}\\\to \mathbb{Z}_{2}[1,1]\hspace*{5mm}\end{gathered}}$ & ${\begin{gathered}[t]\mathbb{Z}\oplus\mathbb{Z}\\\to \mathbb{Z}[2,0]\oplus\mathbb{Z}[1,-1]\hspace*{1.6mm}\end{gathered}}$ & $0$ \\ 
        \multirow{1}{*}{$\text{D}$}&{$\mathcal{S}_{-}$}
        & ${\begin{gathered}[t]\mathbb{Z}_{2}\oplus\mathbb{Z}_{2}\\\to \mathbb{Z}_{2}[1,1]\hspace*{5mm}\end{gathered}}$ & ${\begin{gathered}[t]\mathbb{Z}_{2}\oplus\mathbb{Z}_{2}\\\to \mathbb{Z}_{2}[1,1]\hspace*{5mm}\end{gathered}}$ & ${\begin{gathered}[t]\mathbb{Z}\oplus\mathbb{Z}\\\to \mathbb{Z}[2,0]\oplus\mathbb{Z}[1,-1]\hspace*{1.6mm}\end{gathered}}$  \\ 
        \multirow{1}{*}{$\text{DIII}$}&{$\mathcal{S}_{+-},\eta_{+-}$}
        &  $0$ & $\mathbb{Z}_{2}\oplus\mathbb{Z}_{2}$ & $\mathbb{Z}_{2}\oplus\mathbb{Z}_{2}$ \\ 
        \multirow{1}{*}{$\text{AII}$}&{$\mathcal{S}_{+}$}
        & ${\begin{gathered}[t]2\mathbb{Z}\oplus2\mathbb{Z}\\\to 2\mathbb{Z}[1,-1]\hspace*{3mm}\end{gathered}}$ & $0$ & $\mathbb{Z}_{2}\oplus\mathbb{Z}_{2}$ \\ 
        \multirow{1}{*}{$\text{CII}$}&{$\mathcal{S}_{+-},\eta_{-+}$}
        & $0$ & $2\mathbb{Z}\oplus2\mathbb{Z}$ & $0$ \\ 
        \multirow{1}{*}{$\text{C}$}&{$\mathcal{S}_{-}$}
        &  $0$ & $0$ & $2\mathbb{Z}\oplus2\mathbb{Z}$\\ 
        \multirow{1}{*}{$\text{CI}$}&{$\mathcal{S}_{+-},\eta_{+-}$}
        & $0$ & $0$ & $0$  \\ \hline %\hline
        \multirow{1}{*}{$\text{AI}$}&{$\mathcal{S}_{-}$}
        & $\mathbb{Z}\to 0$ & $0$ & \Z \\ 
        \multirow{1}{*}{$\text{BDI}$}&{$\mathcal{S}_{-+},\eta_{+-}$}
        & $0$ & \Z & $0$ \\ 
        \multirow{1}{*}{$\text{D}$}&{$\mathcal{S}_{+}$}
        & \Z & $0$ & \Z \\ 
        \multirow{1}{*}{$\text{DIII}$}&{$\mathcal{S}_{-+},\eta_{-+}$}
        &  $0$ & \Z & $0$ \\ 
        \multirow{1}{*}{$\text{AII}$}&{$\mathcal{S}_{-}$}
        & $\mathbb{Z}\to 0$ & $0$ & \Z \\ 
        \multirow{1}{*}{$\text{CII}$}&{$\mathcal{S}_{-+},\eta_{+-}$}
        & $0$ & \Z & $0$ \\ 
        \multirow{1}{*}{$\text{C}$}&{$\mathcal{S}_{+}$}
        & $\mathbb{Z}\to2\mathbb{Z}$ & $0$ & \Z\\ 
        \multirow{1}{*}{$\text{CI}$}&{$\mathcal{S}_{-+},\eta_{-+}$}
        & $0$ & \Z & $0$ \\ \hline %\hline
        \multirow{1}{*}{$\text{AI}$}&{$\eta_{+}$}
        & $0$ & $0$ & $0$ \\ 
        \multirow{1}{*}{$\text{BDI}$}&{$\mathcal{S}_{++},\eta_{++}$}
        & \Z & $0$ & $0$ \\ 
        \multirow{1}{*}{$\text{D}$}&{$\eta_{+}$}
        & \Zt & \Z & $0$ \\ 
        \multirow{1}{*}{$\text{DIII}$}&{$\mathcal{S}_{--},\eta_{++}$}
        &  \Zt & \Zt & \Z \\ 
        \multirow{1}{*}{$\text{AII}$}&{$\eta_{+}$}
        & $0$ & \Zt & \Zt \\ 
        \multirow{1}{*}{$\text{CII}$}&{$\mathcal{S}_{++},\eta_{++}$}
        & 2\Z & $0$ & \Zt \\
        \multirow{1}{*}{$\text{C}$}&{$\eta_{+}$}
        &  $0$ & 2\Z & $0$ \\
        \multirow{1}{*}{$\text{CI}$}&{$\mathcal{S}_{--},\eta_{++}$}
        & $0$ & $0$ & 2\Z \\ \hline %\hline
        \multirow{1}{*}{$\text{AI}$}&{$\eta_{-}$}
        & $\mathbb{Z}_{2}\to 0$ & $\mathbb{Z}\to2\mathbb{Z}$ & $0$ \\ 
        \multirow{1}{*}{$\text{BDI}$}&{$\mathcal{S}_{--},\eta_{--}$}
        & \Zt & \Zt & \Z \\ 
        \multirow{1}{*}{$\text{D}$}&{$\eta_{-}$}
        & $0$ & \Zt & \Zt \\ 
        \multirow{1}{*}{$\text{DIII}$}&{$\mathcal{S}_{++},\eta_{--}$}
        &  2\Z & $0$ & \Zt \\ 
        \multirow{1}{*}{$\text{AII}$}&{$\eta_{-}$}
        & $0$ & 2\Z & $0$ \\ 
        \multirow{1}{*}{$\text{CII}$}&{$\mathcal{S}_{--},\eta_{--}$}
        & $0$ & $0$ & 2\Z \\ 
        \multirow{1}{*}{$\text{C}$}&{$\eta_{-}$}
        &  $0$ & $0$ & $0$ \\ 
        \multirow{1}{*}{$\text{CI}$}&{$\mathcal{S}_{++},\eta_{--}$}
        & $\mathbb{Z}\to2\mathbb{Z}$ & $0$ & $0$ \\ \hline \hline
    \end{tabular}
    \end{adjustbox}
    \endgroup
    \caption{Classification of point-gap topological phases in the AZ classes without or with SLS or pH. The subscript of ${\cal S}_{\pm}/\eta_{\pm}$ specifies the commutation (+) or anti-commutation (-) relation to TRS or PHS. For ${\cal S}_{\pm\pm}/\eta_{\pm\pm}$, the first subscript specifies the relation to TRS and the second specifies the relation to PHS.  For the topological numbers with arrows, the classification under OBCs changes from that under PBCs, where the left specifies the classification under PBCs and the right specifies that under OBCs. The topological number $\mathbb{Z}[i,j]$ ($\mathbb{Z}_2[i,j]$) under OBCs indicates the Abelian group $\mathbb{Z}$ ($\mathbb{Z}_2$) generated by the element $(i,j)\in \mathbb{Z}\oplus\mathbb{Z}$ [$(i,j)\in \mathbb{Z}_2\oplus\mathbb{Z}_2$] under PBCs. For the topological numbers without arrows, the classification under OBCs coincides with that under PBCs. 
  \label{table:realAZ+SLS} }
\end{table}

\begin{table}[H]
  \centering
  \begingroup
  \renewcommand{\arraystretch}{1.1}
    \begin{adjustbox}{width=\linewidth}
    \begin{tabular}{ccccc} \hline \hline
        $\text{AZ}^{\dagger}$ class & ~Add. sym.~ & ~$d=1$~ & ~$d=2$~ & ~$d=3$~ \\ \hline %\hline
        \multirow{1}{*}{$\text{AI}^{\dag}$}
        & -&$0$ & $0$ & 2\Z \\ 
        \multirow{1}{*}{$\text{BDI}^{\dag}$}
        & -&$0$ & $0$ & $0$ \\ 
        \multirow{1}{*}{$\text{D}^{\dag}$}
        & -&$\mathbb{Z}\to 0$ & $0$ & $0$ \\ 
        \multirow{1}{*}{$\text{DIII}^{\dag}$}
        & -&$\mathbb{Z}_{2}\to 0$ & $\mathbb{Z}\to2\mathbb{Z}$ & $0$ \\ 
        \multirow{1}{*}{$\text{AII}^{\dag}$}
        & -&$\mathbb{Z}_{2}\to 0$ & $\mathbb{Z}_{2}\to 0$ & $\mathbb{Z}\to2\mathbb{Z}$ \\ 
        \multirow{1}{*}{$\text{CII}^{\dag}$}
        & -&$0$ & \Zt & \Zt \\ 
        \multirow{1}{*}{$\text{C}^{\dag}$}
        & -&$2\mathbb{Z}\to 0$ & $0$ & \Zt \\ 
        \multirow{1}{*}{$\text{CI}^{\dag}$}
        & -&$0$ & $2\mathbb{Z}$ & $0$ \\ \hline %\hline
        \multirow{1}{*}{$\text{AI}^{\dag}$}&{$\mathcal{S}_{+}$}
        & \Z & $0$ & \Z \\ 
        \multirow{1}{*}{$\text{BDI}^{\dag}$}&{$\mathcal{S}_{+-},\eta_{-+}$}
        & $0$ & \Z & $0$ \\ 
        \multirow{1}{*}{$\text{D}^{\dag}$}&{$\mathcal{S}_{-}$}
        & $\mathbb{Z}\to 0$ & $0$ & \Z  \\ 
        \multirow{1}{*}{$\text{DIII}^{\dag}$}&{$\mathcal{S}_{+-},\eta_{+-}$}
        &  $0$ & \Z & $0$ \\ 
        \multirow{1}{*}{$\text{AII}^{\dag}$}&{$\mathcal{S}_{+}$}
        &  $\mathbb{Z}\to2\mathbb{Z}$ & $0$ & \Z \\ 
        \multirow{1}{*}{$\text{CII}^{\dag}$}&{$\mathcal{S}_{+-},\eta_{-+}$}
        & $0$ & \Z & $0$ \\ 
        \multirow{1}{*}{$\text{C}^{\dag}$}&{$\mathcal{S}_{-}$}
        & $\mathbb{Z}\to 0$ & $0$ & \Z\\ 
        \multirow{1}{*}{$\text{CI}^{\dag}$}&{$\mathcal{S}_{+-},\eta_{+-}$}
        & $0$ & \Z & $0$  \\ \hline %\hline
        \multirow{1}{*}{$\text{AI}^{\dag}$}&{$\mathcal{S}_{-}$}
        & $0$ & $0$ & $2\mathbb{Z}\oplus2\mathbb{Z}$ \\ 
        \multirow{1}{*}{$\text{BDI}^{\dag}$}&{$\mathcal{S}_{-+},\eta_{+-}$}
        & $0$ & $0$ & $0$ \\ 
        \multirow{1}{*}{$\text{D}^{\dag}$}&{$\mathcal{S}_{+}$}
        & ${\begin{aligned}[t]&\mathbb{Z}\oplus\mathbb{Z}\\\to &\mathbb{Z}[1,-1]\end{aligned}}$ & $0$ & $0$ \\ 
        \multirow{1}{*}{$\text{DIII}^{\dag}$}&{$\mathcal{S}_{-+},\eta_{-+}$}
        & ${\begin{gathered}[t]\mathbb{Z}_{2}\oplus\mathbb{Z}_{2}\\\to \mathbb{Z}_{2}[1,1]\hspace*{5mm}\end{gathered}}$ & ${\begin{gathered}[t]\mathbb{Z}\oplus\mathbb{Z}\\\to \mathbb{Z}[2,0]\oplus\mathbb{Z}[1,-1]\hspace*{1.6mm}\end{gathered}}$  & $0$  \\ 
        \multirow{1}{*}{$\text{AII}^{\dag}$}&{$\mathcal{S}_{-}$}
        & ${\begin{gathered}[t]\mathbb{Z}_{2}\oplus\mathbb{Z}_{2}\\\to \mathbb{Z}_{2}[1,1]\hspace*{5mm}\end{gathered}}$ & ${\begin{gathered}[t]\mathbb{Z}_{2}\oplus\mathbb{Z}_{2}\\\to \mathbb{Z}_{2}[1,1]\hspace*{5mm}\end{gathered}}$ & ${\begin{gathered}[t]\mathbb{Z}\oplus\mathbb{Z}\\\to \mathbb{Z}[2,0]\oplus\mathbb{Z}[1,-1]\hspace*{1.6mm}\end{gathered}}$  \\ 
        \multirow{1}{*}{$\text{CII}^{\dag}$}&{$\mathcal{S}_{-+},\eta_{+-}$}
        & $0$ & $\mathbb{Z}_{2}\oplus\mathbb{Z}_{2}$ & $\mathbb{Z}_{2}\oplus\mathbb{Z}_{2}$ \\ 
        \multirow{1}{*}{$\text{C}^{\dag}$}&{$\mathcal{S}_{+}$}
        & ${\begin{gathered}[t]2\mathbb{Z}\oplus2\mathbb{Z}\\\to 2\mathbb{Z}[1,-1]\hspace*{3mm}\end{gathered}}$ & $0$ & $\mathbb{Z}_{2}\oplus\mathbb{Z}_{2}$ \\ 
        \multirow{1}{*}{$\text{CI}^{\dag}$}&{$\mathcal{S}_{-+},\eta_{-+}$}
        & $0$ & $2\mathbb{Z}\oplus2\mathbb{Z}$ & $0$ \\ \hline %\hline
        \multirow{1}{*}{$\text{AI}^{\dag}$}&{$\eta_{+}$}
        & \Zt & \Z & $0$ \\ 
        \multirow{1}{*}{$\text{BDI}^{\dag}$}&{$\mathcal{S}_{++},\eta_{++}$}
        & \Z & $0$ & $0$ \\ 
        \multirow{1}{*}{$\text{D}^{\dag}$}&{$\eta_{+}$}
        & $0$ & $0$ & $0$ \\ 
        \multirow{1}{*}{$\text{DIII}^{\dag}$}&{$\mathcal{S}_{--},\eta_{++}$}
        &  \Zt & \Zt & \Z \\ 
        \multirow{1}{*}{$\text{AII}^{\dag}$}&{$\eta_{+}$}
        & $0$ & 2\Z & $0$ \\ 
        \multirow{1}{*}{$\text{CII}^{\dag}$}&{$\mathcal{S}_{++},\eta_{++}$}
        & 2\Z & $0$ & \Zt \\ 
        \multirow{1}{*}{$\text{C}^{\dag}$}&{$\eta_{+}$}
        &  $0$ & \Zt & \Zt \\ 
        \multirow{1}{*}{$\text{CI}^{\dag}$}&{$\mathcal{S}_{--},\eta_{++}$}
        & $0$ & $0$ & 2\Z \\ \hline %\hline
        \multirow{1}{*}{$\text{AI}^{\dag}$}&{$\eta_{-}$}
        & $0$ & \Zt & \Zt \\
        \multirow{1}{*}{$\text{BDI}^{\dag}$}&{$\mathcal{S}_{--},\eta_{--}$}
        & $0$ & $0$ & 2\Z \\
        \multirow{1}{*}{$\text{D}^{\dag}$}&{$\eta_{-}$}
        & $\mathbb{Z}_{2}\to 0$ & $\mathbb{Z}\to2\mathbb{Z}$ & $0$ \\ 
        \multirow{1}{*}{$\text{DIII}^{\dag}$}&{$\mathcal{S}_{++},\eta_{--}$}
        & $\mathbb{Z}\to2\mathbb{Z}$ & $0$ & $0$ \\
        \multirow{1}{*}{$\text{AII}^{\dag}$}&{$\eta_{-}$}
        & $0$ & $0$ & $0$ \\
        \multirow{1}{*}{$\text{CII}^{\dag}$}&{$\mathcal{S}_{--},\eta_{--}$}
        & \Zt & \Zt & \Z \\
        \multirow{1}{*}{$\text{C}^{\dag}$}&{$\eta_{-}$}
        &  $0$ & 2\Z & $0$ \\
        \multirow{1}{*}{$\text{CI}^{\dag}$}&{$\mathcal{S}_{++},\eta_{--}$}
        & 2\Z & $0$ & \Zt \\ \hline \hline
    \end{tabular}
    \end{adjustbox}
    \endgroup
    \caption{Classification of point-gap topological phases in the real AZ$^\dagger$ classes without or with SLS or pH. The subscript of ${\cal S}_{\pm}/\eta_{\pm}$ specifies the commutation (+) or anti-commutation (-) relation to TRS$^\dagger$ or PHS$^\dagger$. For ${\cal S}_{\pm\pm}/\eta_{\pm\pm}$, the first subscript specifies the relation to TRS$^\dagger$ and the second specifies the relation to PHS$^\dagger$. For the topological numbers with arrows, the classification under OBCs changes from that under PBCs, where the left specifies the classification under PBCs and the right specifies that under OBCs. The topological number $\mathbb{Z}[i,j]$ ($\mathbb{Z}_2[i,j]$) under OBCs indicates the Abelian group $\mathbb{Z}$ ($\mathbb{Z}_2$) generated by the element $(i,j)\in \mathbb{Z}\oplus\mathbb{Z}$ [$(i,j)\in \mathbb{Z}_2\oplus\mathbb{Z}_2$] under PBCs. For the topological numbers without arrows, the classification under OBCs coincides with that under PBCs.
    \label{table:realAZdag+SLS} 
}
\end{table}
\clearpage
\bibliography{main}

\begin{thebibliography}{136}%
\makeatletter
\providecommand \@ifxundefined [1]{%
 \@ifx{#1\undefined}
}%
\providecommand \@ifnum [1]{%
 \ifnum #1\expandafter \@firstoftwo
 \else \expandafter \@secondoftwo
 \fi
}%
\providecommand \@ifx [1]{%
 \ifx #1\expandafter \@firstoftwo
 \else \expandafter \@secondoftwo
 \fi
}%
\providecommand \natexlab [1]{#1}%
\providecommand \enquote  [1]{``#1''}%
\providecommand \bibnamefont  [1]{#1}%
\providecommand \bibfnamefont [1]{#1}%
\providecommand \citenamefont [1]{#1}%
\providecommand \href@noop [0]{\@secondoftwo}%
\providecommand \href [0]{\begingroup \@sanitize@url \@href}%
\providecommand \@href[1]{\@@startlink{#1}\@@href}%
\providecommand \@@href[1]{\endgroup#1\@@endlink}%
\providecommand \@sanitize@url [0]{\catcode `\\12\catcode `\$12\catcode `\&12\catcode `\#12\catcode `\^12\catcode `\_12\catcode `\%12\relax}%
\providecommand \@@startlink[1]{}%
\providecommand \@@endlink[0]{}%
\providecommand \url  [0]{\begingroup\@sanitize@url \@url }%
\providecommand \@url [1]{\endgroup\@href {#1}{\urlprefix }}%
\providecommand \urlprefix  [0]{URL }%
\providecommand \Eprint [0]{\href }%
\providecommand \doibase [0]{http://dx.doi.org/}%
\providecommand \selectlanguage [0]{\@gobble}%
\providecommand \bibinfo  [0]{\@secondoftwo}%
\providecommand \bibfield  [0]{\@secondoftwo}%
\providecommand \translation [1]{[#1]}%
\providecommand \BibitemOpen [0]{}%
\providecommand \bibitemStop [0]{}%
\providecommand \bibitemNoStop [0]{.\EOS\space}%
\providecommand \EOS [0]{\spacefactor3000\relax}%
\providecommand \BibitemShut  [1]{\csname bibitem#1\endcsname}%
\let\auto@bib@innerbib\@empty
%</preamble>
\bibitem [{\citenamefont {Rudner}\ and\ \citenamefont {Levitov}(2009)}]{Rudner-09}%
  \BibitemOpen
  \bibfield  {author} {\bibinfo {author} {\bibfnamefont {M.~S.}\ \bibnamefont {Rudner}}\ and\ \bibinfo {author} {\bibfnamefont {L.~S.}\ \bibnamefont {Levitov}},\ }\href {\doibase 10.1103/PhysRevLett.102.065703} {\bibfield  {journal} {\bibinfo  {journal} {Phys. Rev. Lett.}\ }\textbf {\bibinfo {volume} {102}},\ \bibinfo {pages} {065703} (\bibinfo {year} {2009})}\BibitemShut {NoStop}%
\bibitem [{\citenamefont {Sato}\ \emph {et~al.}(2012)\citenamefont {Sato}, \citenamefont {Hasebe}, \citenamefont {Esaki},\ and\ \citenamefont {Kohmoto}}]{Sato-11}%
  \BibitemOpen
  \bibfield  {author} {\bibinfo {author} {\bibfnamefont {M.}~\bibnamefont {Sato}}, \bibinfo {author} {\bibfnamefont {K.}~\bibnamefont {Hasebe}}, \bibinfo {author} {\bibfnamefont {K.}~\bibnamefont {Esaki}}, \ and\ \bibinfo {author} {\bibfnamefont {M.}~\bibnamefont {Kohmoto}},\ }\href {\doibase 10.1143/PTP.127.937} {\bibfield  {journal} {\bibinfo  {journal} {Prog. Theor. Phys.}\ }\textbf {\bibinfo {volume} {127}},\ \bibinfo {pages} {937} (\bibinfo {year} {2012})}\BibitemShut {NoStop}%
\bibitem [{\citenamefont {Esaki}\ \emph {et~al.}(2011)\citenamefont {Esaki}, \citenamefont {Sato}, \citenamefont {Hasebe},\ and\ \citenamefont {Kohmoto}}]{Esaki-11}%
  \BibitemOpen
  \bibfield  {author} {\bibinfo {author} {\bibfnamefont {K.}~\bibnamefont {Esaki}}, \bibinfo {author} {\bibfnamefont {M.}~\bibnamefont {Sato}}, \bibinfo {author} {\bibfnamefont {K.}~\bibnamefont {Hasebe}}, \ and\ \bibinfo {author} {\bibfnamefont {M.}~\bibnamefont {Kohmoto}},\ }\href {\doibase 10.1103/PhysRevB.84.205128} {\bibfield  {journal} {\bibinfo  {journal} {Phys. Rev. B}\ }\textbf {\bibinfo {volume} {84}},\ \bibinfo {pages} {205128} (\bibinfo {year} {2011})}\BibitemShut {NoStop}%
\bibitem [{\citenamefont {Hu}\ and\ \citenamefont {Hughes}(2011)}]{Hu-11}%
  \BibitemOpen
  \bibfield  {author} {\bibinfo {author} {\bibfnamefont {Y.~C.}\ \bibnamefont {Hu}}\ and\ \bibinfo {author} {\bibfnamefont {T.~L.}\ \bibnamefont {Hughes}},\ }\href {\doibase 10.1103/PhysRevB.84.153101} {\bibfield  {journal} {\bibinfo  {journal} {Phys. Rev. B}\ }\textbf {\bibinfo {volume} {84}},\ \bibinfo {pages} {153101} (\bibinfo {year} {2011})}\BibitemShut {NoStop}%
\bibitem [{\citenamefont {Schomerus}(2013)}]{Schomerus-13}%
  \BibitemOpen
  \bibfield  {author} {\bibinfo {author} {\bibfnamefont {H.}~\bibnamefont {Schomerus}},\ }\href {\doibase 10.1364/OL.38.001912} {\bibfield  {journal} {\bibinfo  {journal} {Opt. Lett.}\ }\textbf {\bibinfo {volume} {38}},\ \bibinfo {pages} {1912} (\bibinfo {year} {2013})}\BibitemShut {NoStop}%
\bibitem [{\citenamefont {Malzard}\ \emph {et~al.}(2015)\citenamefont {Malzard}, \citenamefont {Poli},\ and\ \citenamefont {Schomerus}}]{Malzard-15}%
  \BibitemOpen
  \bibfield  {author} {\bibinfo {author} {\bibfnamefont {S.}~\bibnamefont {Malzard}}, \bibinfo {author} {\bibfnamefont {C.}~\bibnamefont {Poli}}, \ and\ \bibinfo {author} {\bibfnamefont {H.}~\bibnamefont {Schomerus}},\ }\href {\doibase 10.1103/PhysRevLett.115.200402} {\bibfield  {journal} {\bibinfo  {journal} {Phys. Rev. Lett.}\ }\textbf {\bibinfo {volume} {115}},\ \bibinfo {pages} {200402} (\bibinfo {year} {2015})}\BibitemShut {NoStop}%
\bibitem [{\citenamefont {Lee}(2016)}]{Lee-16}%
  \BibitemOpen
  \bibfield  {author} {\bibinfo {author} {\bibfnamefont {T.~E.}\ \bibnamefont {Lee}},\ }\href {\doibase 10.1103/PhysRevLett.116.133903} {\bibfield  {journal} {\bibinfo  {journal} {Phys. Rev. Lett.}\ }\textbf {\bibinfo {volume} {116}},\ \bibinfo {pages} {133903} (\bibinfo {year} {2016})}\BibitemShut {NoStop}%
\bibitem [{\citenamefont {Leykam}\ \emph {et~al.}(2017)\citenamefont {Leykam}, \citenamefont {Bliokh}, \citenamefont {Huang}, \citenamefont {Chong},\ and\ \citenamefont {Nori}}]{Leykam-17}%
  \BibitemOpen
  \bibfield  {author} {\bibinfo {author} {\bibfnamefont {D.}~\bibnamefont {Leykam}}, \bibinfo {author} {\bibfnamefont {K.~Y.}\ \bibnamefont {Bliokh}}, \bibinfo {author} {\bibfnamefont {C.}~\bibnamefont {Huang}}, \bibinfo {author} {\bibfnamefont {Y.~D.}\ \bibnamefont {Chong}}, \ and\ \bibinfo {author} {\bibfnamefont {F.}~\bibnamefont {Nori}},\ }\href {\doibase 10.1103/PhysRevLett.118.040401} {\bibfield  {journal} {\bibinfo  {journal} {Phys. Rev. Lett.}\ }\textbf {\bibinfo {volume} {118}},\ \bibinfo {pages} {040401} (\bibinfo {year} {2017})}\BibitemShut {NoStop}%
\bibitem [{\citenamefont {Xu}\ \emph {et~al.}(2017)\citenamefont {Xu}, \citenamefont {Wang},\ and\ \citenamefont {Duan}}]{Xu-17}%
  \BibitemOpen
  \bibfield  {author} {\bibinfo {author} {\bibfnamefont {Y.}~\bibnamefont {Xu}}, \bibinfo {author} {\bibfnamefont {S.-T.}\ \bibnamefont {Wang}}, \ and\ \bibinfo {author} {\bibfnamefont {L.-M.}\ \bibnamefont {Duan}},\ }\href {\doibase 10.1103/PhysRevLett.118.045701} {\bibfield  {journal} {\bibinfo  {journal} {Phys. Rev. Lett.}\ }\textbf {\bibinfo {volume} {118}},\ \bibinfo {pages} {045701} (\bibinfo {year} {2017})}\BibitemShut {NoStop}%
\bibitem [{\citenamefont {Xiong}(2018)}]{Xiong-18}%
  \BibitemOpen
  \bibfield  {author} {\bibinfo {author} {\bibfnamefont {Y.}~\bibnamefont {Xiong}},\ }\href {\doibase 10.1088/2399-6528/aab64a} {\bibfield  {journal} {\bibinfo  {journal} {J. Phys. Commun.}\ }\textbf {\bibinfo {volume} {2}},\ \bibinfo {pages} {035043} (\bibinfo {year} {2018})}\BibitemShut {NoStop}%
\bibitem [{\citenamefont {Shen}\ \emph {et~al.}(2018)\citenamefont {Shen}, \citenamefont {Zhen},\ and\ \citenamefont {Fu}}]{Shen-18}%
  \BibitemOpen
  \bibfield  {author} {\bibinfo {author} {\bibfnamefont {H.}~\bibnamefont {Shen}}, \bibinfo {author} {\bibfnamefont {B.}~\bibnamefont {Zhen}}, \ and\ \bibinfo {author} {\bibfnamefont {L.}~\bibnamefont {Fu}},\ }\href {\doibase 10.1103/PhysRevLett.120.146402} {\bibfield  {journal} {\bibinfo  {journal} {Phys. Rev. Lett.}\ }\textbf {\bibinfo {volume} {120}},\ \bibinfo {pages} {146402} (\bibinfo {year} {2018})}\BibitemShut {NoStop}%
\bibitem [{\citenamefont {Kozii}\ and\ \citenamefont {Fu}()}]{Kozii-17}%
  \BibitemOpen
  \bibfield  {author} {\bibinfo {author} {\bibfnamefont {V.}~\bibnamefont {Kozii}}\ and\ \bibinfo {author} {\bibfnamefont {L.}~\bibnamefont {Fu}},\ }\href@noop {} {}\Eprint {http://arxiv.org/abs/1708.05841} {arXiv:1708.05841} \BibitemShut {NoStop}%
\bibitem [{\citenamefont {Takata}\ and\ \citenamefont {Notomi}(2018)}]{Takata-18}%
  \BibitemOpen
  \bibfield  {author} {\bibinfo {author} {\bibfnamefont {K.}~\bibnamefont {Takata}}\ and\ \bibinfo {author} {\bibfnamefont {M.}~\bibnamefont {Notomi}},\ }\href {\doibase 10.1103/PhysRevLett.121.213902} {\bibfield  {journal} {\bibinfo  {journal} {Phys. Rev. Lett.}\ }\textbf {\bibinfo {volume} {121}},\ \bibinfo {pages} {213902} (\bibinfo {year} {2018})}\BibitemShut {NoStop}%
\bibitem [{\citenamefont {Martinez~Alvarez}\ \emph {et~al.}(2018)\citenamefont {Martinez~Alvarez}, \citenamefont {Barrios~Vargas},\ and\ \citenamefont {Foa~Torres}}]{MartinezAlvarez-18}%
  \BibitemOpen
  \bibfield  {author} {\bibinfo {author} {\bibfnamefont {V.~M.}\ \bibnamefont {Martinez~Alvarez}}, \bibinfo {author} {\bibfnamefont {J.~E.}\ \bibnamefont {Barrios~Vargas}}, \ and\ \bibinfo {author} {\bibfnamefont {L.~E.~F.}\ \bibnamefont {Foa~Torres}},\ }\href {\doibase 10.1103/PhysRevB.97.121401} {\bibfield  {journal} {\bibinfo  {journal} {Phys. Rev. B}\ }\textbf {\bibinfo {volume} {97}},\ \bibinfo {pages} {121401} (\bibinfo {year} {2018})}\BibitemShut {NoStop}%
\bibitem [{\citenamefont {Kawabata}\ \emph {et~al.}(2018{\natexlab{a}})\citenamefont {Kawabata}, \citenamefont {Ashida}, \citenamefont {Katsura},\ and\ \citenamefont {Ueda}}]{KAKU-18}%
  \BibitemOpen
  \bibfield  {author} {\bibinfo {author} {\bibfnamefont {K.}~\bibnamefont {Kawabata}}, \bibinfo {author} {\bibfnamefont {Y.}~\bibnamefont {Ashida}}, \bibinfo {author} {\bibfnamefont {H.}~\bibnamefont {Katsura}}, \ and\ \bibinfo {author} {\bibfnamefont {M.}~\bibnamefont {Ueda}},\ }\href {\doibase 10.1103/PhysRevB.98.085116} {\bibfield  {journal} {\bibinfo  {journal} {Phys. Rev. B}\ }\textbf {\bibinfo {volume} {98}},\ \bibinfo {pages} {085116} (\bibinfo {year} {2018}{\natexlab{a}})}\BibitemShut {NoStop}%
\bibitem [{\citenamefont {Gong}\ \emph {et~al.}(2018)\citenamefont {Gong}, \citenamefont {Ashida}, \citenamefont {Kawabata}, \citenamefont {Takasan}, \citenamefont {Higashikawa},\ and\ \citenamefont {Ueda}}]{Gong-18}%
  \BibitemOpen
  \bibfield  {author} {\bibinfo {author} {\bibfnamefont {Z.}~\bibnamefont {Gong}}, \bibinfo {author} {\bibfnamefont {Y.}~\bibnamefont {Ashida}}, \bibinfo {author} {\bibfnamefont {K.}~\bibnamefont {Kawabata}}, \bibinfo {author} {\bibfnamefont {K.}~\bibnamefont {Takasan}}, \bibinfo {author} {\bibfnamefont {S.}~\bibnamefont {Higashikawa}}, \ and\ \bibinfo {author} {\bibfnamefont {M.}~\bibnamefont {Ueda}},\ }\href {\doibase 10.1103/PhysRevX.8.031079} {\bibfield  {journal} {\bibinfo  {journal} {Phys. Rev. X}\ }\textbf {\bibinfo {volume} {8}},\ \bibinfo {pages} {031079} (\bibinfo {year} {2018})}\BibitemShut {NoStop}%
\bibitem [{\citenamefont {Kawabata}\ \emph {et~al.}(2019{\natexlab{a}})\citenamefont {Kawabata}, \citenamefont {Higashikawa}, \citenamefont {Gong}, \citenamefont {Ashida},\ and\ \citenamefont {Ueda}}]{Kawabata-19}%
  \BibitemOpen
  \bibfield  {author} {\bibinfo {author} {\bibfnamefont {K.}~\bibnamefont {Kawabata}}, \bibinfo {author} {\bibfnamefont {S.}~\bibnamefont {Higashikawa}}, \bibinfo {author} {\bibfnamefont {Z.}~\bibnamefont {Gong}}, \bibinfo {author} {\bibfnamefont {Y.}~\bibnamefont {Ashida}}, \ and\ \bibinfo {author} {\bibfnamefont {M.}~\bibnamefont {Ueda}},\ }\href {\doibase 10.1038/s41467-018-08254-y} {\bibfield  {journal} {\bibinfo  {journal} {Nat. Commun.}\ }\textbf {\bibinfo {volume} {10}},\ \bibinfo {pages} {297} (\bibinfo {year} {2019}{\natexlab{a}})}\BibitemShut {NoStop}%
\bibitem [{\citenamefont {Yao}\ and\ \citenamefont {Wang}(2018)}]{YW-18-SSH}%
  \BibitemOpen
  \bibfield  {author} {\bibinfo {author} {\bibfnamefont {S.}~\bibnamefont {Yao}}\ and\ \bibinfo {author} {\bibfnamefont {Z.}~\bibnamefont {Wang}},\ }\href {\doibase 10.1103/PhysRevLett.121.086803} {\bibfield  {journal} {\bibinfo  {journal} {Phys. Rev. Lett.}\ }\textbf {\bibinfo {volume} {121}},\ \bibinfo {pages} {086803} (\bibinfo {year} {2018})}\BibitemShut {NoStop}%
\bibitem [{\citenamefont {Yao}\ \emph {et~al.}(2018)\citenamefont {Yao}, \citenamefont {Song},\ and\ \citenamefont {Wang}}]{YSW-18-Chern}%
  \BibitemOpen
  \bibfield  {author} {\bibinfo {author} {\bibfnamefont {S.}~\bibnamefont {Yao}}, \bibinfo {author} {\bibfnamefont {F.}~\bibnamefont {Song}}, \ and\ \bibinfo {author} {\bibfnamefont {Z.}~\bibnamefont {Wang}},\ }\href {\doibase 10.1103/PhysRevLett.121.136802} {\bibfield  {journal} {\bibinfo  {journal} {Phys. Rev. Lett.}\ }\textbf {\bibinfo {volume} {121}},\ \bibinfo {pages} {136802} (\bibinfo {year} {2018})}\BibitemShut {NoStop}%
\bibitem [{\citenamefont {Kunst}\ \emph {et~al.}(2018)\citenamefont {Kunst}, \citenamefont {Edvardsson}, \citenamefont {Budich},\ and\ \citenamefont {Bergholtz}}]{Kunst-18}%
  \BibitemOpen
  \bibfield  {author} {\bibinfo {author} {\bibfnamefont {F.~K.}\ \bibnamefont {Kunst}}, \bibinfo {author} {\bibfnamefont {E.}~\bibnamefont {Edvardsson}}, \bibinfo {author} {\bibfnamefont {J.~C.}\ \bibnamefont {Budich}}, \ and\ \bibinfo {author} {\bibfnamefont {E.~J.}\ \bibnamefont {Bergholtz}},\ }\href {\doibase 10.1103/PhysRevLett.121.026808} {\bibfield  {journal} {\bibinfo  {journal} {Phys. Rev. Lett.}\ }\textbf {\bibinfo {volume} {121}},\ \bibinfo {pages} {026808} (\bibinfo {year} {2018})}\BibitemShut {NoStop}%
\bibitem [{\citenamefont {Kawabata}\ \emph {et~al.}(2018{\natexlab{b}})\citenamefont {Kawabata}, \citenamefont {Shiozaki},\ and\ \citenamefont {Ueda}}]{KSU-18}%
  \BibitemOpen
  \bibfield  {author} {\bibinfo {author} {\bibfnamefont {K.}~\bibnamefont {Kawabata}}, \bibinfo {author} {\bibfnamefont {K.}~\bibnamefont {Shiozaki}}, \ and\ \bibinfo {author} {\bibfnamefont {M.}~\bibnamefont {Ueda}},\ }\href {\doibase 10.1103/PhysRevB.98.165148} {\bibfield  {journal} {\bibinfo  {journal} {Phys. Rev. B}\ }\textbf {\bibinfo {volume} {98}},\ \bibinfo {pages} {165148} (\bibinfo {year} {2018}{\natexlab{b}})}\BibitemShut {NoStop}%
\bibitem [{\citenamefont {McDonald}\ \emph {et~al.}(2018)\citenamefont {McDonald}, \citenamefont {Pereg-Barnea},\ and\ \citenamefont {Clerk}}]{McDonald-18}%
  \BibitemOpen
  \bibfield  {author} {\bibinfo {author} {\bibfnamefont {A.}~\bibnamefont {McDonald}}, \bibinfo {author} {\bibfnamefont {T.}~\bibnamefont {Pereg-Barnea}}, \ and\ \bibinfo {author} {\bibfnamefont {A.~A.}\ \bibnamefont {Clerk}},\ }\href {\doibase 10.1103/PhysRevX.8.041031} {\bibfield  {journal} {\bibinfo  {journal} {Phys. Rev. X}\ }\textbf {\bibinfo {volume} {8}},\ \bibinfo {pages} {041031} (\bibinfo {year} {2018})}\BibitemShut {NoStop}%
\bibitem [{\citenamefont {Carlstr\"om}\ and\ \citenamefont {Bergholtz}(2018)}]{Carlstrom-18}%
  \BibitemOpen
  \bibfield  {author} {\bibinfo {author} {\bibfnamefont {J.}~\bibnamefont {Carlstr\"om}}\ and\ \bibinfo {author} {\bibfnamefont {E.~J.}\ \bibnamefont {Bergholtz}},\ }\href {\doibase 10.1103/PhysRevA.98.042114} {\bibfield  {journal} {\bibinfo  {journal} {Phys. Rev. A}\ }\textbf {\bibinfo {volume} {98}},\ \bibinfo {pages} {042114} (\bibinfo {year} {2018})}\BibitemShut {NoStop}%
\bibitem [{\citenamefont {Carlstr\"om}\ \emph {et~al.}(2019)\citenamefont {Carlstr\"om}, \citenamefont {St\aa{}lhammar}, \citenamefont {Budich},\ and\ \citenamefont {Bergholtz}}]{Carlstrom-19}%
  \BibitemOpen
  \bibfield  {author} {\bibinfo {author} {\bibfnamefont {J.}~\bibnamefont {Carlstr\"om}}, \bibinfo {author} {\bibfnamefont {M.}~\bibnamefont {St\aa{}lhammar}}, \bibinfo {author} {\bibfnamefont {J.~C.}\ \bibnamefont {Budich}}, \ and\ \bibinfo {author} {\bibfnamefont {E.~J.}\ \bibnamefont {Bergholtz}},\ }\href {\doibase 10.1103/PhysRevB.99.161115} {\bibfield  {journal} {\bibinfo  {journal} {Phys. Rev. B}\ }\textbf {\bibinfo {volume} {99}},\ \bibinfo {pages} {161115} (\bibinfo {year} {2019})}\BibitemShut {NoStop}%
\bibitem [{\citenamefont {Lee}\ and\ \citenamefont {Thomale}(2019)}]{Lee-19}%
  \BibitemOpen
  \bibfield  {author} {\bibinfo {author} {\bibfnamefont {C.~H.}\ \bibnamefont {Lee}}\ and\ \bibinfo {author} {\bibfnamefont {R.}~\bibnamefont {Thomale}},\ }\href {\doibase 10.1103/PhysRevB.99.201103} {\bibfield  {journal} {\bibinfo  {journal} {Phys. Rev. B}\ }\textbf {\bibinfo {volume} {99}},\ \bibinfo {pages} {201103} (\bibinfo {year} {2019})}\BibitemShut {NoStop}%
\bibitem [{\citenamefont {Jin}\ and\ \citenamefont {Song}(2019)}]{Jin-19}%
  \BibitemOpen
  \bibfield  {author} {\bibinfo {author} {\bibfnamefont {L.}~\bibnamefont {Jin}}\ and\ \bibinfo {author} {\bibfnamefont {Z.}~\bibnamefont {Song}},\ }\href {\doibase 10.1103/PhysRevB.99.081103} {\bibfield  {journal} {\bibinfo  {journal} {Phys. Rev. B}\ }\textbf {\bibinfo {volume} {99}},\ \bibinfo {pages} {081103} (\bibinfo {year} {2019})}\BibitemShut {NoStop}%
\bibitem [{\citenamefont {Budich}\ \emph {et~al.}(2019)\citenamefont {Budich}, \citenamefont {Carlstr\"om}, \citenamefont {Kunst},\ and\ \citenamefont {Bergholtz}}]{Budich-19}%
  \BibitemOpen
  \bibfield  {author} {\bibinfo {author} {\bibfnamefont {J.~C.}\ \bibnamefont {Budich}}, \bibinfo {author} {\bibfnamefont {J.}~\bibnamefont {Carlstr\"om}}, \bibinfo {author} {\bibfnamefont {F.~K.}\ \bibnamefont {Kunst}}, \ and\ \bibinfo {author} {\bibfnamefont {E.~J.}\ \bibnamefont {Bergholtz}},\ }\href {\doibase 10.1103/PhysRevB.99.041406} {\bibfield  {journal} {\bibinfo  {journal} {Phys. Rev. B}\ }\textbf {\bibinfo {volume} {99}},\ \bibinfo {pages} {041406} (\bibinfo {year} {2019})}\BibitemShut {NoStop}%
\bibitem [{\citenamefont {Okugawa}\ and\ \citenamefont {Yokoyama}(2019)}]{Okugawa-19}%
  \BibitemOpen
  \bibfield  {author} {\bibinfo {author} {\bibfnamefont {R.}~\bibnamefont {Okugawa}}\ and\ \bibinfo {author} {\bibfnamefont {T.}~\bibnamefont {Yokoyama}},\ }\href {\doibase 10.1103/PhysRevB.99.041202} {\bibfield  {journal} {\bibinfo  {journal} {Phys. Rev. B}\ }\textbf {\bibinfo {volume} {99}},\ \bibinfo {pages} {041202} (\bibinfo {year} {2019})}\BibitemShut {NoStop}%
\bibitem [{\citenamefont {Liu}\ \emph {et~al.}(2019)\citenamefont {Liu}, \citenamefont {Zhang}, \citenamefont {Ai}, \citenamefont {Gong}, \citenamefont {Kawabata}, \citenamefont {Ueda},\ and\ \citenamefont {Nori}}]{Liu-19}%
  \BibitemOpen
  \bibfield  {author} {\bibinfo {author} {\bibfnamefont {T.}~\bibnamefont {Liu}}, \bibinfo {author} {\bibfnamefont {Y.-R.}\ \bibnamefont {Zhang}}, \bibinfo {author} {\bibfnamefont {Q.}~\bibnamefont {Ai}}, \bibinfo {author} {\bibfnamefont {Z.}~\bibnamefont {Gong}}, \bibinfo {author} {\bibfnamefont {K.}~\bibnamefont {Kawabata}}, \bibinfo {author} {\bibfnamefont {M.}~\bibnamefont {Ueda}}, \ and\ \bibinfo {author} {\bibfnamefont {F.}~\bibnamefont {Nori}},\ }\href {\doibase 10.1103/PhysRevLett.122.076801} {\bibfield  {journal} {\bibinfo  {journal} {Phys. Rev. Lett.}\ }\textbf {\bibinfo {volume} {122}},\ \bibinfo {pages} {076801} (\bibinfo {year} {2019})}\BibitemShut {NoStop}%
\bibitem [{\citenamefont {Yoshida}\ \emph {et~al.}(2019{\natexlab{a}})\citenamefont {Yoshida}, \citenamefont {Peters}, \citenamefont {Kawakami},\ and\ \citenamefont {Hatsugai}}]{Yoshida-19}%
  \BibitemOpen
  \bibfield  {author} {\bibinfo {author} {\bibfnamefont {T.}~\bibnamefont {Yoshida}}, \bibinfo {author} {\bibfnamefont {R.}~\bibnamefont {Peters}}, \bibinfo {author} {\bibfnamefont {N.}~\bibnamefont {Kawakami}}, \ and\ \bibinfo {author} {\bibfnamefont {Y.}~\bibnamefont {Hatsugai}},\ }\href {\doibase 10.1103/PhysRevB.99.121101} {\bibfield  {journal} {\bibinfo  {journal} {Phys. Rev. B}\ }\textbf {\bibinfo {volume} {99}},\ \bibinfo {pages} {121101} (\bibinfo {year} {2019}{\natexlab{a}})}\BibitemShut {NoStop}%
\bibitem [{\citenamefont {Zhou}\ \emph {et~al.}(2019)\citenamefont {Zhou}, \citenamefont {Lee}, \citenamefont {Liu},\ and\ \citenamefont {Zhen}}]{Zhou-19}%
  \BibitemOpen
  \bibfield  {author} {\bibinfo {author} {\bibfnamefont {H.}~\bibnamefont {Zhou}}, \bibinfo {author} {\bibfnamefont {J.~Y.}\ \bibnamefont {Lee}}, \bibinfo {author} {\bibfnamefont {S.}~\bibnamefont {Liu}}, \ and\ \bibinfo {author} {\bibfnamefont {B.}~\bibnamefont {Zhen}},\ }\href {\doibase 10.1364/OPTICA.6.000190} {\bibfield  {journal} {\bibinfo  {journal} {Optica}\ }\textbf {\bibinfo {volume} {6}},\ \bibinfo {pages} {190} (\bibinfo {year} {2019})}\BibitemShut {NoStop}%
\bibitem [{\citenamefont {Lee}\ \emph {et~al.}(2019{\natexlab{a}})\citenamefont {Lee}, \citenamefont {Li},\ and\ \citenamefont {Gong}}]{Lee-Li-Gong-19}%
  \BibitemOpen
  \bibfield  {author} {\bibinfo {author} {\bibfnamefont {C.~H.}\ \bibnamefont {Lee}}, \bibinfo {author} {\bibfnamefont {L.}~\bibnamefont {Li}}, \ and\ \bibinfo {author} {\bibfnamefont {J.}~\bibnamefont {Gong}},\ }\href {\doibase 10.1103/PhysRevLett.123.016805} {\bibfield  {journal} {\bibinfo  {journal} {Phys. Rev. Lett.}\ }\textbf {\bibinfo {volume} {123}},\ \bibinfo {pages} {016805} (\bibinfo {year} {2019}{\natexlab{a}})}\BibitemShut {NoStop}%
\bibitem [{\citenamefont {Kunst}\ and\ \citenamefont {Dwivedi}(2019)}]{Kunst-19}%
  \BibitemOpen
  \bibfield  {author} {\bibinfo {author} {\bibfnamefont {F.~K.}\ \bibnamefont {Kunst}}\ and\ \bibinfo {author} {\bibfnamefont {V.}~\bibnamefont {Dwivedi}},\ }\href {\doibase 10.1103/PhysRevB.99.245116} {\bibfield  {journal} {\bibinfo  {journal} {Phys. Rev. B}\ }\textbf {\bibinfo {volume} {99}},\ \bibinfo {pages} {245116} (\bibinfo {year} {2019})}\BibitemShut {NoStop}%
\bibitem [{\citenamefont {Longhi}(2019)}]{Longhi-19}%
  \BibitemOpen
  \bibfield  {author} {\bibinfo {author} {\bibfnamefont {S.}~\bibnamefont {Longhi}},\ }\href {\doibase 10.1103/PhysRevLett.122.237601} {\bibfield  {journal} {\bibinfo  {journal} {Phys. Rev. Lett.}\ }\textbf {\bibinfo {volume} {122}},\ \bibinfo {pages} {237601} (\bibinfo {year} {2019})}\BibitemShut {NoStop}%
\bibitem [{\citenamefont {Edvardsson}\ \emph {et~al.}(2019)\citenamefont {Edvardsson}, \citenamefont {Kunst},\ and\ \citenamefont {Bergholtz}}]{Edvardsson-19}%
  \BibitemOpen
  \bibfield  {author} {\bibinfo {author} {\bibfnamefont {E.}~\bibnamefont {Edvardsson}}, \bibinfo {author} {\bibfnamefont {F.~K.}\ \bibnamefont {Kunst}}, \ and\ \bibinfo {author} {\bibfnamefont {E.~J.}\ \bibnamefont {Bergholtz}},\ }\href {\doibase 10.1103/PhysRevB.99.081302} {\bibfield  {journal} {\bibinfo  {journal} {Phys. Rev. B}\ }\textbf {\bibinfo {volume} {99}},\ \bibinfo {pages} {081302} (\bibinfo {year} {2019})}\BibitemShut {NoStop}%
\bibitem [{\citenamefont {Kawabata}\ \emph {et~al.}(2019{\natexlab{b}})\citenamefont {Kawabata}, \citenamefont {Shiozaki}, \citenamefont {Ueda},\ and\ \citenamefont {Sato}}]{KSUS-19}%
  \BibitemOpen
  \bibfield  {author} {\bibinfo {author} {\bibfnamefont {K.}~\bibnamefont {Kawabata}}, \bibinfo {author} {\bibfnamefont {K.}~\bibnamefont {Shiozaki}}, \bibinfo {author} {\bibfnamefont {M.}~\bibnamefont {Ueda}}, \ and\ \bibinfo {author} {\bibfnamefont {M.}~\bibnamefont {Sato}},\ }\href {\doibase 10.1103/PhysRevX.9.041015} {\bibfield  {journal} {\bibinfo  {journal} {Phys. Rev. X}\ }\textbf {\bibinfo {volume} {9}},\ \bibinfo {pages} {041015} (\bibinfo {year} {2019}{\natexlab{b}})}\BibitemShut {NoStop}%
\bibitem [{\citenamefont {Zhou}\ and\ \citenamefont {Lee}(2019)}]{ZL-19}%
  \BibitemOpen
  \bibfield  {author} {\bibinfo {author} {\bibfnamefont {H.}~\bibnamefont {Zhou}}\ and\ \bibinfo {author} {\bibfnamefont {J.~Y.}\ \bibnamefont {Lee}},\ }\href {\doibase 10.1103/PhysRevB.99.235112} {\bibfield  {journal} {\bibinfo  {journal} {Phys. Rev. B}\ }\textbf {\bibinfo {volume} {99}},\ \bibinfo {pages} {235112} (\bibinfo {year} {2019})}\BibitemShut {NoStop}%
\bibitem [{\citenamefont {Herviou}\ \emph {et~al.}(2019{\natexlab{a}})\citenamefont {Herviou}, \citenamefont {Bardarson},\ and\ \citenamefont {Regnault}}]{Herviou-19}%
  \BibitemOpen
  \bibfield  {author} {\bibinfo {author} {\bibfnamefont {L.}~\bibnamefont {Herviou}}, \bibinfo {author} {\bibfnamefont {J.~H.}\ \bibnamefont {Bardarson}}, \ and\ \bibinfo {author} {\bibfnamefont {N.}~\bibnamefont {Regnault}},\ }\href {\doibase 10.1103/PhysRevA.99.052118} {\bibfield  {journal} {\bibinfo  {journal} {Phys. Rev. A}\ }\textbf {\bibinfo {volume} {99}},\ \bibinfo {pages} {052118} (\bibinfo {year} {2019}{\natexlab{a}})}\BibitemShut {NoStop}%
\bibitem [{\citenamefont {Zeng}\ \emph {et~al.}(2020)\citenamefont {Zeng}, \citenamefont {Yang},\ and\ \citenamefont {Xu}}]{Zeng-20}%
  \BibitemOpen
  \bibfield  {author} {\bibinfo {author} {\bibfnamefont {Q.-B.}\ \bibnamefont {Zeng}}, \bibinfo {author} {\bibfnamefont {Y.-B.}\ \bibnamefont {Yang}}, \ and\ \bibinfo {author} {\bibfnamefont {Y.}~\bibnamefont {Xu}},\ }\href {\doibase 10.1103/PhysRevB.101.020201} {\bibfield  {journal} {\bibinfo  {journal} {Phys. Rev. B}\ }\textbf {\bibinfo {volume} {101}},\ \bibinfo {pages} {020201} (\bibinfo {year} {2020})}\BibitemShut {NoStop}%
\bibitem [{\citenamefont {Hirsbrunner}\ \emph {et~al.}(2019)\citenamefont {Hirsbrunner}, \citenamefont {Philip},\ and\ \citenamefont {Gilbert}}]{Hirsbrunner-19}%
  \BibitemOpen
  \bibfield  {author} {\bibinfo {author} {\bibfnamefont {M.~R.}\ \bibnamefont {Hirsbrunner}}, \bibinfo {author} {\bibfnamefont {T.~M.}\ \bibnamefont {Philip}}, \ and\ \bibinfo {author} {\bibfnamefont {M.~J.}\ \bibnamefont {Gilbert}},\ }\href {\doibase 10.1103/PhysRevB.100.081104} {\bibfield  {journal} {\bibinfo  {journal} {Phys. Rev. B}\ }\textbf {\bibinfo {volume} {100}},\ \bibinfo {pages} {081104} (\bibinfo {year} {2019})}\BibitemShut {NoStop}%
\bibitem [{\citenamefont {Zirnstein}\ \emph {et~al.}(2021)\citenamefont {Zirnstein}, \citenamefont {Refael},\ and\ \citenamefont {Rosenow}}]{Zirnstein-19}%
  \BibitemOpen
  \bibfield  {author} {\bibinfo {author} {\bibfnamefont {H.-G.}\ \bibnamefont {Zirnstein}}, \bibinfo {author} {\bibfnamefont {G.}~\bibnamefont {Refael}}, \ and\ \bibinfo {author} {\bibfnamefont {B.}~\bibnamefont {Rosenow}},\ }\href {\doibase 10.1103/PhysRevLett.126.216407} {\bibfield  {journal} {\bibinfo  {journal} {Phys. Rev. Lett.}\ }\textbf {\bibinfo {volume} {126}},\ \bibinfo {pages} {216407} (\bibinfo {year} {2021})}\BibitemShut {NoStop}%
\bibitem [{\citenamefont {Pocock}\ \emph {et~al.}(2019)\citenamefont {Pocock}, \citenamefont {Huidobro},\ and\ \citenamefont {Giannini}}]{Pocock-19}%
  \BibitemOpen
  \bibfield  {author} {\bibinfo {author} {\bibfnamefont {S.~R.}\ \bibnamefont {Pocock}}, \bibinfo {author} {\bibfnamefont {P.~A.}\ \bibnamefont {Huidobro}}, \ and\ \bibinfo {author} {\bibfnamefont {V.}~\bibnamefont {Giannini}},\ }\href {\doibase doi:10.1515/nanoph-2019-0033} {\bibfield  {journal} {\bibinfo  {journal} {Nanophotonics}\ }\textbf {\bibinfo {volume} {8}},\ \bibinfo {pages} {1337} (\bibinfo {year} {2019})}\BibitemShut {NoStop}%
\bibitem [{\citenamefont {Kimura}\ \emph {et~al.}(2019)\citenamefont {Kimura}, \citenamefont {Yoshida},\ and\ \citenamefont {Kawakami}}]{Kimura-19}%
  \BibitemOpen
  \bibfield  {author} {\bibinfo {author} {\bibfnamefont {K.}~\bibnamefont {Kimura}}, \bibinfo {author} {\bibfnamefont {T.}~\bibnamefont {Yoshida}}, \ and\ \bibinfo {author} {\bibfnamefont {N.}~\bibnamefont {Kawakami}},\ }\href {\doibase 10.1103/PhysRevB.100.115124} {\bibfield  {journal} {\bibinfo  {journal} {Phys. Rev. B}\ }\textbf {\bibinfo {volume} {100}},\ \bibinfo {pages} {115124} (\bibinfo {year} {2019})}\BibitemShut {NoStop}%
\bibitem [{\citenamefont {Borgnia}\ \emph {et~al.}(2020)\citenamefont {Borgnia}, \citenamefont {Kruchkov},\ and\ \citenamefont {Slager}}]{Borgnia-19}%
  \BibitemOpen
  \bibfield  {author} {\bibinfo {author} {\bibfnamefont {D.~S.}\ \bibnamefont {Borgnia}}, \bibinfo {author} {\bibfnamefont {A.~J.}\ \bibnamefont {Kruchkov}}, \ and\ \bibinfo {author} {\bibfnamefont {R.-J.}\ \bibnamefont {Slager}},\ }\href {\doibase 10.1103/PhysRevLett.124.056802} {\bibfield  {journal} {\bibinfo  {journal} {Phys. Rev. Lett.}\ }\textbf {\bibinfo {volume} {124}},\ \bibinfo {pages} {056802} (\bibinfo {year} {2020})}\BibitemShut {NoStop}%
\bibitem [{\citenamefont {Kawabata}\ \emph {et~al.}(2019{\natexlab{c}})\citenamefont {Kawabata}, \citenamefont {Bessho},\ and\ \citenamefont {Sato}}]{KBS-19}%
  \BibitemOpen
  \bibfield  {author} {\bibinfo {author} {\bibfnamefont {K.}~\bibnamefont {Kawabata}}, \bibinfo {author} {\bibfnamefont {T.}~\bibnamefont {Bessho}}, \ and\ \bibinfo {author} {\bibfnamefont {M.}~\bibnamefont {Sato}},\ }\href {\doibase 10.1103/PhysRevLett.123.066405} {\bibfield  {journal} {\bibinfo  {journal} {Phys. Rev. Lett.}\ }\textbf {\bibinfo {volume} {123}},\ \bibinfo {pages} {066405} (\bibinfo {year} {2019}{\natexlab{c}})}\BibitemShut {NoStop}%
\bibitem [{\citenamefont {Yokomizo}\ and\ \citenamefont {Murakami}(2019)}]{YM-19}%
  \BibitemOpen
  \bibfield  {author} {\bibinfo {author} {\bibfnamefont {K.}~\bibnamefont {Yokomizo}}\ and\ \bibinfo {author} {\bibfnamefont {S.}~\bibnamefont {Murakami}},\ }\href {\doibase 10.1103/PhysRevLett.123.066404} {\bibfield  {journal} {\bibinfo  {journal} {Phys. Rev. Lett.}\ }\textbf {\bibinfo {volume} {123}},\ \bibinfo {pages} {066404} (\bibinfo {year} {2019})}\BibitemShut {NoStop}%
\bibitem [{\citenamefont {Song}\ \emph {et~al.}(2019{\natexlab{a}})\citenamefont {Song}, \citenamefont {Yao},\ and\ \citenamefont {Wang}}]{Song-real-space-19}%
  \BibitemOpen
  \bibfield  {author} {\bibinfo {author} {\bibfnamefont {F.}~\bibnamefont {Song}}, \bibinfo {author} {\bibfnamefont {S.}~\bibnamefont {Yao}}, \ and\ \bibinfo {author} {\bibfnamefont {Z.}~\bibnamefont {Wang}},\ }\href {\doibase 10.1103/PhysRevLett.123.246801} {\bibfield  {journal} {\bibinfo  {journal} {Phys. Rev. Lett.}\ }\textbf {\bibinfo {volume} {123}},\ \bibinfo {pages} {246801} (\bibinfo {year} {2019}{\natexlab{a}})}\BibitemShut {NoStop}%
\bibitem [{\citenamefont {McClarty}\ and\ \citenamefont {Rau}(2019)}]{McClarty-19}%
  \BibitemOpen
  \bibfield  {author} {\bibinfo {author} {\bibfnamefont {P.~A.}\ \bibnamefont {McClarty}}\ and\ \bibinfo {author} {\bibfnamefont {J.~G.}\ \bibnamefont {Rau}},\ }\href {\doibase 10.1103/PhysRevB.100.100405} {\bibfield  {journal} {\bibinfo  {journal} {Phys. Rev. B}\ }\textbf {\bibinfo {volume} {100}},\ \bibinfo {pages} {100405} (\bibinfo {year} {2019})}\BibitemShut {NoStop}%
\bibitem [{\citenamefont {Okuma}\ and\ \citenamefont {Sato}(2019)}]{Okuma-19}%
  \BibitemOpen
  \bibfield  {author} {\bibinfo {author} {\bibfnamefont {N.}~\bibnamefont {Okuma}}\ and\ \bibinfo {author} {\bibfnamefont {M.}~\bibnamefont {Sato}},\ }\href {\doibase 10.1103/PhysRevLett.123.097701} {\bibfield  {journal} {\bibinfo  {journal} {Phys. Rev. Lett.}\ }\textbf {\bibinfo {volume} {123}},\ \bibinfo {pages} {097701} (\bibinfo {year} {2019})}\BibitemShut {NoStop}%
\bibitem [{\citenamefont {Song}\ \emph {et~al.}(2019{\natexlab{b}})\citenamefont {Song}, \citenamefont {Yao},\ and\ \citenamefont {Wang}}]{Song-19-Lindblad}%
  \BibitemOpen
  \bibfield  {author} {\bibinfo {author} {\bibfnamefont {F.}~\bibnamefont {Song}}, \bibinfo {author} {\bibfnamefont {S.}~\bibnamefont {Yao}}, \ and\ \bibinfo {author} {\bibfnamefont {Z.}~\bibnamefont {Wang}},\ }\href {\doibase 10.1103/PhysRevLett.123.170401} {\bibfield  {journal} {\bibinfo  {journal} {Phys. Rev. Lett.}\ }\textbf {\bibinfo {volume} {123}},\ \bibinfo {pages} {170401} (\bibinfo {year} {2019}{\natexlab{b}})}\BibitemShut {NoStop}%
\bibitem [{\citenamefont {Bergholtz}\ and\ \citenamefont {Budich}(2019)}]{Bergholtz-19}%
  \BibitemOpen
  \bibfield  {author} {\bibinfo {author} {\bibfnamefont {E.~J.}\ \bibnamefont {Bergholtz}}\ and\ \bibinfo {author} {\bibfnamefont {J.~C.}\ \bibnamefont {Budich}},\ }\href {\doibase 10.1103/PhysRevResearch.1.012003} {\bibfield  {journal} {\bibinfo  {journal} {Phys. Rev. Res.}\ }\textbf {\bibinfo {volume} {1}},\ \bibinfo {pages} {012003} (\bibinfo {year} {2019})}\BibitemShut {NoStop}%
\bibitem [{\citenamefont {Lee}\ \emph {et~al.}(2019{\natexlab{b}})\citenamefont {Lee}, \citenamefont {Ahn}, \citenamefont {Zhou},\ and\ \citenamefont {Vishwanath}}]{Lee-Vishwanath-19}%
  \BibitemOpen
  \bibfield  {author} {\bibinfo {author} {\bibfnamefont {J.~Y.}\ \bibnamefont {Lee}}, \bibinfo {author} {\bibfnamefont {J.}~\bibnamefont {Ahn}}, \bibinfo {author} {\bibfnamefont {H.}~\bibnamefont {Zhou}}, \ and\ \bibinfo {author} {\bibfnamefont {A.}~\bibnamefont {Vishwanath}},\ }\href {\doibase 10.1103/PhysRevLett.123.206404} {\bibfield  {journal} {\bibinfo  {journal} {Phys. Rev. Lett.}\ }\textbf {\bibinfo {volume} {123}},\ \bibinfo {pages} {206404} (\bibinfo {year} {2019}{\natexlab{b}})}\BibitemShut {NoStop}%
\bibitem [{\citenamefont {Guo}\ \emph {et~al.}(2020)\citenamefont {Guo}, \citenamefont {Wang}, \citenamefont {Wang},\ and\ \citenamefont {Kou}}]{Guo-19}%
  \BibitemOpen
  \bibfield  {author} {\bibinfo {author} {\bibfnamefont {C.-X.}\ \bibnamefont {Guo}}, \bibinfo {author} {\bibfnamefont {X.-R.}\ \bibnamefont {Wang}}, \bibinfo {author} {\bibfnamefont {C.}~\bibnamefont {Wang}}, \ and\ \bibinfo {author} {\bibfnamefont {S.-P.}\ \bibnamefont {Kou}},\ }\href {\doibase 10.1103/PhysRevB.101.144439} {\bibfield  {journal} {\bibinfo  {journal} {Phys. Rev. B}\ }\textbf {\bibinfo {volume} {101}},\ \bibinfo {pages} {144439} (\bibinfo {year} {2020})}\BibitemShut {NoStop}%
\bibitem [{\citenamefont {Yoshida}\ \emph {et~al.}(2019{\natexlab{b}})\citenamefont {Yoshida}, \citenamefont {Kudo},\ and\ \citenamefont {Hatsugai}}]{Yoshida-19-FQH}%
  \BibitemOpen
  \bibfield  {author} {\bibinfo {author} {\bibfnamefont {T.}~\bibnamefont {Yoshida}}, \bibinfo {author} {\bibfnamefont {K.}~\bibnamefont {Kudo}}, \ and\ \bibinfo {author} {\bibfnamefont {Y.}~\bibnamefont {Hatsugai}},\ }\href {\doibase 10.1038/s41598-019-53253-8} {\bibfield  {journal} {\bibinfo  {journal} {Sci. Rep.}\ }\textbf {\bibinfo {volume} {9}},\ \bibinfo {pages} {16895} (\bibinfo {year} {2019}{\natexlab{b}})}\BibitemShut {NoStop}%
\bibitem [{\citenamefont {Rui}\ \emph {et~al.}(2019)\citenamefont {Rui}, \citenamefont {Hirschmann},\ and\ \citenamefont {Schnyder}}]{Rui-19}%
  \BibitemOpen
  \bibfield  {author} {\bibinfo {author} {\bibfnamefont {W.~B.}\ \bibnamefont {Rui}}, \bibinfo {author} {\bibfnamefont {M.~M.}\ \bibnamefont {Hirschmann}}, \ and\ \bibinfo {author} {\bibfnamefont {A.~P.}\ \bibnamefont {Schnyder}},\ }\href {\doibase 10.1103/PhysRevB.100.245116} {\bibfield  {journal} {\bibinfo  {journal} {Phys. Rev. B}\ }\textbf {\bibinfo {volume} {100}},\ \bibinfo {pages} {245116} (\bibinfo {year} {2019})}\BibitemShut {NoStop}%
\bibitem [{\citenamefont {Brzezicki}\ and\ \citenamefont {Hyart}(2019)}]{Brzezicki-19}%
  \BibitemOpen
  \bibfield  {author} {\bibinfo {author} {\bibfnamefont {W.}~\bibnamefont {Brzezicki}}\ and\ \bibinfo {author} {\bibfnamefont {T.}~\bibnamefont {Hyart}},\ }\href {\doibase 10.1103/PhysRevB.100.161105} {\bibfield  {journal} {\bibinfo  {journal} {Phys. Rev. B}\ }\textbf {\bibinfo {volume} {100}},\ \bibinfo {pages} {161105} (\bibinfo {year} {2019})}\BibitemShut {NoStop}%
\bibitem [{\citenamefont {Schomerus}(2020)}]{Schomerus-20}%
  \BibitemOpen
  \bibfield  {author} {\bibinfo {author} {\bibfnamefont {H.}~\bibnamefont {Schomerus}},\ }\href {\doibase 10.1103/PhysRevResearch.2.013058} {\bibfield  {journal} {\bibinfo  {journal} {Phys. Rev. Res.}\ }\textbf {\bibinfo {volume} {2}},\ \bibinfo {pages} {013058} (\bibinfo {year} {2020})}\BibitemShut {NoStop}%
\bibitem [{\citenamefont {Imura}\ and\ \citenamefont {Takane}(2019)}]{Imura-19}%
  \BibitemOpen
  \bibfield  {author} {\bibinfo {author} {\bibfnamefont {K.-I.}\ \bibnamefont {Imura}}\ and\ \bibinfo {author} {\bibfnamefont {Y.}~\bibnamefont {Takane}},\ }\href {\doibase 10.1103/PhysRevB.100.165430} {\bibfield  {journal} {\bibinfo  {journal} {Phys. Rev. B}\ }\textbf {\bibinfo {volume} {100}},\ \bibinfo {pages} {165430} (\bibinfo {year} {2019})}\BibitemShut {NoStop}%
\bibitem [{\citenamefont {Herviou}\ \emph {et~al.}(2019{\natexlab{b}})\citenamefont {Herviou}, \citenamefont {Regnault},\ and\ \citenamefont {Bardarson}}]{Herviou-19-ES}%
  \BibitemOpen
  \bibfield  {author} {\bibinfo {author} {\bibfnamefont {L.}~\bibnamefont {Herviou}}, \bibinfo {author} {\bibfnamefont {N.}~\bibnamefont {Regnault}}, \ and\ \bibinfo {author} {\bibfnamefont {J.~H.}\ \bibnamefont {Bardarson}},\ }\href {\doibase 10.21468/SciPostPhys.7.5.069} {\bibfield  {journal} {\bibinfo  {journal} {SciPost Phys.}\ }\textbf {\bibinfo {volume} {7}},\ \bibinfo {pages} {069} (\bibinfo {year} {2019}{\natexlab{b}})}\BibitemShut {NoStop}%
\bibitem [{\citenamefont {Chang}\ \emph {et~al.}(2020)\citenamefont {Chang}, \citenamefont {You}, \citenamefont {Wen},\ and\ \citenamefont {Ryu}}]{Chang-19}%
  \BibitemOpen
  \bibfield  {author} {\bibinfo {author} {\bibfnamefont {P.-Y.}\ \bibnamefont {Chang}}, \bibinfo {author} {\bibfnamefont {J.-S.}\ \bibnamefont {You}}, \bibinfo {author} {\bibfnamefont {X.}~\bibnamefont {Wen}}, \ and\ \bibinfo {author} {\bibfnamefont {S.}~\bibnamefont {Ryu}},\ }\href {\doibase 10.1103/PhysRevResearch.2.033069} {\bibfield  {journal} {\bibinfo  {journal} {Phys. Rev. Res.}\ }\textbf {\bibinfo {volume} {2}},\ \bibinfo {pages} {033069} (\bibinfo {year} {2020})}\BibitemShut {NoStop}%
\bibitem [{\citenamefont {Zhang}\ and\ \citenamefont {Franz}(2020)}]{Zhang-20}%
  \BibitemOpen
  \bibfield  {author} {\bibinfo {author} {\bibfnamefont {X.-X.}\ \bibnamefont {Zhang}}\ and\ \bibinfo {author} {\bibfnamefont {M.}~\bibnamefont {Franz}},\ }\href {\doibase 10.1103/PhysRevLett.124.046401} {\bibfield  {journal} {\bibinfo  {journal} {Phys. Rev. Lett.}\ }\textbf {\bibinfo {volume} {124}},\ \bibinfo {pages} {046401} (\bibinfo {year} {2020})}\BibitemShut {NoStop}%
\bibitem [{\citenamefont {Zhang}\ \emph {et~al.}(2020)\citenamefont {Zhang}, \citenamefont {Yang},\ and\ \citenamefont {Fang}}]{Zhang-19}%
  \BibitemOpen
  \bibfield  {author} {\bibinfo {author} {\bibfnamefont {K.}~\bibnamefont {Zhang}}, \bibinfo {author} {\bibfnamefont {Z.}~\bibnamefont {Yang}}, \ and\ \bibinfo {author} {\bibfnamefont {C.}~\bibnamefont {Fang}},\ }\href {\doibase 10.1103/PhysRevLett.125.126402} {\bibfield  {journal} {\bibinfo  {journal} {Phys. Rev. Lett.}\ }\textbf {\bibinfo {volume} {125}},\ \bibinfo {pages} {126402} (\bibinfo {year} {2020})}\BibitemShut {NoStop}%
\bibitem [{\citenamefont {Yang}\ \emph {et~al.}(2020)\citenamefont {Yang}, \citenamefont {Zhang}, \citenamefont {Fang},\ and\ \citenamefont {Hu}}]{Yang-19}%
  \BibitemOpen
  \bibfield  {author} {\bibinfo {author} {\bibfnamefont {Z.}~\bibnamefont {Yang}}, \bibinfo {author} {\bibfnamefont {K.}~\bibnamefont {Zhang}}, \bibinfo {author} {\bibfnamefont {C.}~\bibnamefont {Fang}}, \ and\ \bibinfo {author} {\bibfnamefont {J.}~\bibnamefont {Hu}},\ }\href {\doibase 10.1103/PhysRevLett.125.226402} {\bibfield  {journal} {\bibinfo  {journal} {Phys. Rev. Lett.}\ }\textbf {\bibinfo {volume} {125}},\ \bibinfo {pages} {226402} (\bibinfo {year} {2020})}\BibitemShut {NoStop}%
\bibitem [{\citenamefont {Okuma}\ \emph {et~al.}(2020)\citenamefont {Okuma}, \citenamefont {Kawabata}, \citenamefont {Shiozaki},\ and\ \citenamefont {Sato}}]{OKSS-20}%
  \BibitemOpen
  \bibfield  {author} {\bibinfo {author} {\bibfnamefont {N.}~\bibnamefont {Okuma}}, \bibinfo {author} {\bibfnamefont {K.}~\bibnamefont {Kawabata}}, \bibinfo {author} {\bibfnamefont {K.}~\bibnamefont {Shiozaki}}, \ and\ \bibinfo {author} {\bibfnamefont {M.}~\bibnamefont {Sato}},\ }\href {\doibase 10.1103/PhysRevLett.124.086801} {\bibfield  {journal} {\bibinfo  {journal} {Phys. Rev. Lett.}\ }\textbf {\bibinfo {volume} {124}},\ \bibinfo {pages} {086801} (\bibinfo {year} {2020})}\BibitemShut {NoStop}%
\bibitem [{\citenamefont {Longhi}(2020)}]{Longhi-20}%
  \BibitemOpen
  \bibfield  {author} {\bibinfo {author} {\bibfnamefont {S.}~\bibnamefont {Longhi}},\ }\href {\doibase 10.1103/PhysRevLett.124.066602} {\bibfield  {journal} {\bibinfo  {journal} {Phys. Rev. Lett.}\ }\textbf {\bibinfo {volume} {124}},\ \bibinfo {pages} {066602} (\bibinfo {year} {2020})}\BibitemShut {NoStop}%
\bibitem [{\citenamefont {Wang}\ \emph {et~al.}(2020)\citenamefont {Wang}, \citenamefont {Guo},\ and\ \citenamefont {Kou}}]{Wang-19}%
  \BibitemOpen
  \bibfield  {author} {\bibinfo {author} {\bibfnamefont {X.-R.}\ \bibnamefont {Wang}}, \bibinfo {author} {\bibfnamefont {C.-X.}\ \bibnamefont {Guo}}, \ and\ \bibinfo {author} {\bibfnamefont {S.-P.}\ \bibnamefont {Kou}},\ }\href {\doibase 10.1103/PhysRevB.101.121116} {\bibfield  {journal} {\bibinfo  {journal} {Phys. Rev. B}\ }\textbf {\bibinfo {volume} {101}},\ \bibinfo {pages} {121116} (\bibinfo {year} {2020})}\BibitemShut {NoStop}%
\bibitem [{\citenamefont {Matsumoto}\ \emph {et~al.}(2020)\citenamefont {Matsumoto}, \citenamefont {Kawabata}, \citenamefont {Ashida}, \citenamefont {Furukawa},\ and\ \citenamefont {Ueda}}]{Matsumoto-19}%
  \BibitemOpen
  \bibfield  {author} {\bibinfo {author} {\bibfnamefont {N.}~\bibnamefont {Matsumoto}}, \bibinfo {author} {\bibfnamefont {K.}~\bibnamefont {Kawabata}}, \bibinfo {author} {\bibfnamefont {Y.}~\bibnamefont {Ashida}}, \bibinfo {author} {\bibfnamefont {S.}~\bibnamefont {Furukawa}}, \ and\ \bibinfo {author} {\bibfnamefont {M.}~\bibnamefont {Ueda}},\ }\href {\doibase 10.1103/PhysRevLett.125.260601} {\bibfield  {journal} {\bibinfo  {journal} {Phys. Rev. Lett.}\ }\textbf {\bibinfo {volume} {125}},\ \bibinfo {pages} {260601} (\bibinfo {year} {2020})}\BibitemShut {NoStop}%
\bibitem [{\citenamefont {Yoshida}\ \emph {et~al.}(2020{\natexlab{a}})\citenamefont {Yoshida}, \citenamefont {Mizoguchi},\ and\ \citenamefont {Hatsugai}}]{Yoshida-19-mirror}%
  \BibitemOpen
  \bibfield  {author} {\bibinfo {author} {\bibfnamefont {T.}~\bibnamefont {Yoshida}}, \bibinfo {author} {\bibfnamefont {T.}~\bibnamefont {Mizoguchi}}, \ and\ \bibinfo {author} {\bibfnamefont {Y.}~\bibnamefont {Hatsugai}},\ }\href {\doibase 10.1103/PhysRevResearch.2.022062} {\bibfield  {journal} {\bibinfo  {journal} {Phys. Rev. Res.}\ }\textbf {\bibinfo {volume} {2}},\ \bibinfo {pages} {022062} (\bibinfo {year} {2020}{\natexlab{a}})}\BibitemShut {NoStop}%
\bibitem [{\citenamefont {Yokomizo}\ and\ \citenamefont {Murakami}(2020)}]{Yokomizo-20}%
  \BibitemOpen
  \bibfield  {author} {\bibinfo {author} {\bibfnamefont {K.}~\bibnamefont {Yokomizo}}\ and\ \bibinfo {author} {\bibfnamefont {S.}~\bibnamefont {Murakami}},\ }\href {\doibase 10.1103/PhysRevResearch.2.043045} {\bibfield  {journal} {\bibinfo  {journal} {Phys. Rev. Res.}\ }\textbf {\bibinfo {volume} {2}},\ \bibinfo {pages} {043045} (\bibinfo {year} {2020})}\BibitemShut {NoStop}%
\bibitem [{\citenamefont {Li}\ \emph {et~al.}(2020)\citenamefont {Li}, \citenamefont {Lee}, \citenamefont {Mu},\ and\ \citenamefont {Gong}}]{Li-20}%
  \BibitemOpen
  \bibfield  {author} {\bibinfo {author} {\bibfnamefont {L.}~\bibnamefont {Li}}, \bibinfo {author} {\bibfnamefont {C.~H.}\ \bibnamefont {Lee}}, \bibinfo {author} {\bibfnamefont {S.}~\bibnamefont {Mu}}, \ and\ \bibinfo {author} {\bibfnamefont {J.}~\bibnamefont {Gong}},\ }\href {\doibase 10.1038/s41467-020-18917-4} {\bibfield  {journal} {\bibinfo  {journal} {Nat. Commun.}\ }\textbf {\bibinfo {volume} {11}},\ \bibinfo {pages} {5491} (\bibinfo {year} {2020})}\BibitemShut {NoStop}%
\bibitem [{\citenamefont {Kawabata}\ \emph {et~al.}(2020{\natexlab{a}})\citenamefont {Kawabata}, \citenamefont {Okuma},\ and\ \citenamefont {Sato}}]{KOS-20}%
  \BibitemOpen
  \bibfield  {author} {\bibinfo {author} {\bibfnamefont {K.}~\bibnamefont {Kawabata}}, \bibinfo {author} {\bibfnamefont {N.}~\bibnamefont {Okuma}}, \ and\ \bibinfo {author} {\bibfnamefont {M.}~\bibnamefont {Sato}},\ }\href {\doibase 10.1103/PhysRevB.101.195147} {\bibfield  {journal} {\bibinfo  {journal} {Phys. Rev. B}\ }\textbf {\bibinfo {volume} {101}},\ \bibinfo {pages} {195147} (\bibinfo {year} {2020}{\natexlab{a}})}\BibitemShut {NoStop}%
\bibitem [{\citenamefont {Terrier}\ and\ \citenamefont {Kunst}(2020)}]{Terrier-20}%
  \BibitemOpen
  \bibfield  {author} {\bibinfo {author} {\bibfnamefont {F.}~\bibnamefont {Terrier}}\ and\ \bibinfo {author} {\bibfnamefont {F.~K.}\ \bibnamefont {Kunst}},\ }\href {\doibase 10.1103/PhysRevResearch.2.023364} {\bibfield  {journal} {\bibinfo  {journal} {Phys. Rev. Res.}\ }\textbf {\bibinfo {volume} {2}},\ \bibinfo {pages} {023364} (\bibinfo {year} {2020})}\BibitemShut {NoStop}%
\bibitem [{\citenamefont {Bessho}\ and\ \citenamefont {Sato}(2021)}]{Bessho-Sato-20}%
  \BibitemOpen
  \bibfield  {author} {\bibinfo {author} {\bibfnamefont {T.}~\bibnamefont {Bessho}}\ and\ \bibinfo {author} {\bibfnamefont {M.}~\bibnamefont {Sato}},\ }\href {\doibase 10.1103/PhysRevLett.127.196404} {\bibfield  {journal} {\bibinfo  {journal} {Phys. Rev. Lett.}\ }\textbf {\bibinfo {volume} {127}},\ \bibinfo {pages} {196404} (\bibinfo {year} {2021})}\BibitemShut {NoStop}%
\bibitem [{\citenamefont {Claes}\ and\ \citenamefont {Hughes}(2021)}]{Claes-20}%
  \BibitemOpen
  \bibfield  {author} {\bibinfo {author} {\bibfnamefont {J.}~\bibnamefont {Claes}}\ and\ \bibinfo {author} {\bibfnamefont {T.~L.}\ \bibnamefont {Hughes}},\ }\href {\doibase 10.1103/PhysRevB.103.L140201} {\bibfield  {journal} {\bibinfo  {journal} {Phys. Rev. B}\ }\textbf {\bibinfo {volume} {103}},\ \bibinfo {pages} {L140201} (\bibinfo {year} {2021})}\BibitemShut {NoStop}%
\bibitem [{\citenamefont {Zirnstein}\ and\ \citenamefont {Rosenow}(2021)}]{Zirnstein-20}%
  \BibitemOpen
  \bibfield  {author} {\bibinfo {author} {\bibfnamefont {H.-G.}\ \bibnamefont {Zirnstein}}\ and\ \bibinfo {author} {\bibfnamefont {B.}~\bibnamefont {Rosenow}},\ }\href {\doibase 10.1103/PhysRevB.103.195157} {\bibfield  {journal} {\bibinfo  {journal} {Phys. Rev. B}\ }\textbf {\bibinfo {volume} {103}},\ \bibinfo {pages} {195157} (\bibinfo {year} {2021})}\BibitemShut {NoStop}%
\bibitem [{\citenamefont {Denner}\ \emph {et~al.}(2021)\citenamefont {Denner}, \citenamefont {Skurativska}, \citenamefont {Schindler}, \citenamefont {Fischer}, \citenamefont {Thomale}, \citenamefont {Bzdu{\v{s}}ek},\ and\ \citenamefont {Neupert}}]{Denner-20}%
  \BibitemOpen
  \bibfield  {author} {\bibinfo {author} {\bibfnamefont {M.~M.}\ \bibnamefont {Denner}}, \bibinfo {author} {\bibfnamefont {A.}~\bibnamefont {Skurativska}}, \bibinfo {author} {\bibfnamefont {F.}~\bibnamefont {Schindler}}, \bibinfo {author} {\bibfnamefont {M.~H.}\ \bibnamefont {Fischer}}, \bibinfo {author} {\bibfnamefont {R.}~\bibnamefont {Thomale}}, \bibinfo {author} {\bibfnamefont {T.}~\bibnamefont {Bzdu{\v{s}}ek}}, \ and\ \bibinfo {author} {\bibfnamefont {T.}~\bibnamefont {Neupert}},\ }\href {\doibase 10.1038/s41467-021-25947-z} {\bibfield  {journal} {\bibinfo  {journal} {Nat. Commun.}\ }\textbf {\bibinfo {volume} {12}},\ \bibinfo {pages} {5681} (\bibinfo {year} {2021})}\BibitemShut {NoStop}%
\bibitem [{\citenamefont {Okugawa}\ \emph {et~al.}(2020)\citenamefont {Okugawa}, \citenamefont {Takahashi},\ and\ \citenamefont {Yokomizo}}]{Okugawa-takahashi-yokomizo}%
  \BibitemOpen
  \bibfield  {author} {\bibinfo {author} {\bibfnamefont {R.}~\bibnamefont {Okugawa}}, \bibinfo {author} {\bibfnamefont {R.}~\bibnamefont {Takahashi}}, \ and\ \bibinfo {author} {\bibfnamefont {K.}~\bibnamefont {Yokomizo}},\ }\href {\doibase 10.1103/PhysRevB.102.241202} {\bibfield  {journal} {\bibinfo  {journal} {Phys. Rev. B}\ }\textbf {\bibinfo {volume} {102}},\ \bibinfo {pages} {241202(R)} (\bibinfo {year} {2020})}\BibitemShut {NoStop}%
\bibitem [{\citenamefont {Kawabata}\ \emph {et~al.}(2020{\natexlab{b}})\citenamefont {Kawabata}, \citenamefont {Sato},\ and\ \citenamefont {Shiozaki}}]{KSS-20}%
  \BibitemOpen
  \bibfield  {author} {\bibinfo {author} {\bibfnamefont {K.}~\bibnamefont {Kawabata}}, \bibinfo {author} {\bibfnamefont {M.}~\bibnamefont {Sato}}, \ and\ \bibinfo {author} {\bibfnamefont {K.}~\bibnamefont {Shiozaki}},\ }\href {\doibase 10.1103/PhysRevB.102.205118} {\bibfield  {journal} {\bibinfo  {journal} {Phys. Rev. B}\ }\textbf {\bibinfo {volume} {102}},\ \bibinfo {pages} {205118} (\bibinfo {year} {2020}{\natexlab{b}})}\BibitemShut {NoStop}%
\bibitem [{\citenamefont {Fu}\ \emph {et~al.}(2021)\citenamefont {Fu}, \citenamefont {Hu},\ and\ \citenamefont {Wan}}]{Fu-20}%
  \BibitemOpen
  \bibfield  {author} {\bibinfo {author} {\bibfnamefont {Y.}~\bibnamefont {Fu}}, \bibinfo {author} {\bibfnamefont {J.}~\bibnamefont {Hu}}, \ and\ \bibinfo {author} {\bibfnamefont {S.}~\bibnamefont {Wan}},\ }\href {\doibase 10.1103/PhysRevB.103.045420} {\bibfield  {journal} {\bibinfo  {journal} {Phys. Rev. B}\ }\textbf {\bibinfo {volume} {103}},\ \bibinfo {pages} {045420} (\bibinfo {year} {2021})}\BibitemShut {NoStop}%
\bibitem [{\citenamefont {Okuma}\ and\ \citenamefont {Sato}(2021)}]{Okuma-Sato-21}%
  \BibitemOpen
  \bibfield  {author} {\bibinfo {author} {\bibfnamefont {N.}~\bibnamefont {Okuma}}\ and\ \bibinfo {author} {\bibfnamefont {M.}~\bibnamefont {Sato}},\ }\href {\doibase 10.1103/PhysRevB.103.085428} {\bibfield  {journal} {\bibinfo  {journal} {Phys. Rev. B}\ }\textbf {\bibinfo {volume} {103}},\ \bibinfo {pages} {085428} (\bibinfo {year} {2021})}\BibitemShut {NoStop}%
\bibitem [{\citenamefont {Kawabata}\ \emph {et~al.}(2021)\citenamefont {Kawabata}, \citenamefont {Shiozaki},\ and\ \citenamefont {Ryu}}]{KSR-21}%
  \BibitemOpen
  \bibfield  {author} {\bibinfo {author} {\bibfnamefont {K.}~\bibnamefont {Kawabata}}, \bibinfo {author} {\bibfnamefont {K.}~\bibnamefont {Shiozaki}}, \ and\ \bibinfo {author} {\bibfnamefont {S.}~\bibnamefont {Ryu}},\ }\href {\doibase 10.1103/PhysRevLett.126.216405} {\bibfield  {journal} {\bibinfo  {journal} {Phys. Rev. Lett.}\ }\textbf {\bibinfo {volume} {126}},\ \bibinfo {pages} {216405} (\bibinfo {year} {2021})}\BibitemShut {NoStop}%
\bibitem [{\citenamefont {Yi}\ and\ \citenamefont {Yang}(2020)}]{Yi-Yang-20}%
  \BibitemOpen
  \bibfield  {author} {\bibinfo {author} {\bibfnamefont {Y.}~\bibnamefont {Yi}}\ and\ \bibinfo {author} {\bibfnamefont {Z.}~\bibnamefont {Yang}},\ }\href {\doibase 10.1103/PhysRevLett.125.186802} {\bibfield  {journal} {\bibinfo  {journal} {Phys. Rev. Lett.}\ }\textbf {\bibinfo {volume} {125}},\ \bibinfo {pages} {186802} (\bibinfo {year} {2020})}\BibitemShut {NoStop}%
\bibitem [{\citenamefont {Zhang}\ \emph {et~al.}(2022)\citenamefont {Zhang}, \citenamefont {Yang},\ and\ \citenamefont {Fang}}]{Zhang-21}%
  \BibitemOpen
  \bibfield  {author} {\bibinfo {author} {\bibfnamefont {K.}~\bibnamefont {Zhang}}, \bibinfo {author} {\bibfnamefont {Z.}~\bibnamefont {Yang}}, \ and\ \bibinfo {author} {\bibfnamefont {C.}~\bibnamefont {Fang}},\ }\href {\doibase 10.1038/s41467-022-30161-6} {\bibfield  {journal} {\bibinfo  {journal} {Nat. Commun.}\ }\textbf {\bibinfo {volume} {13}},\ \bibinfo {pages} {2496} (\bibinfo {year} {2022})}\BibitemShut {NoStop}%
\bibitem [{\citenamefont {Sun}\ \emph {et~al.}(2021)\citenamefont {Sun}, \citenamefont {Zhu},\ and\ \citenamefont {Hughes}}]{Sun-21}%
  \BibitemOpen
  \bibfield  {author} {\bibinfo {author} {\bibfnamefont {X.-Q.}\ \bibnamefont {Sun}}, \bibinfo {author} {\bibfnamefont {P.}~\bibnamefont {Zhu}}, \ and\ \bibinfo {author} {\bibfnamefont {T.~L.}\ \bibnamefont {Hughes}},\ }\href {\doibase 10.1103/PhysRevLett.127.066401} {\bibfield  {journal} {\bibinfo  {journal} {Phys. Rev. Lett.}\ }\textbf {\bibinfo {volume} {127}},\ \bibinfo {pages} {066401} (\bibinfo {year} {2021})}\BibitemShut {NoStop}%
\bibitem [{\citenamefont {Shiozaki}\ and\ \citenamefont {Ono}(2021)}]{Shiozaki-Ono-21}%
  \BibitemOpen
  \bibfield  {author} {\bibinfo {author} {\bibfnamefont {K.}~\bibnamefont {Shiozaki}}\ and\ \bibinfo {author} {\bibfnamefont {S.}~\bibnamefont {Ono}},\ }\href {\doibase 10.1103/PhysRevB.104.035424} {\bibfield  {journal} {\bibinfo  {journal} {Phys. Rev. B}\ }\textbf {\bibinfo {volume} {104}},\ \bibinfo {pages} {035424} (\bibinfo {year} {2021})}\BibitemShut {NoStop}%
\bibitem [{\citenamefont {Okugawa}\ \emph {et~al.}(2021)\citenamefont {Okugawa}, \citenamefont {Takahashi},\ and\ \citenamefont {Yokomizo}}]{Okugawa-21}%
  \BibitemOpen
  \bibfield  {author} {\bibinfo {author} {\bibfnamefont {R.}~\bibnamefont {Okugawa}}, \bibinfo {author} {\bibfnamefont {R.}~\bibnamefont {Takahashi}}, \ and\ \bibinfo {author} {\bibfnamefont {K.}~\bibnamefont {Yokomizo}},\ }\href {\doibase 10.1103/PhysRevB.103.205205} {\bibfield  {journal} {\bibinfo  {journal} {Phys. Rev. B}\ }\textbf {\bibinfo {volume} {103}},\ \bibinfo {pages} {205205} (\bibinfo {year} {2021})}\BibitemShut {NoStop}%
\bibitem [{\citenamefont {Vecsei}\ \emph {et~al.}(2021)\citenamefont {Vecsei}, \citenamefont {Denner}, \citenamefont {Neupert},\ and\ \citenamefont {Schindler}}]{Vecsei-21}%
  \BibitemOpen
  \bibfield  {author} {\bibinfo {author} {\bibfnamefont {P.~M.}\ \bibnamefont {Vecsei}}, \bibinfo {author} {\bibfnamefont {M.~M.}\ \bibnamefont {Denner}}, \bibinfo {author} {\bibfnamefont {T.}~\bibnamefont {Neupert}}, \ and\ \bibinfo {author} {\bibfnamefont {F.}~\bibnamefont {Schindler}},\ }\href {\doibase 10.1103/PhysRevB.103.L201114} {\bibfield  {journal} {\bibinfo  {journal} {Phys. Rev. B}\ }\textbf {\bibinfo {volume} {103}},\ \bibinfo {pages} {L201114} (\bibinfo {year} {2021})}\BibitemShut {NoStop}%
\bibitem [{\citenamefont {Hu}\ \emph {et~al.}(2022)\citenamefont {Hu}, \citenamefont {Zhao},\ and\ \citenamefont {Liu}}]{Hu-21}%
  \BibitemOpen
  \bibfield  {author} {\bibinfo {author} {\bibfnamefont {H.}~\bibnamefont {Hu}}, \bibinfo {author} {\bibfnamefont {E.}~\bibnamefont {Zhao}}, \ and\ \bibinfo {author} {\bibfnamefont {W.~V.}\ \bibnamefont {Liu}},\ }\href {\doibase 10.1103/PhysRevB.106.094305} {\bibfield  {journal} {\bibinfo  {journal} {Phys. Rev. B}\ }\textbf {\bibinfo {volume} {106}},\ \bibinfo {pages} {094305} (\bibinfo {year} {2022})}\BibitemShut {NoStop}%
\bibitem [{\citenamefont {Ghosh}\ and\ \citenamefont {Nag}(2022)}]{Ghosh-22}%
  \BibitemOpen
  \bibfield  {author} {\bibinfo {author} {\bibfnamefont {A.~K.}\ \bibnamefont {Ghosh}}\ and\ \bibinfo {author} {\bibfnamefont {T.}~\bibnamefont {Nag}},\ }\href {\doibase 10.1103/PhysRevB.106.L140303} {\bibfield  {journal} {\bibinfo  {journal} {Phys. Rev. B}\ }\textbf {\bibinfo {volume} {106}},\ \bibinfo {pages} {L140303} (\bibinfo {year} {2022})}\BibitemShut {NoStop}%
\bibitem [{\citenamefont {Poli}\ \emph {et~al.}(2015)\citenamefont {Poli}, \citenamefont {Bellec}, \citenamefont {Kuhl}, \citenamefont {Mortessagne},\ and\ \citenamefont {Schomerus}}]{Poli-15}%
  \BibitemOpen
  \bibfield  {author} {\bibinfo {author} {\bibfnamefont {C.}~\bibnamefont {Poli}}, \bibinfo {author} {\bibfnamefont {M.}~\bibnamefont {Bellec}}, \bibinfo {author} {\bibfnamefont {U.}~\bibnamefont {Kuhl}}, \bibinfo {author} {\bibfnamefont {F.}~\bibnamefont {Mortessagne}}, \ and\ \bibinfo {author} {\bibfnamefont {H.}~\bibnamefont {Schomerus}},\ }\href {\doibase 10.1038/ncomms7710} {\bibfield  {journal} {\bibinfo  {journal} {Nat. Commun.}\ }\textbf {\bibinfo {volume} {6}},\ \bibinfo {pages} {6710} (\bibinfo {year} {2015})}\BibitemShut {NoStop}%
\bibitem [{\citenamefont {Zeuner}\ \emph {et~al.}(2015)\citenamefont {Zeuner}, \citenamefont {Rechtsman}, \citenamefont {Plotnik}, \citenamefont {Lumer}, \citenamefont {Nolte}, \citenamefont {Rudner}, \citenamefont {Segev},\ and\ \citenamefont {Szameit}}]{Zeuner-15}%
  \BibitemOpen
  \bibfield  {author} {\bibinfo {author} {\bibfnamefont {J.~M.}\ \bibnamefont {Zeuner}}, \bibinfo {author} {\bibfnamefont {M.~C.}\ \bibnamefont {Rechtsman}}, \bibinfo {author} {\bibfnamefont {Y.}~\bibnamefont {Plotnik}}, \bibinfo {author} {\bibfnamefont {Y.}~\bibnamefont {Lumer}}, \bibinfo {author} {\bibfnamefont {S.}~\bibnamefont {Nolte}}, \bibinfo {author} {\bibfnamefont {M.~S.}\ \bibnamefont {Rudner}}, \bibinfo {author} {\bibfnamefont {M.}~\bibnamefont {Segev}}, \ and\ \bibinfo {author} {\bibfnamefont {A.}~\bibnamefont {Szameit}},\ }\href {\doibase 10.1103/PhysRevLett.115.040402} {\bibfield  {journal} {\bibinfo  {journal} {Phys. Rev. Lett.}\ }\textbf {\bibinfo {volume} {115}},\ \bibinfo {pages} {040402} (\bibinfo {year} {2015})}\BibitemShut {NoStop}%
\bibitem [{\citenamefont {Zhen}\ \emph {et~al.}(2015)\citenamefont {Zhen}, \citenamefont {Hsu}, \citenamefont {Igarashi}, \citenamefont {Lu}, \citenamefont {Kaminer}, \citenamefont {Pick}, \citenamefont {Chua}, \citenamefont {Joannopoulos},\ and\ \citenamefont {Solja\u{c}i\'c}}]{Zhen-15}%
  \BibitemOpen
  \bibfield  {author} {\bibinfo {author} {\bibfnamefont {B.}~\bibnamefont {Zhen}}, \bibinfo {author} {\bibfnamefont {C.~W.}\ \bibnamefont {Hsu}}, \bibinfo {author} {\bibfnamefont {Y.}~\bibnamefont {Igarashi}}, \bibinfo {author} {\bibfnamefont {L.}~\bibnamefont {Lu}}, \bibinfo {author} {\bibfnamefont {I.}~\bibnamefont {Kaminer}}, \bibinfo {author} {\bibfnamefont {A.}~\bibnamefont {Pick}}, \bibinfo {author} {\bibfnamefont {S.-L.}\ \bibnamefont {Chua}}, \bibinfo {author} {\bibfnamefont {J.~D.}\ \bibnamefont {Joannopoulos}}, \ and\ \bibinfo {author} {\bibfnamefont {M.}~\bibnamefont {Solja\u{c}i\'c}},\ }\href {\doibase 10.1038/nature14889} {\bibfield  {journal} {\bibinfo  {journal} {Nature}\ }\textbf {\bibinfo {volume} {525}},\ \bibinfo {pages} {354} (\bibinfo {year} {2015})}\BibitemShut {NoStop}%
\bibitem [{\citenamefont {Weimann}\ \emph {et~al.}(2017)\citenamefont {Weimann}, \citenamefont {Kremer}, \citenamefont {Plotnik}, \citenamefont {Lumer}, \citenamefont {Nolte}, \citenamefont {Makris}, \citenamefont {Segev}, \citenamefont {Rechtsman},\ and\ \citenamefont {Szameit}}]{Weimann-17}%
  \BibitemOpen
  \bibfield  {author} {\bibinfo {author} {\bibfnamefont {S.}~\bibnamefont {Weimann}}, \bibinfo {author} {\bibfnamefont {M.}~\bibnamefont {Kremer}}, \bibinfo {author} {\bibfnamefont {Y.}~\bibnamefont {Plotnik}}, \bibinfo {author} {\bibfnamefont {Y.}~\bibnamefont {Lumer}}, \bibinfo {author} {\bibfnamefont {S.}~\bibnamefont {Nolte}}, \bibinfo {author} {\bibfnamefont {K.~G.}\ \bibnamefont {Makris}}, \bibinfo {author} {\bibfnamefont {M.}~\bibnamefont {Segev}}, \bibinfo {author} {\bibfnamefont {M.~C.}\ \bibnamefont {Rechtsman}}, \ and\ \bibinfo {author} {\bibfnamefont {A.}~\bibnamefont {Szameit}},\ }\href {\doibase 10.1038/nmat4811} {\bibfield  {journal} {\bibinfo  {journal} {Nat. Mater.}\ }\textbf {\bibinfo {volume} {16}},\ \bibinfo {pages} {433} (\bibinfo {year} {2017})}\BibitemShut {NoStop}%
\bibitem [{\citenamefont {Xiao}\ \emph {et~al.}(2017)\citenamefont {Xiao}, \citenamefont {Zhan}, \citenamefont {Bian}, \citenamefont {Wang}, \citenamefont {Zhang}, \citenamefont {Wang}, \citenamefont {Li}, \citenamefont {Mochizuki}, \citenamefont {Kim}, \citenamefont {Kawakami}, \citenamefont {Yi}, \citenamefont {Obuse}, \citenamefont {Sanders},\ and\ \citenamefont {Xue}}]{Xiao-17}%
  \BibitemOpen
  \bibfield  {author} {\bibinfo {author} {\bibfnamefont {L.}~\bibnamefont {Xiao}}, \bibinfo {author} {\bibfnamefont {X.}~\bibnamefont {Zhan}}, \bibinfo {author} {\bibfnamefont {Z.~H.}\ \bibnamefont {Bian}}, \bibinfo {author} {\bibfnamefont {K.~K.}\ \bibnamefont {Wang}}, \bibinfo {author} {\bibfnamefont {X.}~\bibnamefont {Zhang}}, \bibinfo {author} {\bibfnamefont {X.~P.}\ \bibnamefont {Wang}}, \bibinfo {author} {\bibfnamefont {J.}~\bibnamefont {Li}}, \bibinfo {author} {\bibfnamefont {K.}~\bibnamefont {Mochizuki}}, \bibinfo {author} {\bibfnamefont {D.}~\bibnamefont {Kim}}, \bibinfo {author} {\bibfnamefont {N.}~\bibnamefont {Kawakami}}, \bibinfo {author} {\bibfnamefont {W.}~\bibnamefont {Yi}}, \bibinfo {author} {\bibfnamefont {H.}~\bibnamefont {Obuse}}, \bibinfo {author} {\bibfnamefont {B.~C.}\ \bibnamefont {Sanders}}, \ and\ \bibinfo {author} {\bibfnamefont {P.}~\bibnamefont {Xue}},\ }\href {\doibase 10.1038/nphys4204} {\bibfield  {journal} {\bibinfo  {journal} {Nat. Phys.}\ }\textbf {\bibinfo {volume} {13}},\
  \bibinfo {pages} {1117} (\bibinfo {year} {2017})}\BibitemShut {NoStop}%
\bibitem [{\citenamefont {St-Jean}\ \emph {et~al.}(2017)\citenamefont {St-Jean}, \citenamefont {Goblot}, \citenamefont {Galopin}, \citenamefont {Lema\^itre}, \citenamefont {Ozawa}, \citenamefont {Gratiet}, \citenamefont {Sagnes}, \citenamefont {Bloch},\ and\ \citenamefont {Amo}}]{St-Jean-17}%
  \BibitemOpen
  \bibfield  {author} {\bibinfo {author} {\bibfnamefont {P.}~\bibnamefont {St-Jean}}, \bibinfo {author} {\bibfnamefont {V.}~\bibnamefont {Goblot}}, \bibinfo {author} {\bibfnamefont {E.}~\bibnamefont {Galopin}}, \bibinfo {author} {\bibfnamefont {A.}~\bibnamefont {Lema\^itre}}, \bibinfo {author} {\bibfnamefont {T.}~\bibnamefont {Ozawa}}, \bibinfo {author} {\bibfnamefont {L.~L.}\ \bibnamefont {Gratiet}}, \bibinfo {author} {\bibfnamefont {I.}~\bibnamefont {Sagnes}}, \bibinfo {author} {\bibfnamefont {J.}~\bibnamefont {Bloch}}, \ and\ \bibinfo {author} {\bibfnamefont {A.}~\bibnamefont {Amo}},\ }\href {\doibase 10.1038/s41566-017-0006-2} {\bibfield  {journal} {\bibinfo  {journal} {Nat. Photonics}\ }\textbf {\bibinfo {volume} {11}},\ \bibinfo {pages} {651} (\bibinfo {year} {2017})}\BibitemShut {NoStop}%
\bibitem [{\citenamefont {Parto}\ \emph {et~al.}(2018)\citenamefont {Parto}, \citenamefont {Wittek}, \citenamefont {Hodaei}, \citenamefont {Harari}, \citenamefont {Bandres}, \citenamefont {Ren}, \citenamefont {Rechtsman}, \citenamefont {Segev}, \citenamefont {Christodoulides},\ and\ \citenamefont {Khajavikhan}}]{Parto-17}%
  \BibitemOpen
  \bibfield  {author} {\bibinfo {author} {\bibfnamefont {M.}~\bibnamefont {Parto}}, \bibinfo {author} {\bibfnamefont {S.}~\bibnamefont {Wittek}}, \bibinfo {author} {\bibfnamefont {H.}~\bibnamefont {Hodaei}}, \bibinfo {author} {\bibfnamefont {G.}~\bibnamefont {Harari}}, \bibinfo {author} {\bibfnamefont {M.~A.}\ \bibnamefont {Bandres}}, \bibinfo {author} {\bibfnamefont {J.}~\bibnamefont {Ren}}, \bibinfo {author} {\bibfnamefont {M.~C.}\ \bibnamefont {Rechtsman}}, \bibinfo {author} {\bibfnamefont {M.}~\bibnamefont {Segev}}, \bibinfo {author} {\bibfnamefont {D.~N.}\ \bibnamefont {Christodoulides}}, \ and\ \bibinfo {author} {\bibfnamefont {M.}~\bibnamefont {Khajavikhan}},\ }\href {\doibase 10.1103/PhysRevLett.120.113901} {\bibfield  {journal} {\bibinfo  {journal} {Phys. Rev. Lett.}\ }\textbf {\bibinfo {volume} {120}},\ \bibinfo {pages} {113901} (\bibinfo {year} {2018})}\BibitemShut {NoStop}%
\bibitem [{\citenamefont {Bahari}\ \emph {et~al.}(2017)\citenamefont {Bahari}, \citenamefont {Ndao}, \citenamefont {Vallini}, \citenamefont {Amili}, \citenamefont {Fainman},\ and\ \citenamefont {Kanté}}]{Bahari-17}%
  \BibitemOpen
  \bibfield  {author} {\bibinfo {author} {\bibfnamefont {B.}~\bibnamefont {Bahari}}, \bibinfo {author} {\bibfnamefont {A.}~\bibnamefont {Ndao}}, \bibinfo {author} {\bibfnamefont {F.}~\bibnamefont {Vallini}}, \bibinfo {author} {\bibfnamefont {A.~E.}\ \bibnamefont {Amili}}, \bibinfo {author} {\bibfnamefont {Y.}~\bibnamefont {Fainman}}, \ and\ \bibinfo {author} {\bibfnamefont {B.}~\bibnamefont {Kanté}},\ }\href {\doibase 10.1126/science.aao4551} {\bibfield  {journal} {\bibinfo  {journal} {Science}\ }\textbf {\bibinfo {volume} {358}},\ \bibinfo {pages} {636} (\bibinfo {year} {2017})}\BibitemShut {NoStop}%
\bibitem [{\citenamefont {Zhao}\ \emph {et~al.}(2018)\citenamefont {Zhao}, \citenamefont {Miao}, \citenamefont {Teimourpour}, \citenamefont {Malzard}, \citenamefont {El-Ganainy}, \citenamefont {Schomerus},\ and\ \citenamefont {Feng}}]{Zhao-18}%
  \BibitemOpen
  \bibfield  {author} {\bibinfo {author} {\bibfnamefont {H.}~\bibnamefont {Zhao}}, \bibinfo {author} {\bibfnamefont {P.}~\bibnamefont {Miao}}, \bibinfo {author} {\bibfnamefont {M.~H.}\ \bibnamefont {Teimourpour}}, \bibinfo {author} {\bibfnamefont {S.}~\bibnamefont {Malzard}}, \bibinfo {author} {\bibfnamefont {R.}~\bibnamefont {El-Ganainy}}, \bibinfo {author} {\bibfnamefont {H.}~\bibnamefont {Schomerus}}, \ and\ \bibinfo {author} {\bibfnamefont {L.}~\bibnamefont {Feng}},\ }\href {\doibase 10.1038/s41467-018-03434-2} {\bibfield  {journal} {\bibinfo  {journal} {Nat. Commun.}\ }\textbf {\bibinfo {volume} {9}},\ \bibinfo {pages} {981} (\bibinfo {year} {2018})}\BibitemShut {NoStop}%
\bibitem [{\citenamefont {Zhou}\ \emph {et~al.}(2018)\citenamefont {Zhou}, \citenamefont {Peng}, \citenamefont {Yoon}, \citenamefont {Hsu}, \citenamefont {Nelson}, \citenamefont {Fu}, \citenamefont {Joannopoulos}, \citenamefont {Soljačić},\ and\ \citenamefont {Zhen}}]{Zhou-18}%
  \BibitemOpen
  \bibfield  {author} {\bibinfo {author} {\bibfnamefont {H.}~\bibnamefont {Zhou}}, \bibinfo {author} {\bibfnamefont {C.}~\bibnamefont {Peng}}, \bibinfo {author} {\bibfnamefont {Y.}~\bibnamefont {Yoon}}, \bibinfo {author} {\bibfnamefont {C.~W.}\ \bibnamefont {Hsu}}, \bibinfo {author} {\bibfnamefont {K.~A.}\ \bibnamefont {Nelson}}, \bibinfo {author} {\bibfnamefont {L.}~\bibnamefont {Fu}}, \bibinfo {author} {\bibfnamefont {J.~D.}\ \bibnamefont {Joannopoulos}}, \bibinfo {author} {\bibfnamefont {M.}~\bibnamefont {Soljačić}}, \ and\ \bibinfo {author} {\bibfnamefont {B.}~\bibnamefont {Zhen}},\ }\href {\doibase 10.1126/science.aap9859} {\bibfield  {journal} {\bibinfo  {journal} {Science}\ }\textbf {\bibinfo {volume} {359}},\ \bibinfo {pages} {1009} (\bibinfo {year} {2018})}\BibitemShut {NoStop}%
\bibitem [{\citenamefont {Harari}\ \emph {et~al.}(2018)\citenamefont {Harari}, \citenamefont {Bandres}, \citenamefont {Lumer}, \citenamefont {Rechtsman}, \citenamefont {Chong}, \citenamefont {Khajavikhan}, \citenamefont {Christodoulides},\ and\ \citenamefont {Segev}}]{Harari-18}%
  \BibitemOpen
  \bibfield  {author} {\bibinfo {author} {\bibfnamefont {G.}~\bibnamefont {Harari}}, \bibinfo {author} {\bibfnamefont {M.~A.}\ \bibnamefont {Bandres}}, \bibinfo {author} {\bibfnamefont {Y.}~\bibnamefont {Lumer}}, \bibinfo {author} {\bibfnamefont {M.~C.}\ \bibnamefont {Rechtsman}}, \bibinfo {author} {\bibfnamefont {Y.~D.}\ \bibnamefont {Chong}}, \bibinfo {author} {\bibfnamefont {M.}~\bibnamefont {Khajavikhan}}, \bibinfo {author} {\bibfnamefont {D.~N.}\ \bibnamefont {Christodoulides}}, \ and\ \bibinfo {author} {\bibfnamefont {M.}~\bibnamefont {Segev}},\ }\href {\doibase 10.1126/science.aar4003} {\bibfield  {journal} {\bibinfo  {journal} {Science}\ }\textbf {\bibinfo {volume} {359}},\ \bibinfo {pages} {eaar4003} (\bibinfo {year} {2018})}\BibitemShut {NoStop}%
\bibitem [{\citenamefont {Bandres}\ \emph {et~al.}(2018)\citenamefont {Bandres}, \citenamefont {Wittek}, \citenamefont {Harari}, \citenamefont {Parto}, \citenamefont {Ren}, \citenamefont {Segev}, \citenamefont {Christodoulides},\ and\ \citenamefont {Khajavikhan}}]{Bandres-18}%
  \BibitemOpen
  \bibfield  {author} {\bibinfo {author} {\bibfnamefont {M.~A.}\ \bibnamefont {Bandres}}, \bibinfo {author} {\bibfnamefont {S.}~\bibnamefont {Wittek}}, \bibinfo {author} {\bibfnamefont {G.}~\bibnamefont {Harari}}, \bibinfo {author} {\bibfnamefont {M.}~\bibnamefont {Parto}}, \bibinfo {author} {\bibfnamefont {J.}~\bibnamefont {Ren}}, \bibinfo {author} {\bibfnamefont {M.}~\bibnamefont {Segev}}, \bibinfo {author} {\bibfnamefont {D.~N.}\ \bibnamefont {Christodoulides}}, \ and\ \bibinfo {author} {\bibfnamefont {M.}~\bibnamefont {Khajavikhan}},\ }\href {\doibase 10.1126/science.aar4005} {\bibfield  {journal} {\bibinfo  {journal} {Science}\ }\textbf {\bibinfo {volume} {359}},\ \bibinfo {pages} {eaar4005} (\bibinfo {year} {2018})}\BibitemShut {NoStop}%
\bibitem [{\citenamefont {Cerjan}\ \emph {et~al.}(2019)\citenamefont {Cerjan}, \citenamefont {Huang}, \citenamefont {Chen}, \citenamefont {Chong},\ and\ \citenamefont {Rechtsman}}]{Cerjan-19}%
  \BibitemOpen
  \bibfield  {author} {\bibinfo {author} {\bibfnamefont {A.}~\bibnamefont {Cerjan}}, \bibinfo {author} {\bibfnamefont {S.}~\bibnamefont {Huang}}, \bibinfo {author} {\bibfnamefont {K.~P.}\ \bibnamefont {Chen}}, \bibinfo {author} {\bibfnamefont {Y.}~\bibnamefont {Chong}}, \ and\ \bibinfo {author} {\bibfnamefont {M.~C.}\ \bibnamefont {Rechtsman}},\ }\href {\doibase 10.1038/s41566-019-0453-z} {\bibfield  {journal} {\bibinfo  {journal} {Nat. Photonics}\ }\textbf {\bibinfo {volume} {13}},\ \bibinfo {pages} {623} (\bibinfo {year} {2019})}\BibitemShut {NoStop}%
\bibitem [{\citenamefont {Zhao}\ \emph {et~al.}(2019)\citenamefont {Zhao}, \citenamefont {Qiao}, \citenamefont {Wu}, \citenamefont {Midya}, \citenamefont {Longhi},\ and\ \citenamefont {Feng}}]{Zhao-19}%
  \BibitemOpen
  \bibfield  {author} {\bibinfo {author} {\bibfnamefont {H.}~\bibnamefont {Zhao}}, \bibinfo {author} {\bibfnamefont {X.}~\bibnamefont {Qiao}}, \bibinfo {author} {\bibfnamefont {T.}~\bibnamefont {Wu}}, \bibinfo {author} {\bibfnamefont {B.}~\bibnamefont {Midya}}, \bibinfo {author} {\bibfnamefont {S.}~\bibnamefont {Longhi}}, \ and\ \bibinfo {author} {\bibfnamefont {L.}~\bibnamefont {Feng}},\ }\href {\doibase 10.1126/science.aay1064} {\bibfield  {journal} {\bibinfo  {journal} {Science}\ }\textbf {\bibinfo {volume} {365}},\ \bibinfo {pages} {1163} (\bibinfo {year} {2019})}\BibitemShut {NoStop}%
\bibitem [{\citenamefont {Brandenbourger}\ \emph {et~al.}(2019)\citenamefont {Brandenbourger}, \citenamefont {Locsin},\ and\ \citenamefont {E.~Lerner}}]{Brandenbourger-19-skin-exp}%
  \BibitemOpen
  \bibfield  {author} {\bibinfo {author} {\bibfnamefont {M.}~\bibnamefont {Brandenbourger}}, \bibinfo {author} {\bibfnamefont {X.}~\bibnamefont {Locsin}}, \ and\ \bibinfo {author} {\bibfnamefont {C.~C.}\ \bibnamefont {E.~Lerner}},\ }\href {\doibase 10.1038/s41467-019-12599-3} {\bibfield  {journal} {\bibinfo  {journal} {Nat. Commun.}\ }\textbf {\bibinfo {volume} {10}},\ \bibinfo {pages} {4608} (\bibinfo {year} {2019})}\BibitemShut {NoStop}%
\bibitem [{\citenamefont {Ghatak}\ \emph {et~al.}(2020)\citenamefont {Ghatak}, \citenamefont {Brandenbourger}, \citenamefont {van Wezel},\ and\ \citenamefont {Coulais}}]{Ghatak-19-skin-exp}%
  \BibitemOpen
  \bibfield  {author} {\bibinfo {author} {\bibfnamefont {A.}~\bibnamefont {Ghatak}}, \bibinfo {author} {\bibfnamefont {M.}~\bibnamefont {Brandenbourger}}, \bibinfo {author} {\bibfnamefont {J.}~\bibnamefont {van Wezel}}, \ and\ \bibinfo {author} {\bibfnamefont {C.}~\bibnamefont {Coulais}},\ }\href {\doibase 10.1073/pnas.2010580117} {\bibfield  {journal} {\bibinfo  {journal} {Proc. Natl. Acad. Sci. U.S.A.}\ }\textbf {\bibinfo {volume} {117}},\ \bibinfo {pages} {29561} (\bibinfo {year} {2020})}\BibitemShut {NoStop}%
\bibitem [{\citenamefont {Helbig}\ \emph {et~al.}(2020)\citenamefont {Helbig}, \citenamefont {Hofmann}, \citenamefont {Imhof}, \citenamefont {Abdelghany}, \citenamefont {Kiessling}, \citenamefont {Molenkamp}, \citenamefont {Lee}, \citenamefont {Szameit}, \citenamefont {Greiter},\ and\ \citenamefont {Thomale}}]{Helbig-19-skin-exp}%
  \BibitemOpen
  \bibfield  {author} {\bibinfo {author} {\bibfnamefont {T.}~\bibnamefont {Helbig}}, \bibinfo {author} {\bibfnamefont {T.}~\bibnamefont {Hofmann}}, \bibinfo {author} {\bibfnamefont {S.}~\bibnamefont {Imhof}}, \bibinfo {author} {\bibfnamefont {M.}~\bibnamefont {Abdelghany}}, \bibinfo {author} {\bibfnamefont {T.}~\bibnamefont {Kiessling}}, \bibinfo {author} {\bibfnamefont {L.}~\bibnamefont {Molenkamp}}, \bibinfo {author} {\bibfnamefont {C.}~\bibnamefont {Lee}}, \bibinfo {author} {\bibfnamefont {A.}~\bibnamefont {Szameit}}, \bibinfo {author} {\bibfnamefont {M.}~\bibnamefont {Greiter}}, \ and\ \bibinfo {author} {\bibfnamefont {R.}~\bibnamefont {Thomale}},\ }\href {\doibase 10.1038/s41567-020-0922-9} {\bibfield  {journal} {\bibinfo  {journal} {Nat. Phys.}\ }\textbf {\bibinfo {volume} {16}},\ \bibinfo {pages} {747} (\bibinfo {year} {2020})}\BibitemShut {NoStop}%
\bibitem [{\citenamefont {Hofmann}\ \emph {et~al.}(2020)\citenamefont {Hofmann}, \citenamefont {Helbig}, \citenamefont {Schindler}, \citenamefont {Salgo}, \citenamefont {Brzezi\ifmmode~\acute{n}\else \'{n}\fi{}ska}, \citenamefont {Greiter}, \citenamefont {Kiessling}, \citenamefont {Wolf}, \citenamefont {Vollhardt}, \citenamefont {Kaba\ifmmode~\check{s}\else \v{s}\fi{}i}, \citenamefont {Lee}, \citenamefont {Bilu\ifmmode \check{s}\else \v{s}\fi{}i\ifmmode~\acute{c}\else \'{c}\fi{}}, \citenamefont {Thomale},\ and\ \citenamefont {Neupert}}]{Hofmann-19-skin-exp}%
  \BibitemOpen
  \bibfield  {author} {\bibinfo {author} {\bibfnamefont {T.}~\bibnamefont {Hofmann}}, \bibinfo {author} {\bibfnamefont {T.}~\bibnamefont {Helbig}}, \bibinfo {author} {\bibfnamefont {F.}~\bibnamefont {Schindler}}, \bibinfo {author} {\bibfnamefont {N.}~\bibnamefont {Salgo}}, \bibinfo {author} {\bibfnamefont {M.}~\bibnamefont {Brzezi\ifmmode~\acute{n}\else \'{n}\fi{}ska}}, \bibinfo {author} {\bibfnamefont {M.}~\bibnamefont {Greiter}}, \bibinfo {author} {\bibfnamefont {T.}~\bibnamefont {Kiessling}}, \bibinfo {author} {\bibfnamefont {D.}~\bibnamefont {Wolf}}, \bibinfo {author} {\bibfnamefont {A.}~\bibnamefont {Vollhardt}}, \bibinfo {author} {\bibfnamefont {A.}~\bibnamefont {Kaba\ifmmode~\check{s}\else \v{s}\fi{}i}}, \bibinfo {author} {\bibfnamefont {C.~H.}\ \bibnamefont {Lee}}, \bibinfo {author} {\bibfnamefont {A.}~\bibnamefont {Bilu\ifmmode \check{s}\else \v{s}\fi{}i\ifmmode~\acute{c}\else \'{c}\fi{}}}, \bibinfo {author} {\bibfnamefont {R.}~\bibnamefont {Thomale}}, \ and\ \bibinfo {author} {\bibfnamefont
  {T.}~\bibnamefont {Neupert}},\ }\href {\doibase 10.1103/PhysRevResearch.2.023265} {\bibfield  {journal} {\bibinfo  {journal} {Phys. Rev. Res.}\ }\textbf {\bibinfo {volume} {2}},\ \bibinfo {pages} {023265} (\bibinfo {year} {2020})}\BibitemShut {NoStop}%
\bibitem [{\citenamefont {Xiao}\ \emph {et~al.}(2020)\citenamefont {Xiao}, \citenamefont {Deng}, \citenamefont {Wang}, \citenamefont {Zhu}, \citenamefont {Wang}, \citenamefont {Yi},\ and\ \citenamefont {Xue}}]{Xiao-19-skin-exp}%
  \BibitemOpen
  \bibfield  {author} {\bibinfo {author} {\bibfnamefont {L.}~\bibnamefont {Xiao}}, \bibinfo {author} {\bibfnamefont {T.}~\bibnamefont {Deng}}, \bibinfo {author} {\bibfnamefont {K.}~\bibnamefont {Wang}}, \bibinfo {author} {\bibfnamefont {G.}~\bibnamefont {Zhu}}, \bibinfo {author} {\bibfnamefont {Z.}~\bibnamefont {Wang}}, \bibinfo {author} {\bibfnamefont {W.}~\bibnamefont {Yi}}, \ and\ \bibinfo {author} {\bibfnamefont {P.}~\bibnamefont {Xue}},\ }\href {\doibase 10.1038/s41567-020-0836-6} {\bibfield  {journal} {\bibinfo  {journal} {Nat. Phys.}\ }\textbf {\bibinfo {volume} {16}},\ \bibinfo {pages} {761} (\bibinfo {year} {2020})}\BibitemShut {NoStop}%
\bibitem [{\citenamefont {Weidemann}\ \emph {et~al.}(2020)\citenamefont {Weidemann}, \citenamefont {Kremer}, \citenamefont {Helbig}, \citenamefont {Hofmann}, \citenamefont {Stegmaier}, \citenamefont {Greiter}, \citenamefont {Thomale},\ and\ \citenamefont {Szameit}}]{Weidemann-20-skin-exp}%
  \BibitemOpen
  \bibfield  {author} {\bibinfo {author} {\bibfnamefont {S.}~\bibnamefont {Weidemann}}, \bibinfo {author} {\bibfnamefont {M.}~\bibnamefont {Kremer}}, \bibinfo {author} {\bibfnamefont {T.}~\bibnamefont {Helbig}}, \bibinfo {author} {\bibfnamefont {T.}~\bibnamefont {Hofmann}}, \bibinfo {author} {\bibfnamefont {A.}~\bibnamefont {Stegmaier}}, \bibinfo {author} {\bibfnamefont {M.}~\bibnamefont {Greiter}}, \bibinfo {author} {\bibfnamefont {R.}~\bibnamefont {Thomale}}, \ and\ \bibinfo {author} {\bibfnamefont {A.}~\bibnamefont {Szameit}},\ }\href {\doibase 10.1126/science.aaz8727} {\bibfield  {journal} {\bibinfo  {journal} {Science}\ }\textbf {\bibinfo {volume} {368}},\ \bibinfo {pages} {311} (\bibinfo {year} {2020})}\BibitemShut {NoStop}%
\bibitem [{\citenamefont {Wang}\ \emph {et~al.}(2021{\natexlab{a}})\citenamefont {Wang}, \citenamefont {Dutt}, \citenamefont {Yang}, \citenamefont {Wojcik}, \citenamefont {Vučković},\ and\ \citenamefont {Fan}}]{Wang-21-1Dwind-exp}%
  \BibitemOpen
  \bibfield  {author} {\bibinfo {author} {\bibfnamefont {K.}~\bibnamefont {Wang}}, \bibinfo {author} {\bibfnamefont {A.}~\bibnamefont {Dutt}}, \bibinfo {author} {\bibfnamefont {K.~Y.}\ \bibnamefont {Yang}}, \bibinfo {author} {\bibfnamefont {C.~C.}\ \bibnamefont {Wojcik}}, \bibinfo {author} {\bibfnamefont {J.}~\bibnamefont {Vučković}}, \ and\ \bibinfo {author} {\bibfnamefont {S.}~\bibnamefont {Fan}},\ }\href {\doibase 10.1126/science.abf6568} {\bibfield  {journal} {\bibinfo  {journal} {Science}\ }\textbf {\bibinfo {volume} {371}},\ \bibinfo {pages} {1240} (\bibinfo {year} {2021}{\natexlab{a}})}\BibitemShut {NoStop}%
\bibitem [{\citenamefont {Zhang}\ \emph {et~al.}(2021{\natexlab{a}})\citenamefont {Zhang}, \citenamefont {Ouyang}, \citenamefont {Huang}, \citenamefont {Wang}, \citenamefont {Zhang}, \citenamefont {Yu}, \citenamefont {Chang}, \citenamefont {Liu}, \citenamefont {Deng},\ and\ \citenamefont {Duan}}]{Zhang-21-1Dwind-exp}%
  \BibitemOpen
  \bibfield  {author} {\bibinfo {author} {\bibfnamefont {W.}~\bibnamefont {Zhang}}, \bibinfo {author} {\bibfnamefont {X.}~\bibnamefont {Ouyang}}, \bibinfo {author} {\bibfnamefont {X.}~\bibnamefont {Huang}}, \bibinfo {author} {\bibfnamefont {X.}~\bibnamefont {Wang}}, \bibinfo {author} {\bibfnamefont {H.}~\bibnamefont {Zhang}}, \bibinfo {author} {\bibfnamefont {Y.}~\bibnamefont {Yu}}, \bibinfo {author} {\bibfnamefont {X.}~\bibnamefont {Chang}}, \bibinfo {author} {\bibfnamefont {Y.}~\bibnamefont {Liu}}, \bibinfo {author} {\bibfnamefont {D.-L.}\ \bibnamefont {Deng}}, \ and\ \bibinfo {author} {\bibfnamefont {L.-M.}\ \bibnamefont {Duan}},\ }\href {\doibase 10.1103/PhysRevLett.127.090501} {\bibfield  {journal} {\bibinfo  {journal} {Phys. Rev. Lett.}\ }\textbf {\bibinfo {volume} {127}},\ \bibinfo {pages} {090501} (\bibinfo {year} {2021}{\natexlab{a}})}\BibitemShut {NoStop}%
\bibitem [{\citenamefont {Zhang}\ \emph {et~al.}(2021{\natexlab{b}})\citenamefont {Zhang}, \citenamefont {Tian}, \citenamefont {Jiang}, \citenamefont {Lu},\ and\ \citenamefont {Chen}}]{Zhang-21-skin-exp}%
  \BibitemOpen
  \bibfield  {author} {\bibinfo {author} {\bibfnamefont {X.}~\bibnamefont {Zhang}}, \bibinfo {author} {\bibfnamefont {Y.}~\bibnamefont {Tian}}, \bibinfo {author} {\bibfnamefont {J.-H.}\ \bibnamefont {Jiang}}, \bibinfo {author} {\bibfnamefont {M.-H.}\ \bibnamefont {Lu}}, \ and\ \bibinfo {author} {\bibfnamefont {Y.-F.}\ \bibnamefont {Chen}},\ }\href {\doibase 10.1038/s41467-021-25716-y} {\bibfield  {journal} {\bibinfo  {journal} {Nat. Commun.}\ }\textbf {\bibinfo {volume} {12}},\ \bibinfo {pages} {5377} (\bibinfo {year} {2021}{\natexlab{b}})}\BibitemShut {NoStop}%
\bibitem [{\citenamefont {Zou}\ \emph {et~al.}(2021)\citenamefont {Zou}, \citenamefont {Chen}, \citenamefont {He}, \citenamefont {Bao}, \citenamefont {Lee}, \citenamefont {Sun},\ and\ \citenamefont {Zhang}}]{Zou-21}%
  \BibitemOpen
  \bibfield  {author} {\bibinfo {author} {\bibfnamefont {D.}~\bibnamefont {Zou}}, \bibinfo {author} {\bibfnamefont {T.}~\bibnamefont {Chen}}, \bibinfo {author} {\bibfnamefont {W.}~\bibnamefont {He}}, \bibinfo {author} {\bibfnamefont {J.}~\bibnamefont {Bao}}, \bibinfo {author} {\bibfnamefont {C.~H.}\ \bibnamefont {Lee}}, \bibinfo {author} {\bibfnamefont {H.}~\bibnamefont {Sun}}, \ and\ \bibinfo {author} {\bibfnamefont {X.}~\bibnamefont {Zhang}},\ }\href {\doibase 10.1038/s41467-021-26414-5} {\bibfield  {journal} {\bibinfo  {journal} {Nat. Commun.}\ }\textbf {\bibinfo {volume} {12}},\ \bibinfo {pages} {7201} (\bibinfo {year} {2021})}\BibitemShut {NoStop}%
\bibitem [{\citenamefont {Palacios}\ \emph {et~al.}(2021)\citenamefont {Palacios}, \citenamefont {Tchoumakov}, \citenamefont {Guix}, \citenamefont {Pagonabarraga}, \citenamefont {S{\'a}nchez},\ and\ \citenamefont {G~Grushin}}]{Palacios-21-skin-exp}%
  \BibitemOpen
  \bibfield  {author} {\bibinfo {author} {\bibfnamefont {L.~S.}\ \bibnamefont {Palacios}}, \bibinfo {author} {\bibfnamefont {S.}~\bibnamefont {Tchoumakov}}, \bibinfo {author} {\bibfnamefont {M.}~\bibnamefont {Guix}}, \bibinfo {author} {\bibfnamefont {I.}~\bibnamefont {Pagonabarraga}}, \bibinfo {author} {\bibfnamefont {S.}~\bibnamefont {S{\'a}nchez}}, \ and\ \bibinfo {author} {\bibfnamefont {A.}~\bibnamefont {G~Grushin}},\ }\href {\doibase 10.1038/s41467-021-24948-2} {\bibfield  {journal} {\bibinfo  {journal} {Nat. Commun.}\ }\textbf {\bibinfo {volume} {12}},\ \bibinfo {pages} {4691} (\bibinfo {year} {2021})}\BibitemShut {NoStop}%
\bibitem [{\citenamefont {Wang}\ \emph {et~al.}(2021{\natexlab{b}})\citenamefont {Wang}, \citenamefont {Dutt}, \citenamefont {Wojcik},\ and\ \citenamefont {Fan}}]{Wang-21-Braid-exp}%
  \BibitemOpen
  \bibfield  {author} {\bibinfo {author} {\bibfnamefont {K.}~\bibnamefont {Wang}}, \bibinfo {author} {\bibfnamefont {A.}~\bibnamefont {Dutt}}, \bibinfo {author} {\bibfnamefont {C.~C.}\ \bibnamefont {Wojcik}}, \ and\ \bibinfo {author} {\bibfnamefont {S.}~\bibnamefont {Fan}},\ }\href {\doibase 10.1038/s41586-021-03848-x} {\bibfield  {journal} {\bibinfo  {journal} {Nature}\ }\textbf {\bibinfo {volume} {598}},\ \bibinfo {pages} {59} (\bibinfo {year} {2021}{\natexlab{b}})}\BibitemShut {NoStop}%
\bibitem [{\citenamefont {Hatsugai}(1993)}]{Hatsugai-93}%
  \BibitemOpen
  \bibfield  {author} {\bibinfo {author} {\bibfnamefont {Y.}~\bibnamefont {Hatsugai}},\ }\href {\doibase 10.1103/PhysRevLett.71.3697} {\bibfield  {journal} {\bibinfo  {journal} {Phys. Rev. Lett.}\ }\textbf {\bibinfo {volume} {71}},\ \bibinfo {pages} {3697} (\bibinfo {year} {1993})}\BibitemShut {NoStop}%
\bibitem [{\citenamefont {Tang}\ \emph {et~al.}(2020)\citenamefont {Tang}, \citenamefont {Zhang}, \citenamefont {Zhang},\ and\ \citenamefont {Zhang}}]{Tang-20}%
  \BibitemOpen
  \bibfield  {author} {\bibinfo {author} {\bibfnamefont {L.-Z.}\ \bibnamefont {Tang}}, \bibinfo {author} {\bibfnamefont {L.-F.}\ \bibnamefont {Zhang}}, \bibinfo {author} {\bibfnamefont {G.-Q.}\ \bibnamefont {Zhang}}, \ and\ \bibinfo {author} {\bibfnamefont {D.-W.}\ \bibnamefont {Zhang}},\ }\href {\doibase 10.1103/PhysRevA.101.063612} {\bibfield  {journal} {\bibinfo  {journal} {Phys. Rev. A}\ }\textbf {\bibinfo {volume} {101}},\ \bibinfo {pages} {063612} (\bibinfo {year} {2020})}\BibitemShut {NoStop}%
\bibitem [{\citenamefont {Liu}\ \emph {et~al.}(2020)\citenamefont {Liu}, \citenamefont {Su}, \citenamefont {Zhang},\ and\ \citenamefont {Jiang}}]{Liu-20}%
  \BibitemOpen
  \bibfield  {author} {\bibinfo {author} {\bibfnamefont {H.}~\bibnamefont {Liu}}, \bibinfo {author} {\bibfnamefont {Z.}~\bibnamefont {Su}}, \bibinfo {author} {\bibfnamefont {Z.-Q.}\ \bibnamefont {Zhang}}, \ and\ \bibinfo {author} {\bibfnamefont {H.}~\bibnamefont {Jiang}},\ }\href {\doibase 10.1088/1674-1056/ab8201} {\bibfield  {journal} {\bibinfo  {journal} {Chin. Phys. B}\ }\textbf {\bibinfo {volume} {29}},\ \bibinfo {pages} {050502} (\bibinfo {year} {2020})}\BibitemShut {NoStop}%
\bibitem [{\citenamefont {Liu}\ \emph {et~al.}(2021)\citenamefont {Liu}, \citenamefont {Zhou}, \citenamefont {Wu}, \citenamefont {Zhang},\ and\ \citenamefont {Jiang}}]{Liu-21}%
  \BibitemOpen
  \bibfield  {author} {\bibinfo {author} {\bibfnamefont {H.}~\bibnamefont {Liu}}, \bibinfo {author} {\bibfnamefont {J.-K.}\ \bibnamefont {Zhou}}, \bibinfo {author} {\bibfnamefont {B.-L.}\ \bibnamefont {Wu}}, \bibinfo {author} {\bibfnamefont {Z.-Q.}\ \bibnamefont {Zhang}}, \ and\ \bibinfo {author} {\bibfnamefont {H.}~\bibnamefont {Jiang}},\ }\href {\doibase 10.1103/PhysRevB.103.224203} {\bibfield  {journal} {\bibinfo  {journal} {Phys. Rev. B}\ }\textbf {\bibinfo {volume} {103}},\ \bibinfo {pages} {224203} (\bibinfo {year} {2021})}\BibitemShut {NoStop}%
\bibitem [{\citenamefont {Ghorashi}\ \emph {et~al.}(2021{\natexlab{a}})\citenamefont {Ghorashi}, \citenamefont {Li}, \citenamefont {Sato},\ and\ \citenamefont {Hughes}}]{Ghorashi-HOD-21}%
  \BibitemOpen
  \bibfield  {author} {\bibinfo {author} {\bibfnamefont {S.~A.~A.}\ \bibnamefont {Ghorashi}}, \bibinfo {author} {\bibfnamefont {T.}~\bibnamefont {Li}}, \bibinfo {author} {\bibfnamefont {M.}~\bibnamefont {Sato}}, \ and\ \bibinfo {author} {\bibfnamefont {T.~L.}\ \bibnamefont {Hughes}},\ }\href {\doibase 10.1103/PhysRevB.104.L161116} {\bibfield  {journal} {\bibinfo  {journal} {Phys. Rev. B}\ }\textbf {\bibinfo {volume} {104}},\ \bibinfo {pages} {L161116} (\bibinfo {year} {2021}{\natexlab{a}})}\BibitemShut {NoStop}%
\bibitem [{\citenamefont {Ghorashi}\ \emph {et~al.}(2021{\natexlab{b}})\citenamefont {Ghorashi}, \citenamefont {Li},\ and\ \citenamefont {Sato}}]{Ghorashi-HOW-21}%
  \BibitemOpen
  \bibfield  {author} {\bibinfo {author} {\bibfnamefont {S.~A.~A.}\ \bibnamefont {Ghorashi}}, \bibinfo {author} {\bibfnamefont {T.}~\bibnamefont {Li}}, \ and\ \bibinfo {author} {\bibfnamefont {M.}~\bibnamefont {Sato}},\ }\href {\doibase 10.1103/PhysRevB.104.L161117} {\bibfield  {journal} {\bibinfo  {journal} {Phys. Rev. B}\ }\textbf {\bibinfo {volume} {104}},\ \bibinfo {pages} {L161117} (\bibinfo {year} {2021}{\natexlab{b}})}\BibitemShut {NoStop}%
\bibitem [{Note1()}]{Note1}%
  \BibitemOpen
  \bibinfo {note} {Skin modes are obtained using GBZ and thus bulk modes \cite {YW-18-SSH}, while surface states are not obtained using GBZ and thus not bulk modes}\BibitemShut {NoStop}%
\bibitem [{\citenamefont {Mondragon-Shem}\ \emph {et~al.}(2014)\citenamefont {Mondragon-Shem}, \citenamefont {Hughes}, \citenamefont {Song},\ and\ \citenamefont {Prodan}}]{Mondragon-Shem-13}%
  \BibitemOpen
  \bibfield  {author} {\bibinfo {author} {\bibfnamefont {I.}~\bibnamefont {Mondragon-Shem}}, \bibinfo {author} {\bibfnamefont {T.~L.}\ \bibnamefont {Hughes}}, \bibinfo {author} {\bibfnamefont {J.}~\bibnamefont {Song}}, \ and\ \bibinfo {author} {\bibfnamefont {E.}~\bibnamefont {Prodan}},\ }\href {\doibase 10.1103/PhysRevLett.113.046802} {\bibfield  {journal} {\bibinfo  {journal} {Phys. Rev. Lett.}\ }\textbf {\bibinfo {volume} {113}},\ \bibinfo {pages} {046802} (\bibinfo {year} {2014})}\BibitemShut {NoStop}%
\bibitem [{\citenamefont {Song}\ and\ \citenamefont {Prodan}(2014)}]{Prodan-14}%
  \BibitemOpen
  \bibfield  {author} {\bibinfo {author} {\bibfnamefont {J.}~\bibnamefont {Song}}\ and\ \bibinfo {author} {\bibfnamefont {E.}~\bibnamefont {Prodan}},\ }\href {\doibase 10.1103/PhysRevB.89.224203} {\bibfield  {journal} {\bibinfo  {journal} {Phys. Rev. B}\ }\textbf {\bibinfo {volume} {89}},\ \bibinfo {pages} {224203} (\bibinfo {year} {2014})}\BibitemShut {NoStop}%
\bibitem [{\citenamefont {Prodan}\ and\ \citenamefont {Schulz-Baldes}(2016)}]{Prodan-16}%
  \BibitemOpen
  \bibfield  {author} {\bibinfo {author} {\bibfnamefont {E.}~\bibnamefont {Prodan}}\ and\ \bibinfo {author} {\bibfnamefont {H.}~\bibnamefont {Schulz-Baldes}},\ }\href {\doibase https://doi.org/10.1016/j.jfa.2016.06.001} {\bibfield  {journal} {\bibinfo  {journal} {J. Funct. Anal.}\ }\textbf {\bibinfo {volume} {271}},\ \bibinfo {pages} {1150} (\bibinfo {year} {2016})}\BibitemShut {NoStop}%
\bibitem [{\citenamefont {Katsura}\ and\ \citenamefont {Koma}(2018)}]{Katsura-Koma-18}%
  \BibitemOpen
  \bibfield  {author} {\bibinfo {author} {\bibfnamefont {H.}~\bibnamefont {Katsura}}\ and\ \bibinfo {author} {\bibfnamefont {T.}~\bibnamefont {Koma}},\ }\href {https://doi.org/10.1063/1.5026964} {\bibfield  {journal} {\bibinfo  {journal} {J. Math. Phys.}\ }\textbf {\bibinfo {volume} {59}},\ \bibinfo {pages} {031903} (\bibinfo {year} {2018})}\BibitemShut {NoStop}%
\bibitem [{\citenamefont {Bellissard}\ \emph {et~al.}(1994)\citenamefont {Bellissard}, \citenamefont {van Elst},\ and\ \citenamefont {Schulz‐~Baldes}}]{Bellissard-94}%
  \BibitemOpen
  \bibfield  {author} {\bibinfo {author} {\bibfnamefont {J.}~\bibnamefont {Bellissard}}, \bibinfo {author} {\bibfnamefont {A.}~\bibnamefont {van Elst}}, \ and\ \bibinfo {author} {\bibfnamefont {H.}~\bibnamefont {Schulz‐~Baldes}},\ }\href {\doibase 10.1063/1.530758} {\bibfield  {journal} {\bibinfo  {journal} {J. Math. Phys.}\ }\textbf {\bibinfo {volume} {35}},\ \bibinfo {pages} {5373} (\bibinfo {year} {1994})}\BibitemShut {NoStop}%
\bibitem [{\citenamefont {Kitaev}(2006)}]{Kitaev-06}%
  \BibitemOpen
  \bibfield  {author} {\bibinfo {author} {\bibfnamefont {A.~Y.}\ \bibnamefont {Kitaev}},\ }\href {https://doi.org/10.1016/j.aop.2005.10.005} {\bibfield  {journal} {\bibinfo  {journal} {Ann. Phys. (N.Y.)}\ }\textbf {\bibinfo {volume} {321}},\ \bibinfo {pages} {2} (\bibinfo {year} {2006})}\BibitemShut {NoStop}%
\bibitem [{Note2()}]{Note2}%
  \BibitemOpen
  \bibinfo {note} {Note that the isotropic structure of $\mp e^{\pm ik_y}$ in the complex energy plane causes highly degenerated in-gap skin modes at $E=0$.}\BibitemShut {Stop}%
\bibitem [{Note3()}]{Note3}%
  \BibitemOpen
  \bibinfo {note} {See Supplemental Material, which includes Refs.~\cite {Prodan-10,Katsura-Koma-16,Akagi-17,Yang-Schnyder-Hu-Chiu}, for a demonstration of how the real-space 3D winding number changes with different boundary conditions, a proof of the BBC for point-gap topological phases in all 38-fold symmetry classes, and a connection between intrinsic point-gap topological phases in the OBC and a single exceptional point.}\BibitemShut {Stop}%
\bibitem [{\citenamefont {Yoshida}\ \emph {et~al.}(2020{\natexlab{b}})\citenamefont {Yoshida}, \citenamefont {Peters}, \citenamefont {Kawakami},\ and\ \citenamefont {Hatsugai}}]{Yoshida-dagger-symmetry}%
  \BibitemOpen
  \bibfield  {author} {\bibinfo {author} {\bibfnamefont {T.}~\bibnamefont {Yoshida}}, \bibinfo {author} {\bibfnamefont {R.}~\bibnamefont {Peters}}, \bibinfo {author} {\bibfnamefont {N.}~\bibnamefont {Kawakami}}, \ and\ \bibinfo {author} {\bibfnamefont {Y.}~\bibnamefont {Hatsugai}},\ }\href {\doibase 10.1093/ptep/ptaa059} {\bibfield  {journal} {\bibinfo  {journal} {Prog. Theor. Exp. Phys.}\ }\textbf {\bibinfo {volume} {2020}},\ \bibinfo {pages} {12A109} (\bibinfo {year} {2020}{\natexlab{b}})}\BibitemShut {NoStop}%
\bibitem [{\citenamefont {Qi}\ \emph {et~al.}(2008)\citenamefont {Qi}, \citenamefont {Hughes},\ and\ \citenamefont {Zhang}}]{QHZ-08}%
  \BibitemOpen
  \bibfield  {author} {\bibinfo {author} {\bibfnamefont {X.-L.}\ \bibnamefont {Qi}}, \bibinfo {author} {\bibfnamefont {T.~L.}\ \bibnamefont {Hughes}}, \ and\ \bibinfo {author} {\bibfnamefont {S.-C.}\ \bibnamefont {Zhang}},\ }\href {\doibase 10.1103/PhysRevB.78.195424} {\bibfield  {journal} {\bibinfo  {journal} {Phys. Rev. B}\ }\textbf {\bibinfo {volume} {78}},\ \bibinfo {pages} {195424} (\bibinfo {year} {2008})}\BibitemShut {NoStop}%
\bibitem [{\citenamefont {Prodan}\ \emph {et~al.}(2010)\citenamefont {Prodan}, \citenamefont {Hughes},\ and\ \citenamefont {Bernevig}}]{Prodan-10}%
  \BibitemOpen
  \bibfield  {author} {\bibinfo {author} {\bibfnamefont {E.}~\bibnamefont {Prodan}}, \bibinfo {author} {\bibfnamefont {T.~L.}\ \bibnamefont {Hughes}}, \ and\ \bibinfo {author} {\bibfnamefont {B.~A.}\ \bibnamefont {Bernevig}},\ }\href {\doibase 10.1103/PhysRevLett.105.115501} {\bibfield  {journal} {\bibinfo  {journal} {Phys. Rev. Lett.}\ }\textbf {\bibinfo {volume} {105}},\ \bibinfo {pages} {115501} (\bibinfo {year} {2010})}\BibitemShut {NoStop}%
\bibitem [{\citenamefont {Katsura}\ and\ \citenamefont {Koma}(2016)}]{Katsura-Koma-16}%
  \BibitemOpen
  \bibfield  {author} {\bibinfo {author} {\bibfnamefont {H.}~\bibnamefont {Katsura}}\ and\ \bibinfo {author} {\bibfnamefont {T.}~\bibnamefont {Koma}},\ }\href {\doibase 10.1063/1.4942494} {\bibfield  {journal} {\bibinfo  {journal} {J. Math. Phys.}\ }\textbf {\bibinfo {volume} {57}},\ \bibinfo {pages} {021903} (\bibinfo {year} {2016})}\BibitemShut {NoStop}%
\bibitem [{\citenamefont {Akagi}\ \emph {et~al.}(2017)\citenamefont {Akagi}, \citenamefont {Katsura},\ and\ \citenamefont {Koma}}]{Akagi-17}%
  \BibitemOpen
  \bibfield  {author} {\bibinfo {author} {\bibfnamefont {Y.}~\bibnamefont {Akagi}}, \bibinfo {author} {\bibfnamefont {H.}~\bibnamefont {Katsura}}, \ and\ \bibinfo {author} {\bibfnamefont {T.}~\bibnamefont {Koma}},\ }\href {https://doi.org/10.7566/JPSJ.86.123710} {\bibfield  {journal} {\bibinfo  {journal} {J. Phys. Soc. Jpn.}\ }\textbf {\bibinfo {volume} {86}},\ \bibinfo {pages} {123710} (\bibinfo {year} {2017})}\BibitemShut {NoStop}%
\bibitem [{\citenamefont {Yang}\ \emph {et~al.}(2021)\citenamefont {Yang}, \citenamefont {Schnyder}, \citenamefont {Hu},\ and\ \citenamefont {Chiu}}]{Yang-Schnyder-Hu-Chiu}%
  \BibitemOpen
  \bibfield  {author} {\bibinfo {author} {\bibfnamefont {Z.}~\bibnamefont {Yang}}, \bibinfo {author} {\bibfnamefont {A.~P.}\ \bibnamefont {Schnyder}}, \bibinfo {author} {\bibfnamefont {J.}~\bibnamefont {Hu}}, \ and\ \bibinfo {author} {\bibfnamefont {C.-K.}\ \bibnamefont {Chiu}},\ }\href {\doibase 10.1103/PhysRevLett.126.086401} {\bibfield  {journal} {\bibinfo  {journal} {Phys. Rev. Lett.}\ }\textbf {\bibinfo {volume} {126}},\ \bibinfo {pages} {086401} (\bibinfo {year} {2021})}\BibitemShut {NoStop}%
\end{thebibliography}%

%%%%%appendix
\clearpage
\onecolumngrid 

%%%%
\newtheorem{theorem}{Theorem}[section]
\newtheorem{lemma}{Lemma}
\newtheorem{cor}{Corollary}

\begin{center}
{\bf \large Supplemental Material}
\end{center}

\makeatletter
\renewcommand{\theequation}{S\arabic{equation}}
\renewcommand{\thefigure}{S\arabic{figure}}
\renewcommand{\thetable}{S\arabic{table}}
\renewcommand{\thesection}{S\arabic{section}}
\renewcommand{\thesubsection}{S\arabic{section}.\arabic{subsection}}
\renewcommand*{\p@subsection}{}
\renewcommand{\thesubsubsection}{S\arabic{section}.\arabic{subsection}.\arabic{subsubsection}}
\renewcommand*{\p@subsubsection}{}
\makeatother

\renewcommand{\theenumi}{\alph{enumi}}

\setcounter{secnumdepth}{3}
\setcounter{equation}{0}
\setcounter{figure}{0}
\setcounter{table}{0}
\setcounter{section}{0}

\section{From PBC to \lowercase{y}OBC in Eq.~(8)}

\begin{figure}[ht]
\centering
\includegraphics[keepaspectratio,scale=0.25]{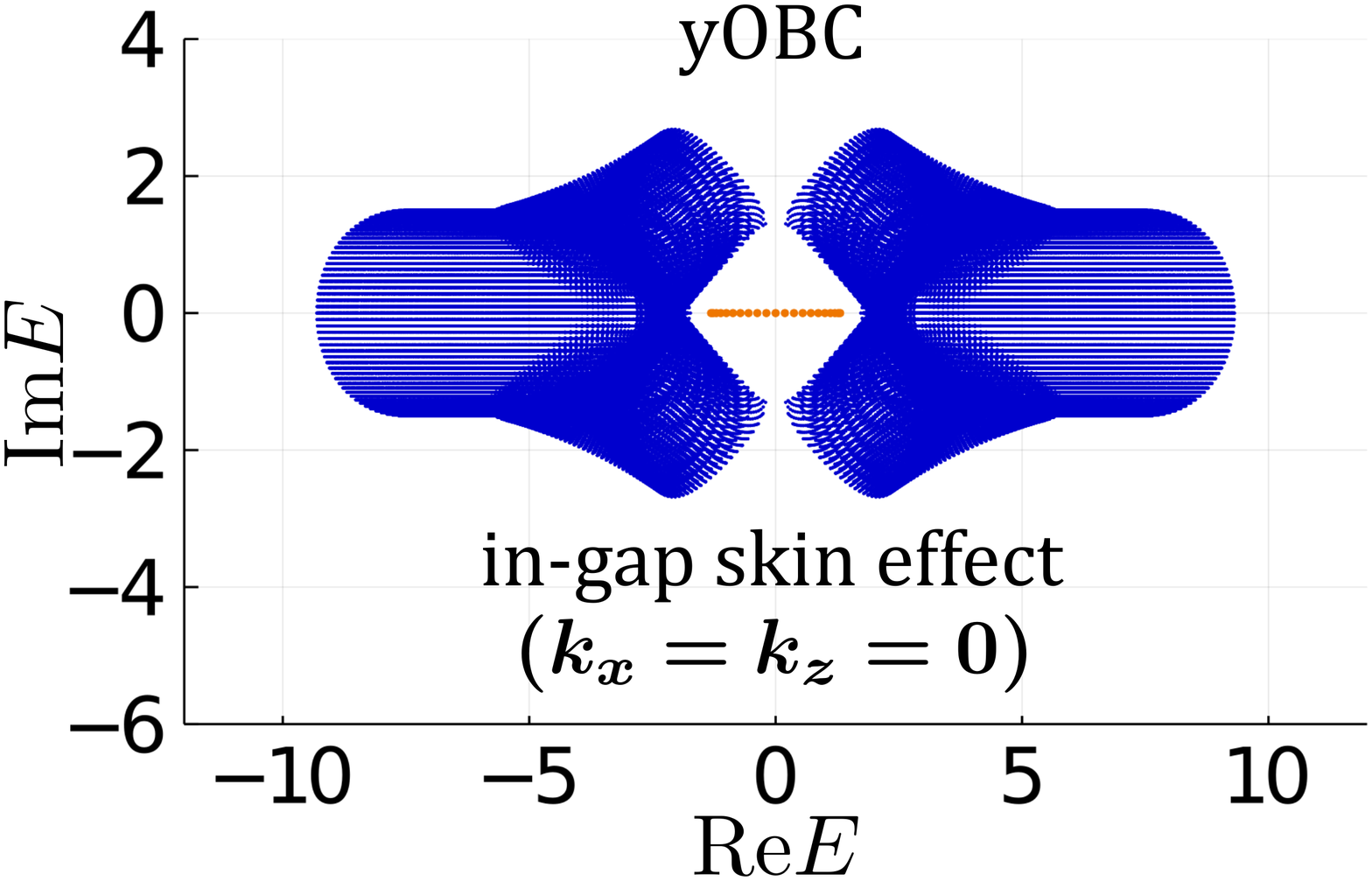}
\caption{The yOBC spectrum of the model in Eq.~(\ref{eq:ETI2-SM}). The system size in the $y$ direction is $L_y=21$ and the momentum resolution in the $k_{x,z}$-directions is taken as $\Delta k_{x,z} = 2\pi/100$. No surface states appear, and instead, the skin modes appear.}
\label{fig:in-gap_skin_mode-SM}
\end{figure}

We here demonstrate that the real-space 3D winding number detects a topological phase transition under continuous change of the boundary condition. Let us start with the Hamiltonian in Eq.~(8) of the main text:
\begin{align}
  H(\bm{k}) = \sin{k_x}\sigma_{x} + 2\sin{k_z}\sigma_{y} + 2\left(2 - \sum_{i=x,y,z}\cos{k_i}\right)\sigma_{z}
+ \dfrac{3}{2}i(\sin{k_y}+\sin{k_z})\sigma_{0}.
\label{eq:ETI2-SM}
\end{align}
While this model has the non-trivial 3D winding number $-1$ in a region including $E=0$ under the full PBCs, in-gap skin modes arise under the yOBC, instead of  surface states, as shown in Fig.~\ref{fig:in-gap_skin_mode-SM}. In the following, we will reveal that the absence of surface states can be understood as a topological phase transition under a continuous deformation of the boundary condition by using the real-space 3D winding number.

First, we note that the in-gap skin modes originate from modes with $(k_x, k_z)=(0,0)$, where the Hamiltonian in Eq.~(\ref{eq:ETI2-SM}) becomes
\begin{align}
  H(k_x=0,k_y,k_z=0) = -2\cos{k_y}\sigma_{z} +\dfrac{3}{2}i\sin{k_y}\sigma_{0}.
\label{eq:ETI2-1d_sector-SM}
\end{align}
This Hamiltonian gives elliptical complex spectra in the eigensector of $\sigma_z=\pm 1$, which have the energy winding numbers $\mp 1$ along the $k_y$ direction.
According to the general theory of NHSE~\cite{Zhang-19,OKSS-20}, the energy winding results in skin modes in Fig.~\ref{fig:in-gap_skin_mode-SM}. %Also note that, as explained in the main text, the in-gap skin modes are always accompanied by bulk modes covering the point-gapped region under high momentum resolutions (Fig.~\ref{fig:3dwinding-SM} (c)). Therefore, we cannot have a well-defined 3D winding number under the yOBC, as shown in Fig.~\ref{fig:3dwinding-SM} (c).

\begin{figure}[ht]
\RawFloats
\begin{minipage}[t]{0.32\hsize}
\centering
\includegraphics[keepaspectratio,width=\linewidth]{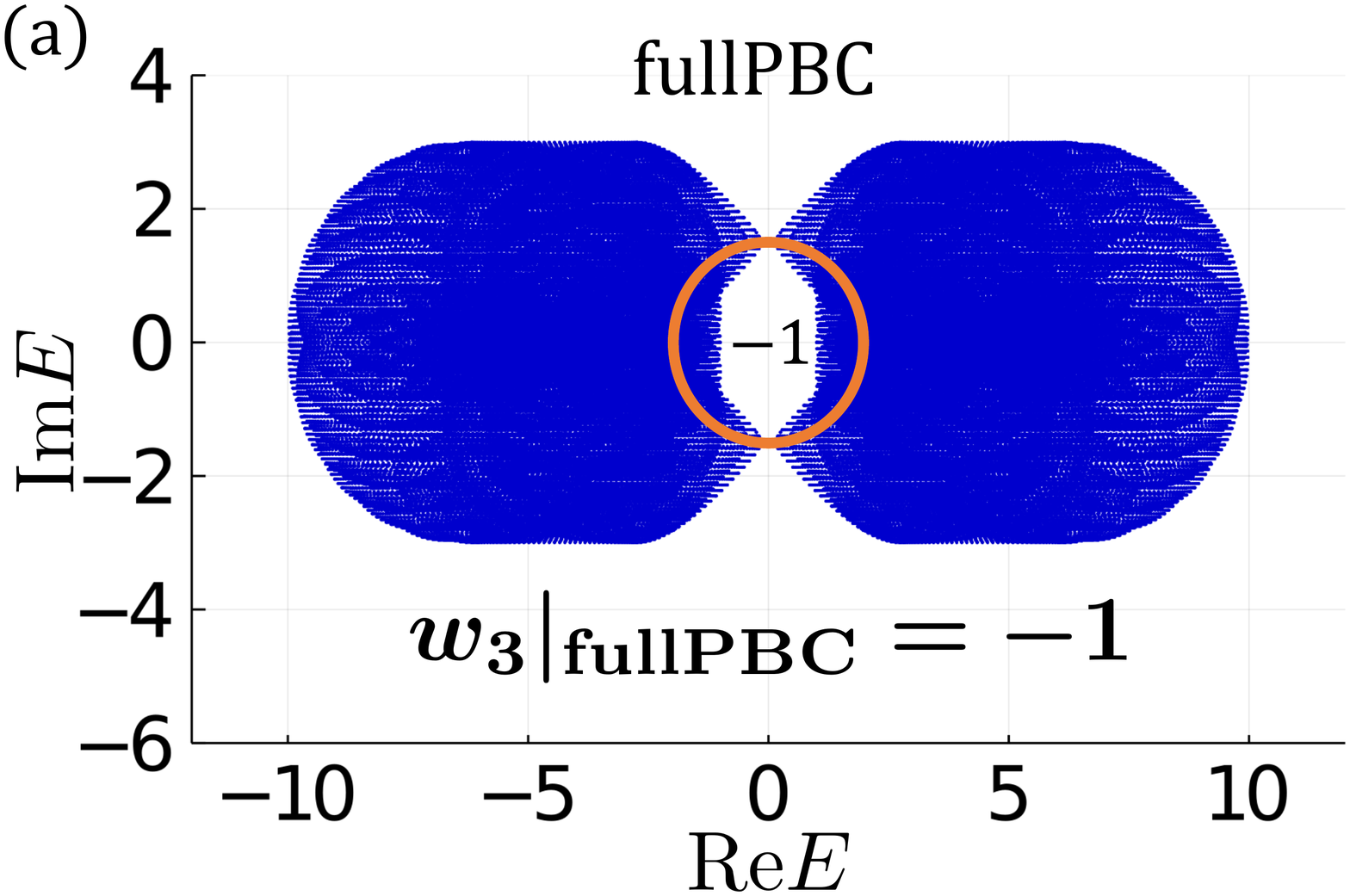}
\end{minipage}\hspace{0mm}
\begin{minipage}[t]{0.32\hsize}
\centering
\includegraphics[keepaspectratio,width=\linewidth]{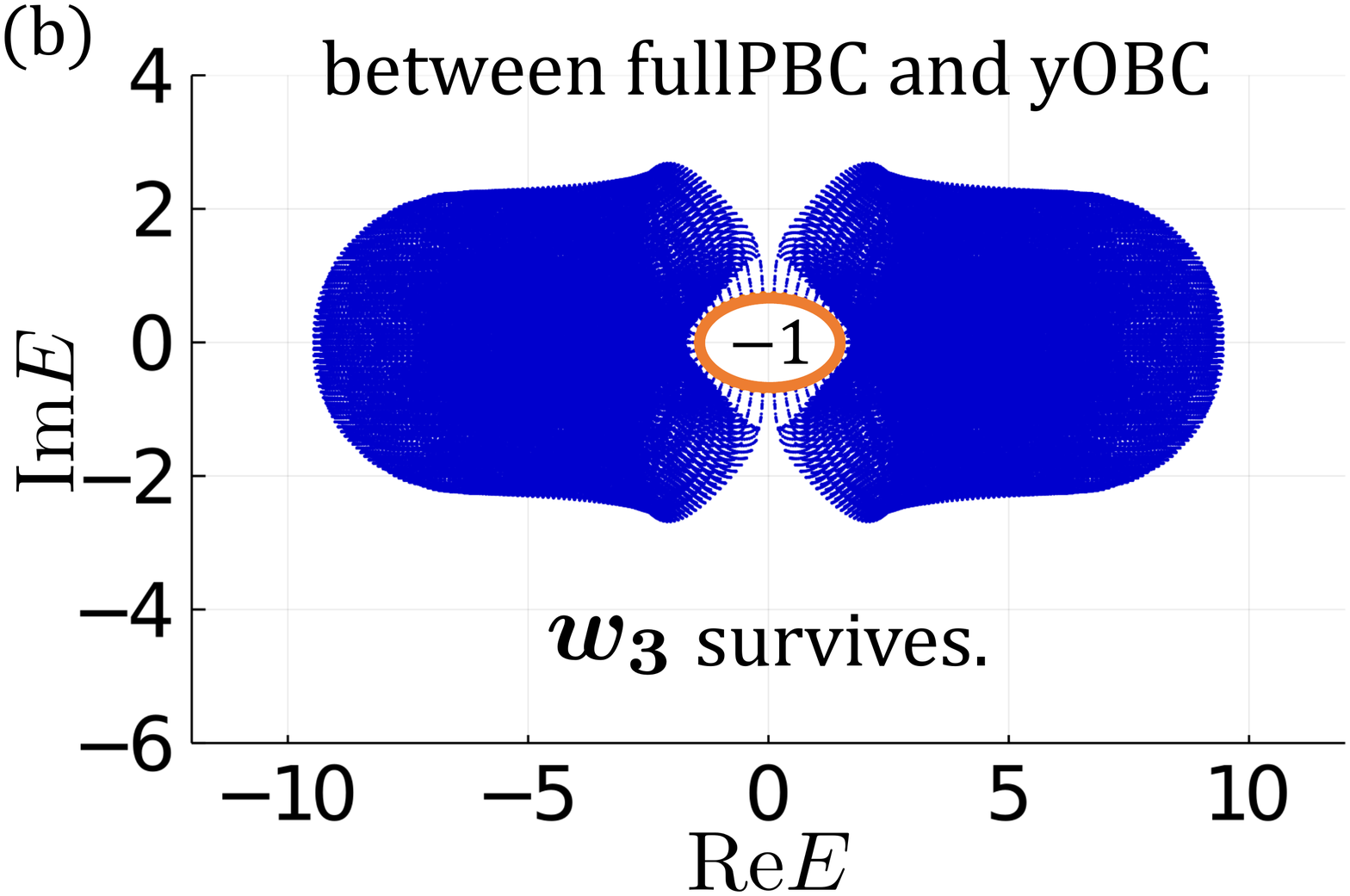}
\end{minipage}\hspace{0mm}
\begin{minipage}[t]{0.32\hsize}
\centering
\includegraphics[keepaspectratio,width=\linewidth]{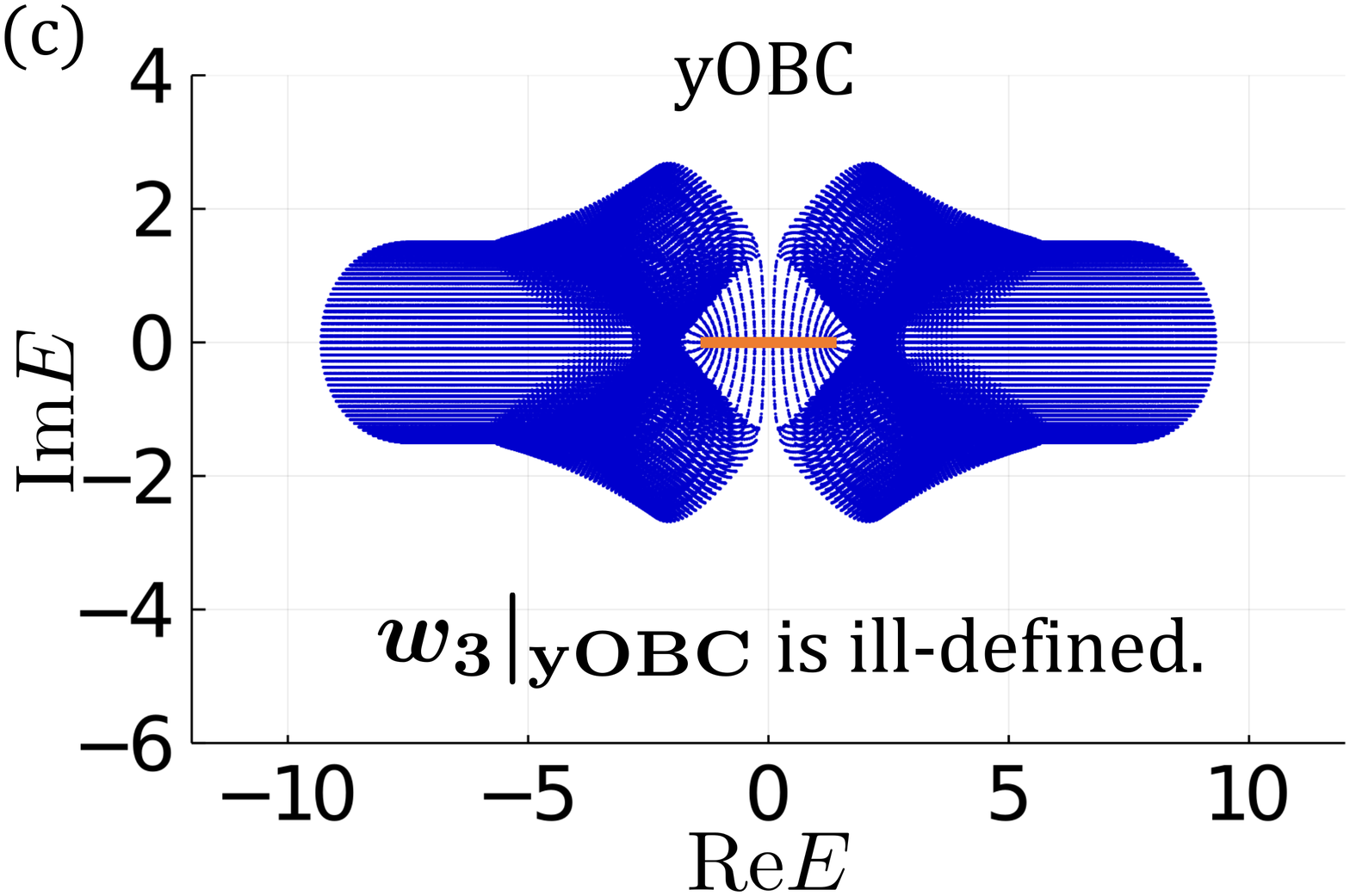}
\end{minipage}\hspace{0mm}
\begin{minipage}[t]{0.32\hsize}
\centering
\includegraphics[keepaspectratio,width=\linewidth]{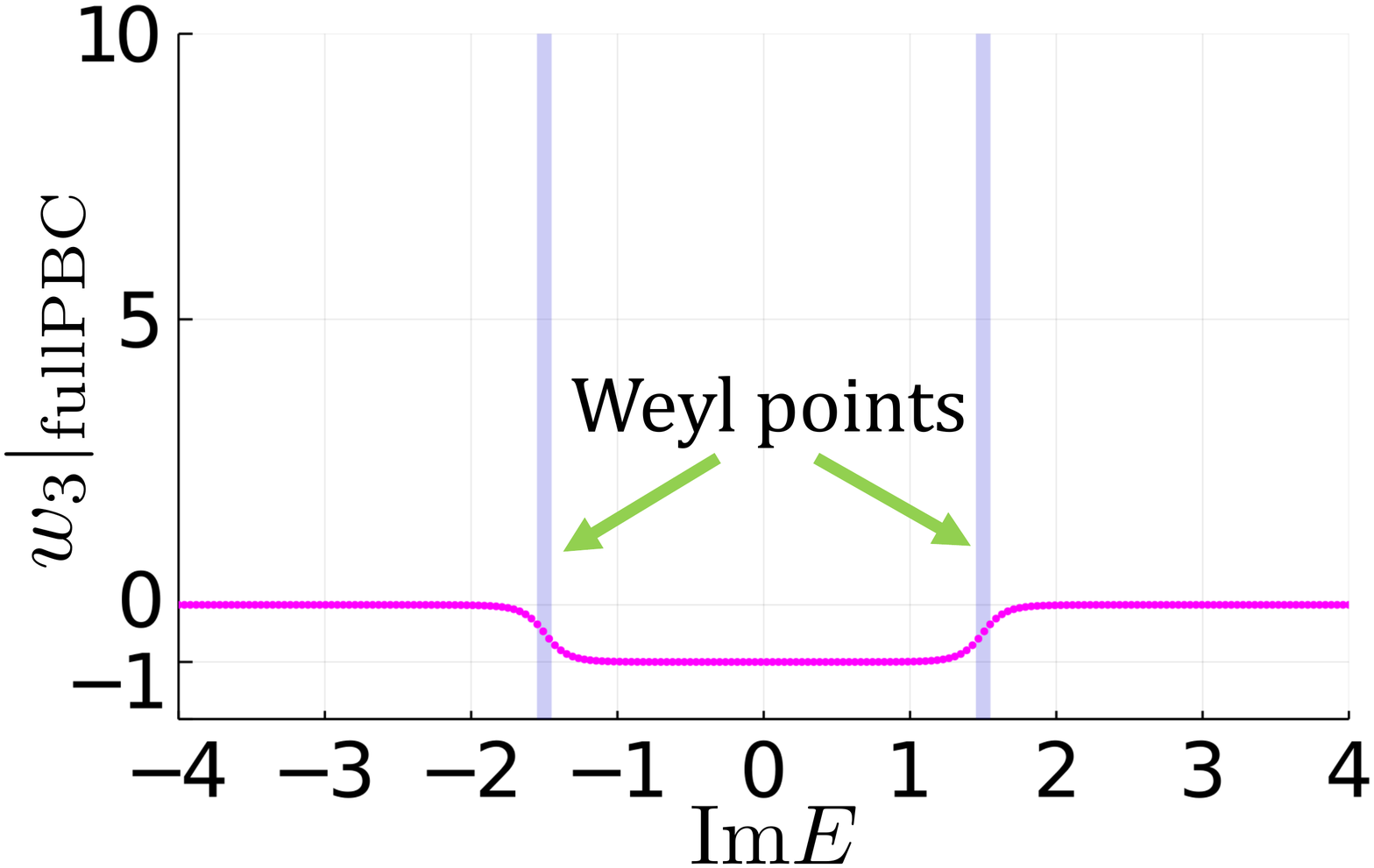}
\end{minipage}\hspace{0mm}
\begin{minipage}[t]{0.32\hsize}
\centering
\includegraphics[keepaspectratio,width=\linewidth]{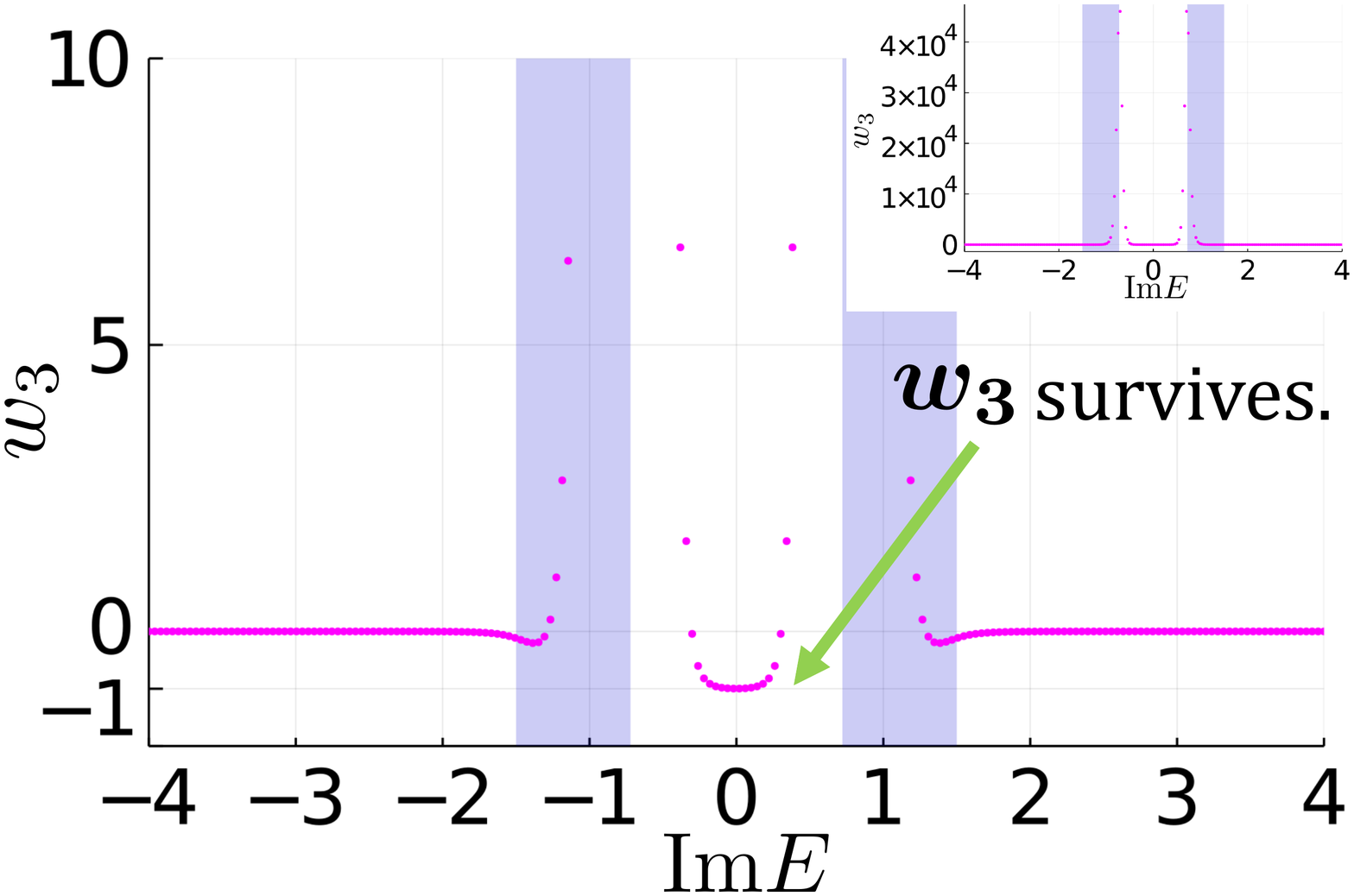}
\end{minipage}\hspace{0mm}
\begin{minipage}[t]{0.32\hsize}
\centering
\includegraphics[keepaspectratio,width=\linewidth]{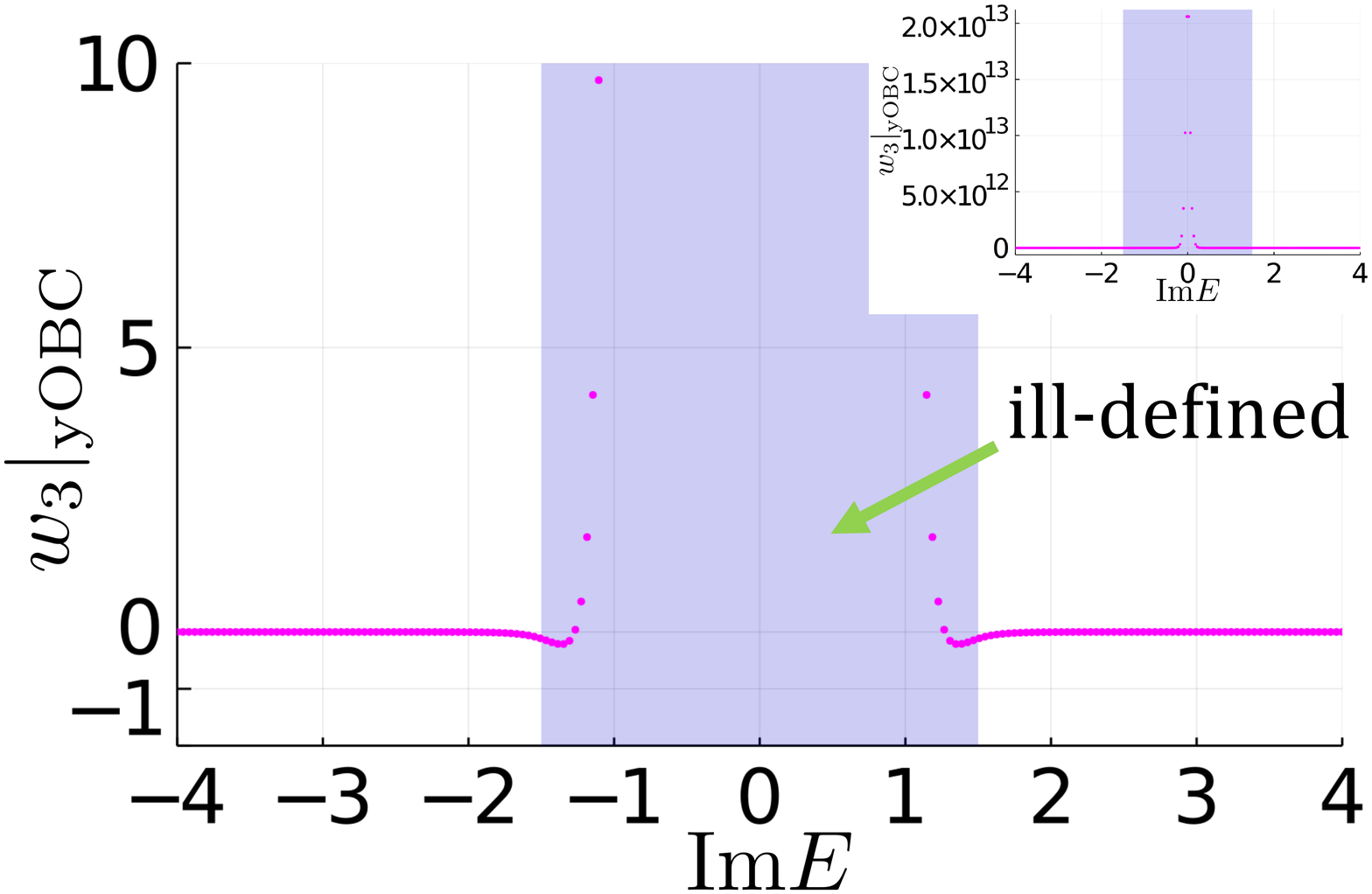}
\end{minipage}\hspace{0mm}
\caption{The changes of the spectrum~(top) and the real-space 3D winding number~(bottom) of the model in Eq.~(\ref{eq:ETI2-SM}) from under the full PBC to the yOBC. %The system size in y direction
$L_y$ is the same as that in Fig. \ref{fig:in-gap_skin_mode-SM}, but the momentum resolution around $(k_x,k_z)=(0,0)$ is much finer in the energy spectrum. The orange circles in the top figures represent the modes with $(k_x,k_z)=(0,0)$. 
In the bottom figures, the 3D winding numbers are calculated at $E=0+i{\rm Im}E$ with $-4<{\rm Im}E<4$. The purple shaded regions in the bottom figures correspond to point-gapless regions in the top figures.
%along the energy reference line.
(a) The full PBC. A point gap is open in the region containing $E=0$ with the nontrivial 3D winding number $-1$. The orange modes at $(k_x,k_z)=(0,0)$ form a loop and have a nontrivial 1D winding number in each eigensector of $\sigma_z=\pm 1$. 
(b) A boundary condition between the full PBC and the yOBC. The hopping amplitude between the $y=1$ sites and the $y=L_y$ sites is $0.00005$. The point gapped region containing $E=0$ shrinks, but the 3D winding number is $-1$ in the point gapped region. 
(c) The yOBC. The point-gapped region including $E=0$ is completely closed by the in-gap skin effect with %the in-gap skin modes at $(k_x,k_z)=(0,0)$ and 
the orange modes near $(k_x,k_z)=(0,0)$. The in-gap skin effect makes the 3D winding number ill-defined.
%The changes of the 3D winding number of the ETI in Eq.~(\ref{eq:ETI2-SM}) during the change from the full PBC to the yOBC. The energy references are $E_\text{p}=i\mathrm{Im}E_\text{p}$ with $-4<\mathrm{Im}E_\text{p}<4$. The labels (a)-(c) correspond to those in Fig. \ref{fig:yOBC2-SM}. (a) The 3D winding number under the full PBC. The red thin regions represent the energies at the Weyl points $(k_x,k_y,k_z)=(0,\pm\dfrac{\pi}{2},0)$. 
%(b) The 3D winding number under the condition between the full PBC and the yOBC. The red shaded regions represent the point-gap closed regions caused by the shrinking of spectra at $(k_x,k_z)=(0,0)$. The 3D winding number is still well-defined in a region where the point gap remains open. 
%(c) The 3D winding number under the yOBC. The in-gap skin modes make the 3D winding number ill-defined throughout the originally point-gapped region.} where that is non-trivial under the full PBC.
}
\label{fig:3dwinding-SM}
\end{figure}

In Fig.~\ref{fig:3dwinding-SM}, we show how the complex spectrum and the real-space 3D winding number changes under the continuous deformation of the boundary conditions.
Here we define the real-space 3D winding number as follows.
Taking the real-space representation of $H({\bm k})$ in the $y$-direction and performing the projection to the bulk, we have the bulk Hamiltonian ${\cal H}_{\rm bulk}(k_x,k_z)$.
Then, the real-space 3D winding number at $E$ is
\begin{align}
w_{3}=-\frac{i}{12\pi}\int_{BZ} d^2{\bm k}{\cal T}_y
[\varepsilon^{ijk}Q_i Q_j Q_k]    
\end{align}
with $Q_{i=x,z}=i({\cal H}_{\rm bulk}
-E)^{-1}\partial_{k_i}({\cal H}_{\rm bulk}-E)$ and 
$Q_y=({\cal H}_{\rm bulk}-E)^{-1}[Y, {\cal H}_{\rm bulk}-E]$, where $Y$ is the position operator of the $y$-coordinate, and ${\cal T}_y$ stands for the trace per unit length in the $y$-direction. 
Decreasing the hopping amplitudes between the $y=1$ sites and $y=L_y$ sites, we can change the boundary conditions from the full PBC to the yOBC smoothly. 
%As one decreases the hopping amplitudes,
During the process, the point-gapped region becomes smaller because of the shrinking of the modes with $(k_x,k_z)=(0,0)$ in the complex energy plane. Accordingly, we can confirm that the real-space 3D winding number changes from the well-defined value $-1$ to ill-defined ones in the original point-gapped region around $E=0$, as illustrated in Fig.~\ref{fig:3dwinding-SM}.  Therefore, we conclude that no surface states appear under the yOBC in Eq.~(\ref{eq:ETI2-SM}) as a result of the topological phase transition.

%As these calculations show, the real-space 3D winding number correctly captures the bulk topology in the imposed boundary condition. 

\section{Proof of the BBC for point-gap topological phases in AZ$^\dagger$ symmetry classes}
\label{sec:S2}

In this section, we give a proof of the BBC for point-gap topological phases with AZ$^\dagger$ symmetry. 
Below, we assume that the spatial dimension $d$ is $d=2,3$, but the generalization to higher dimensions is straightforward.

Our basic strategy of the proof is to use the additivity of topological phases in the framework of $K$-theory.
Since we can add the topological number of $H_1$ to that of $H_2$ by stacking $H_1$ and $H_2$ as
\begin{align}
\begin{pmatrix}
H_1 & 0 \\
0 & H_2
\end{pmatrix},
\end{align}
we can generate any topological phases by considering a proper stacks of a model with the minimal topological number.
From $K$-theory, we also know that  models with the same topological numbers can deform into each other without gap-closing, up to addition of topologically trivial Hamiltonians. 
Thus, once we can show the BBC for a model with the minimal topological number, the BBC holds generally. 

We can construct a model with the minimal topological number, which we call the minimal model,  as follows.
Except for 3D class CII$^\dagger$, we consider the following form of Hamiltonian, 
\begin{align}
  H_{\text{minimal}}(\bm{k}) &=  H_\text{SM}(\bm{k}) + i\sin{k_{d}}\Gamma_d,
\quad
  H_\text{SM}(\bm{k}) = \sum_{i=1}^{d-1}\sin{k_i}\Gamma_{i} + \left[(d-1) - \sum_{i=1}^{d}\cos{k_i}\right]\Gamma_{0},
\label{eq:simple model}
  \\
  E_\text{minimal}(\bm{k}) &= \pm \sqrt{\sum_{i=1}^{d-1}\left(\sin k_i\right)^2 + \left[(d-1)-\sum_{i=1}^{d} \cos k_i\right]^2} 
  +s\, i \sin k_{d},
  \label{eq:simple-energy}
\end{align}
where $d\ge 2$ is the spatial dimension, $\Gamma_{i=0,1,2,\dots,d-1}$ are Gamma
matrices that anti-commute with each other, and $\Gamma_d$ is a matrice that commutes with all $\Gamma_{i=0,1,2,\dots,d-1}$, and $\Gamma_{\mu}^2=1$ ($\mu=0,1,\dots,d$).
The sign factor $s$ in Eq.~(\ref{eq:simple-energy}) takes $s=+1$ ($s=\pm 1$) for $\Gamma_d=\hat{1}$ ($\Gamma_d\neq \hat{1}$).
Note that $H_\text{SM}(\bm{k})$ is the Hamiltonian of a Hermitian Weyl/Dirac semimetal, and 
$H_{\rm minimal}$ is straightforward generalization of Eq.~($1$) in the main text.
Since $H_{\text{minimal}}(\bm{k})$ commutes with $i\sin{k_d}\Gamma_{d}$, the non-Hermitian term $i\sin{k_d}\Gamma_{d}$ only gives a complex energy shift if we impose the PBC in the $x_d$-direction. % and other terms are all Hermitian.
Therefore, skin effects do not occur when we keep the PBC in the $x_d$-direction.

%Then, we see that the model show the point-gap bulk-boundary correspondence.
%The point-gap topological number is easily obtained by 

The point-gap topological number of $H_{\rm minimal}({\bm k})$ under the full PBC is given by the topological number of the doubled Hermitian Hamiltonian \cite{KSUS-19},
\begin{align}
\tilde{H}({\bm k})=
\begin{pmatrix}
0 & H_{\rm minimal}({\bm k})\\
H^\dagger_{\rm minimal}({\bm k}) & 0
\end{pmatrix}
=\sum_{i=1}^d\sin k_i\tilde{\Gamma}_i+ \left[(d-1)-\sum_{i=1}^{d} \cos k_i\right]\tilde{\Gamma}_0.
\label{eq:doubled_TI}
\end{align}
Here $\tilde{\Gamma}_{\mu=0,\dots, d}$ are the Gamma matrices in the doubled space, 
\begin{align}
\tilde{\Gamma}_{i=0,\dots,d-1}=
\begin{pmatrix}
0 &\Gamma_i\\
\Gamma_i & 0
\end{pmatrix},
\quad
\tilde{\Gamma}_d
=
\begin{pmatrix}
0 & i\Gamma_d\\
-i\Gamma_d & 0
\end{pmatrix},
\end{align}
which satisfy $\{\tilde{\Gamma}_\mu, \tilde{\Gamma}_\nu\}=\delta_{\mu,\nu}$.
We can easily evaluate the topological number of 
Eq.~(\ref{eq:doubled_TI}), namely the point-gap topological number of $H_{\rm minimal}({\bm k})$, because Eq.~(\ref{eq:doubled_TI}) is the standard massive Dirac Hamiltonian for topological insulators.
The obtained point-gap topological number coincides with that under the OBC in other than the $x_d$-direction, say the OBC in the $x_1$-direction (x$_1$OBC), as no skin effects occur. 
%Because of the absence of the skin effects, 
%Using the extended Nielsen-Ninomiya theorem~\cite{Bessho-Sato-20}, one can easily evaluate the point-gap topological number of $H_{\rm minimal}({\bm k})$ in Eq.(\ref{eq:simple model}).
%Futhermore, we can show that $H_{\text{minimal}}(\bm{k})$ has nontrivial point-gap topological number.
%The Hamiltonian has Weyl/Dirac points at the time-reversal momenta $\bm{k}=(0,0,...,\pm\dfrac{\pi}{2})$ with pure imaginary energy $E(\bm{k})=s \, i \sin k_d$.
%, which open a point-gap at $E=0$ by $i\sin{k_d}\Gamma_{0}=\pm i\Gamma_{0}$. 
%According to the theorem, 
%the net topological number of Weyl/Dirac points with positive imaginary energy equals to the point-gap topological number under the full PBCs for $\text{AZ}^{\dag}$ symmetry classes~\cite{Bessho-Sato-20}. A
One can also show that this model has surface states under the x$_1$OBC at the same time.
Since the Weyl/Dirac semimetal model $H_\text{SM}$ has Fermi arc surface states between two Weyl/Dirac points, $H_{\text{minimal}}(\bm{k})$ also has surface states under the x$_1$OBC. 
As a result, we have the BBC between the surface states and the bulk point-gap topological number.

%These surface states can be understood as surface states of Weyl/Dirac semimetals, $H_{\text{WSM/DSM}}(\bm{k}) = H_{\text{simple}}(\bm{k}) - i\sin{k_d}\Gamma_{0}$ as in th

For 3D class CII$^\dagger$, the minimal model has the form
\begin{align}
H_{\rm minimal}({\bm k})=H_{\rm SM}({\bm k})+i\sin k_2 \Gamma_2+i\sin k_3 \Gamma_3,     \quad 
H_{\rm SM}({\bm k})=\sin k_1 \Gamma_1 +\left[2-\sum_{i=1}^3 \cos k_i\right]\Gamma_0,
\end{align}
where $\Gamma_\mu$ satisfy
\begin{align}
\{\Gamma_0,\Gamma_1\}=\{\Gamma_2,\Gamma_3\}=[\Gamma_{i=0,1}, \Gamma_{j=2,3}]=0,
\quad \Gamma_{\mu=0,1,2,3}^2=1.
\end{align}
The double Hamiltonian of this model also has the form of the standard massive Dirac Hamiltonian, and thus we can easily evaluate the point-gap topological number. We also find that $H_{\rm SM}$ has $E=0$ surface states, which give surface states with ${\rm Re}E=0$ of $H_{\rm minimal}$. Thus, we have the BBC again.

%\textcolor{green}{We note that 2D class AIII$^\dagger$ also has the model in Eq.~(\ref{eq:simple model}), but the above argument does not hold. This is because the skin effect occurs under the OBCs for any directions.} \textcolor{red}{2d AIIdag dato Eq.S5 niha dekinai ninshikidesu. nakamura}

In Table \ref{table:ClassificationTable}, we summarize possible point-gap topological phases under the OBC for AZ$^\dagger$ classes with the spatial dimension $d=1,2,3$.
Below, we present the minimal model for each point-gap topological phase in Table \ref{table:ClassificationTable}.
In the following, $\tau_{i=x,y,z}$ and $\sigma_{i=x,y,z}$ represent the Pauli
matrices, and $\tau_{0}$ and  $\sigma_{0}$ are the 2×2 identity matrix.

\begin{center}
\begin{table}[h]
    \caption{Classification of point-gap topological phases under the OBC. We consider the $\mathrm{AZ}^{\dag}$ symmetry classes with the spatial dimension $d =1,2$ and $3$. The section numbers for the minimal models are shown for each topological number.}
  \begingroup
  \renewcommand{\arraystretch}{1.1}
     \begin{tabular}{ccccccc} \hline \hline
    Symmetry class & ~TRS$^\dagger$~ & ~PHS$^\dagger$~ & ~~CS~~ & ~$d=1$~ & ~$d=2$~ & ~$d=3$~ \\ \hline
		\multirow{1}{*}{A}
		& $0$ & $0$ & $0$ &\color{red}0\color{black} & $0$ & \color{green}\Z\color{black}~\color{black}[Sec.\ref{proof:A}] \\ 
		\multirow{1}{*}{AIII}
		& $0$ & $0$ & $1$ & $0$ & \color{green}\Z\color{black}~\color{black}[Sec.\ref{proof:AIII}] & $0$ \\ \hline
    \multirow{1}{*}{$\text{AI}^{\dag}$}
    & $+1$ & $0$ & $0$ &$0$ & $0$ & \color{green}2\Z\color{black}~\color{black}[Sec.\ref{proof:AIdag}] \\ 
    \multirow{1}{*}{$\text{BDI}^{\dag}$}
    & $+1$ & $+1$ & $1$ &$0$ & $0$ & $0$ \\ 
    \multirow{1}{*}{$\text{D}^{\dag}$}
    & $0$ & $+1$ & $0$ &\color{red}$0$\color{black} & $0$ & $0$ \\
    \multirow{1}{*}{$\text{DIII}^{\dag}$}
    & $-1$ & $+1$ & $1$ &\color{red}$0$\color{black} & \color{blue}$2\mathbb{Z}$\color{black}~\color{black}[Sec.\ref{proof:DIIIdag}] & $0$ \\ 
    \multirow{1}{*}{$\text{AII}^{\dag}$}
    & $-1$ & $0$ & $0$ &\color{red}0\color{black} & \color{red}0\color{black} & \color{blue}$2\mathbb{Z}$\color{black}~\color{black}[Sec.\ref{proof:AIIdag}] \\ 
    \multirow{1}{*}{$\text{CII}^{\dag}$}
    & $-1$ & $-1$ & $1$ &$0$ & \color{green}\Zt\color{black}~\color{black}[Sec.\ref{proof:CIIdag}] & \color{green}\Zt\color{black}~\color{black}[Sec.\ref{proof:CIIdag}] \\ 
    \multirow{1}{*}{$\text{C}^{\dag}$}
    & $0$ & $-1$ & $0$ &\color{red}0\color{black} & $0$ & \color{green}\Zt\color{black}~\color{black}[Sec.\ref{proof:Cdag}] \\ 
    \multirow{1}{*}{$\text{CI}^{\dag}$}
    & $+1$ & $-1$ & $1$ &$0$ & \color{green}2\Z\color{black}~\color{black}[Sec.\ref{proof:CIdag}] & $0$ \\ \hline \hline
  \end{tabular}
  \endgroup
 \label{table:ClassificationTable} 
\end{table}
\end{center}

\subsection{class A}\label{proof:A}
A non-Hermitian system in class A has a nontrivial point-gap topological phase in $d=3$. The minimal model is
\begin{align}
  H(\bm{k}) &= \sin{k_x}\sigma_{x} + \sin{k_y}\sigma_{y} + \left(2 - \sum_{i=x,y,z}\cos{k_i}\right)\sigma_{z} + i\sin{k_z}\sigma_{0}, 
\label{eq:3D-A}
\end{align}
and the point-gap topological number under the full PBC is the 3D winding number in Eq.~(2), which takes $+1$ for Eq.~(\ref{eq:3D-A}). As shown in the main text, the corresponding point-gap topological number under the xOBC is 

\begin{align}
  w_{3}|_{\rm xOBC} = -\dfrac{i}{12\pi}\int_{\text{BZ}}d^{2}\bm{k}\hspace{0.5mm}\mathcal{T}_x[\varepsilon^{ijk}Q_iQ_jQ_k],
\label{eq:w3xOBC}
\end{align}
with $Q_x = \mathcal{H}_{\text{bulk}}^{-1}[X,\mathcal{H}_{\text{bulk}}]$ and $Q_{i=y,z} = i\mathcal{H}_{\text{bulk}}^{-1}\partial_{k_{i}}\mathcal{H}_{\text{bulk}}$, where $X$ is the position operator in the $x$-coordinate, $\mathcal{T}_x$
stands for the trace per unit length in the $x$-direction, and  $\mathcal{H}_{\text{bulk}}(k_y,k_z)$
is the 3D bulk Hamiltonian with momentum-space representation in the $y$- and $z$-directions and real-space representation in the $x$-direction. 

\subsection{class AIII}\label{proof:AIII}
A non-Hermitian system in class AIII has a nontrivial point-gap topological phase in $d=2$. The minimal model is
\begin{align}
  H(\bm{k}) &= \sin{k_x}\sigma_{x} + \left(1 - \sum_{i=x,y}\cos{k_i}\right)\sigma_{y} + i\sin{k_y}\sigma_{0},
\label{eq:2D-AIII}
\end{align}
which hosts CS 
\begin{align}
  \Gamma H^{\dag}(\bm{k}) \Gamma^{-1} = - H(\bm{k}), \quad \Gamma = \sigma_{z}.
\label{eq:sym-AIII}
\end{align}
The point-gap topological number under the full PBC is the first Chern number of the Hermitian matrix $iH(\bm{k})\Gamma$ \cite{KSUS-19}:
\begin{align}
  iH(\bm{k})\Gamma &= \sin{k_x}\sigma_{y} - \sin{k_y}\sigma_{z} - \left(1 - \sum_{i=x,y}\cos{k_i}\right)\sigma_{x},
\label{eq:2D-AIII-chern}
\end{align}
which takes $+1$ for Eq.~(\ref{eq:2D-AIII}). 
The corresponding point-gap topological number under the xOBC is
\begin{align}
  Ch_{1}|_{\rm xOBC} = 2\pi i\mathcal{T}_{xy}\left(P_{\alpha}[[X,P_{\alpha}],[Y,P_{\alpha}]]\right),
\label{eq:Ch1xOBC}
\end{align}
where $P_{\alpha}$ is the bulk-band projection operator of $i{\cal H}\Gamma$ in a band $\alpha$, $X$ is the position operator in the $x$-coordinate, $Y$ is the position operator in the $y$-coordinate, $\mathcal{T}_{xy}$ stands for the trace per unit area in the $xy$-plane, and $i{\cal H}\Gamma$ is the bulk Hamiltonian of the real-space representation of $iH(\bm{k})\Gamma$ \cite{Bellissard-94,Kitaev-06,Prodan-10,Katsura-Koma-18,Song-real-space-19}.

\subsection{class $\text{AI}^{\dag}$}\label{proof:AIdag}
In class $\text{AI}^{\dag}$, we have a nontrivial point-gap topological phase in $d=3$. The required symmetry is
\begin{align}
  {\cal C} H^T({\bm k}) {\cal C}^{-1}=H(-{\bm k}),
  \quad {\cal C}{\cal C}^{*}=+1,
  \quad {\cal C} = \tau_{x}\sigma_{x},
\label{eq:sym-AIdag}
\end{align}
and the minimal model is
\begin{align}
  H(\bm{k}) &= \left[\sin{k_x}\tau_{x} + \sin{k_y}\tau_{y} + \left(2 - \sum_{i=x,y,z}\cos{k_i}\right)\tau_{z} + i\sin{k_z}\tau_{0}\right]\sigma_{z}.
\label{eq:3D-AIdag}
\end{align}
The point-gap topological number in 3D class $\text{AI}^{\dag}$ is the same as that in 3D class A.
Since Eq.~(\ref{eq:3D-AIdag}) is a stacking of the class A model in Eq.~(\ref{eq:3D-A}) to the eigensector of $\sigma_{z}=\pm1$, the 3D winding number takes $+2\in2$\Z~for Eq.~(\ref{eq:3D-AIdag}).

\subsection{class $\text{DIII}^{\dag}$}\label{proof:DIIIdag}
In class $\text{DIII}^{\dag}$, there exists a nontrivial point-gap topological phase in $d=2$. The minimal model is
\begin{align}
  H(\bm{k}) &= \sin{k_x}\tau_{x}\sigma_{x} + \left(1 - \sum_{i=x,y}\cos{k_i}\right)\tau_{x}\sigma_{y} - i\sin{k_y}\tau_{y}\sigma_{z},
\label{eq:2D-DIIIdag}
\end{align}
which has TRS$^\dagger$,  PHS$^\dagger$ and their combination CS
\begin{align}
  {\cal C} H^T({\bm k}) {\cal C}^{-1} &= H(-{\bm k}),\quad {\cal C}{\cal C}^{*}=-1, 
  \quad {\cal C} = i\tau_{y}\sigma_{0}\nonumber,\\
  {\cal T} H^{*}({\bm k}) {\cal T}^{-1} &= -H(-{\bm k}),\quad {\cal T}{\cal T}^{*}=+1,
  \quad {\cal T} = \tau_{0}\sigma_{0}\label{eq:sym-DIIIdag},\\
  \Gamma H^\dagger({\bm k})\Gamma^{-1}&=-H({\bm k}), \quad
  \Gamma = i{\cal C}{\cal T}^*\nonumber.
\end{align}
The point-gap topological number in 2D class $\text{DIII}^{\dag}$ is the same as that in 2D class AIII.
$iH(\bm{k})\Gamma$ for Eq.~(\ref{eq:2D-DIIIdag}) is 
\begin{align}
  iH(\bm{k})\Gamma &= \sin{k_x}\tau_{z}\sigma_{x}  - \sin{k_y}\tau_{0}\sigma_{z} + \left(1 - \sum_{i=x,y}\cos{k_i}\right)\tau_{z}\sigma_{y},
\label{eq:2D-DIIIdag-chern}
\end{align}
which is a stacking of Chern insulators to the eigensector of $\tau_{z}=\pm1$.
Thus, the first Chern number takes $+2\in2$\Z~for Eq.~(\ref{eq:2D-DIIIdag}).

\subsection{class $\text{AII}^{\dag}$}\label{proof:AIIdag}
In class $\text{AII}^{\dag}$, there exists a nontrivial point-gap topological phase in $d=3$. The minimal model is
\begin{align}
  H(\bm{k}) &= \tau_{y}\left[\sin{k_x}\sigma_{z} + \sin{k_y}\sigma_{x} + \left(2 - \sum_{i=x,y,z}\cos{k_i}\right)\sigma_{y} + i\sin{k_z}\sigma_{0}\right],
\label{eq:3D-AIIdag}
\end{align}
which has TRS$^\dagger$, 
\begin{align}
  {\cal C} H^T({\bm k}) {\cal C}^{-1}=H(-{\bm k}),
  \quad {\cal C}{\cal C}^{*}=-1,\quad
  {\cal C} = i\tau_{y}\sigma_{0}.
\label{eq:sym-AIIdag}
\end{align}
The point-gap topological number in 3D class $\text{AII}^{\dag}$ is the same as that in 3D class A.
Since Eq.~(\ref{eq:3D-AIIdag}) is a stacking of the minimal model of 3D class A to the eigensector of $\tau_{y}=\pm1$, the 3D winding number takes $+2\in2$\Z~for Eq.~(\ref{eq:3D-AIIdag}).

\subsection{class $\text{CII}^{\dag}$}\label{proof:CIIdag}
In class $\text{CII}^{\dag}$, there exist nontrivial point-gap topological phases in $d=2,3$.
We impose the following AZ$^\dagger$ symmetry, 
\begin{align}
  \label{eq:sym-CIIdag-2D}
  {\cal C} H^T({\bm k}) {\cal C}^{-1} &= H(-{\bm k}), \quad {\cal C}{\cal C}^{*}=-1, \quad {\cal C}=i\tau_{y}\sigma_{0},\nonumber
  \\
  {\cal T} H^{*}({\bm k}) {\cal T}^{-1} &= -H(-{\bm k}),\quad {\cal T}{\cal T}^{*}=-1, \quad {\cal T} = i\tau_{y}\sigma_{z},
  \\
\Gamma H^\dagger({\bm k})\Gamma^{-1}&=-H({\bm k}), \quad \Gamma = {\cal C}{\cal T}^*.
  \nonumber  
\end{align}
For $d=2$, the minimal model is 
\begin{align}
  H(\bm{k}) &= \sin k_x \tau_x \sigma_x
%  + \sin k_y \tau_y \sigma_y s_x
  + \left(1 - \sum_{i=x,y}\cos{k_i}\right) \tau_x\sigma_y
  + i \sin k_y \tau_z \sigma_z.
\label{eq:CIIdag-2D}
\end{align}
The point-gap topological number under the full PBCs is the Kane-Male invariant for 2D class AII Hermitian matrix $iH(\bm{k})\Gamma$ \cite{KSUS-19}, which takes $1\in$~\Zt~for Eq.~(\ref{eq:CIIdag-2D}). The corresponding point-gap topological number under the xOBC is
\begin{align}
  \eta|_{\rm xOBC} &= \text{dim ker}(A-1)~\text{mod}~2
\label{eq:Z2xOBC}
\end{align}
with
\begin{align}
  A = P_{F} - D^{\dag}_{\bm{a}}P_{F}D_{\bm{a}},\quad
  D_{\bm{a}}(\bm{r}) = \dfrac{\hat{x}+i\hat{y}-(a_{x}+i a_{y})}{|\hat{x}+i\hat{y}-(a_{x}+i a_{y})|}\nonumber.
\end{align}
Here $P_{F}$ is the projection operator of $i{\cal H}\Gamma$ 
onto the bulk states below the Fermi energy, where $i{\cal H}\Gamma$ is the real-space representation of $iH(\bm{k})\Gamma$,  $\bm{r}=(\hat{x},\hat{y})\in\mathbb{Z}^{2}$ represents the position operator in the $xy$-plane, and $\bm{a}=(a_x,a_y)\in\mathbb{R}^{2}\backslash\mathbb{Z}^{2}$ is a two dimensional vector in the $xy$-plane \cite{Katsura-Koma-16,Katsura-Koma-18,Akagi-17}. 

%, and $\bm{\gamma}=(\gamma_{x},\gamma_{y},\gamma_{z})$ stands for the Pauli matrices

For $d=3$, the minimal model is
\begin{align}
  H(\bm{k}) &= \sin k_x \tau_x \sigma_x
  + \left(2 - \sum_{i=x,y,z}\cos{k_i}\right) \tau_x\sigma_y
  + i\sin k_y \tau_x \sigma_0
  + i \sin k_z \tau_z \sigma_z.
\label{eq:CIIdag-3D}
\end{align}
The point-gap topological number under the full PBC is the $\mathbb{Z}_2$ invariant of the  class AII Hermitian matrix $iH(\bm{k})\Gamma$ \cite{KSUS-19}, which takes $1\in$~\Zt~for Eq.~(\ref{eq:CIIdag-3D}). The corresponding point-gap topological number under the xOBC is the 3D version of Eq.~(\ref{eq:Z2xOBC}) \cite{Katsura-Koma-18,Akagi-17}.

%\begin{align}
  %{\cal C} H^T({\bm k}) {\cal C}^{-1} &= H(-{\bm k}),~{\cal C}{\cal C}^{*}=-1,~{\cal C}=i\tau_{z}\sigma_{y}s_{0}\\
  %{\cal T} H^{*}({\bm k}) {\cal T}^{-1} &= -H(-{\bm k}),~{\cal T}{\cal T}^{*}=-1,~{\cal T} = i\tau_{0}\sigma_{y}s_{0},\\
  %\Gamma &= {\cal C}{\cal T},
%\label{eq:sym-CIIdag}
%\end{align}

%\begin{align}
  %H(\bm{k}) &= i\dfrac{\tau_{0}+\tau_{z}}{2}\left[\sin{k_x}\sigma_{x}s_{x} + \sin{k_y}\sigma_{y}s_{x} + \sin{k_z}\sigma_{z}s_{x} + \sin{k_w}\sigma_{0}s_{y} + \left(3 - \sum_{i=x,y,z,w}\cos{k_i}\right)\sigma_{0}s_{z}\right]\nonumber\\
  %&+ i\dfrac{\tau_{0}-\tau_{z}}{2}\sigma_{0}s_{0},
%\label{eq:CIIdag}
%\end{align}

\subsection{class $\text{C}^{\dag}$}\label{proof:Cdag}
In class $\text{C}^{\dag}$, there exists a nontrivial point-gap topological phase in $d=3$. The minimal model is

\begin{align}
  H(\bm{k}) &= 
  \sin k_x \tau_x \sigma_x 
  +\sin k_y \tau_x\sigma_y 
  + \left(2 - \sum_{i=x,y,z}\cos{k_i}\right)\tau_0\sigma_z
  + i\sin k_z  \tau_z\sigma_z,
\label{eq:Cdag}
\end{align}
which has the following PHS$^\dagger$,
\begin{align}
  {\cal T} H^{*}({\bm k}) {\cal T}^{-1} &= -H(-{\bm k}),
  \quad {\cal T}{\cal T}^{*}=-1,
  \quad {\cal T} = i\tau_{z}\sigma_{y}.
\label{eq:sym-Cdag}
\end{align}
The point-gap topological number under the full PBCs is the \Zt~invariant of the Hermitian Hamiltonian
\begin{align}
\begin{pmatrix}
0 & H({\bm k})\\
H^{\dagger}({\bm k}) & 0
\end{pmatrix}    
\end{align}
in class CII \cite{KSUS-19}, which takes $1\in$~\Zt~for Eq.~(\ref{eq:Cdag}). The corresponding point-gap topological number under the xOBC is its real-space representation given in Ref.~\cite{Katsura-Koma-18}.

\subsection{class $\text{CI}^{\dag}$}\label{proof:CIdag}
In class $\text{CI}^{\dag}$, there exists a nontrivial point-gap topological phase in $d=2$. The minimal model is 

\begin{align}
  H(\bm{k}) &= \sin k_x \tau_y\sigma_x 
  + \left(1 - \sum_{i=x,y} \cos{k_i} \right)\tau_{x}\sigma_{x} 
  + i\sin k_y \tau_z\sigma_y,
\label{eq:2D-CIdag}
\end{align}
which obeys
\begin{align}
  {\cal C} H^T({\bm k}) {\cal C}^{-1} &= H(-{\bm k}),\quad {\cal C}{\cal C}^{*}=+1,\quad~{\cal C} = \tau_{0}\sigma_{0}\nonumber,
  \\
  {\cal T} H^{*}({\bm k}) {\cal T}^{-1} &= -H(-{\bm k}),\quad {\cal T}{\cal T}^{*}=-1,\quad~{\cal T} = i\tau_{0}\sigma_{y}\label{eq:sym-CIdag},
  \\
  \Gamma H^\dagger ({\bm k})\Gamma&=-H({\bm k}), \quad
  \Gamma = i{\cal C}{\cal T}^*\nonumber.
\end{align}
The point-gap topological number under the full PBC  is the same as that in 2D class AIII.
Since $iH(\bm{k})\Gamma$ for Eq.~(\ref{eq:2D-CIdag}) is a stacking of two Chern insulators to the eigensectors of $\sigma_{z}=\pm1$:
\begin{align}
  iH(\bm{k})\Gamma &= \sin{k_x}\tau_{y}\sigma_{z} 
  + \sin{k_y}\tau_{z}\sigma_{0} + \left(1 - \sum_{i=x,y}\cos{k_i}\right)\tau_{x}\sigma_{z},
\label{eq:2D-CIdag-chern}
\end{align}
the first Chern number takes $+2\in2$\Z~for Eq.~(\ref{eq:2D-CIdag}).

\section{Bulk-boundary correspondence in 2D class AIII}
\begin{figure}[h]
\RawFloats
\begin{minipage}[t]{0.49\hsize}
\centering
\includegraphics[keepaspectratio,width=\linewidth]{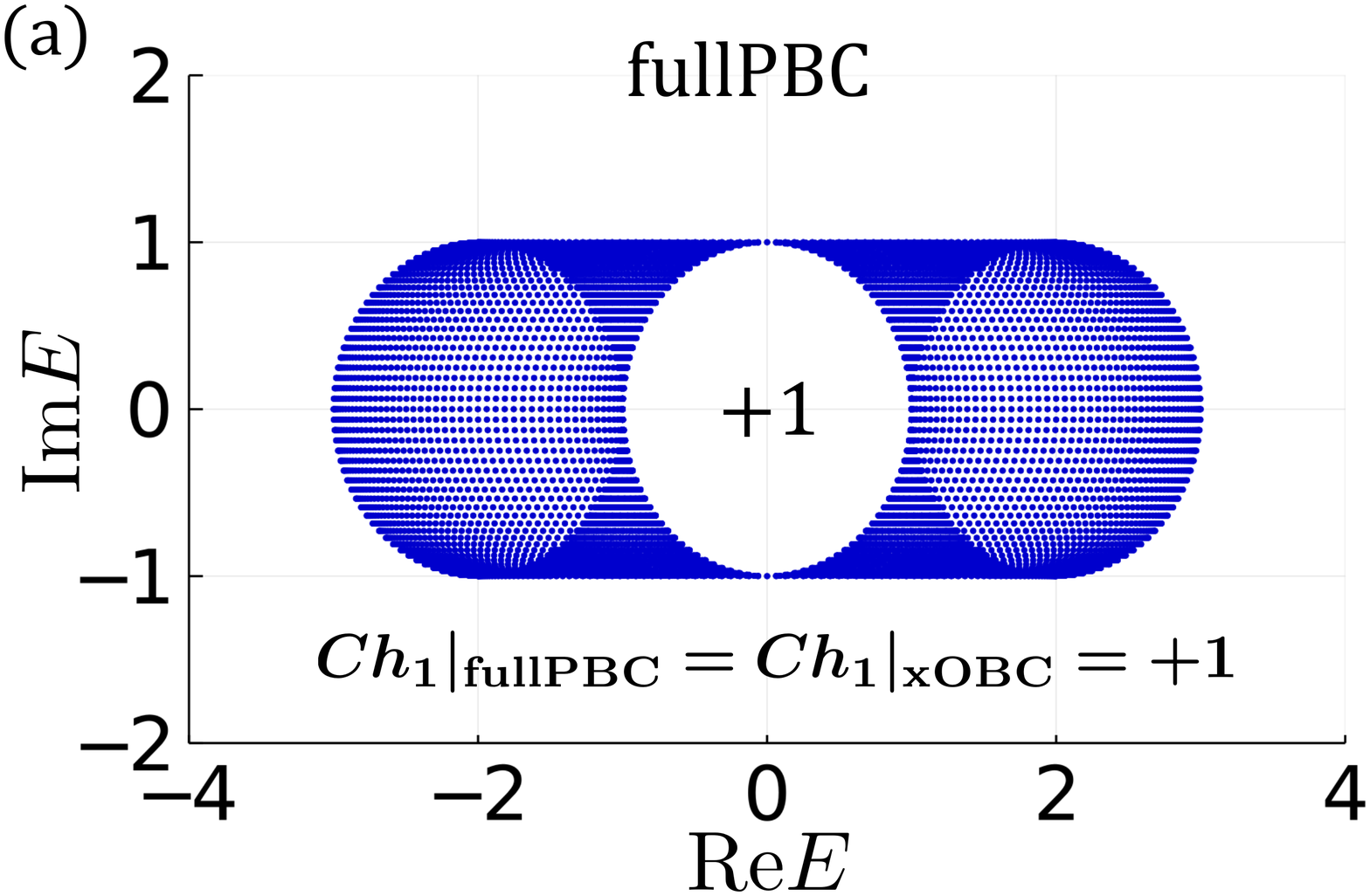}
\end{minipage}\hspace{0mm}
\begin{minipage}[t]{0.49\hsize}
\centering
\includegraphics[keepaspectratio,width=\linewidth]{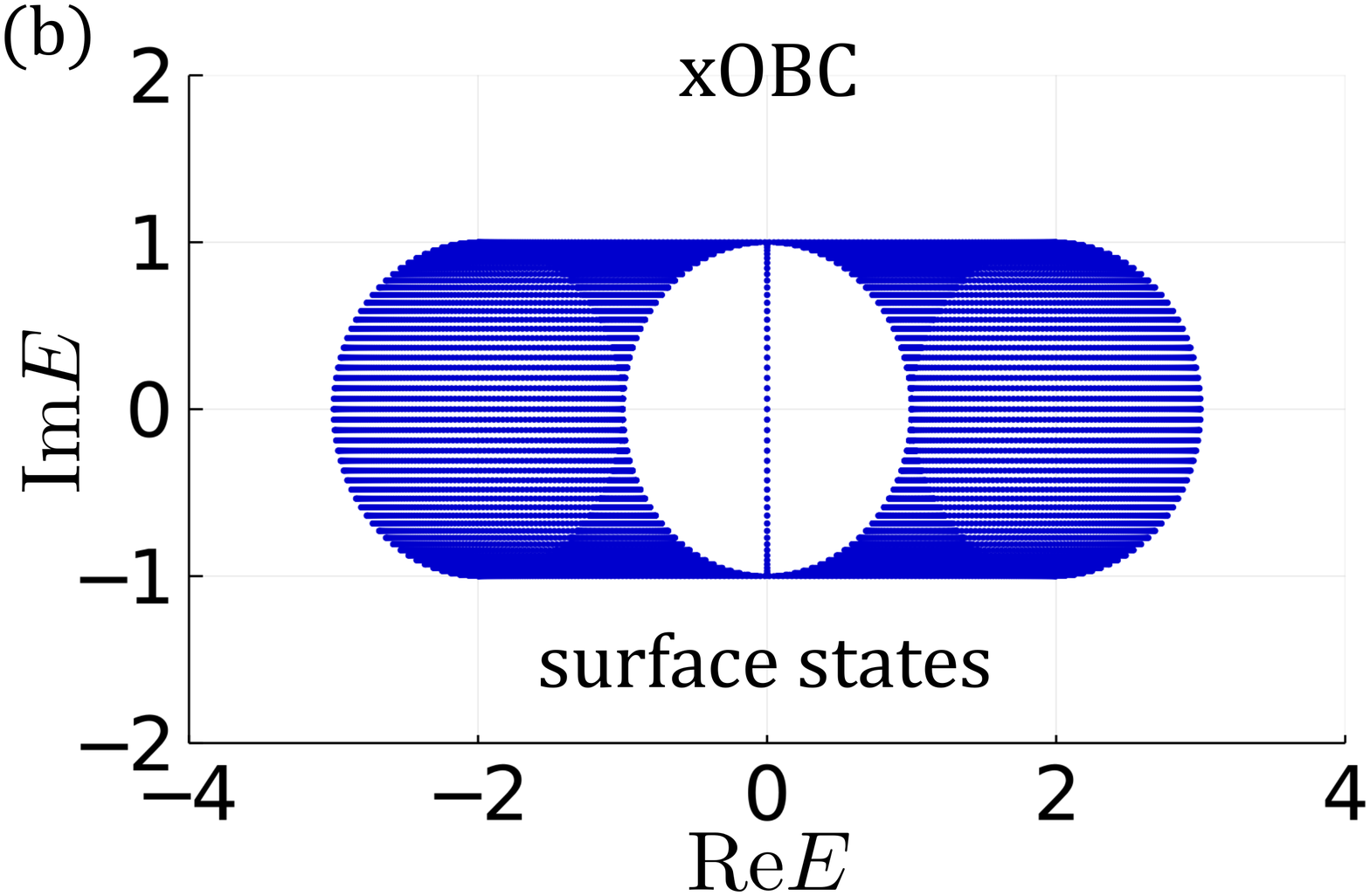}
\end{minipage}\hspace{0mm}
\caption{The energy spectra of the model in Eq.~(\ref{eq:2D-AIII-primitive}). The system sizes are $L_x=L_y=100$. (a) The full PBC spectrum. A point gap is open in the region containing $E=0$, with the nontrivial first Chern number $+1$. No NHSE occurs and the spectra coincide with the bulk spectra under the xOBC. (b) The xOBC spectrum. Surface states appear along $\mathrm{Re}E=0$ where the first Chern number takes +1. Note that the first Chern number is well-defined only on the line $\mathrm{Re}E=0$ in class AIII, since the system has chiral symmetry only when the reference energy of the point gap is purely imaginary.
\label{fig:2dAIII-SM}}
\end{figure}

We here explain more details of the BBC in point-gap topological phases for 2D class AIII. 
We start with the model in Eq.~(\ref{eq:2D-AIII}):
\begin{align}
  H_{\rm AIII}(\bm{k}) &= \sin{k_x}\sigma_{x} + \left(1 - \sum_{i=x,y}\cos{k_i}\right)\sigma_{y} + i\sin{k_y}\sigma_{0}.
\label{eq:2D-AIII-primitive}
\end{align}
As explained in Sec.\ref{proof:AIII}, this model has the first Chern number $+1$ of $iH_{\rm AIII}\Gamma$ under the xOBC. Since $H_{\rm AIII}(\bm{k})$ commutes with $i\sin{k_y}\sigma_{0}$, the surface state of this model with a fixed $k_y$ is that of the Su-Schrieffer–Heeger (SSH) model,
$H_{\text{SSH}}(\bm{k}) = H_{\rm AIII}(\bm{k})-i\sin{k_y}\sigma_{0}=\sin{k_x}\sigma_{x} + \left(1 - \sum_{i=x,y}\cos{k_i}\right)\sigma_{y}$. For $-\pi/2<k_y<\pi/2$, the SSH model supports the 1D winding number $+1$, and thus it has corresponding zero modes $\ket{\psi_{+}(k_y)}$ with chirality, $\Gamma\ket{\psi_{+}(k_y)}=+1\ket{\psi_{+}(k_y)}$, under the xOBC. By taking into account the complex energy shift from the non-Hermitian term $i\sin {k_y}\sigma_0$, the zero modes give surface states in the point-gapped region, as shown in Fig.~\ref{fig:2dAIII-SM}(b).

These surface states have a topological number with the same value as the bulk first Chern number. The effective Hamiltonian of the surface states around $E=0$ takes the form of $h_{\rm surface}(k_y)=ik_y\ket{\psi_{+}(k_y)}\bra{\psi_{+}(k_y)}$ as the zero mode of the SSH model shifted by the non-Hermitian term of $H_{\rm AIII}$, $ik_y$. Then, we can define the occupation number ${\cal N}(k_y)$ (the number of negative eigenvalues) for $ih_{\rm surface}(k_y)\Gamma$ on the surface BZ. Since $\Gamma\ket{\psi_{+}(k_y)}=+1\ket{\psi_{+}(k_y)}$, $ih_{\rm surface}(k_y)\Gamma$ is equal to $-k_y\ket{\psi_{+}(k_y)}\bra{\psi_{+}(k_y)}$ and we get 
\begin{align}
  {\cal N}(k_y) &=     
    \begin{cases}
        1 & \text{for $k_y>0$}\\
        0 & \text{for $k_y<0$}
    \end{cases}.
\end{align}
The difference of the occupation number $\nu_{\rm AIII}$:
\begin{align}
  \nu_{\rm AIII} &= {\cal N}(k_y>0)-{\cal N}(k_y<0)=+1
\end{align}
protects the surface energy $E=0$ at $k_y=0$ and thus measures the topology of the surface states, of which value coincides with the bulk first Chern number. Consequently, we have the BBC under the xOBC for Eq.~(\ref{eq:2D-AIII-primitive}):
\begin{align}
  Ch_{1}|_{\rm xOBC} = \nu_{\rm AIII}.
\label{eq:2D-AIII-BBC}
\end{align}
In a general case, we can get the BBC by stacking the model in Eq.~(\ref{eq:2D-AIII-primitive}) up to continuous deformations. The BBC for other symmetry classes also can be proved in a similar manner.

\section{The BBC for point-gap topological phases in 38-fold symmetry classes}

According to a general theory in Ref.\cite{KSUS-19}, there are 38-fold symmetry classes in non-Hermitian systems.
In this section, we extend our arguments to all the 38-fold symmetry classes.

First, we briefly explain the 38-fold symmetry classes. 
In addition to AZ$^\dagger$ symmetries discussed in the main text, non-Hermitian systems may host the original AZ symmetries defined by the following equations:
\begin{align}
  \label{eq:sym-AZ}
  {\cal T} H^{*}({\bm k}) {\cal T}^{-1} &= H(-{\bm k}),\quad {\cal T}{\cal T}^{*}=\pm 1,\nonumber
  \\
  {\cal C} H^T({\bm k}) {\cal C}^{-1} &= -H(-{\bm k}), \quad {\cal C}{\cal C}^{*}=\pm 1,
  \\
\Gamma H^\dagger({\bm k})\Gamma^{-1}&=-H({\bm k}),\quad \Gamma^{2}=1,
  \nonumber  
\end{align}
where ${\cal T},{\cal C}$ and $\Gamma$ are unitary operators corresponding to time-reversal symmetry (TRS), particle-hole symmetry (PHS), and chiral symmetry (CS), respectively. Furthermore, one can introduce sublattice symmetry (SLS),
%\begin{align}
%{\cal S}H{\cal S}^{-1}=-H, \quad {\cal S}^2=1,    
%\end{align}
\begin{align}
  {\cal S}H(\bm{k}){\cal S}^{-1} = -H(\bm{k}), \quad {\cal S}^{2}=1,
  \label{sym:SLS}
\end{align}
with a unitary operator ${\cal S}$, which is distinct from CS in non-Hermitian systems. The presence and/or absence of these symmetries give the 38-fold independent symmetry classes, which is a natural generalization of the Hermitian 10-fold AZ symmetry classes to non-Hermitian systems \cite{KSUS-19}.

To specify the 38-fold symmetry classes, we introduce the convention used in Ref.\cite{KSUS-19} as follows:
(i) The AZ and AZ$^\dagger$ symmetries define 18-fold symmetry classes in Table \ref{table:AZ_AZdag_class}. 
(ii) Furthermore, each of them can host SLS additionally, which defines AZ class + ${\cal S}$ and AZ$^\dagger$ class + ${\cal S}$. We also introduce the subindex $+(-)$ of ${\cal S}$ specifying the commutation (anti-commutation) relation between SLS and AZ or AZ$^\dagger$ symmetries. For an AZ (AZ$^\dagger$) class having both TRS (TRS$^\dagger$) and PHS (PHS$^\dagger$), ${\cal S}$ has a double subindex, where the first index specifies the commutation or anticommutation relation between SLS and TRS (TRS$^\dagger$), and the second one specifies those between SLS and PHS (PHS$^\dagger$), respectively.

\begin{center}
\begin{table}[h]
	\caption{AZ and AZ$^\dagger$ symmetry classes for non-Hermitian Hamiltonians.}
     \begin{tabular}{lcccccc} \hline \hline
    Sym. class & &TRS & PHS & CS & TRS$^\dagger$ & PHS$^\dagger$ \\ \hline 
	Complex AZ &\multirow{1}{*}{A}
		& $0$ & $0$ & $0$ & $0$ & $0$ \\
		&\multirow{1}{*}{AIII}
		& $0$ & $0$ & $1$ & $0$ & $0$  \\ \hline
    Real AZ &\multirow{1}{*}{$\text{AI}$}
    & $+1$ & $0$ & $0$ &$0$ & $0$  \\ 
    &\multirow{1}{*}{$\text{BDI}$}
    & $+1$ & $+1$ & $1$ &0 & 0 \\ 
    &\multirow{1}{*}{$\text{D}$}
    & $0$ & $+1$ & $0$ &0 & 0  \\ 
    &\multirow{1}{*}{$\text{DIII}$}
    & $-1$ & $+1$ & $1$ & $0$ & 0  \\ 
    &\multirow{1}{*}{$\text{AII}$}
    & $-1$ & $0$ & $0$ &0 & $0$  \\ 
    &\multirow{1}{*}{$\text{CII}$}
    & $-1$ & $-1$ & $1$ &$0$ & 0  \\ 
    &\multirow{1}{*}{$\text{C}$}
    & $0$ & $-1$ & $0$ & $0$ & $0$  \\ 
    &\multirow{1}{*}{$\text{CI}$}
    & $+1$ & $-1$ & $1$ &$0$ & $0$  \\ \hline
    Real AZ$^\dagger$ &\multirow{1}{*}{$\text{AI}^\dagger$}
    & $0$ & $0$ & $0$ &+1 & $0$  \\ 
    &\multirow{1}{*}{$\text{BDI}^\dagger$}
    & $0$ & $0$ & $1$ &+1 & +1 \\ 
    &\multirow{1}{*}{$\text{D}^\dagger$}
    & $0$ & $0$ & $0$ &0 & +1  \\ 
    &\multirow{1}{*}{$\text{DIII}^\dagger$}
    & $0$ & $0$ & $1$ & -1 & +1  \\ 
    &\multirow{1}{*}{$\text{AII}^\dagger$}
    & $0$ & $0$ & $0$ &-1 & $0$  \\ 
    &\multirow{1}{*}{$\text{CII}^\dagger$}
    & $0$ & $0$ & $1$ &-1 & -1  \\ 
    &\multirow{1}{*}{$\text{C}^\dagger$}
    & $0$ & $0$ & $0$ & $0$ & $-1$  \\ 
    &\multirow{1}{*}{$\text{CI}^\dagger$}
    & $0$ & $0$ & $1$ &$+1$ & $-1$  \\ \hline \hline  
    \end{tabular}
 \label{table:AZ_AZdag_class} 
\end{table}
\end{center}

Let us now identify the independent 38-fold symmetry classes. 
As was shown in Ref.\cite{KSUS-19}, we can divide the 38-fold symmetry classes according to the number $N \le 3$ of their generators TRS, PHS, TRS$^\dagger$ and PHS$^\dagger$. (i) For $N$=0, we have 5 classes, which are classes A, AIII, A + ${\cal S}$, AIII + ${\cal S}_+$, and AIII + ${\cal S}_-$.
(ii) For $N=1$, we have 6 classes, AI, AII, AI$^\dagger$, AII$^\dagger$, D, and C.
(iii) For $N=2$, there are 15 classes, BDI, DIII, CI, CII, BDI$^\dagger$, DIII$^\dagger$, CI$^\dagger$, CII$^\dagger$, AI + ${\cal S}_\pm$, AII + ${\cal S}_+$, D + ${\cal S}_\pm$, and C + ${\cal S}_\pm$. 
(iv) For $N=3$, we have 12 classes, BDI + ${\cal S}_{\pm\pm}$, DIII + ${\cal S}_{+\pm}$, CI + ${\cal S}_{\pm\pm}$, and CII + ${\cal S}_{+\pm}$.
Therefore, we have, in total, 5+6+15+12=38 classes.
Here we have several remarks: First,
AZ$^\dagger$ + ${\cal S}$ classes do not appear above because the combination between TRS$^\dagger$ (PHS$^\dagger$) and SLS gives PHS (TRS) so they are not independent of AZ + ${\cal S}$ classes. 
Second, AII + ${\cal S}_-$ coincides with AI + ${\cal S}_-$ by multiplying $i$ to the Hamiltonian and replacing the original TRS with the combination of TRS and SLS.
In a similar manner, DIII + ${\cal S}_{-\pm}$ and CII + ${\cal S}_{-\pm}$ are equal to BDI + ${\cal S}_{-\pm}$ and CI + ${\cal S}_{-\pm}$, respectively. 
Thus, the above list does not include them.

Among these 38-fold symmetry classes, we have proved the BBC in point-gap topological phases for 10-fold AZ$^\dagger$ classes in Sec.\ref{sec:S2}. 
The AZ$^\dagger$ symmetry classes correspond to A, AIII, AI$^\dagger$, BDI$^\dagger$, AI, DIII$^\dagger$, AII$^\dagger$, CII$^\dagger$, AII, and CI$^\dagger$ in the above 38 classes since one can identify D$^\dagger$ and C$^\dagger$ with AI and AII by multiplying the imaginary unit $i$ to the Hamiltonian. 
We argue below the BBC in point-gap topological phases for the remaining 28 symmetry classes.

\subsection{Classes without SLS}

In the remaining 28 classes, 6 classes (BDI, D, DIII, CII, C, and CI) in Table \ref{table:without SLS} do not support SLS. 
We first argue these classes in $d=1,2,3$. In a manner similar to Sec.\ref{sec:S2}, on the basis of $K$-theory, we can prove the BBC by constructing a model with a minimal topological number.
Below we present such minimal models. 
Except for $d=1,2$ in class BDI and $d=1$ in class D, the minimal model has the form of Eq.(\ref{eq:simple model}), and for $d=1,2$ in class BDI and $d=1$ in class D, the minimal model is either anti-Hermitian or Hermitian.
Therefore, all the minimal models can avoid skin effects, so have the common value of the point-gap topological numbers between OBC and PBC. 
Furthermore, using the conventional BBC for Hermitian systems, each of these models has a boundary state, so we have the BBC.

\begin{table}[H]
  \centering
  \begingroup
  \renewcommand{\arraystretch}{1.1}
    \begin{tabular}{cccc} \hline \hline
        AZ class & ~$d=1$~ & ~$d=2$~ & ~$d=3$~ \\ \hline
        \multirow{1}{*}{$\text{BDI}$}
        &\color{green}\Zt~\color{black}[Sec.\ref{proof:BDI}] & \color{green}\Z~\color{black}[Sec.\ref{proof:BDI}] & $0$ \\ 
        \multirow{1}{*}{$\text{D}$}
        &\color{green}\Zt~\color{black}[Sec.\ref{proof:D}] & \color{green}\Zt~\color{black}[Sec.\ref{proof:D}] & \color{green}\Z~\color{black}[Sec.\ref{proof:D}] \\ 
        \multirow{1}{*}{$\text{DIII}$}
        &$0$ &\color{green}\Zt~\color{black}[Sec.\ref{proof:DIII}] & \color{green}\Zt~\color{black}[Sec.\ref{proof:DIII}] \\ 
        \multirow{1}{*}{$\text{CII}$}
        &$0$ & \color{green}2\Z~\color{black}[Sec.\ref{proof:CII}] & $0$ \\ 
        \multirow{1}{*}{$\text{C}$}
        &$0$ & $0$ & \color{green}2\Z~\color{black}[Sec.\ref{proof:C}] \\ 
        \multirow{1}{*}{$\text{CI}$}
        &$0$ & $0$ & $0$ \\ \hline \hline
    \end{tabular}
    \endgroup
    \caption{Point-gap topological phases in 6 classes without SLS. These classes do not show the skin effects, and thus the classification under the OBCs coincides with that under the PBC. The section numbers for the proof of the classification under OBCs and the BBC are shown for each topological number.}
    \label{table:without SLS} 
\end{table}

\subsubsection{class $\text{BDI}$}\label{proof:BDI}
In class $\text{BDI}$, there exist nontrivial point-gap topological phases in $d=1,2$ \cite{KSUS-19}.
%We impose the following AZ symmetry, 
%\begin{align}
%  \label{eq:sym-BDI}
%  {\cal T} H^{*}({\bm k}) {\cal T}^{-1} &= H(-{\bm k}),\quad {\cal T}{\cal T}^{*}=+1, \quad {\cal T}=\sigma_{0},\nonumber
%  \\
%  {\cal C} H^T({\bm k}) {\cal C}^{-1} &= -H(-{\bm k}), \quad {\cal C}{\cal C}^{*}=+1, \quad {\cal C}=\sigma_{0},
%  \\
%\Gamma H^\dagger({\bm k})\Gamma^{-1}&=-H({\bm k}), \quad \Gamma = {\cal C}{\cal T}^*.
%  \nonumber  
%\end{align}
For $d=1$, the minimal model is
\begin{align}
  H(k_x) &= i\left(\sin k_x \sigma_x - \cos{k_x} \sigma_y\right),
\label{eq:BDI-1D}
\end{align}
which has TRS and PHS of class BDI with ${\cal T}=\sigma_0$ and ${\cal C}=\sigma_0$.
This model has a non-zero \Zt~point-gap topological number of BDI class in $d=1$, which is defined by $(-1)^\nu={\rm sgn}({\rm Pf}[H(\pi){\cal C}]/{\rm Pf}[H(0){\cal C}])$ \cite{KSUS-19}. 
Furthermore, as Eq.(\ref{eq:BDI-1D}) is anti-Hermitian so does not show skin effects, this model has the same non-zero point-gap topological number under OBC, which is defined by
\begin{align}
  \eta|_{\rm OBC} &= \text{dim ker}(A-1)~\text{mod.}~2
\label{eq:Z2OBC}
\end{align}
with
\begin{align}
  A = P_{F} - D_{a_x}P_{F}D_{a_x},\quad
  D_{a_x}(\hat{x}) = \dfrac{\hat{x}-a_{x}}{|\hat{x}-a_{x}|}.\nonumber
\end{align}
Here $P_{F}$ is the projection operator of the Hertmian Hamltonian $i{\cal H}{\cal T}{\cal C}^*$ 
onto the its negative energy states, where ${\cal H}$ is the real-space representation of $H(k_x)$, $\hat{x}\in\mathbb{Z}$ is the position operator in the real space lattice $x$, and $a_x\in\mathbb{R}\backslash\mathbb{Z}$ \cite{Katsura-Koma-18}. 
One can easily check that this model has a zero energy boundary state under OBC, so the BBC holds. 

For $d=2$, the minimal model is 
\begin{align}
  H(\bm{k}) &= i\left[\sin k_x \sigma_x
  + \sin k_y \sigma_z
  + \left(1 - \sum_{i=x,y}\cos{k_i}\right) \sigma_y\right],
\label{eq:BDI-2D}
\end{align}
with ${\cal T}=\sigma_0$ and ${\cal C}=\sigma_0$.
The point-gap topological number in 2D class $\text{BDI}$ is the first Chern number of the Hermitian Hamiltonian
$iH(\bm{k}){\cal T}{\cal C}^*$. For Eq.~(\ref{eq:BDI-2D}), we have 
\begin{align}
  iH(\bm{k})\Gamma &= -\sin k_x \sigma_x - \sin k_y \sigma_z - \left(1 - \sum_{i=x,y}\cos{k_i}\right) \sigma_y,
\label{eq:BDI-2D-chern}
\end{align}
which gives the first Chern number $+1\in$\Z.
This model does not show skin effects because of the anti-Hermiticity and shows a chiral edge state closing the point gap at $E=0$ under OBCs. Thus we have the BBC.

\subsubsection{class $\text{D}$}\label{proof:D}
In class $\text{D}$, there exist nontrivial point-gap topological phases in $d=1,2$ and $3$.
%The required symmetry is
%\begin{align}
%  {\cal C} H^T({\bm k}) {\cal C}^{-1} &= -H(-{\bm k}), \quad {\cal C}{\cal C}^{*}=+1, \quad {\cal C}=\sigma_{0}. 
%  \label{eq:sym-D}
%\end{align}
For $d=1$, the minimal model is
\begin{align}
  H(k_x) &= \sin k_x \sigma_x + \cos{k_x} \sigma_y,
\label{eq:D-1D}
\end{align}
which has PHS of class D with ${\cal C}=\sigma_0$.
The point-gap topological number is given by 
\begin{align}
(-1)^{\nu_1[H]}=\text{sgn}
\left\{
\dfrac{\operatorname{Pf}[H(\pi)\mathcal{C}]}{\operatorname{Pf}[H(0)\mathcal{C}]}
~\times~\text{exp}\left[-\dfrac{1}{2}\int_{k_x=0}^{k_x=\pi}\operatorname{\it{d}}\operatorname{log}\operatorname{det}[H(k_x)\mathcal{C}]
\right]\right\},
\label{eq:Z21D-D}
\end{align}
which becomes non-trivial for Eq.(\ref{eq:D-1D}). This model is Hermitian and thus does not show skin effects.
Therefore, this model supports the same non-trivial point-gap topological number under OBC.
Because the minimal model has a boundary zero mode under OBC, we have the BBC.
%the same formula as that for 1D class \text{BDI} \cite{Katsura-Koma-18}.

For $d=2$, the minimal model is 
\begin{align}
  H(\bm{k}) &= \sin k_x \sigma_x
  + \left(1 - \sum_{i=x,y}\cos{k_i}\right) \sigma_y + i\sin k_y \sigma_0, \quad {\cal C}=\sigma_0.
\label{eq:D-2D}
\end{align}
This model has the same form as Eq.(\ref{eq:simple model}), and thus we can prove the BBC in a similar manner.
The \Zt~point-gap topological number is defined by \cite{KSUS-19}
\begin{align}
    (-1)^{\nu_2[H]} = \prod_{\text{X=I,II}}\text{sgn}\left\{\dfrac{\operatorname{Pf}[H(\bm{k}_{\text{X}+})\mathcal{C}]}{\operatorname{Pf}[H(\bm{k}_{\text{X}-})\mathcal{C}]}~\times~\text{exp}\left[-\dfrac{1}{2}\int_{\bm{k}=\bm{k}_{\text{X}-}}^{\bm{k}=\bm{k}_{\text{X}+}}\operatorname{\it{d}}\operatorname{log}\operatorname{det}[H(\bm{k})\mathcal{C}]\right]\right\},
    \label{eq:Z22D-D}
\end{align}
where $(\bm{k}_{\text{I}+},\bm{k}_{\text{I}-})$ and $(\bm{k}_{\text{II}+},\bm{k}_{\text{II}-})$ are two pairs of particle-hole symmetric momenta.

%The corresponding point-gap topological number under the xOBC is the 2D version of Eq.~(\ref{eq:Z2OBC}) \cite{Katsura-Koma-18}, which takes non-trivial value $1\in\mathbb{Z}_2$ for Eq.~(\ref{eq:D-2D}).

For $d=3$, the minimal model is 
\begin{align}
  H(\bm{k}) &= \sin{k_x}\sigma_{x} + \sin{k_y}\sigma_{z} + \left(2 - \sum_{i=x,y,z}\cos{k_i}\right)\sigma_{y} - i\sin{k_z}\sigma_{0}, \quad {\cal C}=\sigma_0,
\label{eq:D-3D}
\end{align}
which also has the form of Eq.(\ref{eq:simple model}).
The point-gap topological number in 3D class $\text{D}$ coincides with the winding number for 3D class A, which takes $+1\in$~\Z~for Eq.~(\ref{eq:D-3D}).

\subsubsection{class $\text{DIII}$}\label{proof:DIII}
In class $\text{DIII}$, there exist nontrivial point-gap topological phases in $d=2,3$.
%We impose the following AZ symmetry, 
%\begin{align}
%  \label{eq:sym-DIII}
%  {\cal T} H^{*}({\bm k}) {\cal T}^{-1} &= H(-{\bm k}),\quad {\cal T}{\cal T}^{*}=-1, \quad 
%  {\cal T}=i\tau_{y}\sigma_{0},\nonumber
%  \\
%  {\cal C} H^T({\bm k}) {\cal C}^{-1} &= -H(-{\bm k}), \quad {\cal C}{\cal C}^{*}=+1, \quad {\cal C}=\tau_{0}\sigma_{0},
%  \\
%\Gamma H^\dagger({\bm k})\Gamma^{-1}&=-H({\bm k}), \quad \Gamma = i{\cal C}{\cal T}^*.
%  \nonumber  
%\end{align}
For $d=2$, the minimal model is 
\begin{align}
  H(\bm{k}) &= \sin{k_x}\tau_{x}\sigma_{x} + \left(1 - \sum_{i=x,y}\cos{k_i}\right)\tau_{x}\sigma_{y} + i\sin{k_y}\tau_{0}\sigma_{0},
  \quad   {\cal T}=i\tau_{y}\sigma_{0},
  \quad {\cal C}=\tau_0\sigma_0.
\label{eq:DIII-2D}
\end{align}
The point-gap topological number is the Kane-Male \Zt~invariant for 2D class AII Hermitian matrix $iH(\bm{k}){\cal T}{\cal C}^*$ \cite{KSUS-19}, which is non-trivial for Eq.~(\ref{eq:DIII-2D}). 
%The corresponding point-gap topological number under the xOBC is given by Eq.~(\ref{eq:Z2xOBC}).
For $d=3$, the minimal model is 
\begin{align}
  H(\bm{k}) &= \sin{k_x}\tau_{x}\sigma_{x} + \sin{k_y}\tau_{x}\sigma_{z} + \left(2 - \sum_{i=x,y,z}\cos{k_i}\right)\tau_{x}\sigma_{y} + i\sin{k_z}\tau_{0}\sigma_{0},
  \quad   {\cal T}=i\tau_{y}\sigma_{0},
  \quad {\cal C}=\tau_0\sigma_0.
  \label{eq:DIII-3D}
\end{align}
The point-gap topological number under the full PBC is the $\mathbb{Z}_2$ invariant of the class AII Hermitian matrix $iH(\bm{k}){\cal T}{\cal C}^*$ \cite{KSUS-19}, which is non-trivial for Eq.~(\ref{eq:DIII-3D}). 
These models have the form of Eq.(\ref{eq:simple model}), and thus we have the BBC.
%The corresponding point-gap topological number under the xOBC is the 3D version of Eq.~(\ref{eq:Z2xOBC}) \cite{Katsura-Koma-18,Akagi-17}.

%\subsection{class $\text{AII}$}
%A Hamiltonian $H_{\text{AII}}$ in class $\text{AII}$ is related to a Hamiltonian $H_{\text{C}^\dagger}$ in class $\text{C}^\dagger$ via $H_{\text{AII}}=iH_{\text{C}^\dagger}$. Thus, from Sec.S2, there exist nontrivial point-gap topological phases only in $d=3$. From Eq.~(\ref{eq:Cdag}), the minimal model is given by

%\begin{align}
%  H(\bm{k}) &= 
%  i\left[\sin k_x \tau_x \sigma_x 
%  + \sin k_y \tau_x\sigma_y 
%  + \left(2 - \sum_{i=x,y,z}\cos{k_i}\right)\tau_0\sigma_z\right]
%  - \sin k_z  \tau_z\sigma_z,
%\label{eq:AII-3D}
%\end{align}
%which has the following TRS,
%\begin{align}
%  {\cal T} H^{*}({\bm k}) {\cal T}^{-1} &= H(-{\bm k}),
%  \quad {\cal T}{\cal T}^{*}=-1,
%  \quad {\cal T} = i\tau_{z}\sigma_{y}.
%\label{eq:sym-AII}
%\end{align}
%The point-gap topological number under the full PBCs is the \Zt~invariant of the Hermitian Hamiltonian
%\begin{align}
%\begin{pmatrix}
%0 & H({\bm k})\\
%H^{\dagger}({\bm k}) & 0
%\end{pmatrix}    
%\end{align}
%in class CII \cite{KSUS-19}, which takes $1\in$~\Zt~for Eq.~(\ref{eq:AII-3D}). The corresponding point-gap topological number under the xOBC is its %real-space representation given in Ref.~\cite{Katsura-Koma-18}.

\subsubsection{class $\text{CII}$}\label{proof:CII}
In class $\text{CII}$, there exist nontrivial point-gap topological phases in $d=2$.
%We impose the following AZ symmetry, 
%\begin{align}
%  \label{eq:sym-CII}
%  {\cal T} H^{*}({\bm k}) {\cal T}^{-1} &= H(-{\bm k}),\quad {\cal T}{\cal T}^{*}=-1, 
%  \quad {\cal T}=i\tau_{y}\sigma_{z},\nonumber
%  \\
%  {\cal C} H^T({\bm k}) {\cal C}^{-1} &= -H(-{\bm k}), \quad {\cal C}{\cal C}^{*}=-1, 
%  \quad {\cal C}=i\tau_{y}\sigma_{0},
%  \\
%\Gamma H^\dagger({\bm k})\Gamma^{-1}&=-H({\bm k}), \quad \Gamma = {\cal C}{\cal T}^*.
%  \nonumber  
%\end{align}
For $d=2$, the minimal model is 
\begin{align}
  H(\bm{k}) &= \tau_{0}\left[\sin{k_x}\sigma_{x} + \left(1 - \sum_{i=x,y}\cos{k_i}\right)\sigma_{y} + i\sin{k_y}\sigma_{0}\right],
\label{eq:CII-2D}
\end{align}
with ${\cal T}=i\tau_{y}\sigma_{z}$ and ${\cal C}=i\tau_{y}\sigma_{0}$.
 The point-gap topological number is the first Chern number of $iH({\bm k}){\cal T}{\cal C}^*$.
Since the above model is a stacking of Eq.~(\ref{eq:2D-AIII}), this model has the first Chern number $+2\in2$\Z~ and the BBC holds.
%~for Eq.~(\ref{eq:CII-2D}).

\subsubsection{class $\text{C}$}\label{proof:C}
In class $\text{C}$, there exists a non-trivial point-gap topological phase in $d=3$.
%The required symmetry is
%\begin{align}
%  {\cal C} H^T({\bm k}) {\cal C}^{-1} &= -H(-{\bm k}), \quad {\cal C}{\cal C}^{*}=-1, \quad {\cal %C}=i\tau_{y}\sigma_{0}. 
% \label{eq:sym-C}
%\end{align}
%
The minimal model is
\begin{align}
  H(\bm{k}) &= \tau_{0}\left[\sin{k_x}\sigma_{x} + \sin{k_y}\sigma_{z} + \left(1 - \sum_{i=x,y,z}\cos{k_i}\right)\sigma_{y} - i\sin{k_z}\sigma_{0}\right], \quad {\cal C}=i\tau_{y}\sigma_{0}. 
  \label{eq:C-3D}
\end{align}
The point-gap topological number is equal to the 3D winding number in $d=3$ class A.
Since Eq.~(\ref{eq:C-3D}) is a stacking of Eq.(\ref{eq:3D-A}), the 3D winding number for Eq.~(\ref{eq:C-3D}) takes $+2\in2$\Z, and the BBC holds.

\subsubsection{class \text{CI}}
The point gap topological phase is trivial for class CI in $d=1,2,3$. 

\subsection{Classes with SLS}

So far, we have argued the BBC for point-gap topological phases in 16 classes, all of which do not support SLS. 
Below we discuss the remaining 22 classes supporting SLS in Eq.(\ref{sym:SLS}). See Table \ref{table:situation with SLS}.
Without loss of generality, we choose ${\cal S}$ as $\mu_{z}\otimes \hat{1}$ such that $H(\bm{k})$ takes the form of
\begin{align}
H(\bm{k})=
\begin{pmatrix}
0 & h_{+}({\bm k})\\
h_{-}({\bm k}) & 0
\end{pmatrix}, \label{eq:SLS}
\end{align}
where $\mu_{i=x,y,z}$ are the Pauli matrices, $\mu_{0}$ is the 2×2 identity matrix and $\hat{1}$ is an identity matrix acting on $h_{\pm}(\bm{k})$. 

In a manner similar to the above,  we prove the BBC by constructing models with minimal point-gap topological numbers.
To show the existence of boundary states for these models under OBCs, we use the following lemmas:
%From Eq.~(\ref{eq:SLS}), we have
%\begin{align}
%\text{det}[H(\bm{k})-E]=E^2 - h_{+}h_{-}=0,
%\label{eq:eigenEq-SLS}
%\end{align}
%and thus, the eigenvalues of $H(\bm{k})$ coincide with the square root of those of $h_{+}h_{-}$. 

%\begin{itemize}
%    \item 
%Let $H$ and $h_{\pm}$ be the Hamiltonian and its off-diagonal components in Eq.(\ref{eq:SLS}). 
\begin{lemma}

Suppose $h_\pm$ is diagonalizable and $[h_{+},h_-]=0$, and let 
$|\phi_n\rangle$ be an eigenvector 
 diagonalizing $h_+$ and $h_-$ simultaneously,
\begin{align}
h_\pm |\phi_n\rangle=E^{\pm}_{n}|\phi_n\rangle.    
\end{align}
Then, we obtain the eigenvectors and eigenenergies of $H$ from those of $h_{\pm}$,
\begin{align}
H
\begin{pmatrix}
c_n^\pm\sqrt{E^+_{n}}\ket{\phi_n} \\
\pm c_n^\pm \sqrt{E^-_{n}}\ket{\phi_n} 
\end{pmatrix}
= \pm \sqrt{E^+_n E^-_n}
\begin{pmatrix}
c_n^\pm \sqrt{E^+_{n}}\ket{\phi_n} \\
\pm c_n^\pm\sqrt{E^-_{n}}\ket{\phi_n} 
\end{pmatrix},
\label{eq:eigenvec-SLS}
\end{align}
where $c_n^\pm$ is a constant.
%where $\ket{\psi}$ is a common eigenvector of $h_\pm$ with eigenenergies $E_\pm$, {\it i.e.} $h_{\pm}\ket{\psi}=E_{\pm}\ket{\psi}$.
%Because we have ${\rm det}(H^2-E^2)={\rm det}(h_+h_--E^2)$ from $[h_+,h_-]=0$, 
%the above equation gives the complete sets of spectrum of $H$.
\end{lemma}
\begin{lemma}
\label{lemma:2}
Suppose $h_-h_+$ is diagonalizable and $h_-$ has its inverse $h_-^{-1}$, and let $|\psi_n\rangle$ be an eigenvector of $h_-h_+$,
\begin{align}
h_-h_+|\psi_n\rangle=\lambda_n|\psi_n\rangle.    
\end{align}
Then, we obtain the eigenvectors and eigenenergies of $H$ as,
\begin{align}
H
\begin{pmatrix}
\sqrt{\lambda_n}h_-^{-1}|\psi_n\rangle\\
\pm |\psi_n\rangle
\end{pmatrix}
=\pm \sqrt{\lambda_n}
\begin{pmatrix}
\sqrt{\lambda_n}h_-^{-1}|\psi_n\rangle\\
\pm |\psi_n\rangle
\end{pmatrix}.
\end{align}
\end{lemma}
These lemmas imply that if $h_\pm$ ($h_-h_+$) in the former (latter) lemma does not show skin effects and supports a boundary state, then $H$ also does. 
Here it should be noted that the commutation relation $[h_-,h_+]=0$ and the invertibility $h_-^{-1}$ may depend on the boundary condition. For instance, whereas $e^{ik_x}$ and $e^{-ik_x}$ commute with each other, their matrix representations under the OBC do not. The invertibility of $e^{ik_x}$ also depends on the boundary condition because $e^{ik_x}$ has a zero mode under the OBC.
%The 16 of 38-fold classification are provided in Sec.S2 and S4, which consists of the 10 AZ$^{\dagger}$ symmetry classes and the 6 of 10 AZ symmetry classes excluding the 4 overlapping with AZ$^{\dagger}$ symmetry classes (A, AIII, AI and AII) \cite{KSUS-19}. The remaining 22 classes are in AZ symmetry classes with SLS \cite{KSUS-19} as detailed below.

For a system satisfying the assumption for Lemma \ref{lemma:2}, we also have the following corollary.

\begin{cor}
If $h_+$ has a (right) zero eigenvalue, $h_+|\psi_0\rangle=0$, then we have
\begin{align}
H|\Psi_0^1\rangle=0, 
\quad
H|\Psi_0^2\rangle=|\Psi_0^1\rangle,
\end{align}
with
\begin{align}
|\Psi_0^1\rangle=
\begin{pmatrix}
0\\
|\psi_0\rangle
\end{pmatrix},
\quad
|\Psi_0^2\rangle
=
\begin{pmatrix}
h_-^{-1}|\psi_0\rangle\\
0
\end{pmatrix}.
\end{align}
\label{cor:1}
\end{cor}
The equation in Corollary implies that $H$ has a $2\times 2$ Jordan block with zero eigenenergy
\begin{align}
H|\Psi_0^i\rangle=\sum_{j=1,2}|\Psi_0^j\rangle J_{ji},    
\quad J=
\begin{pmatrix}
0 & 1\\
0 & 0
\end{pmatrix},
\end{align}
and thus it has an exceptional point.
We use this result in Sec.\ref{S:intrinsic}.

Once we choose the basis in Eq.(\ref{eq:SLS}), we only need to consider how AZ symmetries act on $h_\pm$.   
For complex AZ + ${\cal S}$ classes, we have 3 different situations:
\begin{enumerate}
\item $H$ does not have any AZ symmetries, which corresponds to A + ${\cal S}$.
In this case, $h_{\pm}$ does not have any AZ symmetries.
\item CS commutes with SLS, {\it i.e.} AIII + ${\cal S}_+$.
Without loss of generality, we take 
\begin{align}
\Gamma=\mu_0\otimes \gamma
\end{align}
with $\gamma$ a unitary operator acting on $h_{\pm}$, which leads to
\begin{align}
\gamma h^\dagger_{\pm}({\bm k})\gamma^{-1}=-h_{\mp}({\bm k}), \quad \gamma^2=1.
\end{align}
Therefore, $h_\pm$ does not have CS, but they are not independent.
\item CS anti-commutes with SLS, {\it i.e.} AIII + ${\cal S}_-$.
Without loss of generality, we take 
\begin{align}
\Gamma=\mu_x\otimes \gamma
\end{align}
with $\gamma$ a unitary operator acting on $h_{\pm}$, which leads to
\begin{align}
\gamma h^\dagger_{\pm}({\bm k})\gamma^{-1}=-h_{\pm}({\bm k}), \quad \gamma^2=1.
\end{align}
Therefore, $h_\pm$ has its own CS, and they are independent.

\end{enumerate}
For real AZ + ${\cal S}$ classes, we have 4 different situations:
\begin{enumerate}
\setcounter{enumi}{3}
\item 
%$h_\pm$ supports the same AZ symmetries as $H$ does. This happens when
TRS (if exists) commutes with SLS, while PHS (if exists) anti-commutes with SLS:
We can take generally
\begin{align}
  \label{sym:SLS+-}
  {\cal T} = \mu_{0}\otimes T, \quad
  {\cal C} = \mu_{x}\otimes C, 
  %\nonumber
  %\\
  %\Gamma = \mu_{x}\otimes \gamma,\quad
  %{\cal S} = \mu_{z}\otimes \hat{1}, 
\end{align}
where $T$ and $C$ are unitary operators acting on $h_{\pm}(\bm{k})$.
In this basis, TRS and PHS in Eq.(\ref{eq:sym-AZ}) reduce to
\begin{align}
  \label{eq:hSLS+-}
  T h^{*}_{\pm}({\bm k}) T^{-1} &= h_{\pm}(-{\bm k}),\quad TT^{*}=\pm 1,\nonumber
  \\
  C h^{T}_{\pm}({\bm k}) C^{-1} &= -h_{\pm}(-{\bm k}), \quad CC^{*}=\pm 1.
%  \\
%  \gamma h^{\dagger}_{\pm}({\bm k})\gamma^{-1}&=-h_{\pm}({\bm k}),\quad \gamma^{2}=1. \nonumber
\end{align}
Therefore, $h_{+}(\bm{k})$ and $h_{-}(\bm{k})$ are independent and belong to the same real AZ class as $H(\bm{k})$. 
We have 8 such classes, AI + ${\cal S}_+$, BDI + ${\cal S}_{+-}$, D + ${\cal S}_-$, DIII + ${\cal S}_{+-}$, AII + ${\cal S}_+$, CII + ${\cal S}_{+-}$, C + ${\cal S}_-$, and CI + ${\cal S}_{+-}$.

\item 
%$h_\pm$ has neither TRS nor PHS. We have this situation when 
TRS (if exists) anti-commutes with SLS, but PHS (if exists) commutes with SLS:
Generally, we can take the basis, 
\begin{align}
  \label{sym:SLS-+}
  {\cal T} = \mu_{x}\otimes T, \quad
  {\cal C} = \mu_{0}\otimes C, 
%  \nonumber\\
%  \Gamma = \mu_{x}\otimes \gamma,\quad
%  {\cal S} = \mu_{z}\otimes \hat{1}, 
\end{align}
which leads to 
\begin{align}
  \label{eq:hSLS-+}
  T h^{*}_{\pm}({\bm k}) T^{-1} &= h_{\mp}(-{\bm k}),\quad TT^{*}=+ 1,\nonumber
  \\
  C h^{T}_{\pm}({\bm k}) C^{-1} &= -h_{\mp}(-{\bm k}), \quad CC^{*}=\pm 1.
%  \\
%  \gamma h^{\dagger}_{\pm}({\bm k})\gamma^{-1}&=-h_{\pm}({\bm k}),\quad \gamma^{2}=1.\nonumber 
\end{align}
Thus, $h_\pm$ supports neither TRS nor PHS, but it retains the combination of TRS and PHS, {\it i.e.} CS.
There are 5 such independent classes, AI + ${\cal S}_-$, BDI + ${\cal S}_{-+}$, D + ${\cal S}_+$, C + ${\cal S}_+$, and CI + ${\cal S}_{-+}$.

\item 
$H$ has TRS and PHS both commuting with SLS:
In this case, we can take the basis
\begin{align}
  \label{sym:SLS++}
  {\cal T} = \mu_{0}\otimes T, \quad
  {\cal C} = \mu_{0}\otimes C, 
%  \Gamma = \mu_{0}\otimes \gamma,\quad
%  {\cal S} = \mu_{z}\otimes \hat{1},
\end{align}
from which we have
\begin{align}
  \label{eq:hSLS++}
  T h^{*}_{\pm}({\bm k}) T^{-1} &= h_{\pm}(-{\bm k}),\quad TT^{*}=\pm 1,\nonumber
  \\
  C h^{T}_{\pm}({\bm k}) C^{-1} &= -h_{\mp}(-{\bm k}), \quad CC^{*}=\pm 1.
%  \\
%  \gamma h^{\dagger}_{\pm}({\bm k})\gamma^{-1}&=-h_{\mp}({\bm k}),\quad \gamma^{2}=1.\nonumber 
\end{align}
Thus, $h_\pm$ may retain only TRS. (If $H$ does not have TRS, $h_\pm$ has no symmetry.)
We have 4 such independent classes: BDI + ${\cal S}_{++}$, DIII + ${\cal S}_{++}$, CII + ${\cal S}_{++}$, and CI + ${\cal S}_{++}$.

\item 
$H$ has TRS and PHS both anti-commuting with SLS:
Taking the basis
\begin{align}
  \label{sym:SLS--}
  {\cal T} = \mu_{x}\otimes T, \quad
  {\cal C} = \mu_{x}\otimes C, 
%  \nonumber\\
%  \Gamma = \mu_{0}\otimes \gamma,\quad
%  {\cal S} = \mu_{z}\otimes \hat{1},
\end{align}
we have
\begin{align}
  \label{eq:hSLS--}
  T h^{*}_{\pm}({\bm k}) T^{-1} &= h_{\mp}(-{\bm k}),\quad TT^{*}=\pm 1,\nonumber
  \\
  C h^{T}_{\pm}({\bm k}) C^{-1} &= -h_{\pm}(-{\bm k}), \quad CC^{*}=\pm 1,
%  \\
%  \gamma h^{\dagger}_{\pm}({\bm k})\gamma^{-1}&=-h_{\mp}({\bm k}),\quad \gamma^{2}=1.\nonumber 
\end{align}
so $h_\pm$ supports only PHS. 
We have such 2 independent classes: BDI + ${\cal S}_{--}$, and CI + ${\cal S}_{--}$.
\end{enumerate}

%First, we consider the 10 of 22 symmetry classes listed in Table \ref{table:AZ+SLS+-}. In this case, we have SLS which commutes with TRS and anticommutes with both PHS and CS,
%\begin{align}
%  \label{eq:SLS+-}
%  {\cal T} H^{*}({\bm k}) {\cal T}^{-1} &= H(-{\bm k}),\quad {\cal T}{\cal T}^{*}=\pm 1,\nonumber
%  \\
%  {\cal C} H^T({\bm k}) {\cal C}^{-1} &= -H(-{\bm k}), \quad {\cal C}{\cal C}^{*}=\pm 1,\nonumber 
%  \\
%  \Gamma H^\dagger({\bm k})\Gamma^{-1}&=-H({\bm k}),\quad \Gamma^{2}=1, 
%  \\
%  {\cal S}H(\bm{k}){\cal S}^{-1} &= -H(\bm{k}), \quad {\cal S}^{2}=1,\nonumber 
%  \\
%  {\cal T}{\cal S}^{*}={\cal S}{\cal T}, \quad &{\cal C}{\cal S}^{*}=-{\cal S}{\cal C}, \quad \Gamma{\cal S}=-{\cal S}\Gamma. %\nonumber
%\end{align}
%Without loss of generality, we can take
%\begin{align}
%  \label{sym:SLS+-}
%%  {\cal T} &= \mu_{0}\otimes T, \nonumber \\
 % {\cal C} &= \mu_{x}\otimes C, \\
 % \Gamma &= \mu_{x}\otimes \gamma,\nonumber \\
 % {\cal S} &= \mu_{z}\otimes \hat{1}, \nonumber
%\end{align}
%where $T, C$ and $\gamma$ are unitary operators acting on $h_{\pm}(\bm{k})$.
%\begin{align}
%  \label{eq:hSLS+-}
%  T h^{*}_{\pm}({\bm k}) T^{-1} &= h_{\pm}(-{\bm k}),\quad TT^{*}=\pm 1,\nonumber
%  \\
%  C h^{T}_{\pm}({\bm k}) C^{-1} &= -h_{\pm}(-{\bm k}), \quad CC^{*}=\pm 1,
%  \\
%  \gamma h^{\dagger}_{\pm}({\bm k})\gamma^{-1}&=-h_{\pm}({\bm k}),\quad \gamma^{2}=1.\nonumber 
%\end{align}

Below, we construct the minimal models for these 22 classes.

\begin{table}[H]
  \centering
  \begingroup
  \renewcommand{\arraystretch}{1.1}
  \hspace*{-2.6em}
    \begin{tabular}{ccccccccc} \hline \hline
        Case && AZ class && ~SLS~ && ~$d=1$~ & ~$d=2$~ & ~$d=3$~ \\ \hline
        a && \multirow{1}{*}{$\text{A}$}&&{$\mathcal{S}$}&
        &$\color{blue}\mathbb{Z}\oplus\mathbb{Z} \to \mathbb{Z}[1,-1]$~\color{black}[Sec.\ref{proof:A+S}] & $0$ & $\color{green}\mathbb{Z}\oplus\mathbb{Z}$~\color{black}[Sec.\ref{proof:A+S}] \\
        b && \multirow{1}{*}{$\text{AIII}$}&&{$\mathcal{S}_{+}$}&
        &  $\color{green}\mathbb{Z}$~\color{black}[Sec.\ref{proof:AIII+S+}] & $0$ & $\color{green}\mathbb{Z}$~\color{black}[Sec.\ref{proof:AIII+S+}] \\
        c && \multirow{1}{*}{$\text{AIII}$}&&{$\mathcal{S}_{-}$}&
        &  $0$ & $\color{green}\mathbb{Z}\oplus\mathbb{Z}$~\color{black}[Sec.\ref{proof:AIII+S-}] & $0$ \\ \hline 
        d && \multirow{1}{*}{$\text{AI}$}&&{$\mathcal{S}_{+}$}&
        & $\color{blue}\mathbb{Z}\oplus\mathbb{Z} \to \mathbb{Z}[1,-1]$~\color{black}[Sec.\ref{proof:AI+S+}] & $0$ & $0$ \\ 
        && \multirow{1}{*}{$\text{BDI}$}&&{$\mathcal{S}_{+-}$}&
        & $\color{blue}\mathbb{Z}_{2}\oplus\mathbb{Z}_{2} \to \mathbb{Z}_{2}[1,1]$~\color{black}[Sec.\ref{proof:BDI+S+-}] & $\color{blue}\mathbb{Z}\oplus\mathbb{Z} \to \mathbb{Z}[2,0]\oplus\mathbb{Z}[1,-1]$~\color{black}[Sec.\ref{proof:BDI+S+-}] & $0$ \\ 
        &&\multirow{1}{*}{$\text{D}$}&&{$\mathcal{S}_{-}$}&
        & $\color{blue}\mathbb{Z}_{2}\oplus\mathbb{Z}_{2} \to \mathbb{Z}_{2}[1,1]$~\color{black}[Sec.\ref{proof:D+S-}] & $\color{blue}\mathbb{Z}_{2}\oplus\mathbb{Z}_{2} \to \mathbb{Z}_{2}[1,1]$~\color{black}[Sec.\ref{proof:D+S-}] & $\color{blue}\mathbb{Z}\oplus\mathbb{Z} \to \mathbb{Z}[2,0]\oplus\mathbb{Z}[1,-1]$~\color{black}[Sec.\ref{proof:D+S-}]  \\ 
        &&\multirow{1}{*}{$\text{DIII}$}&&{$\mathcal{S}_{+-}$}&
        &  $0$ & $\color{green}\mathbb{Z}_{2}\oplus\mathbb{Z}_{2}$~\color{black}[Sec.\ref{proof:DIII+S+-}] & $\color{green}\mathbb{Z}_{2}\oplus\mathbb{Z}_{2}$~\color{black}[Sec.\ref{proof:DIII+S+-}] \\ 
        &&\multirow{1}{*}{$\text{AII}$}&&{$\mathcal{S}_{+}$}&
        & $\color{blue}2\mathbb{Z}\oplus2\mathbb{Z} \to 2\mathbb{Z}[1,-1]$~\color{black}[Sec.\ref{proof:AII+S+}] & $0$ & $\color{green}\mathbb{Z}_{2}\oplus\mathbb{Z}_{2}$~\color{black}[Sec.\ref{proof:AII+S+}] \\ 
        &&\multirow{1}{*}{$\text{CII}$}&&{$\mathcal{S}_{+-}$}&
        & $0$ & $\color{green}2\mathbb{Z}\oplus2\mathbb{Z}$~\color{black}[Sec.\ref{proof:CII+S+-}] & $0$ \\ 
        &&\multirow{1}{*}{$\text{C}$}&&{$\mathcal{S}_{-}$}&
        &  $0$ & $0$ & $\color{green}2\mathbb{Z}\oplus2\mathbb{Z}$~\color{black}[Sec.\ref{proof:C+S-}]\\ 
        &&\multirow{1}{*}{$\text{CI}$}&&{$\mathcal{S}_{+-}$}&
        & $0$ & $0$ & $0$  \\ \hline
        e && \multirow{1}{*}{$\text{AI}$}&&{$\mathcal{S}_{-}$}&
        & $\color{red}\mathbb{Z}\to 0$~\color{black}[Sec.\ref{proof:AI+S-}] & $0$ & \color{green}\Z~\color{black}[Sec.\ref{proof:AI+S-}] \\ 
        &&\multirow{1}{*}{$\text{BDI}$}&&{$\mathcal{S}_{-+}$}&
        & $0$ & \color{green}\Z~\color{black}[Sec.\ref{proof:BDI+S-+}] & $0$ \\ 
        &&\multirow{1}{*}{$\text{D}$}&&{$\mathcal{S}_{+}$}&
        & \color{green}\Z~\color{black}[Sec.\ref{proof:D+S+}] & $0$ & \color{green}\Z~\color{black}[Sec.\ref{proof:D+S+}] \\ 
        &&\multirow{1}{*}{$\text{C}$}&&{$\mathcal{S}_{+}$}&
        & $\color{blue}\mathbb{Z}\to2\mathbb{Z}$~\color{black}[Sec.\ref{proof:C+S+}] & $0$ & \color{green}\Z~\color{black}[Sec.\ref{proof:C+S+}]\\ 
        &&\multirow{1}{*}{$\text{CI}$}&&{$\mathcal{S}_{-+}$}&
        & $0$ & \color{green}\Z~\color{black}[Sec.\ref{proof:CI+S-+}] & $0$ \\ \hline 
        f && \multirow{1}{*}{$\text{BDI}$}&&{$\mathcal{S}_{++}$}&
        & \color{green}\Z~\color{black}[Sec.\ref{proof:BDI+S++}] & $0$ & $0$ \\ 
        &&\multirow{1}{*}{$\text{DIII}$}&&{$\mathcal{S}_{++}$}&
        &  \color{green}2\Z~\color{black}[Sec.\ref{proof:DIII+S++}] & $0$ & \color{green}\Zt~\color{black}[Sec.\ref{proof:DIII+S++}] \\ 
        &&\multirow{1}{*}{$\text{CII}$}&&{$\mathcal{S}_{++}$}&
        & \color{green}2\Z~\color{black}[Sec.\ref{proof:CII+S++}] & $0$ & \color{green}\Zt~\color{black}[Sec.\ref{proof:CII+S++}] \\
        &&\multirow{1}{*}{$\text{CI}$}&&{$\mathcal{S}_{++}$}&
        & $\color{blue}\mathbb{Z}\to2\mathbb{Z}$~\color{black}[Sec.\ref{proof:CI+S++}] & $0$ & $0$ \\ \hline 
        g && \multirow{1}{*}{$\text{BDI}$}&&{$\mathcal{S}_{--}$}&
        & \color{green}\Zt~\color{black}[Sec.\ref{proof:BDI+S--}] & \color{green}\Zt~\color{black}[Sec.\ref{proof:BDI+S--}] & \color{green}\Z~\color{black}[Sec.\ref{proof:BDI+S--}] \\ 
        &&\multirow{1}{*}{$\text{CI}$}&&{$\mathcal{S}_{--}$}&
        & $0$ & $0$ & \color{green}2\Z~\color{black}[Sec.\ref{proof:CI+S--}] \\ \hline \hline
    \end{tabular}
    \endgroup
    \caption{Point-gap topological phases in 22 classes with SLS. For topological numbers colored red or blue, the left specifies topological numbers under PBCs and the right specifies those under OBCs. For topological numbers colored green, the classification under OBCs coincides with those under PBCs.
    The section numbers for the proof of the classification under OBCs and the BBC are shown for each topological number.
    \label{table:situation with SLS} }
\end{table}

\subsubsection{class A + ${\cal S}$ (case a.)}\label{proof:A+S}
This class has nontrivial point-gap topological phases in $d=1,3$.
%The Hamiltonian takes the form of
%\begin{align}
%H(\bm{k})=
%\begin{pmatrix}
%0 & h_{+}({\bm k})\\
%h_{-}({\bm k}) & 0
%\end{pmatrix}, \label{eq:A+SLS}
%\end{align}
%and there is no constraint on $h_{\pm}(\bm{k})$.
For $d=1$, the point-gap topological number under the PBC is a pair of the 1D winding numbers for $h_{\pm}(k_x)$, $(w_{1}[h_{+}],w_{1}[h_{-}])\in\mathbb{Z}\oplus\mathbb{Z}$ \cite{KSUS-19}, where $w_1[h_\pm]$ is given by
\begin{align}
    w_{1}[h_{\pm}] = \int_{0}^{2\pi}\dfrac{dk_x}{2\pi i}~\text{Tr}[h_{\pm}^{-1}\partial_{k_x}h_{\pm}].
    \label{eq:1Dwinding}
\end{align}
%Since $\text{det}[H(\bm{k})]=-\text{det}[h_{+}({\bm k})]\cdot\text{det}[h_{-}({\bm k})]$, $H(\bm{k})$ has a point-gap at $E=0$. 
We also have the 1D winding number $w_1[H]$ of $H(k_x)$, $w_{1}[H] = w_{1}[h_{+}]+w_{1}[h_{-}]$, which gives skin effects. 
Therefore, we have $w_{1}[h_{+}]+w_{1}[h_{-}]=0$ under the OBC \cite{Zhang-19,OKSS-20}. 
As a result, under the OBC,
the possible point-gap topological number is $(w_1[h_+], -w_1[h_+])$ and the $\mathbb{Z}\oplus\mathbb{Z}$ classification changes to the $\mathbb{Z}$ one.
Here we use the notation $\mathbb{Z}[1,-1]$ to represent the latter $\mathbb{Z}$ classification because its generator is given by the element $(1,-1)\in \mathbb{Z}\oplus\mathbb{Z}$ of the former. 
(We often use the same notation below.)
The minimal model of this phase is 
\begin{align}
\label{eq:A+SLS-1D}
&h_{+}(k_x) = -i\text{e}^{ik_x}, \nonumber\\
&h_{-}(k_x) = i\text{e}^{-ik_x}.
\end{align}
which has $(+1,-1)\in\mathbb{Z}\oplus\mathbb{Z}$. This model gives 
\begin{align}
H(k_x)=\sin k_x \mu_x+\cos k_x \mu_y, \quad {\cal S}=\mu_z,
\end{align}
which supports a boundary zero mode under the OBC. Thus, we have the BBC.

For $d=3$, the classification of the point-gap topological phase is $\mathbb{Z}\oplus \mathbb{Z}$, of which topological numbers are a pair of the 3D winding numbers of $h_{\pm}(\bm{k})$, $(w_{3}[h_{+}],w_{3}[h_{-}])\in\mathbb{Z}\oplus\mathbb{Z}$ \cite{KSUS-19}.
Thus, we have two generators, $(1,0)$ and $(0,1)$.
The minimal model for the generator $(1,0)$ is
\begin{align}
\label{eq:A+SLS-3D}
&h_{+}(\bm{k}) = \sin{k_x}\sigma_{x} + \sin{k_y}\sigma_{y} + \left(2 - \sum_{i=x,y,z}\cos{k_i}\right)\sigma_{z} + i\sin{k_z}\sigma_{0},\nonumber\\ 
&h_{-}(\bm{k}) = \sigma_{0}.
\end{align}
%which takes $(+1,0)\in\mathbb{Z}\oplus\mathbb{Z}$ for Eq.~(\ref{eq:A+SLS-3D}). 
We also have the minimal model for another generator $(0,+1)\in\mathbb{Z}\oplus\mathbb{Z}$, by swapping $h_{+}(\bm{k})$ and $h_{-}(\bm{k})$ in Eq.~(\ref{eq:A+SLS-3D}).
Under the xOBC, we have $[h_+,h_-]=0$ and either $h_+$ or $h_-$ supports boundary states. Thus we have the BBC.

\subsubsection{class AIII + ${\cal S}_{+}$ (case b.)}\label{proof:AIII+S+}
In this class, there exist nontrivial point-gap topological phases characterized by the winding numbers of $h_+$ in $d=1,3$.
For $d=1$, the minimal model is 
\begin{align}
\label{eq:AIII+SLS+-1D}
&h_{+}(k_x) = \text{e}^{ik_x},\nonumber \\
&h_{-}(k_x) = -\gamma h^{\dagger}_{+}(k_x) \gamma^{-1} = -\text{e}^{-ik_x},
\end{align}
where $\gamma=1$ and $w_{1}[h_{+}]=+1\in\mathbb{Z}$. From this, we have $H(k_x)=i(\sin k_x \mu_x+\cos k_x\mu_y)$ with ${\cal S}=\mu_z$ and $\Gamma=1$.
This model has a boundary state under the OBC, and thus we have the BBC.

For $d=3$, we have the minimal model with $\gamma=\sigma_{0}$ and $w_{3}[h_{+}]=+1\in\mathbb{Z}$ as follows.
\begin{align}
\label{eq:AIII+SLS+-3D}
&h_{+}(\bm{k}) = \sin{k_x}\sigma_{x} + \sin{k_y}\sigma_{y} + \left(2 - \sum_{i=x,y,z}\cos{k_i}\right)\sigma_{z} + i\sin{k_z}\sigma_{0},\nonumber\\ 
&h_{-}(\bm{k}) = -\gamma h^{\dagger}_{+}({\bm k}) \gamma^{-1} = -\sin{k_x}\sigma_{x} - \sin{k_y}\sigma_{y} - \left(2 - \sum_{i=x,y,z}\cos{k_i}\right)\sigma_{z} + i\sin{k_z}\sigma_{0}.
\end{align}
Note that $h_+({\bm k})$ concides with Eq.(\ref{eq:3D-A}) so $h_\pm$ has surface states under the xOBC.
Thus, from $[h_+,h_-]=0$, we have the BBC.

\subsubsection{class AIII + ${\cal S}_{-}$ (case c.)}\label{proof:AIII+S-}
In this class, $h_\pm$ has its own CS so belongs to class AIII.  Thus, 
we have a $\mathbb{Z}\oplus\mathbb{Z}$ point-gap topological phase in $d=2$.
A pair of the first Chern numbers for $ih_{\pm}(\bm{k})\gamma$, $(Ch_{1}[ih_{+}\gamma],Ch_{1}[ih_{-}\gamma])$$\in\mathbb{Z}\oplus\mathbb{Z}$, is the point-gap topological number\cite{KSUS-19}.
Using the model in Eq. (\ref{eq:2D-AIII}), we obtain the minimal model with the topological number $(+1,0)\in\mathbb{Z}\oplus\mathbb{Z}$ as
\begin{align}
\label{eq:AIII+SLS--2D}
&h_{+}(\bm{k}) = \sin{k_x}\sigma_{x} + \left(1 - \sum_{i=x,y}\cos{k_i}\right)\sigma_{y} + i\sin{k_y}\sigma_{0},
\nonumber\\ 
&h_{-}(\bm{k}) = i\sigma_{0},
\quad \gamma=\sigma_z.
\end{align}
We also have another minimal model with topological number $(0,+1)\in\mathbb{Z}\oplus\mathbb{Z}$ by exchanging $h_{+}(\bm{k})$ and $h_{-}(\bm{k})$ in Eq.~(\ref{eq:AIII+SLS--2D}).
In both cases, $h_+$ and $h_-$ commute, and either $h_-$ or $h_-$ has a boundary state. Thus, the BBC holds.

\subsubsection{class AI + ${\cal S}_+$ (case d.)}\label{proof:AI+S+}
In this class, $h_\pm$ belongs to class AI, and there is a $\mathbb{Z}\oplus \mathbb{Z}$ point-gap topological phase in $d=1$. The point-gap topological number under the  PBC is a pair of the 1D winding number $(w_1[h_+],w_1[h_-])$, where $w_1[h_\pm]$ is given by Eq.(\ref{eq:1Dwinding}).
In a manner similar to the $d=1$ class A + ${\cal S}$ case, one can avoid skin effects only when $w_1[h_+]+w_1[h_-]=0$, 
so the point-gap topological phase under the OBC is classified as $\mathbb{Z}[1,-1]$.
The minimal model is 
\begin{align}
\label{eq:AI+SLS+-1D}
h_{+}(k_x) = \text{e}^{ik_x} \quad
h_{-}(k_x) = \text{e}^{-ik_x}, \quad T=1.
\end{align}
This model gives $H(k_x)=\cos k_x \mu_x - \sin k_x\mu_y$ with ${\cal T}=\mu_0$ and ${\cal S}=\mu_z$, which has a zero energy boundary state. Thus we have the BBC.

\subsubsection{class BDI + ${\cal S}_{+-}$ (case d.)}\label{proof:BDI+S+-}
In class $\text{BDI}+\mathcal{S}_{+-}$, $h_\pm$ belongs to class BDI, which has nontrivial point-gap topological phases in $d=1,2$. 
%The required symmetry is
%\begin{align}
%  \label{sym:BDI+S+-}
%  T h^{*}_{\pm}({\bm k}) T^{-1} &= h_{\pm}(-{\bm k}),\quad TT^{*}=+ 1,\quad T=\sigma_0\nonumber
%  \\
%  C h^{T}_{\pm}({\bm k}) C^{-1} &= -h_{\pm}(-{\bm k}), \quad CC^{*}=+ 1,\quad C=\sigma_0,
% \\
% \gamma h^{\dagger}_{\pm}({\bm k})\gamma^{-1}&=-h_{\pm}({\bm k}),\quad \gamma^{2}=1,\quad \gamma=CT^{*}.\nonumber 
%\end{align}
For $d=1$, the point-gap topological number under the PBC is a pair of the $\mathbb{Z}_{2}$ invariants for $h_{\pm}(k_x)$, $(\nu[h_{+}],\nu[h_{-}])\in\mathbb{Z}_{2}\oplus\mathbb{Z}_{2}$ \cite{KSUS-19},
\begin{align}
    (-1)^{\nu_1[h_{\pm}]} = \text{sgn}\left\{\dfrac{\operatorname{Pf}[h_{\pm}(\pi)C]}{\operatorname{Pf}[h_{\pm}(0)C]}
%    ~\times~\text{exp}\left[-\dfrac{1}{2}\int_{k_x=0}^{k_x=\pi}\operatorname{\it{d}}\operatorname{log}\operatorname{det}[h_{\pm}C]\right]
    \right\},
    \label{eq:Z2PBC-PHS}
\end{align}
which coincides with Eq.~(\ref{eq:Z2OBC}) in OBCs. 
As the Hamiltonian $H(k_x)$ has TRS$^\dagger$ for class AII$^\dagger$,
\begin{align}
\mathcal{T}_{\text{AII}^{\dagger}}H^{\text{T}}(k_x)\mathcal{T}_{\text{AII}^{\dagger}}^{-1}=H(-k_x),    
\quad
\mathcal{T}_{\text{AII}^{\dagger}}\mathcal{T}_{\text{AII}^{\dagger}}^{*}=-1, 
\end{align}
where $\mathcal{T}_{\text{AII}^{\dagger}}$ is given by the combination of ${\cal C}$ and ${\cal S}$, 
{\it i.e.} $\mathcal{T}_{\text{AII}^{\dagger}}=\mathcal{SC}=i\mu_{y}\otimes C$, we also have the $\mathbb{Z}_{2}$ number for $H(k_x)$ in class AII$^{\dagger}$,
\begin{align}
    (-1)^{\nu_1[H]} = \text{sgn}\left\{\dfrac{\operatorname{Pf}[H(\pi)\mathcal{C}_{\text{AII}^{\dagger}}]}{\operatorname{Pf}[H(0)\mathcal{C}_{\text{AII}^{\dagger}}]}~\times~\text{exp}\left[-\dfrac{1}{2}\int_{k_x=0}^{k_x=\pi}\operatorname{\it{d}}\operatorname{log}\operatorname{det}\left[H(k_x)\mathcal{C}_{\text{AII}^{\dagger}}\right]\right]\right\},
    \label{eq:Z2PBC-TRSdag}
\end{align}
which satisfies  $\nu_1[H]=\nu_1[h_+]+\nu_1[h_-]~(\text{mod.}~2)$. 
Since the $\mathbb{Z}_2$ number results in the symmetry-skin effect \cite{OKSS-20}, 
we have the condition $\nu_1[h_+]+\nu_1[h_-]=0$ (mod.2) to avoid it. 
Therefore, the point-gap topological phase under OBC is $\mathbb{Z}_{2}[1,1]$.
The minimal model is 
\begin{align}
\label{eq:BDI+SLS+--1D}
h_{+}(k_x) = h_{-}(k_x) = i\left(\sin k_x \sigma_x - \cos{k_x} \sigma_y\right),
\end{align}
which shows the BBC.

For $d=2$, the point-gap topological phase is $\mathbb{Z}\oplus\mathbb{Z}$, which topological number is given by a pair of the first Chern numbers, $(Ch_1[ih_+\gamma], Ch_1[ih_-\gamma])$ \cite{KSUS-19}. 
However, using the dimension reduction discussed in the main text, we have the symmetry-protected skin effect when the first Chern number of $iH\Gamma$, $Ch_1[iH\Gamma]=Ch_1[ih_+ \gamma]+Ch_1[ih_-\gamma]$ is odd. 
Therefore, the point-gap topological phase becomes $\mathbb{Z}[1,-1]\oplus \mathbb{Z}[2,0]$ under OBCs.

%In class BDI under full PBCs, however, an odd parity of the non-trivial 2D $\mathbb{Z}$ index yields a non-zero 1D $\mathbb{Z}_2$ index along a high-symmetry line in the BZ, in a similar vein to the dimensional reduction presented in the main text. Thus, the topological phases meeting $Ch_{1}[ih_{+}\gamma]+Ch_{1}[ih_{-}\gamma]=1~(\text{mod}~2)$ in PBCs necessarily show NHSE because of $\eta[h_+]+\eta[h_-]=\eta[H]=1~(\text{mod}~2)$ along a high-symmetry line in the BZ, and only those satisfying $Ch_{1}[ih_{+}\gamma]+Ch_{1}[ih_{-}\gamma]=0~(\text{mod}~2)$ survive under OBCs. As a result, the classification in OBCs consists of the pair of even indices and the pair of odd indices, $2\mathbb{Z}\oplus2\mathbb{Z},(2\mathbb{Z}+1)\oplus(2\mathbb{Z}+1)\in\mathbb{Z}\oplus\mathbb{Z}$.

Using Eq.(\ref{eq:BDI-2D}) for $h_\pm$, we can construct the minimal model with the topological number $(1,-1)$ as
\begin{align}
\label{eq:BDI+SLS+--2D}
h_{+}(\bm{k}) = -h_{-}(\bm{k}) = i\left[\sin k_x \sigma_x
  + \sin k_y \sigma_z
  + \left(1 - \sum_{i=x,y}\cos{k_i}\right) \sigma_y\right], \quad C=T=\sigma_0
\end{align}
In a similar manner, another minimal model with the topological number $(2,0)$ is given by
\begin{align}
&h_+({\bm k})=i\tau_0\left[\sin k_x \sigma_x
  + \sin k_y \sigma_z
  + \left(1 - \sum_{i=x,y}\cos{k_i}\right) \sigma_y\right],
\nonumber\\
&h_-({\bm k})=i\tau_y\sigma_0, \quad T=C=\tau_0\sigma_0.
\end{align}
In both models, $h_+$ and $h_-$ commute with each other and they have boundary states when their first Chern numbers are non-zero.
Therefore, the BBC holds.

\subsubsection{class D + ${\cal S}_{-}$ (case d.)}\label{proof:D+S-}
In class $\text{D}+\mathcal{S}_{-}$, $h_\pm$ belongs to class D, which realizes non-trivial point-gap topological phases in $d=1,2,3$. %We impose the following symmetry on $h_{\pm}(\bm{k})$,
%\begin{align}
%  \label{sym:D+S-}
%  C h^{T}_{\pm}({\bm k}) C^{-1} &= -h_{\pm}(-{\bm k}), \quad CC^{*}=+ 1,\quad C=\sigma_0.
%\end{align}
For $d=1$, $h_\pm$ supports the $\mathbb{Z}_2$ point-gap topological number $\nu_1[h_\pm]$ in Eq.(\ref{eq:Z21D-D}), and thus the classification under the PBC is given by $(\nu_1[h_+],\nu_1[h_-]) \in \mathbb{Z}_2\oplus\mathbb{Z}_2$.
Then, as $H({\bm k})$ has TRS$^\dagger$ of class AII$^\dagger$,
\begin{align}
{\cal T}_{\rm AII^\dagger} H^T({\bm k}){\cal T}_{\rm AII^\dagger}^{-1}=H(-{\bm k})    
\end{align}
with $\mathcal{T}_{\text{AII}^{\dagger}}=\mathcal{SC}$, we have the 1D $\mathbb{Z}_2$ number $\nu_1[H]$ for the symmetry-protected skin effect, so we have the condition $\nu_1[H]=\nu_1[h_+]+\nu_1[h_-]=0$ (mod.2) to avoid the skin effects.
As a result, 
the point-gap topological phase under the OBC becomes $\mathbb{Z}_2[1,1]$.
From Eq.(\ref{eq:D-1D}), the minimal model is 
\begin{align}
\label{eq:D+S--1D}
h_{+}(k_x) = h_{-}(k_x) = \sin k_x \sigma_x + \cos{k_x} \sigma_y, \quad C=\sigma_0,
\end{align}
which has $(1,1)\in\mathbb{Z}_{2}\oplus\mathbb{Z}_{2}$.
Since it supports a boundary state, we have the BBC.

For $d=2$, the point-gap topological number under the PBC is a pair of $(\nu_2[h_+], \nu_2[h_-])$ in Eq.~(\ref{eq:Z22D-D}) \cite{KSUS-19}. 
Then, from TRS$^\dagger$ in the above, the dimension reduction argument in the main text indicates that there arises the symmetry-protected skin effect when $\nu_2[h_+]+\nu_2[h_-]=1$ (mod.2). 
Hence, the topological classification under OBCs becomes $\mathbb{Z}_2[1,1]$.
The minimal model with topological number $(1,1)$ is 
\begin{align}
\label{eq:D+S--2D}
h_{+}(\bm{k}) = h_{-}(\bm{k}) = \sin k_x \sigma_x + \left(1 - \sum_{i=x,y}\cos{k_i}\right) \sigma_y + i\sin k_y \sigma_0 , \quad C=\sigma_0,
\end{align}
which shows the BBC.

In $d=3$, a pair of the 3D winding numbers $(w_3[h_+],w_3[h_-])$ characterize the point-gap topological phase under the PBC \cite{KSUS-19}. In a manner similar to the above, we have the symmetry-protected skin effect when $w_3[h_+]+w_3[h_-]$ is odd.
Consequently, the classification under OBCs becomes $\mathbb{Z}[1,-1]\oplus \mathbb{Z}[2,0]$.
The minimal model with topological number $(1,-1)$ is 
\begin{align}
\label{eq:D+S--3D}
&h_{+}(\bm{k}) = \sin{k_x}\sigma_{x} + \sin{k_y}\sigma_{z} + \left(2 - \sum_{i=x,y,z}\cos{k_i}\right)\sigma_{y} - i\sin{k_z}\sigma_{0},\nonumber\\
&h_{-}(\bm{k}) = \sin{k_x}\sigma_{x} + \sin{k_y}\sigma_{z} + \left(2 - \sum_{i=x,y,z}\cos{k_i}\right)\sigma_{y} + i\sin{k_z}\sigma_{0}, \quad C=\sigma_0,
\end{align}
and the minimal model with topological number $(2,0)$ is
\begin{align}
&h_{+}(\bm{k}) = \tau_0\left(
\sin{k_x}\sigma_{x} + \sin{k_y}\sigma_{z} + \left(2 - \sum_{i=x,y,z}\cos{k_i}\right)\sigma_{y} - i\sin{k_z}\sigma_{0}
\right),\nonumber\\
&h_{-}(\bm{k}) = \tau_y\sigma_0,
\quad C=\tau_0\sigma_0,
\end{align}
both of which show the BBC.

\subsubsection{class DIII + ${\cal S}_{+-}$ (case d.)}\label{proof:DIII+S+-}
The off-diagonal components $h_\pm$ belong to class DIII, so we have nontrivial point-gap topological phases in $d=2,3$.
%In this class, the required symmetry is
%\begin{align}
%  \label{sym:DIII+S+-}
%  T h^{*}_{\pm}({\bm k}) T^{-1} &= h_{\pm}(-{\bm k}),\quad TT^{*}=- 1,\quad T=i\tau_y\sigma_0,\nonumber
%  \\
%  C h^{T}_{\pm}({\bm k}) C^{-1} &= -h_{\pm}(-{\bm k}), \quad CC^{*}=+ 1,\quad C=\tau_0\sigma_0,
%  \\
% \gamma h^{\dagger}_{\pm}({\bm k})\gamma^{-1}&=-h_{\pm}({\bm k}),\quad \gamma^{2}=1,\quad \gamma=iCT^{*}.\nonumber 
%\end{align}
%Since $h_+$ and $h_-$ commutes and they have boundary states, we have the BBC.
Under the PBC, these phases are characterized by a pair of the $\mathbb{Z}_2$ invariants in $d=2,3$ class DIII \cite{KSUS-19} for $h_{\pm}(\bm{k})$. 

For $d=2$, the minimal model is
\begin{align}
\label{eq:DIII+SLS+--2D}
&h_{+}(\bm{k}) = \sin{k_x}\tau_{x}\sigma_{x} + \left(1 - \sum_{i=x,y}\cos{k_i}\right)\tau_{x}\sigma_{y} + i\sin{k_y}\tau_{0}\sigma_{0},\nonumber\\ 
&h_{-}(\bm{k}) = i\tau_{y}\sigma_{z},\quad T=i\tau_y\sigma_0,\quad C=\tau_0\sigma_0,
\end{align}
which takes $(1,0)\in\mathbb{Z}_2\oplus\mathbb{Z}_2$. We can also construct another minimal model with topological number $(0,1)\in\mathbb{Z}_2\oplus\mathbb{Z}_2$ by exchanging $h_{+}(\bm{k})$ and $h_{-}(\bm{k})$ in Eq.~(\ref{eq:DIII+SLS+--2D}).
Since $h_+$ in Eq.(\ref{eq:DIII+SLS+--2D}) has a boundary state and commutes with $h_-$, we have the BBC.

Similarly, for $d=3$, the minimal model with topological number $(1,0)$ is
\begin{align}
\label{eq:DIII+SLS+--3D}
&h_{+}(\bm{k}) = \sin{k_x}\tau_{x}\sigma_{x} + \sin{k_y}\tau_{x}\sigma_{z} + \left(2 - \sum_{i=x,y,z}\cos{k_i}\right)\tau_{x}\sigma_{y} + i\sin{k_z}\tau_{0}\sigma_{0},\nonumber\\ 
&h_{-}(\bm{k}) = i\tau_{y}\sigma_{0}, \quad T=i\tau_y\sigma_0,\quad C=\tau_0\sigma_0,
\end{align}
%
%\color{blue}\begin{align}
%\label{eq:DIII+SLS+--3D}
%&h_{+}(\bm{k}) = \sin{k_x}\tau_{x}\sigma_{x} + \sin{k_y}\tau_{x}\sigma_{z} + %\sin{k_z}\tau_{z}\sigma_{0}+ \left(2 - \sum_{i=x,y,z}\cos{k_i}\right)\tau_{x}\sigma_{y},\nonumber\\ 
%&h_{-}(\bm{k}) = i\tau_{y}\sigma_{0}, \quad T=i\tau_y\sigma_0,\quad C=\tau_0\sigma_0,
%\end{align}\color{black}
We also have another minimal model with topological number $(0,1)$ by swapping $h_{+}(\bm{k})$ and $h_{-}(\bm{k})$ in Eq.~(\ref{eq:DIII+SLS+--3D}).
Both models show the BBC from Lemma \ref{lemma:2}.

\subsubsection{class AII + ${\cal S}_+$ (case d.)}\label{proof:AII+S+}
Since $h_\pm$ belongs to class AII, there are nontrivial point-gap topological phases in $d=1,3$. 
For $d=1$, the point-gap topological phase is $2\mathbb{Z}\oplus 2\mathbb{Z}$ \cite{KSUS-19}, which topological number is given by a pair of the 1D winding numbers $w_1[h_\pm]$.
To avoid skin effects, the 1D winding number $w_1[H]$ of $H$ should vanish, so we have the condition $w_1[H]=w_1[h_+]+w_1[h_-]=0$. Thus, the classification under the OBC is $2\mathbb{Z}[1,-1]$.
The minimal model is 
\begin{align}
\label{eq:AII+SLS+-1D}
h_{+}(k_x) = i\text{e}^{ik_x}\sigma_y, \quad
h_{-}(k_x) = -i\text{e}^{-ik_x}\sigma_y, \quad T=i\sigma_y,
\end{align}
which has $(+2,-2)\in2\mathbb{Z}\oplus2\mathbb{Z}$ and shows the BBC.

For $d=3$, the point-gap topological number under the PBC is a pair of the $\mathbb{Z}_2$ invariants in class AII \cite{KSUS-19} for $h_{\pm}(\bm{k})$. We have the minimal model with $(1,0)\in\mathbb{Z}_2\oplus\mathbb{Z}_2$,
\begin{align}
\label{eq:AII+SLS+--3D}
&h_{+}(k_x) = i\left[\sin k_x \tau_x \sigma_x 
  + \sin k_y \tau_x\sigma_y 
  + \left(2 - \sum_{i=x,y,z}\cos{k_i}\right)\tau_0\sigma_z\right]
  - \sin k_z  \tau_z\sigma_z, \nonumber\\
&h_{-}(k_x) = \tau_0\sigma_0, \quad T=i\tau_z\sigma_y.
\end{align}
We can obtain another minimal model with topological number $(0,1)$ by exchanging $h_{+}(\bm{k})$ and $h_{-}(\bm{k})$ in Eq.~(\ref{eq:AII+SLS+--3D}).
From them, the BBC holds.

\subsubsection{class CII + ${\cal S}_{+-}$ (case d.)}\label{proof:CII+S+-}
In class $\text{CII}+\mathcal{S}_{+-}$, $h_\pm$ belongs to class CII, so there exist nontrivial point-gap topological phases in $d=2$. 
%The required symmetry is
%\begin{align}
%  \label{sym:CII+S+-}
%  T h^{*}_{\pm}({\bm k}) T^{-1} &= h_{\pm}(-{\bm k}),\quad TT^{*}=- 1,\quad T=i\tau_y\sigma_z\nonumber
%  \\
%  C h^{T}_{\pm}({\bm k}) C^{-1} &= -h_{\pm}(-{\bm k}), \quad CC^{*}=- 1,\quad C=i\tau_y\sigma_0,
%  \\
% \gamma h^{\dagger}_{\pm}({\bm k})\gamma^{-1}&=-h_{\pm}({\bm k}),\quad \gamma^{2}=1,\quad \gamma=CT^{*}.\nonumber 
%\end{align}
The point-gap topological number is a pair of the first Chern numbers $(Ch_1[ih_+TC^*], Ch_1[ih_-TC^*]) \in 2\mathbb{Z}\oplus 2\mathbb{Z}$. 
The minimal model with topological number $(2,0)$ is 
\begin{align}
\label{eq:CII+S+--2D}
&h_{+}(\bm{k}) = \tau_{0}\left[\sin{k_x}\sigma_{x} + \left(1 - \sum_{i=x,y}\cos{k_i}\right)\sigma_{y} + i\sin{k_y}\sigma_{0}\right],\nonumber\\ 
&h_{-}(\bm{k}) = i\tau_y\sigma_{0}, \quad T=i\tau_y\sigma_z, \quad C=i\tau_y\sigma_0, 
\end{align}
and another one with topological number $(0,2)$ is obtained by switching $h_{+}(\bm{k})$ and $h_{-}(\bm{k})$ in Eq.~(\ref{eq:CII+S+--2D}).
From them, we have the BBC.

\subsubsection{class C + ${\cal S}_{-}$ (case d.)}\label{proof:C+S-}
$h_\pm$ belongs to class C, and we have nontrivial point-gap topological phases in $d=3$. 
A pair of the 3D winding numbers $(w_3[h_+],w_3[h_-])\in 2\mathbb{Z}\oplus 2\mathbb{Z}$ characterize the point-gap topological phases under the PBC \cite{KSUS-19}. 
The minimal model with topological number $(2.0)$ is
\begin{align}
\label{eq:C+S--3D}
&h_{+}(\bm{k}) = \tau_{0}\left[\sin{k_x}\sigma_{x} + \sin{k_y}\sigma_{z} + \left(2 - \sum_{i=x,y,z}\cos{k_i}\right)\sigma_{y} - i\sin{k_z}\sigma_{0}\right],\\ 
&h_{-}(\bm{k}) = i\tau_y\sigma_{0}, \quad C=i\tau_y\sigma_0,
\end{align} 
and we also have aother minimal model with $(0,+2)\in2\mathbb{Z}\oplus2\mathbb{Z}$ similarly.
These models show the BBC.

\subsubsection{class CI + ${\cal S}_{+-}$ (case d.)}\label{proof:CI+S+-}
Since the point gap topological phases are trivial for class CI in $d=1,2,3$, so are also for this class. 

\subsubsection{class AI + ${\cal S}_-$ (case e.)}\label{proof:AI+S-}
In this class, $h_{+}$ belongs to class A, and TRS exchanges $h_+$ and $h_-$. 
Thus, we have nontrivial point-gap topological phases in $d=1,3$ under the PBC.
For $d=1$, the corresponding point-gap topological number is the 1D winding number $w_1[h_+]$ for $h_{+}(k_x)$ \cite{KSUS-19}. 
We can also define the 1D winding number $w_1[H]$ for $H(k_x)$, which satisfies $w_{1}[H] = w_{1}[h_{+}]+w_{1}[h_{-}]$, where $w_1[h_-]$ is the 1D winding number for $h_-$. 
Then, from  $h_{-}(k_x)=T h^{*}_{+}(-k_x) T^{-1}$, we have $w_{1}[h_{+}]=w_{1}[h_{-}]$.
Thus, to avoid skin effect, we need $w_{1}[h_+]=0$, which implies no point-gap topological phase survives under the OBC.

For $d=3$, we have a nontrivial point-gap topological phase under the OBC.  The minimal model is 
\begin{align}
\label{eq:AI+SLS--3D}
&h_{+}(\bm{k}) = \sin{k_x}\sigma_{x} + \sin{k_y}\sigma_{y} + \left(2 - \sum_{i=x,y,z}\cos{k_i}\right)\sigma_{z} + i\sin{k_z}\sigma_{0},\nonumber\\
&h_{-}(\bm{k}) = T h^{*}_{+}(-{\bm k}) T^{-1}=-\sin{k_x}\sigma_{x} - \sin{k_y}\sigma_{y} - \left(2 - \sum_{i=x,y,z}\cos{k_i}\right)\sigma_{z} + i\sin{k_z}\sigma_{0},
\end{align}
where $T=\sigma_x$.
As $h_\pm$ has a boundary state and $[h_+, h_-]=0$, the BBC holds.

\subsubsection{class BDI + ${\cal S}_{-+}$ (case e.)}\label{proof:BDI+S-+}
In class $\text{BDI}+{\cal S}_{-+}$, $h_{+}(\bm{k})$ has CS, $\gamma h^\dagger_+({\bm k})\gamma^{\dagger}=-h_+({\bm k})$ with $\gamma=TC^*$, and TRS and PHS exchange $h_+$ and $h_-$.
So $h_+$ belongs to class AIII.
Thus, we have a non-trivial point-gap topological phase in $d=2$.
The Chern number for $ih_{+}(\bm{k})\gamma$ characterizes the point-gap topological phase\cite{KSUS-19}. The minimal model is
\begin{align}
\label{eq:BDI+SLS-+-2D}
&h_{+}(\bm{k}) = \sin{k_x}\sigma_{x} + \left(1 - \sum_{i=x,y}\cos{k_i}\right)\sigma_{y} + i\sin{k_y}\sigma_{0}, \nonumber\\
&h_{-}(\bm{k}) = T h^{*}_{+}(-{\bm k}) T^{-1} = -C h^{\text{T}}_{+}(-{\bm k}) C^{-1}= \sin{k_x}\sigma_{x} + \left(1 - \sum_{i=x,y}\cos{k_i}\right)\sigma_{y} + i\sin{k_y}\sigma_{0},
\end{align}
with $T=\gamma=\sigma_z,C=\sigma_0$. 
$h_+$ and $h_-$ in the minimal model commute and they support boundary states.
Thus, we have the BBC.

\subsubsection{class D + ${\cal S}_+$ (case e.)}\label{proof:D+S+}
In this class, $h_{+}(\bm{k})$ belongs to class A and PHS exchanges $h_+$ and $h_-$.
The system has non-trivial topological phases in $d=1,3$.
For $d=1$, the corresponding point-gap topological number under the PBC is the 1D winding number $w_1[h_+]\in \mathbb{Z}$ for $h_{+}(k_x)$. We can also consider the 1D winding number $w_1[H]$ for $H(k_x)$, $w_{1}[H] = w_{1}[h_{+}]+w_{1}[h_{-}]$, where $w_1[h_-]$ is the 1D winding number for $h_-(k_x)$.
However, the relation $h_{-}(k_x)=-C h^{\text{T}}_{+}(-k_x) C^{-1}$ leads $w_{1}[h_{+}]=-w_{1}[h_{-}]$, so $w_{1}[H]$ always vanishes. Thus, no skin effect occurs.
The minimal model with $w_1[h_+]=1$ is
\begin{align}
\label{eq:D+S+-1D}
h_{+}(k_x) = \text{e}^{ik_x}, \quad
h_{-}(k_x) = -C h^{\text{T}}_{+}(-k_x) C^{-1} = -\text{e}^{-ik_x},
\end{align}
where $C=1$. This model gives $H(k_x)=i(\sin k_x \mu_x+\cos k_x \mu_y)$, which supports a boundary state. Thus the BBC holds.

In $d=3$, the point-gap topological number is the 3D winding number $w_3[h_+]\in \mathbb{Z}$.
The minimal model is 
\begin{align}
\label{eq:D+S+-3D}
&h_{+}(\bm{k}) = \sin{k_x}\sigma_{x} + \sin{k_y}\sigma_{y} + \left(2 - \sum_{i=x,y,z}\cos{k_i}\right)\sigma_{z} + i\sin{k_z}\sigma_{0}, \nonumber\\
&h_{-}(\bm{k}) = -C h^{\text{T}}_{+}(-{\bm k}) C^{-1} =\sin{k_x}\sigma_{x} + \sin{k_y}\sigma_{y} + \left(2 - \sum_{i=x,y,z}\cos{k_i}\right)\sigma_{z} + i\sin{k_z}\sigma_{0},
\end{align}
where $C=\sigma_x$ and $w_{3}[h_{+}]=+1$. This model shows the BBC.

\subsubsection{class C + ${\cal S}_+$ (case e.)}\label{proof:C+S+}
For this class, $h_{+}(\bm{k})$ belongs to class A, and PHS exchanges $h_+$ and $h_-$.
Thus, point-gap topological phases under the PBC are $\mathbb{Z}$ in $d=1,3$.
For $d=1$, the point-gap topological number is the 1D winding number $w_1[h_+]$ for $h_{+}(k_x)$.
Because $H({\bm k})$ has TRS for class AII$^\dagger$, 
${\cal T}_{\rm AII^\dagger}H^T({\bm k}){\cal T}_{\rm AII^\dagger}^{-1}=H(-{\bm k})$ with ${\cal T}_{\rm AII^\dagger}={\cal S}{\cal C}$, 
we also have the 1D $\mathbb{Z}_2$ number $\nu_1[H]$ responsible for the symmetry-protected skin effect. 
From the straightforward calculation, we can show that $\nu_1[H]=w_1[h_+]$ (mod.2), thus 
%In $d=1$, since we can regard $h(k_x)\equiv h_{+}(k_x)h_{-}(k_x)=-h_{+}(k_x)C h^{\text{T}}_{+}(-k_x) C^{-1}$ as being in class AII$^{\dagger}$ with $C$, we can define the $\mathbb{Z}_{2}$ invariant for $h(k_x)$,
%\begin{align}
%    (-1)^{\eta[h]} &= \text{sgn}\left\{\dfrac{\operatorname{Pf}[h(\pi)C]}{\operatorname{Pf}[h(0)C]}~\times~\text{exp}\left[-\dfrac{1}{2}\int_{k_x=0}^{k_x=\pi}\operatorname{\it{d}}\operatorname{log}\operatorname{det}\left[h(k_x)C\right]\right]\right\},
%\end{align}
%which leads $\eta[h]=w_{1}[h_{+}]~(\text{mod}~2)$. As a result, 
only the even part of $w_{1}[h_{+}]$  survives under the OBC.
Thus, the classification under the OBC becomes $2\mathbb{Z}$.
%NHSE in $h_{+}(k_x)h_{-}(k_x)$ and the topological classification changes from $\mathbb{Z}$ to $2\mathbb{Z}$ under OBCs, because the square root of $h_{+}(k_x)h_{-}(k_x)$ yields the eigenvalues of $H(k_x)$. Thus, 
The minimal model with $w_{1}[h_{+}]=2$ is
\begin{align}
\label{eq:C+S+-1D}
h_{+}(k_x) = \text{e}^{ik_x}\sigma_0, \quad
h_{-}(k_x) = -C h^{\text{T}}_{+}(-k_x) C^{-1} = -\text{e}^{-ik_x}\sigma_0,
\end{align}
with $C=i\sigma_y$. This model gives $H(k_x)=i(\sin k_x\mu_x+\cos k_x\mu_y)\sigma_0$, which support a boundary state.
Thus, we have the BBC.

For $d=3$, the point-gap topological number is the 3D winding number $w_3[h_+]\in \mathbb{Z}$ for $h_+$. 
Whereas $H({\bm k})$ has TRS for class AII$^\dagger$ in the above, the 3D winding number $w_3[H]$ of $H$ takes an even number because $H({\bm k})$ has PHS for class C at the same time.
Therefore, no symmetry-protected skin effect occurs by the dimensional reduction.
%
%As a result, there is no dimensional reduction even though $H(\bm{k})$ is also in class $\text{AII}^{\dagger}$ with $\mathcal{C}_{\text{AII}^{\dagger}}=\mathcal{SC}$. 
The minimal model with $w_3[h_+]=1$ is
\begin{align}
\label{eq:C+S+-3D}
&h_{+}(\bm{k}) = \dfrac{\tau_0+\tau_z}{2}\left[\sin{k_x}\sigma_{x} + \sin{k_y}\sigma_{y} + \left(2 - \sum_{i=x,y,z}\cos{k_i}\right)\sigma_{z} + i\sin{k_z}\sigma_{0}\right]+\dfrac{\tau_0-\tau_z}{2}\sigma_0, \nonumber\\
&h_{-}(\bm{k}) = -C h^{\text{T}}_{+}(-\bm{k}) C^{-1} = \dfrac{\tau_0-\tau_z}{2}\left[-\sin{k_x}\sigma_{x} - \sin{k_y}\sigma_{y} + \left(2 - \sum_{i=x,y,z}\cos{k_i}\right)\sigma_{z} + i\sin{k_z}\sigma_{0}\right]-\dfrac{\tau_0+\tau_z}{2}\sigma_0,
\end{align}
where $C=i\tau_x\sigma_y$. Note that $h_\pm$ is block-diagonal and one of the blocks has the form of Eq.(\ref{eq:3D-A}).
Therefore, $h_\pm$ has a boundary state under the xOBC. Furthermore, we have $[h_+,h_-]=0$ under the xOBC, so the BBC holds.

\subsubsection{class CI + ${\cal S}_{-+}$ (case e.)}\label{proof:CI+S-+}
In class $\text{CI}+{\cal S}_{-+}$, $h_{+}(\bm{k})$ belongs to class AIII, and thus the system has a $\mathbb{Z}$ point-gap topological phase in $d=2$.
The corresponding point-gap topological number is the first Chern number for $ih_{+}(\bm{k})\gamma$ \cite{KSUS-19}. 
The minimal model with $Ch_{1}[ih_{+}\gamma]=1$ is
\begin{align}
\label{eq:CI+SLS-+-2D}
&h_{+}(\bm{k}) = \dfrac{\tau_0+\tau_z}{2}\left[\sin{k_x}\sigma_{x} + \left(1 - \sum_{i=x,y}\cos{k_i}\right)\sigma_{y} + i\sin{k_y}\sigma_{0}\right]+\dfrac{\tau_0-\tau_z}{2}i\sigma_0, \nonumber\\
&h_{-}(\bm{k}) = T h^{*}_{+}(-{\bm k}) T^{-1} = -C h^{\text{T}}_{+}(-{\bm k}) C^{-1} = \dfrac{\tau_0-\tau_z}{2}\left[-\sin{k_x}\sigma_{x} + \left(1 - \sum_{i=x,y}\cos{k_i}\right)\sigma_{y} + i\sin{k_y}\sigma_{0}\right]-\dfrac{\tau_0+\tau_z}{2}i\sigma_0,
\end{align}
where $T=\tau_x\sigma_x,C=\tau_x\sigma_y,\gamma=iTC^{*}=\tau_0\sigma_z$. This model show the BBC.

\subsubsection{class BDI + ${\cal S}_{++}$ (case f.)}\label{proof:BDI+S++}
In class $\text{BDI}+{\cal S}_{++}$, $h_{+}(\bm{k})$ belongs to class AI and there exist nontrivial point-gap topological phases in $d=1$. The minimal model is the same as Eq.~(\ref{eq:AIII+SLS+-1D}) with $T=C=\gamma=1$.

\subsubsection{class DIII + ${\cal S}_{++}$ (case f.)}\label{proof:DIII+S++}
For this class, $h_{+}(\bm{k})$ belongs to class AII and nontrivial point-gap topological phases exist in $d=1,3$.
For $d=1$, the minimal model is 
\begin{align}
\label{eq:DIII+SLS++-1D}
h_{+}(k_x) = \text{e}^{ik_x}\sigma_0, \quad
h_{-}(k_x) = -C h^{T}_{+}(-k_x) C^{-1} = -\gamma h^{\dagger}_{+}(k_x) \gamma^{-1} = -\text{e}^{-ik_x}\sigma_0,
\end{align}
where $T=i\sigma_y, C=\sigma_0$ and $\gamma=\sigma_0$. This model has $w_{1}[h_{+}]=2\in2\mathbb{Z}$, and shows the BBC.

For $d=3$, the minimal model is
\begin{align}
\label{eq:DIII+SLS++-3D}
&h_{+}({\bm k}) = i\left[\sin{k_x}\tau_{x}\sigma_{x} + \sin{k_y}\tau_{x}\sigma_{y} + \left(2 - \sum_{i=x,y,z}\cos{k_i}\right)\tau_{0}\sigma_{z}\right] +\sin{k_z}\tau_{z}\sigma_{z}, \nonumber\\
&h_{-}({\bm k}) = -C h^{T}_{+}(-{\bm k}) C^{-1} = -\gamma h^{\dagger}_{+}({\bm k}) \gamma^{-1} = i\left[\sin{k_x}\tau_{x}\sigma_{x} + \sin{k_y}\tau_{x}\sigma_{y} + \left(2 - \sum_{i=x,y,z}\cos{k_i}\right)\tau_{0}\sigma_{z}\right] +\sin{k_z}\tau_{z}\sigma_{z},
\end{align}
where $T=i\tau_{z}\sigma_y, C=\tau_{x}\sigma_{x}$ and $\gamma=\tau_{y}\sigma_{z}$. This model hosts the 3D non-trivial $\mathbb{Z}_{2}$ number $\nu_3[h_+]$ in class AII.
Because $h_+$ and $h_-$ commute and support boundary states, the BBC holds.

\subsubsection{class CII + ${\cal S}_{++}$ (case f.)}\label{proof:CII+S++}
In this class, $h_{+}(\bm{k})$ belongs to class AII and nontrivial point-gap topological phases exist in $d=1,3$.
For $d=1$, the point-gap topological number is the 1D winding number $w_1[h_+]\in 2\mathbb{Z}$ and 
the minimal model is the same as Eq.~(\ref{eq:DIII+SLS++-1D}) with $T=C=i\sigma_y, \gamma=\sigma_0$. 
For $d=3$, the point-gap topological number is the 3D $\mathbb{Z}_2$ number $\nu_3[h_+]$ for class AII \cite{KSUS-19}, and the minimal model with $\nu_3[H]=1$ is 
\begin{align}
\label{eq:CII+SLS++-3D}
&h_{+}({\bm k}) = i\left[\sin{k_x}\tau_{x}\sigma_{x} + \sin{k_y}\tau_{x}\sigma_{y} + \left(2 - \sum_{i=x,y,z}\cos{k_i}\right)\tau_{0}\sigma_{z}\right] +\sin{k_z}\tau_{z}\sigma_{z}, \nonumber\\
&h_{-}({\bm k}) = -C h^{T}_{+}(-{\bm k}) C^{-1} = -\gamma h^{\dagger}_{+}({\bm k}) \gamma^{-1} = i\left[\sin{k_x}\tau_{x}\sigma_{x} + \sin{k_y}\tau_{x}\sigma_{y} + \left(2 - \sum_{i=x,y,z}\cos{k_i}\right)\tau_{0}\sigma_{z}\right] - \sin{k_z}\tau_{z}\sigma_{z},
\end{align}
where $T=C=i\tau_{z}\sigma_y$ and $\gamma=\tau_{0}\sigma_{0}$. Since $h_\pm$ supports a boundary state and $[h_+,h_-]=0$ under the xOBC. Thus the BBC holds.

\subsubsection{class CI + ${\cal S}_{++}$ (case f.)}\label{proof:CI+S++}
In this class, $h_{+}(\bm{k})$ belongs to class AI and there exists a $\mathbb{Z}$ point-gap topological phase in $d=1$. In a similar manner to $\text{C}+{\cal S}_{+}$, $H({\bm k})$ has TRS$^\dagger$ for class AII$^\dagger$, thus 
the odd parity of $w_{1}[h_{+}]$ causes the symmetry-protected skin effect, and the topological classification changes from $\mathbb{Z}$ to $2\mathbb{Z}$ under the OBC. The minimal model with $w_{1}[h_{+}]=2$ is
\begin{align}
\label{eq:CI+S++-1D}
h_{+}(k_x) = \text{e}^{ik_x}\sigma_0, \quad
h_{-}(k_x) = -C h^{\text{T}}_{+}(-k_x) C^{-1} = -\gamma h^{\dagger}_{+}(k_x) \gamma^{-1} = -\text{e}^{-ik_x}\sigma_0,
\end{align}
where $T=\sigma_0, C=i\sigma_y$ and $\gamma = \sigma_y$. This model shows the BBC.

\subsubsection{class BDI + ${\cal S}_{--}$ (case g.)}\label{proof:BDI+S--}
In this class, $h_{\pm}(\bm{k})$ belongs to class D, and there exist point-gap topological phases in $d=1,2,3$.  
For $d=1$, the point-gap topological number is the 1D $\mathbb{Z}_2$ number $\nu_1[h_+]$, 
\begin{align}
(-1)^{\nu_1[h_+]}=\text{sgn}
\left\{
\dfrac{\operatorname{Pf}[h_+(\pi)C]}{\operatorname{Pf}[h_+(0)C]}
~\times~\text{exp}\left[-\dfrac{1}{2}\int_{k_x=0}^{k_x=\pi}\operatorname{\it{d}}\operatorname{log}\operatorname{det}[h_+(k_x)C]
\right]\right\}.
\end{align}
The minimal model is
\begin{align}
\label{eq:BDI+S---1D}
h_{+}(k_x) = \sin k_x \sigma_x + \cos{k_x} \sigma_y, \quad
h_{-}(k_x) = Th_{+}^{*}(-k_x)T^{-1} = -\gamma h^{\dagger}_{+}(k_x) \gamma^{-1} = -\sin k_x \sigma_x - \cos{k_x} \sigma_y,
\end{align}
with $C=T=\sigma_0$, and we have the BBC.

For $d=2$, the point-gap topological number is the 2D $\mathbb{Z}_2$ number $\nu_2[h_+]$,
\begin{align}
    (-1)^{\nu_2[h_+]} = \prod_{\text{X=I,II}}\text{sgn}\left\{\dfrac{\operatorname{Pf}[h_+(\bm{k}_{\text{X}+})C
    ]}{\operatorname{Pf}[h_+(\bm{k}_{\text{X}-})C]}~\times~\text{exp}\left[-\dfrac{1}{2}\int_{\bm{k}=\bm{k}_{\text{X}-}}^{\bm{k}=\bm{k}_{\text{X}+}}\operatorname{\it{d}}\operatorname{log}\operatorname{det}[h_+(\bm{k})C]\right]\right\},
\end{align}
where $(\bm{k}_{\text{I}+},\bm{k}_{\text{I}-})$ and $(\bm{k}_{\text{II}+},\bm{k}_{\text{II}-})$ are two pairs of particle-hole symmetric momenta. 
The minimal model is
\begin{align}
\label{eq:BDI+S---2D}
&h_{+}(\bm{k}) = \sin k_x \sigma_x + \sin k_y \sigma_z + \left(1 - \sum_{i=x,y}\cos{k_i}\right) \sigma_y + i\sin k_y \sigma_0, 
\nonumber\\
&h_{-}(\bm{k}) = Th_{+}^{*}(-\bm{k})T^{-1} = -\gamma h^{\dagger}_{+}(\bm{k}) \gamma^{-1} = -\sin k_x \sigma_x - \sin k_y \sigma_z - \left(1 - \sum_{i=x,y}\cos{k_i}\right) \sigma_y + i\sin k_y \sigma_0,
\end{align}
with $C=T=\sigma_0$.
Since $h_\pm$ supports a boundary state and $[h_+,h_-]=0$ under the xOBC, we have the BBC.

For $d=3$, the point-gap topological number is the 3D winding number $w_3[h_+]\in \mathbb{Z}$ for $h_+$.  
The minimal model is
\begin{align}
\label{eq:BDI+S---3D}
&h_{+}(\bm{k}) = \sin{k_x}\sigma_{x} + \sin{k_y}\sigma_{z} + \left(2 - \sum_{i=x,y,z}\cos{k_i}\right)\sigma_{y} - i\sin{k_z}\sigma_{0}, \nonumber\\
&h_{-}(\bm{k}) = Th_{+}^{*}(-\bm{k})T^{-1} = -\gamma h^{\dagger}_{+}(\bm{k}) \gamma^{-1} = -\sin{k_x}\sigma_{x} - \sin{k_y}\sigma_{z} - \left(2 - \sum_{i=x,y,z}\cos{k_i}\right)\sigma_{y} - i\sin{k_z}\sigma_{0},
\end{align}
with $C=T=\sigma_0$, which also shows the BBC.

\subsubsection{class CI + ${\cal S}_{--}$ (case g.)}\label{proof:CI+S--}
In this class, $h_+(\bm{k})$ belongs to class C, and there exists a $2\mathbb{Z}$ point-gap topological phase in $d=3$.
The point-gap topological number is the 3D winding number $w_3[h_+]$ for $h_+$.
The minimal model is
\begin{align}
\label{eq:CI+S---3D}
&h_{+}(\bm{k}) = \tau_{0}\left[\sin{k_x}\sigma_{x} + \sin{k_y}\sigma_{z} + \left(1 - \sum_{i=x,y,z}\cos{k_i}\right)\sigma_{y} - i\sin{k_z}\sigma_{0}\right], \nonumber\\
&h_{-}(\bm{k}) = T h^{*}_{+}({-\bm k}) T^{-1} = -\gamma h^{\dagger}_{+}(\bm{k}) \gamma^{-1} = -\tau_{0}\left[\sin{k_x}\sigma_{x} + \sin{k_y}\sigma_{z} + \left(1 - \sum_{i=x,y,z}\cos{k_i}\right)\sigma_{y} + i\sin{k_z}\sigma_{0}\right],
\end{align}
with $T=\tau_0\sigma_0$ and $C=i\tau_y\sigma_0$. Since $h_\pm$ supports a boundary state and $[h_+,h_-]=0$ under the xOBC, we have the BBC.

%%%%%%test
\clearpage
\section{Classification tables for point-gap topological phases}
In the previous section, we have completed the proof of the BBC for point-gap topological phases in 38 classes.
In this section,  we present classification tables in a more convenient form and summarize how the point-gap topological phases change between the PBC and the OBC.
As an additional symmetry, we introduce here pseudo-hermiticity defined by
\begin{align}
\eta H^{\dagger}({\bm k})\eta^{-1}= H({\bm k}),
\quad \eta^2=1,
\end{align}
with a unitary matrix $\eta$. Since pseudo-hermiticity is equivalent to CS by multiplying the Hamiltonian by $i$, it does not change the classification. However, pseudo-hermiticity serves as a key internal symmetry in non-Hermitian physics, so its inclusion is convenient for application.

To obtain classification tables, we use the equivalence relations between classes.
Tables \ref{tab: SLS - AZ}, \ref{tab: SLS - AZ-dag} and \ref{tab: pH & SLS} summarize the equivalent relations between classes. 
Combining these tables with the results in Tables \ref{table:without SLS} and \ref{table:situation with SLS}, we obtain the classification tables in Tables \ref{table:complexAZ+SLS}, \ref{table:realAZ+SLS} and \ref{table:realAZdag+SLS}.

\begin{table}[htbp]
	\centering
	\caption{Equivalence among the real AZ symmetry classes with SLS. The subscript of ${\cal S}_{\pm}$ specifies the commutation (+) or anti-commutation (-) relation to TRS or PHS in the real AZ class. For ${\cal S}_{\pm\pm}$, the first subscript specifies the relation to TRS and the second specifies the relation to PHS. }
	\label{tab: SLS - AZ}
     \begin{tabular}{ccccccc} \hline \hline
    ~$\text{AZ}$ class~ & ~${\cal S}_{-}$~ & ~${\cal S}_{-+}$~ & ~${\cal S}_{--}$~\\ \hline %\hline
    $\text{AI}$ & ~AII + ${\cal S}_{-}$~ & & \\
    $\text{BDI}$ & & ~DIII + ${\cal S}_{-+}$~ & ~DIII + ${\cal S}_{--}$~ \\
    $\text{CI}$ & & ~CII + ${\cal S}_{-+}$~ & ~CII + ${\cal S}_{--}$~ \\ \hline \hline
  \end{tabular}
\end{table}

\begin{table}[htbp]
	\centering
	\caption{Equivalence between the real AZ$^\dagger$ symmetry class with SLS and the real AZ symmetry class with SLS \cite{KSUS-19}. The subscript of ${\cal S}_{\pm}$ specifies the commutation (+) or anti-commutation (-) relation to TRS/TRS$^\dagger$ or PHS/PHS$^\dagger$. For ${\cal S}_{\pm\pm}$, the first subscript specifies the relation to TRS/TRS$^\dagger$ and the second specifies the relation to PHS/PHS$^\dagger$.}
	\label{tab: SLS - AZ-dag}
     \begin{tabular}{ccccccc} \hline \hline
    ~$\text{AZ}^{\dag}$ class~ & ~${\cal S}_{+}$~ & ~${\cal S}_{-}$~ & ~${\cal S}_{++}$~ & ~${\cal S}_{+-}$~ & ~${\cal S}_{-+}$~ & ~${\cal S}_{--}$~\\ \hline %\hline
    $\text{AI}^{\dag}$ & ~D + ${\cal S}_{+}$~ & ~C + ${\cal S}_{-}$~ & & & & \\
    $\text{BDI}^{\dag}$ & & & ~BDI + ${\cal S}_{++}$~ & ~DIII + ${\cal S}_{-+}$~ & ~CI + ${\cal S}_{+-}$~ & ~CII + ${\cal S}_{--}$~ \\
    $\text{D}^{\dag}$ & ~AI + ${\cal S}_{+}$~ & ~AII + ${\cal S}_{-}$~ & & & & \\
    $\text{DIII}^{\dag}$ & & & ~CI + ${\cal S}_{++}$~ & ~CII + ${\cal S}_{-+}$~ & ~BDI + ${\cal S}_{+-}$~ & ~DIII + ${\cal S}_{--}$~ \\
    $\text{AII}^{\dag}$ & ~C + ${\cal S}_{+}$~ & ~D + ${\cal S}_{-}$~ & & & & \\
    $\text{CII}^{\dag}$ & & & ~CII + ${\cal S}_{++}$~ & ~CI + ${\cal S}_{-+}$~ & ~DIII + ${\cal S}_{+-}$~ & ~BDI + ${\cal S}_{--}$~ \\
    $\text{C}^{\dag}$ & ~AII + ${\cal S}_{+}$~ & ~AI + ${\cal S}_{-}$~ & & & & \\
    $\text{CI}^{\dag}$ & & & ~DIII + ${\cal S}_{++}$~ & ~BDI + ${\cal S}_{-+}$~ & ~CII + ${\cal S}_{+-}$~ & ~CI + ${\cal S}_{--}$~ \\ \hline \hline
  \end{tabular}
\end{table}

\begin{table}[htbp]
	\centering
	\caption{Equivalence between the AZ/AZ$^\dagger$ symmetry class with pseudo-Hermiticity and those with SLS \cite{KSUS-19}. The subscript of $\eta_\pm/{\cal S}_{\pm}$ specifies the commutation (+) or anti-commutation (-) relation to TRS/TRS$^\dagger$ or PHS/PHS$^\dagger$. For $\eta_{\pm\pm}/{\cal S}_{\pm\pm}$, the first subscript specifies the relation to TRS/TRS$^\dagger$ and the second specifies the relation to PHS/PHS$^\dagger$.}
	\label{tab: pH & SLS}
     \begin{tabular}{cccccccc} \hline \hline
    ~Sym. class~ & ~$\eta$~ & ~$\eta_{+}$~ & ~$\eta_{-}$~ & ~$\eta_{++}$~ & ~$\eta_{+-}$~ & ~$\eta_{-+}$~ & ~$\eta_{--}$~\\ \hline %\hline
    A & ~AIII~ & & & & & & \\
    AIII & & ~AIII + ${\cal S}_{+}$~ & ~AIII + ${\cal S}_{-}$~ & & & & \\ \hline %\hline
    AI & & ~$\text{BDI}^{\dag}$~ & ~$\text{DIII}^{\dag}$~ & & & & \\
    BDI & & & & ~BDI + ${\cal S}_{++}$~ & ~BDI + ${\cal S}_{-+}$~ & ~BDI + ${\cal S}_{+-}$~ & ~BDI + ${\cal S}_{--}$~ \\
    D & & ~BDI~ & ~DIII~ & & & & \\
    DIII & & & & ~DIII + ${\cal S}_{--}$~ & ~DIII + ${\cal S}_{+-}$~ & ~DIII + ${\cal S}_{-+}$~ & ~DIII + ${\cal S}_{++}$~ \\
    AII & & ~$\text{CII}^{\dag}$~ & ~$\text{CI}^{\dag}$~ & & & & \\
    CII & & & & ~CII + ${\cal S}_{++}$~ & ~CII + ${\cal S}_{-+}$~ & ~CII + ${\cal S}_{+-}$~ & ~CII + ${\cal S}_{--}$~ \\
    C & & ~CII~ & ~CI~ & & & & \\
    CI & & & & ~CI + ${\cal S}_{--}$~ & ~CI + ${\cal S}_{+-}$~ & ~CI + ${\cal S}_{-+}$~ & ~CI + ${\cal S}_{++}$~ \\ \hline %\hline
    $\text{AI}^{\dag}$ & & ~BDI~ & ~DIII~ & & & & \\
    $\text{BDI}^{\dag}$ & & & & ~$\text{BDI}^{\dag}$ + ${\cal S}_{++}$~ & ~$\text{BDI}^{\dag}$ + ${\cal S}_{-+}$~ & ~$\text{BDI}^{\dag}$ + ${\cal S}_{+-}$~ & ~$\text{BDI}^{\dag}$ + ${\cal S}_{--}$~ \\
    $\text{D}^{\dag}$ & & ~$\text{BDI}^{\dag}$~ & ~$\text{DIII}^{\dag}$~ & & & & \\
    $\text{DIII}^{\dag}$ & & & & ~$\text{DIII}^{\dag}$ + ${\cal S}_{--}$~ & ~$\text{DIII}^{\dag}$ + ${\cal S}_{+-}$~ & ~$\text{DIII}^{\dag}$ + ${\cal S}_{-+}$~ & ~$\text{DIII}^{\dag}$ + ${\cal S}_{++}$~ \\
    $\text{AII}^{\dag}$ & & ~CII~ & ~CI~ & & & & \\
    $\text{CII}^{\dag}$ & & & & ~$\text{CII}^{\dag}$ + ${\cal S}_{++}$~ & ~$\text{CII}^{\dag}$ + ${\cal S}_{-+}$~ & ~$\text{CII}^{\dag}$ + ${\cal S}_{+-}$~ & ~$\text{CII}^{\dag}$ + ${\cal S}_{--}$~ \\
    $\text{C}^{\dag}$ & & ~$\text{CII}^{\dag}$~ & ~$\text{CI}^{\dag}$~ & & & & \\
    $\text{CI}^{\dag}$ & & & & ~$\text{CI}^{\dag}$ + ${\cal S}_{--}$~ & ~$\text{CI}^{\dag}$ + ${\cal S}_{+-}$~ & ~$\text{CI}^{\dag}$ + ${\cal S}_{-+}$~ & ~$\text{CI}^{\dag}$ + ${\cal S}_{++}$~ \\ \hline \hline
  \end{tabular}
\end{table}

\begin{table}[htbp]
  \centering
  \begingroup
  \renewcommand{\arraystretch}{1.1}
    \begin{tabular}{ccccc} \hline \hline
        AZ class & ~Add. sym.~ & ~$d=1$~ & ~$d=2$~ & ~$d=3$~ \\ \hline %\hline
        \multirow{1}{*}{A}
        & -&$\color{red}\mathbb{Z}\to 0$ & $0$ & \color{green}\Z \\ 
        \multirow{1}{*}{AIII}
        &  -&$0$ &\color{green}\Z\color{black} & $0$ \\ \hline %\hline
        \multirow{1}{*}{$\text{A}$}&{$\mathcal{S}$}
        &  $\color{blue}\mathbb{Z}\oplus\mathbb{Z} \to \mathbb{Z}[1,-1]$ & $0$ & $\color{green}\mathbb{Z}\oplus\mathbb{Z}$ \\ 
        \multirow{1}{*}{$\text{AIII}$}&{$\mathcal{S}_{-},\eta_{-}$}
        &  $0$ & $\color{green}\mathbb{Z}\oplus\mathbb{Z}$ & $0$ \\ \hline %\hline
        \multirow{1}{*}{$\text{A}$}&{$\eta$}
        &  $0$ & \color{green}\Z & $0$ \\ 
        \multirow{1}{*}{$\text{AIII}$}&{$\mathcal{S}_{+},\eta_{+}$}
        &  \color{green}\Z & $0$ & \color{green}\Z \\ \hline \hline
    \end{tabular}
    \endgroup
     \caption{Classification of point-gap topological phases in the complex AZ classes without or with SLS or pseudo-Hermiticity. The subscript of ${\cal S}_{\pm}/\eta_{\pm}$ specifies the commutation (+) or anti-commutation (-) relation to CS. For the topological numbers colored red or blue, 
    the classification under OBCs changes from that under PBCs, where
    the left specifies the classification under PBCs and the right specifies that under OBCs. The topological number $\mathbb{Z}[i,j]$ under OBCs indicates the abelian group $\mathbb{Z}$ generated by the element $(i,j)\in \mathbb{Z}\oplus\mathbb{Z}$ under PBCs. For the topological numbers colored green, the classification under OBCs coincides with that under PBCs. 
    \label{table:complexAZ+SLS} 
    }
\end{table}

\begin{table}[htbp]
  \centering
  \begingroup
  \renewcommand{\arraystretch}{1.1}
    \begin{tabular}{ccccc} \hline \hline
        AZ class & ~Add. sym.~ & ~$d=1$~ & ~$d=2$~ & ~$d=3$~ \\ \hline %\hline
        \multirow{1}{*}{$\text{AI}$}
        & -&$\color{red}\mathbb{Z}\to 0$ & $0$ & $0$ \\ 
        \multirow{1}{*}{$\text{BDI}$}
        & -&\color{green}\Zt & \color{green}\Z & $0$ \\ 
        \multirow{1}{*}{$\text{D}$}
        & -&\color{green}\Zt & \color{green}\Zt & \color{green}\Z \\ 
        \multirow{1}{*}{$\text{DIII}$}
        & -&$0$ &\color{green}\Zt & \color{green}\Zt \\ 
        \multirow{1}{*}{$\text{AII}$}
        & -&$\color{red}2\mathbb{Z}\to 0$ & $0$ & \color{green}\Zt \\ 
        \multirow{1}{*}{$\text{CII}$}
        & -&$0$ & \color{green}2\Z & $0$ \\ 
        \multirow{1}{*}{$\text{C}$}
        & -&$0$ & $0$ & \color{green}2\Z \\ 
        \multirow{1}{*}{$\text{CI}$}
        & -&$0$ & $0$ & $0$ \\ \hline %\hline
        \multirow{1}{*}{$\text{AI}$}&{$\mathcal{S}_{+}$}
        & $\color{blue}\mathbb{Z}\oplus\mathbb{Z} \to \mathbb{Z}[1,-1]$ & $0$ & $0$ \\ 
        \multirow{1}{*}{$\text{BDI}$}&{$\mathcal{S}_{+-},\eta_{-+}$}
        & $\color{blue}\mathbb{Z}_{2}\oplus\mathbb{Z}_{2} \to \mathbb{Z}_{2}[1,1]$ & $\color{blue}\mathbb{Z}\oplus\mathbb{Z} \to \mathbb{Z}[2,0]\oplus\mathbb{Z}[1,-1]$ & $0$ \\ 
        \multirow{1}{*}{$\text{D}$}&{$\mathcal{S}_{-}$}
        & $\color{blue}\mathbb{Z}_{2}\oplus\mathbb{Z}_{2} \to \mathbb{Z}_{2}[1,1]$ & $\color{blue}\mathbb{Z}_{2}\oplus\mathbb{Z}_{2} \to \mathbb{Z}_{2}[1,1]$ & $\color{blue}\mathbb{Z}\oplus\mathbb{Z} \to \mathbb{Z}[2,0]\oplus\mathbb{Z}[1,-1]$  \\ 
        \multirow{1}{*}{$\text{DIII}$}&{$\mathcal{S}_{+-},\eta_{+-}$}
        &  $0$ & $\color{green}\mathbb{Z}_{2}\oplus\mathbb{Z}_{2}$ & $\color{green}\mathbb{Z}_{2}\oplus\mathbb{Z}_{2}$ \\ 
        \multirow{1}{*}{$\text{AII}$}&{$\mathcal{S}_{+}$}
        & $\color{blue}2\mathbb{Z}\oplus2\mathbb{Z} \to 2\mathbb{Z}[1,-1]$ & $0$ & $\color{green}\mathbb{Z}_{2}\oplus\mathbb{Z}_{2}$ \\ 
        \multirow{1}{*}{$\text{CII}$}&{$\mathcal{S}_{+-},\eta_{-+}$}
        & $0$ & $\color{green}2\mathbb{Z}\oplus2\mathbb{Z}$ & $0$ \\ 
        \multirow{1}{*}{$\text{C}$}&{$\mathcal{S}_{-}$}
        &  $0$ & $0$ & $\color{green}2\mathbb{Z}\oplus2\mathbb{Z}$\\ 
        \multirow{1}{*}{$\text{CI}$}&{$\mathcal{S}_{+-},\eta_{+-}$}
        & $0$ & $0$ & $0$  \\ \hline %\hline
        \multirow{1}{*}{$\text{AI}$}&{$\mathcal{S}_{-}$}
        & $\color{red}\mathbb{Z}\to 0$ & $0$ & \color{green}\Z \\ 
        \multirow{1}{*}{$\text{BDI}$}&{$\mathcal{S}_{-+},\eta_{+-}$}
        & $0$ & \color{green}\Z & $0$ \\ 
        \multirow{1}{*}{$\text{D}$}&{$\mathcal{S}_{+}$}
        & \color{green}\Z & $0$ & \color{green}\Z \\ 
        \multirow{1}{*}{$\text{DIII}$}&{$\mathcal{S}_{-+},\eta_{-+}$}
        &  $0$ & \color{green}\Z & $0$ \\ 
        \multirow{1}{*}{$\text{AII}$}&{$\mathcal{S}_{-}$}
        & $\color{red}\mathbb{Z}\to 0$ & $0$ & \color{green}\Z \\ 
        \multirow{1}{*}{$\text{CII}$}&{$\mathcal{S}_{-+},\eta_{+-}$}
        & $0$ & \color{green}\Z & $0$ \\ 
        \multirow{1}{*}{$\text{C}$}&{$\mathcal{S}_{+}$}
        & $\color{blue}\mathbb{Z}\to2\mathbb{Z}$ & $0$ & \color{green}\Z\\ 
        \multirow{1}{*}{$\text{CI}$}&{$\mathcal{S}_{-+},\eta_{-+}$}
        & $0$ & \color{green}\Z & $0$ \\ \hline %\hline
        \multirow{1}{*}{$\text{AI}$}&{$\eta_{+}$}
        & $0$ & $0$ & $0$ \\ 
        \multirow{1}{*}{$\text{BDI}$}&{$\mathcal{S}_{++},\eta_{++}$}
        & \color{green}\Z & $0$ & $0$ \\ 
        \multirow{1}{*}{$\text{D}$}&{$\eta_{+}$}
        & \color{green}\Zt & \color{green}\Z & $0$ \\ 
        \multirow{1}{*}{$\text{DIII}$}&{$\mathcal{S}_{--},\eta_{++}$}
        &  \color{green}\Zt & \color{green}\Zt & \color{green}\Z \\ 
        \multirow{1}{*}{$\text{AII}$}&{$\eta_{+}$}
        & $0$ & \color{green}\Zt & \color{green}\Zt \\ 
        \multirow{1}{*}{$\text{CII}$}&{$\mathcal{S}_{++},\eta_{++}$}
        & \color{green}2\Z & $0$ & \color{green}\Zt \\
        \multirow{1}{*}{$\text{C}$}&{$\eta_{+}$}
        &  $0$ & \color{green}2\Z & $0$ \\
        \multirow{1}{*}{$\text{CI}$}&{$\mathcal{S}_{--},\eta_{++}$}
        & $0$ & $0$ & \color{green}2\Z \\ \hline %\hline
        \multirow{1}{*}{$\text{AI}$}&{$\eta_{-}$}
        & $\color{red}\mathbb{Z}_{2}\to 0$ & $\color{blue}\mathbb{Z}\to2\mathbb{Z}$ & $0$ \\ 
        \multirow{1}{*}{$\text{BDI}$}&{$\mathcal{S}_{--},\eta_{--}$}
        & \color{green}\Zt & \color{green}\Zt & \color{green}\Z \\ 
        \multirow{1}{*}{$\text{D}$}&{$\eta_{-}$}
        & $0$ & \color{green}\Zt & \color{green}\Zt \\ 
        \multirow{1}{*}{$\text{DIII}$}&{$\mathcal{S}_{++},\eta_{--}$}
        &  \color{green}2\Z & $0$ & \color{green}\Zt \\ 
        \multirow{1}{*}{$\text{AII}$}&{$\eta_{-}$}
        & $0$ & \color{green}2\Z & $0$ \\ 
        \multirow{1}{*}{$\text{CII}$}&{$\mathcal{S}_{--},\eta_{--}$}
        & $0$ & $0$ & \color{green}2\Z \\ 
        \multirow{1}{*}{$\text{C}$}&{$\eta_{-}$}
        &  $0$ & $0$ & $0$ \\ 
        \multirow{1}{*}{$\text{CI}$}&{$\mathcal{S}_{++},\eta_{--}$}
        & $\color{blue}\mathbb{Z}\to2\mathbb{Z}$ & $0$ & $0$ \\ \hline \hline
    \end{tabular}
    \endgroup
    \caption{Classification of point-gap topological phases in the real AZ classes without or with SLS or pseudo-Hermiticity. The subscript of ${\cal S}_{\pm}/\eta_{\pm}$ specifies the commutation (+) or anti-commutation (-) relation to TRS or PHS. For ${\cal S}_{\pm\pm}/\eta_{\pm\pm}$, the first subscript specifies the relation to TRS and the second specifies the relation to PHS.  For the topological numbers colored red or blue, the classification under OBCs changes from that under PBCs, where the left specifies the classification under PBCs and the right specifies that under OBCs. The topological number $\mathbb{Z}[i,j]$ ($\mathbb{Z}_2[i,j]$) under OBCs indicates the abelian group $\mathbb{Z}$ ($\mathbb{Z}_2$) generated by the element $(i,j)\in \mathbb{Z}\oplus\mathbb{Z}$ ($(i,j)\in \mathbb{Z}_2\oplus\mathbb{Z}_2$) under PBCs. For the topological numbers colored green, the classification under OBCs coincides with that under PBCs. 
  \label{table:realAZ+SLS} }
\end{table}

\clearpage

\begin{table}[H]
  \centering
  \begingroup
  \renewcommand{\arraystretch}{1.1}
    \begin{tabular}{ccccc} \hline \hline
        $\text{AZ}^{\dagger}$ class & ~Add. sym.~ & ~$d=1$~ & ~$d=2$~ & ~$d=3$~ \\ \hline %\hline
        \multirow{1}{*}{$\text{AI}^{\dag}$}
        & -&$0$ & $0$ & \color{green}2\Z \\ 
        \multirow{1}{*}{$\text{BDI}^{\dag}$}
        & -&$0$ & $0$ & $0$ \\ 
        \multirow{1}{*}{$\text{D}^{\dag}$}
        & -&$\color{red}\mathbb{Z}\to 0$ & $0$ & $0$ \\ 
        \multirow{1}{*}{$\text{DIII}^{\dag}$}
        & -&$\color{red}\mathbb{Z}_{2}\to 0$ & $\color{blue}\mathbb{Z}\to2\mathbb{Z}$ & $0$ \\ 
        \multirow{1}{*}{$\text{AII}^{\dag}$}
        & -&$\color{red}\mathbb{Z}_{2}\to 0$ & $\color{red}\mathbb{Z}_{2}\to 0$ & $\color{blue}\mathbb{Z}\to2\mathbb{Z}$ \\ 
        \multirow{1}{*}{$\text{CII}^{\dag}$}
        & -&$0$ & \color{green}\Zt & \color{green}\Zt \\ 
        \multirow{1}{*}{$\text{C}^{\dag}$}
        & -&$\color{red}2\mathbb{Z}\to 0$ & $0$ & \color{green}\Zt \\ 
        \multirow{1}{*}{$\text{CI}^{\dag}$}
        & -&$0$ & $\color{green}2\mathbb{Z}$ & $0$ \\ \hline %\hline
        \multirow{1}{*}{$\text{AI}^{\dag}$}&{$\mathcal{S}_{+}$}
        & \color{green}\Z & $0$ & \color{green}\Z \\ 
        \multirow{1}{*}{$\text{BDI}^{\dag}$}&{$\mathcal{S}_{+-},\eta_{-+}$}
        & $0$ & \color{green}\Z & $0$ \\ 
        \multirow{1}{*}{$\text{D}^{\dag}$}&{$\mathcal{S}_{-}$}
        & $\color{red}\mathbb{Z}\to 0$ & $0$ & \color{green}\Z  \\ 
        \multirow{1}{*}{$\text{DIII}^{\dag}$}&{$\mathcal{S}_{+-},\eta_{+-}$}
        &  $0$ & \color{green}\Z & $0$ \\ 
        \multirow{1}{*}{$\text{AII}^{\dag}$}&{$\mathcal{S}_{+}$}
        &  $\color{blue}\mathbb{Z}\to2\mathbb{Z}$ & $0$ & \color{green}\Z \\ 
        \multirow{1}{*}{$\text{CII}^{\dag}$}&{$\mathcal{S}_{+-},\eta_{-+}$}
        & $0$ & \color{green}\Z & $0$ \\ 
        \multirow{1}{*}{$\text{C}^{\dag}$}&{$\mathcal{S}_{-}$}
        & $\color{red}\mathbb{Z}\to 0$ & $0$ & \color{green}\Z\\ 
        \multirow{1}{*}{$\text{CI}^{\dag}$}&{$\mathcal{S}_{+-},\eta_{+-}$}
        & $0$ & \color{green}\Z & $0$  \\ \hline %\hline
        \multirow{1}{*}{$\text{AI}^{\dag}$}&{$\mathcal{S}_{-}$}
        & $0$ & $0$ & $\color{green}2\mathbb{Z}\oplus2\mathbb{Z}$ \\ 
        \multirow{1}{*}{$\text{BDI}^{\dag}$}&{$\mathcal{S}_{-+},\eta_{+-}$}
        & $0$ & $0$ & $0$ \\ 
        \multirow{1}{*}{$\text{D}^{\dag}$}&{$\mathcal{S}_{+}$}
        & $\color{blue}\mathbb{Z}\oplus\mathbb{Z} \to \mathbb{Z}[1,-1]$ & $0$ & $0$ \\ 
        \multirow{1}{*}{$\text{DIII}^{\dag}$}&{$\mathcal{S}_{-+},\eta_{-+}$}
        & $\color{blue}\mathbb{Z}_{2}\oplus\mathbb{Z}_{2} \to \mathbb{Z}_{2}[1,1]$ & $\color{blue}\mathbb{Z}\oplus\mathbb{Z} \to \mathbb{Z}[2,0]\oplus\mathbb{Z}[1,-1]$  & $0$  \\ 
        \multirow{1}{*}{$\text{AII}^{\dag}$}&{$\mathcal{S}_{-}$}
        & $\color{blue}\mathbb{Z}_{2}\oplus\mathbb{Z}_{2} \to \mathbb{Z}_{2}[1,1]$ & $\color{blue}\mathbb{Z}_{2}\oplus\mathbb{Z}_{2} \to \mathbb{Z}_{2}[1,1]$ & $\color{blue}\mathbb{Z}\oplus\mathbb{Z} \to \mathbb{Z}[2,0]\oplus\mathbb{Z}[1,-1]$  \\ 
        \multirow{1}{*}{$\text{CII}^{\dag}$}&{$\mathcal{S}_{-+},\eta_{+-}$}
        & $0$ & $\color{green}\mathbb{Z}_{2}\oplus\mathbb{Z}_{2}$ & $\color{green}\mathbb{Z}_{2}\oplus\mathbb{Z}_{2}$ \\ 
        \multirow{1}{*}{$\text{C}^{\dag}$}&{$\mathcal{S}_{+}$}
        & $\color{blue}2\mathbb{Z}\oplus2\mathbb{Z} \to 2\mathbb{Z}[1,-1]$ & $0$ & $\color{green}\mathbb{Z}_{2}\oplus\mathbb{Z}_{2}$ \\ 
        \multirow{1}{*}{$\text{CI}^{\dag}$}&{$\mathcal{S}_{-+},\eta_{-+}$}
        & $0$ & $\color{green}2\mathbb{Z}\oplus2\mathbb{Z}$ & $0$ \\ \hline %\hline
        \multirow{1}{*}{$\text{AI}^{\dag}$}&{$\eta_{+}$}
        & \color{green}\Zt & \color{green}\Z & $0$ \\ 
        \multirow{1}{*}{$\text{BDI}^{\dag}$}&{$\mathcal{S}_{++},\eta_{++}$}
        & \color{green}\Z & $0$ & $0$ \\ 
        \multirow{1}{*}{$\text{D}^{\dag}$}&{$\eta_{+}$}
        & $0$ & $0$ & $0$ \\ 
        \multirow{1}{*}{$\text{DIII}^{\dag}$}&{$\mathcal{S}_{--},\eta_{++}$}
        &  \color{green}\Zt & \color{green}\Zt & \color{green}\Z \\ 
        \multirow{1}{*}{$\text{AII}^{\dag}$}&{$\eta_{+}$}
        & $0$ & \color{green}2\Z & $0$ \\ 
        \multirow{1}{*}{$\text{CII}^{\dag}$}&{$\mathcal{S}_{++},\eta_{++}$}
        & \color{green}2\Z & $0$ & \color{green}\Zt \\ 
        \multirow{1}{*}{$\text{C}^{\dag}$}&{$\eta_{+}$}
        &  $0$ & \color{green}\Zt & \color{green}\Zt \\ 
        \multirow{1}{*}{$\text{CI}^{\dag}$}&{$\mathcal{S}_{--},\eta_{++}$}
        & $0$ & $0$ & \color{green}2\Z \\ \hline %\hline
        \multirow{1}{*}{$\text{AI}^{\dag}$}&{$\eta_{-}$}
        & $0$ & \color{green}\Zt & \color{green}\Zt \\
        \multirow{1}{*}{$\text{BDI}^{\dag}$}&{$\mathcal{S}_{--},\eta_{--}$}
        & $0$ & $0$ & \color{green}2\Z \\
        \multirow{1}{*}{$\text{D}^{\dag}$}&{$\eta_{-}$}
        & $\color{red}\mathbb{Z}_{2}\to 0$ & $\color{blue}\mathbb{Z}\to2\mathbb{Z}$ & $0$ \\ 
        \multirow{1}{*}{$\text{DIII}^{\dag}$}&{$\mathcal{S}_{++},\eta_{--}$}
        & $\color{blue}\mathbb{Z}\to2\mathbb{Z}$ & $0$ & $0$ \\
        \multirow{1}{*}{$\text{AII}^{\dag}$}&{$\eta_{-}$}
        & $0$ & $0$ & $0$ \\
        \multirow{1}{*}{$\text{CII}^{\dag}$}&{$\mathcal{S}_{--},\eta_{--}$}
        & \color{green}\Zt & \color{green}\Zt & \color{green}\Z \\
        \multirow{1}{*}{$\text{C}^{\dag}$}&{$\eta_{-}$}
        &  $0$ & \color{green}2\Z & $0$ \\
        \multirow{1}{*}{$\text{CI}^{\dag}$}&{$\mathcal{S}_{++},\eta_{--}$}
        & \color{green}2\Z & $0$ & \color{green}\Zt \\ \hline \hline
    \end{tabular}
    \endgroup
    \caption{Classification of point-gap topological phases in the real AZ$^\dagger$ classes without or with SLS or pseudo-Hermiticity. The subscript of ${\cal S}_{\pm}/\eta_{\pm}$ specifies the commutation (+) or anti-commutation (-) relation to TRS$^\dagger$ or PHS$^\dagger$. For ${\cal S}_{\pm\pm}/\eta_{\pm\pm}$, the first subscript specifies the relation to TRS$^\dagger$ and the second specifies the relation to PHS$^\dagger$.    
    For the topological numbers colored red or blue, the
classification under OBCs changes from that under PBCs, where the left specifies the classification under PBCs and the right specifies that under OBCs. The topological number $\mathbb{Z}[i,j]$ ($\mathbb{Z}_2[i,j]$) under OBCs indicates the abelian group $\mathbb{Z}$ ($\mathbb{Z}_2$) generated by the element $(i,j)\in \mathbb{Z}\oplus\mathbb{Z}$ ($(i,j)\in \mathbb{Z}_2\oplus\mathbb{Z}_2$) under PBCs.   For the topological numbers colored green, the classification under OBCs coincides with that under PBCs.
    \label{table:realAZdag+SLS} 
}
\end{table}

\clearpage
\section{Intrinsic point-gap topological phases}
\label{S:intrinsic}

As discussed in Ref.\cite{OKSS-20}, there exist two types of point-gap topological phases: One is those smoothly connected to conventional Hermitian (or anti-Hermitan) topological phases without point-gap closing, and the other is not.
The latter is called intrinsic point-gap topological phases.
Comparing Tables S7, S8, and S9 in Ref.\cite{OKSS-20} 
for intrinsic point-gap topological phases
with Tables \ref{table:complexAZ+SLS}, \ref{table:realAZ+SLS} and \ref{table:realAZdag+SLS} in the above, we can predict which intrinsic point-gap topological phases result in boundary states. We summarize the results in Tables \ref{table:intrinsic complexAZ+SLS}, \ref{table:intrinsic realAZ+SLS} and \ref{table:intrinsic realAZdag+SLS}.
Here the superscripts "SE" and "BS" of the topological numbers indicate the skin effect and the boundary state, respectively:
If the $\mathbb{Z}^{\rm SE}$ or $\mathbb{Z}_2^{\rm SE}$ ($\mathbb{Z}^{\rm BS}$ or $\mathbb{Z}_2^{\rm BS}$) topological number is nonzero, we have skin effects (boundary states) in the corresponding intrinsic point-gap topological phase.

\begin{figure}[htbp]
\includegraphics[keepaspectratio,scale=0.12]{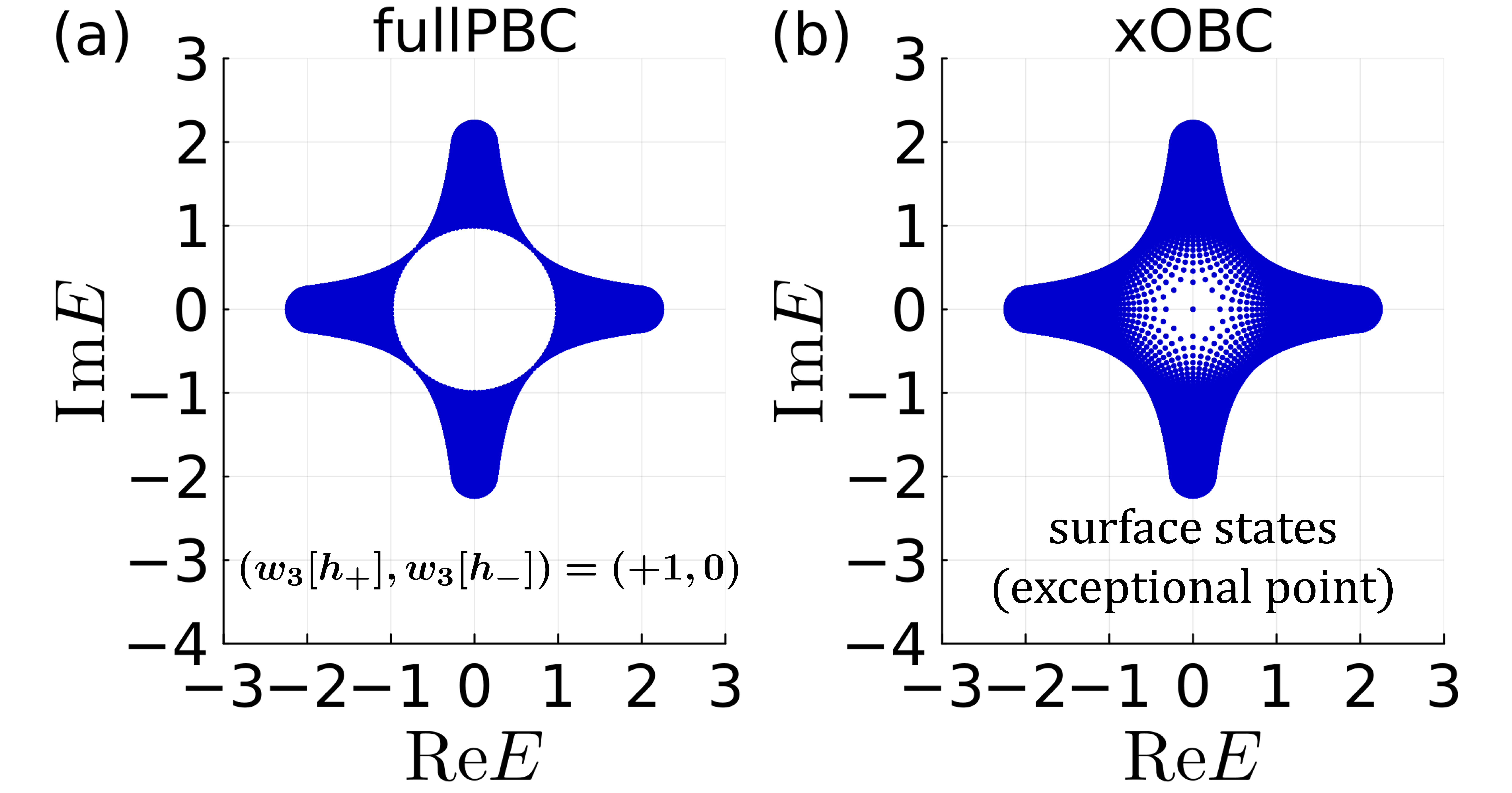}
\caption{The energy spectra of the 3D class A + ${\cal S}$ model in Eq.~(\ref{eq:A+SLS-3D}). The system sizes are $L_x=30,L_y=L_z=60$. This model has an intrinsic point-gap topological number and its boundary state hosts an exceptional point.
\label{fig:3dA+S-SM}}
\end{figure}

\begin{figure}[htbp]
\includegraphics[keepaspectratio,scale=0.12]{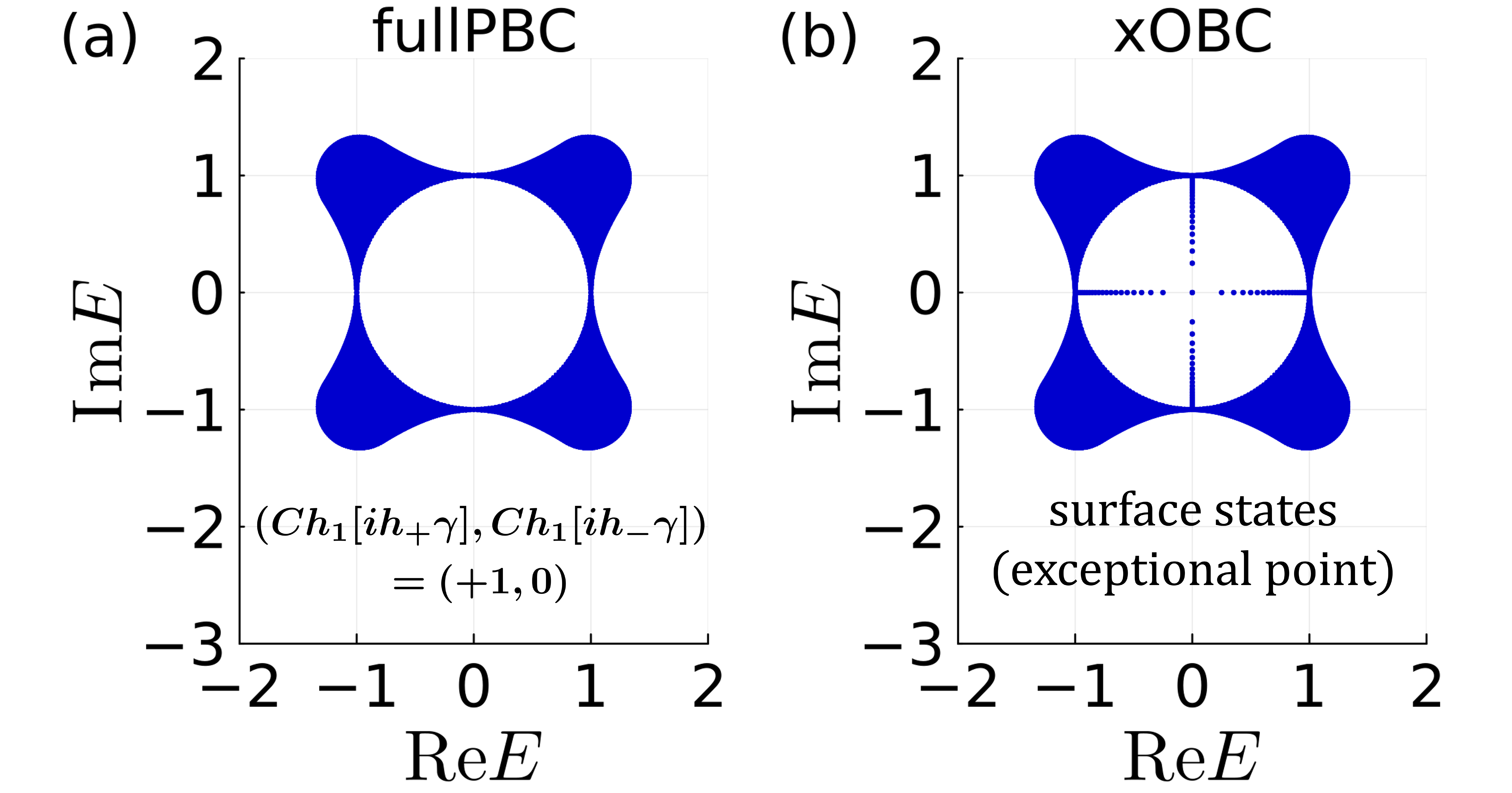}
\caption{The energy spectra of the 2D class AIII + ${\cal S}_-$ model in Eq.~(\ref{eq:AIII+SLS--2D}). The system sizes are $L_x=L_y=100$. This model has an intrinsic point-gap topological number and its boundary state hosts an exceptional point.
\label{fig:2dAIII+S--SM}}
\end{figure}

A remarkable feature of intrinsic point-gap topological phases is that boundary states can avoid the doubling theorem in Ref.\cite{Yang-Schnyder-Hu-Chiu} and have a single exceptional point on a boundary. Whereas such a surface state has been known to appear in an exceptional topological insulator \cite{Denner-20}, the exact condition for the appearance has not been specified before.
Here we would like to point out that the presence of an intrinsic topological phase is necessary for such a boundary state with a single exceptional point:
If the system is not in an intrinsic point-gap topological phase, it is smoothly deformable to a Hermitian or an anti-Hermitian one without point-gap closing. 
Therefore, even if the system supports exceptional points, the exceptional points should appear in a pair. 
Otherwise, the exceptional points can not disappear because they have their own topological numbers \cite{KBS-19}, which contradicts the fact that the system is topologically equivalent to a Hermitian (or anit-Hermitian) one where no exceptional point exists. 
Actually, the exceptional topological insulator in Ref.\cite{Denner-20} has the intrinsic point-gap topological number in 3D class A (see Table \ref{table:intrinsic complexAZ+SLS}), which is the 3D winding number $w_3=1$. 
In Figs. \ref{fig:3dA+S-SM} and \ref{fig:2dAIII+S--SM}, we show another examples of such exceptional boundary states. 
In both cases, the presence of exceptional points immediately follows from Corollary \ref{cor:1}. 
In Fig.\ref{fig:3dA+S-SM}, we consider the 3D class A + ${\cal S}$ model in Eq.(\ref{eq:A+SLS-3D}). As discussed in Sec.\ref{proof:A+S}, a pair of the 3D winding numbers $(w_3[h_+], w_3[h_-])\in \mathbb{Z}\oplus\mathbb{Z}$ characterizes the point-gap topological phase in this class, and the topological number of this model is $(1,0)$.
According to Table S4 in Ref.\cite{OKSS-20}, this number realizes an intrinsic point-gap topological phase, which is consistent with the presence of an exceptional point in Fig.\ref{fig:3dA+S-SM}. In a similar manner, we can check that the 2D class AIII + ${\cal S}_-$ model in Eq.(\ref{eq:AIII+SLS--2D}) realizes an intrinsic point-gap topological phase, which is also consistent with the presence of an exceptional point in Fig.\ref{fig:2dAIII+S--SM}. 
%Note that these types of boundary states never appear in non-intrinsic point-gap topological phases: Since any non-intrinsic point-gap topological phase is smoothly deformable to a Hermitian or an anti-Hermitian one,  
%even if its boundary state supports exceptional points, 
%the exceptional points should appear in a pair. 
%As discussed in Ref.\cite{KBS-19}, these exceptional points have their own point-gap %topological number so the boundary states 

\clearpage

\begin{table}[htbp]
  \centering
  \begingroup
  \renewcommand{\arraystretch}{1.2}
    \begin{tabular}{ccccc} \hline \hline
        AZ class & ~Add. sym.~ & ~$d=1$~ & ~$d=2$~ & ~$d=3$~ \\ \hline %\hline
        \multirow{1}{*}{A}
        & -&\color{red}$\mathbb{Z}^{\rm SE}$ & $0$ & \color{green}$\mathbb{Z}^{\rm BS}$ \\ %\hline
        \multirow{1}{*}{AIII}
        &  -&$0$ & $0$ & $0$ \\ \hline %\hline
        \multirow{1}{*}{$\text{A}$}&{$\mathcal{S}$}
        &\color{red}$\mathbb{Z}^{\rm SE}$ & $0$ & \color{green}$\mathbb{Z}^{\rm BS}$ \\ %\hline
        \multirow{1}{*}{$\text{AIII}$}&{$\mathcal{S}_{-},\eta_{-}$}
        &  $0$ & \color{green}$\mathbb{Z}_{2}^{\rm BS}$ & $0$ \\ \hline %\hline
        \multirow{1}{*}{$\text{A}$}&{$\eta$}
        &  $0$ & $0$ & $0$ \\ %\hline
        \multirow{1}{*}{$\text{AIII}$}&{$\mathcal{S}_{+},\eta_{+}$}
        &  $0$ & $0$ & $0$ \\ \hline \hline
    \end{tabular}
    \endgroup
    \caption{Classification of intrinsic point-gap topological phases for complex AZ classes without or with SLS or pseudo-Hermiticity. The subscript of ${\cal S}_{\pm}/\eta_{\pm}$ specifies the commutation (+) or anti-commutation (-) relation to CS.
    The topological numbers colored red (green) result in non-Hermitian skin effects (boundary states).   \label{table:intrinsic complexAZ+SLS} 
    }
\end{table}
\begin{table}[htbp]
\RawFloats
  \begin{minipage}[t]{.45\textwidth}
  \centering
  \begingroup
  \renewcommand{\arraystretch}{1.2}
    \begin{tabular}{ccccc} \hline \hline
        AZ class & ~Add. sym.~ & ~$d=1$~ & ~$d=2$~ & ~$d=3$~ \\ \hline %\hline
        \multirow{1}{*}{$\text{AI}$}
        & -&\color{red}$\mathbb{Z}^{\rm SE}$ & $0$ & $0$ \\ %\hline
        \multirow{1}{*}{$\text{BDI}$}
        & -&$0$ & $0$ & $0$ \\ %\hline
        \multirow{1}{*}{$\text{D}$}
        & -&$0$ & $0$ & \color{green}$\mathbb{Z}^{\rm BS}$ \\ %\hline
        \multirow{1}{*}{$\text{DIII}$}
        & -&$0$ & $0$ & $0$ \\ %\hline
        \multirow{1}{*}{$\text{AII}$}
        & -&\color{red}$2\mathbb{Z}^{\rm SE}$ & $0$ & $0$ \\ %\hline
        \multirow{1}{*}{$\text{CII}$}
        & -&$0$ & $0$ & $0$ \\ %\hline
        \multirow{1}{*}{$\text{C}$}
        & -&$0$ & $0$ & \color{green}$2\mathbb{Z}^{\rm BS}$ \\ %\hline
        \multirow{1}{*}{$\text{CI}$}
        & -&$0$ & $0$ & $0$ \\ \hline %\hline
        \multirow{1}{*}{$\text{AI}$}&{$\mathcal{S}_{+}$}
        & \color{red}$\mathbb{Z}^{\rm SE}$ & $0$ & $0$ \\ %\hline
        \multirow{1}{*}{$\text{BDI}$}&{$\mathcal{S}_{+-},\eta_{-+}$}
        & \color{red}$\mathbb{Z}_{2}^{\rm SE}$ & \color{red}$\mathbb{Z}_{2}^{\rm SE}$ & $0$ \\ %\hline
        \multirow{1}{*}{$\text{D}$}&{$\mathcal{S}_{-}$}
        & \color{red}$\mathbb{Z}_{2}^{\rm SE}$ & \color{red}$\mathbb{Z}_{2}^{\rm SE}$ & \begin{tabular}{c}\color{red}$(2\mathbb{Z}+1)^{\rm SE}$ \\ \color{green}$2\mathbb{Z}^{\rm BS}$\end{tabular} \\ %\hline
        \multirow{1}{*}{$\text{DIII}$}&{$\mathcal{S}_{+-},\eta_{+-}$}
        &  $0$ & \color{green}$\mathbb{Z}_{2}^{\rm BS}$ & \color{green}$\mathbb{Z}_{2}^{\rm BS}$ \\ %\hline
        \multirow{1}{*}{$\text{AII}$}&{$\mathcal{S}_{+}$}
        &  \color{red}$\mathbb{Z}^{\rm SE}$ & $0$ & \color{green}$\mathbb{Z}_{2}^{\rm BS}$ \\ %\hline
        \multirow{1}{*}{$\text{CII}$}&{$\mathcal{S}_{+-},\eta_{-+}$}
        & $0$ & \color{green}$\mathbb{Z}_{2}^{\rm BS}$ & $0$ \\ %\hline
        \multirow{1}{*}{$\text{C}$}&{$\mathcal{S}_{-}$}
        &  $0$ & $0$ & \color{green}$\mathbb{Z}^{\rm BS}$\\ %\hline
        \multirow{1}{*}{$\text{CI}$}&{$\mathcal{S}_{+-},\eta_{+-}$}
        & $0$ & $0$ & $0$  \\ \hline %\hline
        \multirow{1}{*}{$\text{AI}$}&{$\mathcal{S}_{-}$}
        & \color{red}$\mathbb{Z}^{\rm SE}$ & $0$ & $0$ \\ %\hline
        \multirow{1}{*}{$\text{BDI}$}&{$\mathcal{S}_{-+},\eta_{+-}$}
        & $0$ & $0$ & $0$ \\ %\hline
        \multirow{1}{*}{$\text{D}$}&{$\mathcal{S}_{+}$}
        & $0$& $0$ & \color{green}$\mathbb{Z}^{\rm BS}$ \\ %\hline
        \multirow{1}{*}{$\text{DIII}$}&{$\mathcal{S}_{-+},\eta_{-+}$}
        &  $0$ & $0$ & $0$ \\ %\hline
        \multirow{1}{*}{$\text{AII}$}&{$\mathcal{S}_{-}$}
        &  \color{red}$\mathbb{Z}^{\rm SE}$ & $0$ & $0$ \\ %\hline
        \multirow{1}{*}{$\text{CII}$}&{$\mathcal{S}_{-+},\eta_{+-}$}
        & $0$ & \color{green}$\mathbb{Z}_{2}^{\rm BS}$ & $0$ \\ %\hline
        \multirow{1}{*}{$\text{C}$}&{$\mathcal{S}_{+}$}
        &  \color{red}$\mathbb{Z}_{2}^{\rm SE}$ & $0$ & \color{green}$\mathbb{Z}^{\rm BS}$\\ %\hline
        \multirow{1}{*}{$\text{CI}$}&{$\mathcal{S}_{-+},\eta_{-+}$}
        & $0$ & \color{green}$\mathbb{Z}_{2}^{\rm BS}$ & $0$ \\ \hline %\hline
        \multirow{1}{*}{$\text{AI}$}&{$\eta_{+}$}
        & $0$ & $0$ & $0$ \\ %\hline
        \multirow{1}{*}{$\text{BDI}$}&{$\mathcal{S}_{++},\eta_{++}$}
        & $0$ & $0$ & $0$ \\ %\hline
        \multirow{1}{*}{$\text{D}$}&{$\eta_{+}$}
        & $0$ & $0$ & $0$ \\ %\hline
        \multirow{1}{*}{$\text{DIII}$}&{$\mathcal{S}_{--},\eta_{++}$}
        & $0$ & $0$ & $0$ \\ %\hline
        \multirow{1}{*}{$\text{AII}$}&{$\eta_{+}$}
        & $0$ & $0$ & $0$ \\ %\hline
        \multirow{1}{*}{$\text{CII}$}&{$\mathcal{S}_{++},\eta_{++}$}
        & $0$ & $0$ & $0$ \\ %\hline
        \multirow{1}{*}{$\text{C}$}&{$\eta_{+}$}
        & $0$ & $0$ & $0$ \\ %\hline
        \multirow{1}{*}{$\text{CI}$}&{$\mathcal{S}_{--},\eta_{++}$}
        & $0$ & $0$ & $0$ \\ \hline%\hline
        \multirow{1}{*}{$\text{AI}$}&{$\eta_{-}$}
        & \color{red}$\mathbb{Z}_{2}^{\rm SE}$ & \color{red}$\mathbb{Z}_{2}^{\rm SE}$ & $0$ \\ %\hline
        \multirow{1}{*}{$\text{BDI}$}&{$\mathcal{S}_{--},\eta_{--}$}
        &  $0$ & $0$ & $0$ \\ %\hline
        \multirow{1}{*}{$\text{D}$}&{$\eta_{-}$}
        &  $0$ & $0$ & $0$ \\ %\hline
        \multirow{1}{*}{$\text{DIII}$}&{$\mathcal{S}_{++},\eta_{--}$}
        &  $0$ & $0$ & $0$ \\ %\hline
        \multirow{1}{*}{$\text{AII}$}&{$\eta_{-}$}
        &  $0$ & $0$ & $0$ \\ %\hline
        \multirow{1}{*}{$\text{CII}$}&{$\mathcal{S}_{--},\eta_{--}$}
        &  $0$ & $0$ & $0$ \\ %\hline
        \multirow{1}{*}{$\text{C}$}&{$\eta_{-}$}
        &  $0$ & $0$ & $0$ \\ %\hline
        \multirow{1}{*}{$\text{CI}$}&{$\mathcal{S}_{++},\eta_{--}$}
        & \color{red}$\mathbb{Z}_{2}^{\rm SE}$ & $0$ & $0$ \\ \hline \hline
    \end{tabular}
    \endgroup
   \caption{
    Classification of intrinsic point-gap topological phases for real AZ classes without or with SLS or pseudo-Hermiticity. The subscript of ${\cal S}_{\pm}/\eta_{\pm}$ specifies the commutation (+) or anti-commutation (-) relation to TRS or PHS.
    For ${\cal S}_{\pm\pm}/\eta_{\pm\pm}$, the first subscript specifies the relation to TRS and the second one specifies the relation to PHS.
    The topological numbers colored red (green) result in non-Hermitian skin effects (boundary states).  \label{table:intrinsic realAZ+SLS} }
  \end{minipage}
  \hfill
   \begin{minipage}[t]{.45\textwidth}
  \centering
  \begingroup
  \renewcommand{\arraystretch}{1.2}
    \begin{tabular}{ccccc} \hline \hline
        $\text{AZ}^{\dag}$ class & ~Add. sym.~ & ~$d=1$~ & ~$d=2$~ & ~$d=3$~ \\ \hline %\hline
        \multirow{1}{*}{$\text{AI}^{\dag}$}
        & -&$0$ & $0$ & \color{green}$2\mathbb{Z}^{\rm BS}$ \\ %\hline
        \multirow{1}{*}{$\text{BDI}^{\dag}$}
        & -&$0$ & $0$ & $0$ \\ %\hline
        \multirow{1}{*}{$\text{D}^{\dag}$}
        & -&\color{red}$\mathbb{Z}^{\rm SE}$ & $0$ & $0$ \\ %\hline
        \multirow{1}{*}{$\text{DIII}^{\dag}$}
        & -&\color{red}$\mathbb{Z}_{2}^{\rm SE}$ & \color{red}$\mathbb{Z}_{2}^{\rm SE}$ & $0$ \\ %\hline
        \multirow{1}{*}{$\text{AII}^{\dag}$}
        & -&\color{red}$\mathbb{Z}_{2}^{\rm SE}$ & \color{red}$\mathbb{Z}_{2}^{\rm SE}$ & \begin{tabular}{c}\color{red}$(2\mathbb{Z}+1)^{\rm SE}$ \\ \color{green}$2\mathbb{Z}^{\rm BS}$\end{tabular} \\% \hline
        \multirow{1}{*}{$\text{CII}^{\dag}$}
        & -&$0$ & $0$ & $0$ \\ %\hline
        \multirow{1}{*}{$\text{C}^{\dag}$}
        & -&\color{red}$2\mathbb{Z}^{\rm SE}$ & $0$ & $0$ \\ %\hline
        \multirow{1}{*}{$\text{CI}^{\dag}$}
        & -&$0$ & $0$ & $0$ \\ \hline %\hline
        \multirow{1}{*}{$\text{AI}^{\dag}$}&{$\mathcal{S}_{+}$}
        & $0$ & $0$ & \color{green}$\mathbb{Z}^{\rm BS}$ \\ %\hline
        \multirow{1}{*}{$\text{BDI}^{\dag}$}&{$\mathcal{S}_{+-},\eta_{-+}$}
        & $0$ & $0$ & $0$ \\ %\hline
        \multirow{1}{*}{$\text{D}^{\dag}$}&{$\mathcal{S}_{-}$}
        & \color{red}$\mathbb{Z}^{\rm SE}$ & $0$ & $0$ \\ %\hline
        \multirow{1}{*}{$\text{DIII}^{\dag}$}&{$\mathcal{S}_{+-},\eta_{+-}$}
        &  $0$ & \color{green}$\mathbb{Z}_{2}^{\rm BS}$ & $0$ \\ %\hline
        \multirow{1}{*}{$\text{AII}^{\dag}$}&{$\mathcal{S}_{+}$}
        &  \color{red}$\mathbb{Z}_{2}^{\rm SE}$ & $0$ & \color{green}$\mathbb{Z}^{\rm BS}$ \\ %\hline
        \multirow{1}{*}{$\text{CII}^{\dag}$}&{$\mathcal{S}_{+-},\eta_{-+}$}
        & $0$ & \color{green}$\mathbb{Z}_{2}^{\rm BS}$ & $0$ \\ %\hline
        \multirow{1}{*}{$\text{C}^{\dag}$}&{$\mathcal{S}_{-}$}
        & \color{red}$\mathbb{Z}^{\rm SE}$ & $0$ & $0$ \\ %\hline
        \multirow{1}{*}{$\text{CI}^{\dag}$}&{$\mathcal{S}_{+-},\eta_{+-}$}
        & $0$ & $0$ & $0$  \\ \hline %\hline
        \multirow{1}{*}{$\text{AI}^{\dag}$}&{$\mathcal{S}_{-}$}
        & $0$ & $0$ & \color{green}$\mathbb{Z}^{\rm BS}$ \\ %\hline
        \multirow{1}{*}{$\text{BDI}^{\dag}$}&{$\mathcal{S}_{-+},\eta_{+-}$}
        & $0$ & $0$ & $0$ \\ %\hline
        \multirow{1}{*}{$\text{D}^{\dag}$}&{$\mathcal{S}_{+}$}
        & \color{red}$\mathbb{Z}^{\rm SE}$ & $0$ & $0$ \\ %\hline
        \multirow{1}{*}{$\text{DIII}^{\dag}$}&{$\mathcal{S}_{-+},\eta_{-+}$}
        &  \color{red}$\mathbb{Z}_{2}^{\rm SE}$ & \color{red}$\mathbb{Z}_{2}^{\rm SE}$ & $0$ \\ %\hline
        \multirow{1}{*}{$\text{AII}^{\dag}$}&{$\mathcal{S}_{-}$}
        & \color{red}$\mathbb{Z}_{2}^{\rm SE}$ & \color{red}$\mathbb{Z}_{2}^{\rm SE}$ & \begin{tabular}{c}\color{red}$(2\mathbb{Z}+1)^{\rm SE}$ \\ \color{green}$2\mathbb{Z}^{\rm BS}$\end{tabular} \\ %\hline
        \multirow{1}{*}{$\text{CII}^{\dag}$}&{$\mathcal{S}_{-+},\eta_{+-}$}
        & $0$ & \color{green}$\mathbb{Z}_{2}^{\rm BS}$ & \color{green}$\mathbb{Z}_{2}^{\rm BS}$ \\ %\hline
        \multirow{1}{*}{$\text{C}^{\dag}$}&{$\mathcal{S}_{+}$}
        &  \color{red}$\mathbb{Z}^{\rm SE}$ & $0$ & \color{green}$\mathbb{Z}_{2}^{\rm BS}$\\ %\hline
        \multirow{1}{*}{$\text{CI}^{\dag}$}&{$\mathcal{S}_{-+},\eta_{-+}$}
        & $0$ & \color{green}$\mathbb{Z}_{2}^{\rm BS}$ & $0$ \\ \hline %\hline
        \multirow{1}{*}{$\text{AI}^{\dag}$}&{$\eta_{+}$}
        & $0$ & $0$ & $0$ \\ %\hline
        \multirow{1}{*}{$\text{BDI}^{\dag}$}&{$\mathcal{S}_{++},\eta_{++}$}
        & $0$ & $0$ & $0$ \\ %\hline
        \multirow{1}{*}{$\text{D}^{\dag}$}&{$\eta_{+}$}
        & $0$ & $0$ & $0$ \\ %\hline
        \multirow{1}{*}{$\text{DIII}^{\dag}$}&{$\mathcal{S}_{--},\eta_{++}$}
        & $0$ & $0$ & $0$ \\ %\hline
        \multirow{1}{*}{$\text{AII}^{\dag}$}&{$\eta_{+}$}
        & $0$ & $0$ & $0$ \\ %\hline
        \multirow{1}{*}{$\text{CII}^{\dag}$}&{$\mathcal{S}_{++},\eta_{++}$}
        & $0$ & $0$ & $0$ \\ %\hline
        \multirow{1}{*}{$\text{C}^{\dag}$}&{$\eta_{+}$}
        & $0$ & $0$ & $0$ \\ %\hline
        \multirow{1}{*}{$\text{CI}^{\dag}$}&{$\mathcal{S}_{--},\eta_{++}$}
        & $0$ & $0$ & $0$ \\ \hline%\hline
        \multirow{1}{*}{$\text{AI}^{\dag}$}&{$\eta_{-}$}
        &  $0$ & $0$ & $0$ \\ %\hline
        \multirow{1}{*}{$\text{BDI}^{\dag}$}&{$\mathcal{S}_{--},\eta_{--}$}
        &  $0$ & $0$ & $0$ \\ %\hline
        \multirow{1}{*}{$\text{D}^{\dag}$}&{$\eta_{-}$}
        & \color{red}$\mathbb{Z}_{2}^{\rm SE}$ & \color{red}$\mathbb{Z}_{2}^{\rm SE}$ & $0$ \\ %\hline
        \multirow{1}{*}{$\text{DIII}^{\dag}$}&{$\mathcal{S}_{++},\eta_{--}$}
        & \color{red}$\mathbb{Z}_{2}^{\rm SE}$ & $0$ & $0$ \\ %\hline
        \multirow{1}{*}{$\text{AII}^{\dag}$}&{$\eta_{-}$}
        &  $0$ & $0$ & $0$ \\ %\hline
        \multirow{1}{*}{$\text{CII}^{\dag}$}&{$\mathcal{S}_{--},\eta_{--}$}
        &  $0$ & $0$ & $0$ \\ %\hline
        \multirow{1}{*}{$\text{C}^{\dag}$}&{$\eta_{-}$}
        &  $0$ & $0$ & $0$ \\ %\hline
        \multirow{1}{*}{$\text{CI}^{\dag}$}&{$\mathcal{S}_{++},\eta_{--}$}
        &  $0$ & $0$ & $0$ \\ \hline \hline
    \end{tabular}
    \endgroup
    \caption{Classification of intrinsic point-gap topological phases for real AZ$^\dagger$ classes without or with SLS or pseudo-Hermiticity. The subscript of ${\cal S}_{\pm}/\eta_{\pm}$ specifies the commutation (+) or anti-commutation (-) relation to TRS$^\dagger$ or PHS$^\dagger$.
    For ${\cal S}_{\pm\pm}/\eta_{\pm\pm}$, the first subscript specifies the relation to TRS$^\dagger$ and the second one specifies the relation to PHS$^\dagger$.
    The topological numbers colored red (green) result in non-Hermitian skin effects (boundary states).\label{table:intrinsic realAZdag+SLS} }
  \end{minipage}
\end{table}

\end{document}